\documentclass[useAMS,usenatbib]{mn2e}

\usepackage{graphicx}
\usepackage{natbib}
\usepackage{subscript}
\usepackage{multicol}
\usepackage{float}
\usepackage{amsmath,amssymb,latexsym} 
\usepackage{chngcntr}
\usepackage{color}
\usepackage[T1]{fontenc}
\usepackage{aecompl}

\newcommand{\appropto}{\mathrel{\vcenter{
  \offinterlineskip\halign{\hfil$##$\cr
    \propto\cr\noalign{\kern2pt}\sim\cr\noalign{\kern-2pt}}}}}

\title[The wind speeds, dust content, and mass-loss rates of evolved AGB stars at varying metallicity]{The wind speeds, dust content, and mass-loss rates of evolved AGB and RSG stars at varying metallicity}
\author[S. R. Goldman et al. ]{Steven R. Goldman$^{1}$\thanks{E-mail: s.r.goldman@keele.ac.uk}, Jacco Th. van Loon$^{1}$, Albert A. Zijlstra$^{2}$, James A. Green$^{3,4}$, \newauthor  Peter R. Wood$^{5}$, Ambra Nanni$^{6}$, Hiroshi Imai$^{7}$, Patricia A. Whitelock$^{8,9}$, \newauthor Mikako Matsuura$^{10}$, Martin A. T. Groenewegen$^{11}$, and Jos\'e F. G\'omez$^{12}$ \vspace{0.25cm} \\ 
$^{1}$Astrophysics Group, Lennard-Jones Laboratories, Keele University, ST5 5BG, UK\\
$^{2}$Jodrell Bank Centre for Astrophysics, Alan Turing Building, School of Physics and Astronomy, The University of Manchester,\\ Oxford Road, Manchester, M13 9PL, UK \\
$^{3}$SKA Organisation, Jodrell Bank Observatory, Lower Withington, Macclesfield, Cheshire, SK11 9DL, UK \\
$^{4}$CSIRO Astronomy and Space Science, Australia Telescope National Facility, PO Box 76, Epping, NSW 1710, Australia\\
$^{5}$Research School of Astronomy and Astrophysics, Australian National University, Weston Creek, ACT 2611, Australia\\
$^{6}$Dipartimento di Fisica e Astronomia Galileo Galilei, vicolo dell'Osservatorio 3, 35141 Padova PD, Italy\\
$^{7}$Department of Physics and Astronomy, Kagoshima University, 1-21-35 Korimoto, Kagoshima 890-0065, Japan\\
$^{8}$South African Astronomical Observatory (SAAO), PO Box 9, 7935 Observatory, South Africa\\
$^{9}$Astronomy Department, University of Cape Town, 7701 Rondebosch, South Africa\\
$^{10}$School of Physics and Astronomy, Cardiff University, Queen's Buildings, The Parade, Cardiff, CF24 3AA, UK\\
$^{11}$Royal Observatory of Belgium, Ringlaan 3, B-1180 Brussels, Belgium \\
$^{12}$Instituto de Astrof\'{\i}sica de Andaluc\'{\i}a, CSIC, Glorieta de la Astronom\'{\i}a s/n, 18008 Granada, Spain}

\date{Accepted 2016 XXXXXXX XX. Received 2016   XXXXXXX XX}
\pubyear{2016}

\begin{document}
\pagerange{\pageref{firstpage}--\pageref{lastpage}}
\maketitle
\label{firstpage}

\begin{abstract}
\noindent We present the results of our survey of 1612 MHz circumstellar OH maser emission from asymptotic giant branch (AGB) stars and red supergiants (RSGs) in the Large Magellanic Cloud. We have discovered four new circumstellar maser sources in the LMC, and increased the number of reliable wind speeds from IR stars in the LMC from 5 to 13. Using our new wind speeds, as well as those from Galactic sources, we have derived an updated relation for dust driven winds: $v_{exp} \propto Z L^{0.4}$. We compare the sub-solar metallicity LMC OH/IR stars with carefully selected samples of more metal-rich OH/IR stars, also at known distances, in the Galactic Centre and Galactic Bulge. For 8 of the Bulge stars we derive pulsation periods for the first time, using near-IR photometry from the VVV survey. We have modeled our LMC OH/IR stars and developed an empirical method of deriving gas-to-dust ratios and mass loss rates by scaling the models to the results from maser profiles. We have done this also for samples in the Galactic Centre and Bulge and derived a new mass loss prescription that includes luminosity, pulsation period, and gas-to-dust ratio $\dot{M} = 1.06^{+3.5}_{-0.8} \rm{ \cdot }10^{-5}\,(L/10^4\,\rm{L}_\odot)^{0.9\pm0.1}(P/500\,\rm{d})^{0.75\pm0.3} (r_{gd}/200)^{-0.03\pm0.07}\,\rm{M_{\odot}} yr^{-1}$. The tightest correlation is found between mass loss rate and luminosity. We find that the gas-to-dust ratio has little effect on the mass loss of oxygen-rich AGB stars and RSGs within the Galaxy and the LMC. This suggests that mass loss of oxygen-rich AGB stars and RSGs is (nearly) independent of metallicity between a half and twice solar. 

\end{abstract} 

\begin{keywords}
masers -- stars: AGB and post-AGB -- supergiants -- stars: mass-loss -- stars: winds, outflows -- Magellanic Clouds 
\end{keywords}

\section{Introduction}
A remarkable aspect of the evolution of Asymptotic Giant Branch (AGB) stars (0.8 M\textsubscript{$\odot$} $\lesssim$ M $\lesssim 8$ M\textsubscript{$\odot$}) and some red supergiants (RSGs) (M $\lesssim$ 8 M\textsubscript{$\odot$}) is their period of intense mass loss. Near the end of their lifetime, AGB stars will lose up to 85\% of their initial mass at a rate of up to $10^{-4}$ solar masses per year, and contribute a significant amount of dust and chemically enriched material to the interstellar medium (ISM) (Herwig 2005). They may collectively contribute more dust to the ISM than any other type of celestial body in the Universe. Yet, this is still controversial, as the effects of metallicity and luminosity on the mass loss of these stars remain unclear. Massive AGB stars and RSGs have been instrumental in the ongoing star formation and chemical evolution of massive sphroidals at high redshift, and possibly in the multiple populations found in massive Galactic globular clusters. Constraining this mass loss rate is also hugely important for understanding supernova progenitors and lightcurves. Mass loss models have been developed and refined (Reimers 1975; Baud \& Habing 1983; Schr\"{o}der \& Cuntz 2005; van Loon et al. 2005), but require observations for verification.

During their stage of intense mass loss, AGB and RSG stars lose mass in the form of a stellar ``superwind,'' a concept first introduced by Iben and Renzini (1983). This superwind is accelerated by radiation pressure on circumstellar dust grains and produces velocities ranging from about $5 - 30$ km s\textsuperscript{$-1$}. The dust grains form at a few stellar radii and are composed of either carbonaceous or silicate dust grains, depending on the composition of the star's atmosphere and the efficiency of hot-bottom burning and third dredge up events (Herwig, 2005). Stars with an initial mass $\lesssim$ 1.5 M\textsubscript{$\odot$} or $\gtrsim$ 4 M\textsubscript{$\odot$} and metallicity within a factor of few around solar will end their lives oxygen-rich.

The large quantity of dust surrounding some of these stars leaves many of them obscured at visual wavelengths. However, the stellar light heats dust grains which re-emit energy in the infrared (IR). By modeling the spectral energy distributions (SEDs) and making assumptions we can derive important and hard to obtain stellar parameters such as stellar mass loss rate and wind speed. Dusty oxygen-rich AGB and RSG stars often also exhibit OH, H\textsubscript{2}O, and SiO maser emission (Nyman et al. 1998). From these unique properties, these stars have been aptly named OH/IR stars.
 
OH/IR stars generally emit their strongest OH masers at 1612 MHz, which occurs at several hundred stellar radii from the star. These masers are pumped by infrared radiation and trace terminal velocities of the envelope's expansion due to their large distances from the central star (Elitzur, 1992). The OH molecules are formed when H\textsubscript{2}O molecules at the periphery of the inner envelope are photo-dissociated by interstellar ultra-violet radiation. In addition to OH masers, SiO and H\textsubscript{2}O masers are used to probe the velocities of the inner dust-free zones, and accelerating wind zones, respectively (Richards et al. 2014). All three types of masers are critical to understanding the kinematics of dust-driven winds.  

Circumstellar 1612 MHz OH masers exhibit double-peaked profiles. The double-peak is a result of the expansion of the OH shell and the radially amplification of 1612 MHz photons from this shell (Engels \& Bunzel 2015). From OH masers we can determine the velocity of outflowing material as half the separation of the main twin emission peaks. The peaks represent the blue and redshifted final wind speeds of outflowing material moving toward and away from us, respectively. This expansion velocity can be used to test \hspace{0.5cm} dust-driven wind theory: \begin{equation} v_{exp} \propto r_{gd}^{-1/2}L^{1/4} \end{equation} where $L$ is the luminosity, and $r_{gd}$ is the gas-to-dust ratio; the dust-to-gas ratio has shown strong empirical evidence to scale approximately with the metal content (van Loon 2000; Habing 1994; Elitzur \& Ivezi{\'c}, 2001; van Loon 2012). Often OH masers are the only available tool to derive expansion velocities in extra-galactic systems as they radiate with an intensity much greater than the alternative, thermal CO line emission.
 
We can test and refine dust-driven wind theory by observing OH masers in the Large Magellanic Cloud (LMC), which has a metallicity of a half solar for the ISM and for stars formed in the most recent few Gyr (Choudhury et al. 2016), and is within close proximity. The well known distance to the LMC at 50 kpc (Feast 2013) gives us accurate luminosity measurements and the ability to test the dependence of mass loss on luminosity. What follows is a refined analysis of OH maser searches in the LMC by Wood et al. (1992), and Marshall et al. (2004), as well as new detections from our new observations with the Parkes and Australia Telescope Compact Array (ATCA) radio telescopes. We include samples of OH/IR stars in the Galactic Centre and Bulge to extend our metallicity range to super-solar values, in order to determine the dependence of mass loss on metallicity.\newline

\begin{table*}
\caption[]{The LMC target sample including the detections from Wood et al.\ (1992), van Loon et al. (1998a) and Marshall et al. (2004); J2000 positions are taken from 2MASS and ATCA observations. Also listed are the previously detected OH masers, and whether we detected OH maser emission here (no entry means target has not been attempted).}
\begin{minipage}[c]{\textwidth}
\centering
\begin{tabular}{llccccrr}
\hline\hline
Object           &
Alternative      &
\multicolumn{2}{c}{2MASS position (J2000)} &
\multicolumn{2}{c}{ATCA position (J2000)} &
Previous         &
This work        \\
name             &
names             &
RA                  &
DEC                  &
RA                  &
DEC                  &
OH               &
OH               \\

\hline 
IRAS\,04407$-$7000 &
LI-LMC\,4          &
04 40 28.5         &
$-69$ 55 14           &
&
&
yes\rlap{$^{2,1}$} &
yes                \\
IRAS\,04498$-$6842 &
LI-LMC\,60         &
04 49 41.5         &
$-68$ 37 52          &
04 49 40.9         &
$-68$ 37 50          &
yes\rlap{$^1$}     &
yes                \\
IRAS\,04509$-$6922 &
LI-LMC\,77         &
04 50 40.5         &
$-69$ 17 32          &
04 50 40.3         &
$-69$ 17 35          &
no\rlap{$^{3}$}   &
yes                \\
IRAS\,04516$-$6902 &
LI-LMC\,92         &
04 51 29.0         &
$-68$ 57 50         &
04 51 29.6         &
$-68$ 57 47         &
no\rlap{$^1$}      &
yes                \\
IRAS\,04537$-$6922 &
WOH S 60 &
04 53 30.9 &
$-69$ 17 50 &
&
&
&
no \\
IRAS\,04545$-$7000 &
LI-LMC\,159        &
04 54 10.1        &
$-69$ 55 58          &
&
&
yes\rlap{$^{3}$}  &
yes                \\
IRAS\,04553$-$6825 &
WOH\,G064          &
04 55 10.5         &
$-68$ 20 30          &
&
&
yes\rlap{$^{3,1}$} &
yes                \\
IRAS\,04553$-$6933 &
LI-LMC\,183         &
04 55 03.1         &
$-69$ 29 13          &
&
&
                   &
no                \\
IRAS\,05003$-$6712 &
LI-LMC\,297       &
05 00 19.0         &
$-67$ 07 58          &   
&
&  
yes\rlap{$^1$}     &
yes                \\
IRAS\,05280$-$6910 &
NGC\,1984-IRS1     &
05 27 40.1         &
$-69$ 08 05           &
05 27 41.3         &
$-69$ 08 02           &
yes\rlap{$^{3}$}  &
yes                   \\
IRAS\,05294$-$7104 &
LI-LMC\,1153       &
05 28 48.2         &
$-71$ 02 29          &
05 28 48.9         &
$-71$ 02 32          &
no\rlap{$^1$}      &
yes                   \\
IRAS\,05298$-$6957 &
LI-LMC\,1164       &
05 29 24.6         &
$-69$ 55 14          &
&
&
yes\rlap{$^{3}$}  &
                   \\
IRAS\,05329$-$6957 &
LI-LMC 1286, TRM 60             &
05 32 52.3              &
$-67$ 06 26               &
&
&
 yes\rlap{$^{3}$}        &
 no                  \\                   
IRAS\,05402$-$6956 &
LI-LMC\,1506       &
05 39 44.8              &
$-69$ 55 18               &
&
&
yes\rlap{$^{3}$}  &
                   \\
IRAS\,05558$-$7000 &
LI-LMC 1790, TRM 58       &
05 55 21.0              &
$-70$ 00 03               &
&
&
yes\rlap{$^1$}     &
yes                \\
MSX\,LMC\,807      &
LI-LMC\,1280       &
05 32 37.2              &
$-67$ 06 56           &
05 32 37.3              &
$-67$ 06 57           &
                   &
yes                \\
MSX\,LMC\,811      &
                           &
05 32 51.3              &
$-67$ 06 52               &
05 32 51.3              &
$-67$ 06 52               &
yes\rlap{$^{3}$}           &
yes                \\
MSX\,LMC\,815      &
                   &
05 35 14.1            &
$-67$ 43 56      &
&
&
                &
 no                \\
MSX\,LMC\,1207      &
LI-LMC 182                   &
04 55 06.5            &
$-69$ 17 09      &
&
&
                &
 no                \\
\hline
\end{tabular}\\
References:
$^1$Marshall et al.\ (2004)
$^2$van Loon et al.\ (1998a)
$^3$Wood et al.\ (1992)
\end{minipage}
\end{table*}

\begin{table*}
\caption[]{The LMC sample with bolometric magnitudes $M_{\rm bol}$, pulsation periods $P$, K-band peak-to-peak pulsation amplitudes ${\Delta}K$, mean K-band magnitudes and $J-K$ colours from Whitelock et al.\ (2003). \textit{Spitzer} 8.0 $\mu$m ($F_{8}$) and 24 $\mu$m ($F_{24}$) flux densities are from the SAGE project (Meixner et al. 2006), that use IRAC (Fazio et al 2004) and MIPS (Rieke et al. 2004), respectively on board the Spitzer Space Telescope (Werner et al. 2004). Spectral types are from van Loon et al.\ (2005), except where noted otherwise. We adopt a distance to the LMC of 50 kpc (Feast 2013).}
\begin{minipage}[c]{\textwidth}
\centering
\begin{tabular}{lcccccccl}
\hline\hline
Object           &
$M_{\rm bol}$    &
$P$              &
${\Delta}K$      &
$K$              &
$F_{8}$          &
$F_{24}$         &
$J-K$            &
Spectral         \\
name             &
(mag)            &
(days)           &
(mag)            &
(mag)            &
(Jy)             &
(Jy)             &
                 &
type             \\

\hline 
IRAS\,04407$-$7000 &
$-$7.11            &
1264\rlap{$^5$}    &
1.23               &
8.79               &
0.51               &
0.68               &
2.34               &
M7.5               \\
IRAS\,04498$-$6842 &
$-$7.72            &
1292               &
1.30               &
8.08               &
0.78               &
0.69               &
1.86               &
M10                \\
IRAS\,04509$-$6922 &
$-$7.28            &
1292               &
1.45               &
8.59               &
0.37               &
0.86\rlap{$^8$}&
2.21               &
M10\rlap{$^{11}$}  \\
IRAS\,04516$-$6902 &
$-$7.11            &
1165\rlap{$^5$}    &
1.41               &
8.72               &
0.30               &
0.55\rlap{$^8$}    &
2.32               &
M9\rlap{$^{11}$}   \\
IRAS\,04537$-$6922    &
&
&
&
8.06\rlap{$^2$} &
0.04 &
1.30 &
1.27 \rlap{$^2$} &
M3\rlap{$^3$} \\
IRAS\,04545$-$7000 &
$-$6.56            &
1216               &
1.57               &
\llap{1}0.13       &
0.17               &
0.38               &
5.70\rlap{$^{14}$} &
                   \\
IRAS\,04553$-$6825 &
$-$9.19\rlap{$^{10}$} &
\hspace*{1mm}841   &
0.34               &
7.09               &
5.30               &
\llap{1}3.53\rlap{$^8$}   &
2.60\rlap{$^8$}    &
M7.5               \\
IRAS\,05003$-$6712 &
$-$6.20            &
\hspace*{1mm}942\rlap{$^5$}&
1.59               &
9.95               &
0.14               &
0.19               &
2.95               &
M9                 \\
IRAS\,05280$-$6910 &
$-$7.75\rlap{$^{14}$} &
                   &
                   &
12.87\rlap{$^{13}$} &
1.03               &
\llap{2}4.18\rlap{$^{12}$} &
                   &
                   \\
IRAS\,05294$-$7104 &
$-$6.79            &
1079               &
1.20               &
9.21               &
0.23               &
0.25               &
2.97               &
M8                 \\
IRAS\,05298$-$6957 &
$-$6.72\rlap{$^{14}$} &
1280\rlap{$^{14}$} &
2.00\rlap{$^{14}$} &
\llap{1}0.29\rlap{$^{14}$} &
0.59               &
1.05               &
3.54\rlap{$^{14}$} &
                   \\
IRAS\,05402$-$6956 &
$-$6.77            &
1393               &
1.80               &
\llap{1}0.40       &
0.62               &
1.00               &
4.46\rlap{$^{14}$} &
                   \\
IRAS\,05558$-$7000 &
$-$6.97            &
1220               &
1.42               &
9.25               &
0.33               &
0.49               &
3.27               &
                   \\
MSX\,LMC\,807      &
$-$6.08\rlap{$^4$}                   &
                   &
                   &
\llap{1}1.93\rlap{$^1$}   &
0.33               &
0.45               &
                   &
                   \\
MSX\,LMC\,811      &
$-$6.71\rlap{$^4$}                   &
                   &
                   &
\llap{1}1.12\rlap{$^2$}   &
0.76               &
1.25               &
5.2\rlap{$^7$}   &
                   \\
MSX\,LMC\,815      &
$-$8.00\rlap{$^4$}                   &
590\rlap{$^6$}                   &
                   &
8.14\rlap{$^1$}    &
0.19               &
0.20\rlap{$^9$}    &
1.36\rlap{$^1$}    &
M4                 \\
MSX\,LMC\,1207      &
                   &
                   &
                   &
11.50   &
0.12               &
    &
2.87    &
                \\
\hline
\end{tabular}\\
References:
$^1$Buchanan et al. (2006)
$^2$Cutri et al. (2003)
$^3$Humphreys (1978)
$^4$Jones et al. (2012)
$^5$Kim et al. (2014)
$^6$Pierce et al. (2000)
$^7$Sloan et al. (2008)
$^8$Trams et al. (1999)
$^{9}$van Aarle et al. (2011)
$^{10}$van Loon et al. (1999)
$^{11}$van Loon et al. (1998a)
$^{12}$van Loon et al. (2001b)
$^{13}$van Loon et al. (2005)
$^{14}$Wood et al. (1992)
\end{minipage}
\end{table*}

\begin{table*}
\caption[]{Results of recent maser searches in the Large Magellanic Cloud.}
\begin{minipage}[c]{\textwidth}
\centering
\begin{tabular}{ l c r c c c c c c c c c c }
  \hline\hline
  Object &  $v_{exp}$ & Parkes (P) or &
   Integration & \multicolumn{2}{c}{Velocity at peaks} & \multicolumn{2}{c}{$F_{int}$} & 
  $\sigma$ \\
  name &
  (km s$^{-1}$) & 
  ATCA (A) & 
  time (h) & 
  \multicolumn{2}{c}{(km s$^{-1}$)} & 
  \multicolumn{2}{c}{(mJy km s$^{-1}$)} & 
  (mJy) \\
  &
  &
  epoch &
  & 
  Blue & 
  Red &  
  Blue & 
  Red &  
  &
  \\
  \hline

  IRAS 04407$-$7000 &
  8.35 &
  P 2003 &
  8.6 &
  222.8 &
  239.5, 237 &
  9.6 &
  35.5 & 
  5.9 \\
  & 
   &
  P 2005 &
  1.7 &
   &
  240, 237 &
  &
  89.8 &
  8.8 \\
  & 
  &
  Combined &
  10.2 &
  223 &
  240, 237 &
  12&
  105.8 &
  4.8 \\

  IRAS 04498$-$6842 &
  13.0 &
  P 2003 &
  13.8 &
  246.3 &
  &
  22.9 &
  &
  8.2 \\
  & 
  &
  A 2015 &
  4.6 &
  246.3 &
  272.3 &
  8.2 &
  44.5 &
  10.6 \\
  &
  &
  Combined &
  18.4 &
  246.3 &
  272.3 &
  67.3 &
  6.4 &
  6.3 \\ 
  
  IRAS 04509$-$6922 &
  11.4 &
  P 2003 &
  5.0 &
  254.2 &
  274.7 &
  5.1 &
  3.4 &
  10.1 \\
  & 
  &
  P 2005 &
  22.7 &
  255.9 &
  278.6 &
  53.3 &
  44.2 &
  3.9 \\
  & 
  &
  A 2015 &
  4.1 &
  256 &
  & 
  101.5 &
  &
  26.7 \\ 
  & 
  &
  Combined &
  31.8 &
  255 &
  278.6 &
  75.2 &
  52.1 &
  3.7 \\  

  IRAS 04516$-$6902 &
  10.0 &
  P 2003 &
  8.5 &
   &
  287.3 &
  &
  16.0 &
  7.0 \\
  & 
  &
  A 2015 &
  12.6 &
  &
  287.3 &
  &
  30.6 &
  3.8 \\
  & 
  &
  Combined &
  21.1 &
  267.1&
  287.1& 
  13.3 &
  30.4 & 
  3.4 \\ 

  IRAS 04537$-$6922 &
  ? &
  A 2016 &
  10.8 & 
   &
   &
   &
   &
  3.7 \\
  
  IRAS 04545$-$7000 &
  7.7 &
  P 1992 &
  1 &
  258.7 &
  274 &
   &
   &
  6.1 \\
   &
   &
  P 2005 &
  4.5 &
  258.7 &
  274 &
  191.5 &
  37.6 &
  6.1 \\
  &
  &
  Combined &
  4.5 &
  258.7 &
  274 &
  191.5 &
  37.6 &
  6.1 \\

  IRAS 04553$-$6825 &
  23.8 &
  P 2003 &
  0.8 &
  253.4, 263 &
  281, 300.9 &
  1709, 414 &
  11.7, 8.2 &
  12.7 \\
  & 
  &
  P 2005 &
  0.6 &
  253.4, 263 &
  281, 300.9 &
  1346, 289 &
  74.1, 52.6 &
  14.0 \\
  & 
  &
  Combined &
  1.4 &
  253.4, 263 &
  281, 300.9 &
  1417, 272.6 &
  32.3, 21.9 &
  9.8 \\

  IRAS 04553$-$6933 &
  ? &
  A 2016 &
  10.8 & 
   &
   &
   &
   &
  3.6 \\
  
  IRAS 05003$-$6712 &
  ? &
  P 2003 &
  7.2 &
  287? &
  269.5 &
  87.1 &
  &
  6.7 \\
  & 
  &
  P 2005 &
  3.4 &
   &
  269.5 &
  40.5 &
  &
  5.8 \\
  & 
  &
  Combined &
  10.6 &
  243 &
  269.5 &
  65.8 &
  &
  4.2 \\
  
  IRAS 05280$-$6910 &
  20.6 &
  P 1992 &
  1 & 
  255 &
  289 &
  419 &
  398 &
  17 \\
  &
  &
  A 2016 &
  5.7 &
  258.1 &
  292.6 &
  131.1 &
  147.3  &
  11.5 \\

  IRAS 05294$-$7104 &
  10.3 &
  A 2016 &
  9.8 & 
  269.7 &
  290.3 &
  9.6 &
  34.2 &
  5.5 \\
 
  IRAS 05298$-$6957 &
  10.5 &
  P 1992 &
  1 & 
  271 &
  292 &
  519 &
  448 &
  17 \\

  IRAS 05402$-$6956 &
  10.5 &
  P 1992 &
  1 & 
  263 &
  284 &
  334 &
  181 &
  17 \\

  IRAS 05558$-$7000 &
  8 &
  P 2003 &
  7.7 &
  265 &
  281.5 &
  18.5 &
  11.7 &
  5.7 &
  \\
  & 
  &
  P 2005 &
  3.4 &
  265 &
  281.5 &
  68.7 &
  52.5 &
  5.9 \\
  & 
  &
  Combined &
  11.2 &
  265 &
  281.5 &
  70.1 &
  51.0 &
  4.1 \\
  
  MSX LMC 807 &
  8.15 &
  A 2015 &
  9.3 &
  306.4 &
  322.7 &
  19.8 &
  12.1 &
  4.6 \\

  MSX LMC 811 &
  8.25 &
  A 2015 &
  9.5 &
  302.4 &
  318.9 &
  151 &
  188 &
  4.3 \\
  
  MSX LMC 815 &
  ?&
  A 2015 &
  14.1 &
  &
  &
  &
  &
  4.5 \\

  MSX LMC 1207 &
  ? &
  A 2016 &
  10.8 & 
   &
   &
   &
   &
  3.8 \\
\hline
\end{tabular}
\end{minipage}
\end{table*}

\section{Description of the Sample}

Circumstellar OH masers have been discovered within our Galaxy, and also within the LMC (Table 1). The Hamburg Database (Engels et al. 2015) has compiled thousands of OH maser detections within our Galaxy, yet few have been discovered in the LMC. Previous OH maser searches in the LMC have been completed by Wood et al. (1992), van Loon et al. (1998a), van Loon et al. (2001b) and Marshall et al. (2004) using IRAS data for selection criteria. These searches were biased against objects in molecular cloud regions which require better resolution. LMC searches of other maser types have supplemented these searches including the SiO maser search by van Loon et al. (1996, 2001b) and the water maser searches by van Loon et al. (1998b,2001b) and Imai et al. (2013).

Following the successful selection method adopted by Marshall et al. (2004), with the added benefit of existing mid-IR spectra obtained with the Spitzer Space Telescope (Buchanan et al. 2006; Kemper et al. 2010; Woods et al. 2011), we selected the brightest oxygen-rich (silicate dust) mid-IR sources (which pumps the OH maser) with the reddest optical and near-IR colours known in the LMC. These have luminosities $L\gtrsim 20,000$ L$_\odot$. Despite the superior angular resolution of Spitzer, all targets are exclusively IRAS or MSX sources, suggesting few (if any) luminous OH/IR stars were missed. Indeed, none other were found in a recent search for extremely dusty luminous red supergiants in the LMC (Jones et al. 2015) based on the Herschel Space Observatory far-IR survey HERITAGE (Meixner et al. 2013). \vspace{0.5cm}

Additionally, we targeted stars that pulsate with very long periods and large amplitudes, ignoring normal Mira variables which have shorter periods, thinner shells, and weaker OH emission. Typical LMC sources pulsate with periods $P\gtrsim 1000$ d and with K-band ($\sim 2.2 \micron$) amplitudes $\Delta K \geq 1.2$ mag as is shown in Table 2. As a consequence of their thick envelope of circumstellar dust, these stars are obscured at short wavelengths and are bright in the mid-IR ($F_{25} \gtrsim 0.5$ Jy).

In order to determine how the dusty wind depends on stellar parameters we have collected two comparison samples of OH/IR stars in the inner Milky Way. We describe these samples in detail in Section 4.2.2, but suffice to highlight here some of their relevant similarities and differences. Like the LMC sample, both the Galactic Centre and Galactic Bulge samples have relatively uniform and well-known distances, OH maser detections, and IR photometry. The Galactic Centre sources have reliable pulsation periods; for the Galactic Bulge sources we determine them for the first time here. By virtue of their OH detection these represent the most extreme OH/IR populations in those stellar systems. However, because they are dominated by relatively old, low-mass stars -- and more massive stars are just too rare to yield examples of the briefest, most extreme phases of evolution -- their luminosities are typically lower than those of the LMC sample, $L\sim10,000$ L$_\odot$. The same selection bias will also have predominantly selected metal-rich stars, which are not present in the LMC.

We have detected four new 1612 MHz circumstellar OH masers in the LMC (IRAS 04509$-$6922, IRAS 04516$-$6902, IRAS 05294$-$7104 and MSX LMC 807) and obtained wind speeds for six others. We successfully observed thirteen targets with three discoveries, one serendipitous discovery, and two non-detections; Galactic maser sources (shown in Appendix B) were used to check the pointing. We have used the velocities at the peaks of the maser profile to measure the expansion velocity of our sources with an uncertainty of around 1 km s$^{-1}$. We have also successfully observed one target in the 1665 MHz maser transition, IRAS 05280$-$6910. Before conducting this survey, only five of the known LMC maser profiles were reliable. We now have thirteen reliable maser profiles, doubling the previous LMC sample and bringing us closer to testing stellar wind driving mechanisms. We will describe each in turn in the following subsections. No emission was detected from any source at 1667 or 1720 MHz. 

\section{Observations}

\subsection {The Parkes Radio Telescope}

Targeted observations were done with the 64-m Parkes radio telescope in New South Wales, Australia. The observations were done in 2003 from August 14 to 22 (Marshall et al. 2004), in 2005 (new) from July 8 to 9, and from August 13 to 22, observing the 1612 MHz OH satellite line. The multibeam receiver used a dual polarisation setup with 8-MHz bandwidth and 8192 channels yielding 0.18 km s\textsuperscript{$-1$} channel\textsuperscript{$-1$} velocity resolution. The observations used a $12^{\prime}$ beam and an 8 MHz bandwith. Most observations were done in frequency switching mode except for the 2003 observations of IRAS 04498$-$6842, IRAS 04509$-$6922, and IRAS 04553$-$6922, which were done in position switching mode; integration times for observations are listed in Table 3.

Data were reduced using the \textsc{casa asap} toolkit (McMullin et al. 2007). The data were inspected manually for quality and for radio frequency interference (RFI). Each scan that was deemed reliable was averaged by polarisation and then a line-free baseline was fit and subtracted with a low order polynomial. The scans were then averaged and reduced separately for each epoch. Additionally, the collective data for each target were averaged and reduced. All scans were aligned in velocity, weighted by the noise (T$_{sys}^{-2}$), and fit with another low order polynomial, which was then subtracted. Any bad data spikes were determined visually and removed from within the file. All spectra were hanning smoothed to extenuate any potential maser peaks. This process has yielded spectra for each target from the 2003 and 2005 observing sessions as well as spectra with the combined data. As these sources are highly variable, multi-epoch observations have been critical for confirmation. 

\subsection{The Australia Telescope Compact Array (ATCA)}
The data were taken between March 6 and March 8, 2015, on June 11, 2015, and between February 21 and February 26 2016, using the 6C, 6A, and 6B array configurations of the ATCA, respectively. We did targeted integrations observing all four OH transitions (1612, 1665, 1667, 1720 MHz) using the CFB 1M-0.5k correlator configuration with zoom bands. The observations used a synthesized beam of around $7^{\prime\prime}$ and a velocity range of 465  km s$^{-1}$. During the first two observations we aimed the telescope directly at our primary targets (MSX LMC 807 and MSX LMC 815). The third observation used a position switching technique to get full \textsl{uv} coverage of our third primary target, IRAS 04516$-$6902, while getting partial coverage on two secondary targets, IRAS 04509$-$6922 and IRAS 04498$-$6842. Our two secondary sources could not fit within the primary beam of a single pointing, thus the need to alternate fields to observe with coverage similar to that of the shape of a Venn diagram with pointings separated by $\sim 13^{\prime}$. The fourth observation was split, observing the faint detection of IRAS 04516$-$6902, and the nondetection MSX LMC 815. The fifth, sixth, and seventh observations were aimed directly at our primary targets (IRAS 04553-6933, IRAS 05294-7104 and IRAS 05280-6910). 

The ATCA data were inspected, flagged, and calibrated, using the \textsc{Miriad} package (Sault et al. 1995). The visibility data were then transformed into three-dimensional data cubes. The source-finding package \textsc{Duchamp} (Whiting 2012) was used to search for maser sources within the $28^{\prime}$ full width half maximum of each of the fields. For the resulting peaks, our intended targets, and any other potential targets from Riebel et al. (2012, 2015) or SIMBAD, a spectrum was extracted for a region covering the size of the synthesized beam ($\sim 7^{\prime\prime}$). All spectra for each epoch were combined and weighted by $1/\sigma^2$, where $\sigma$ is the standard deviation determined from the line-free channels of the displayed spectrum.


\begin{figure}
 \centering
 \includegraphics[width=\columnwidth]{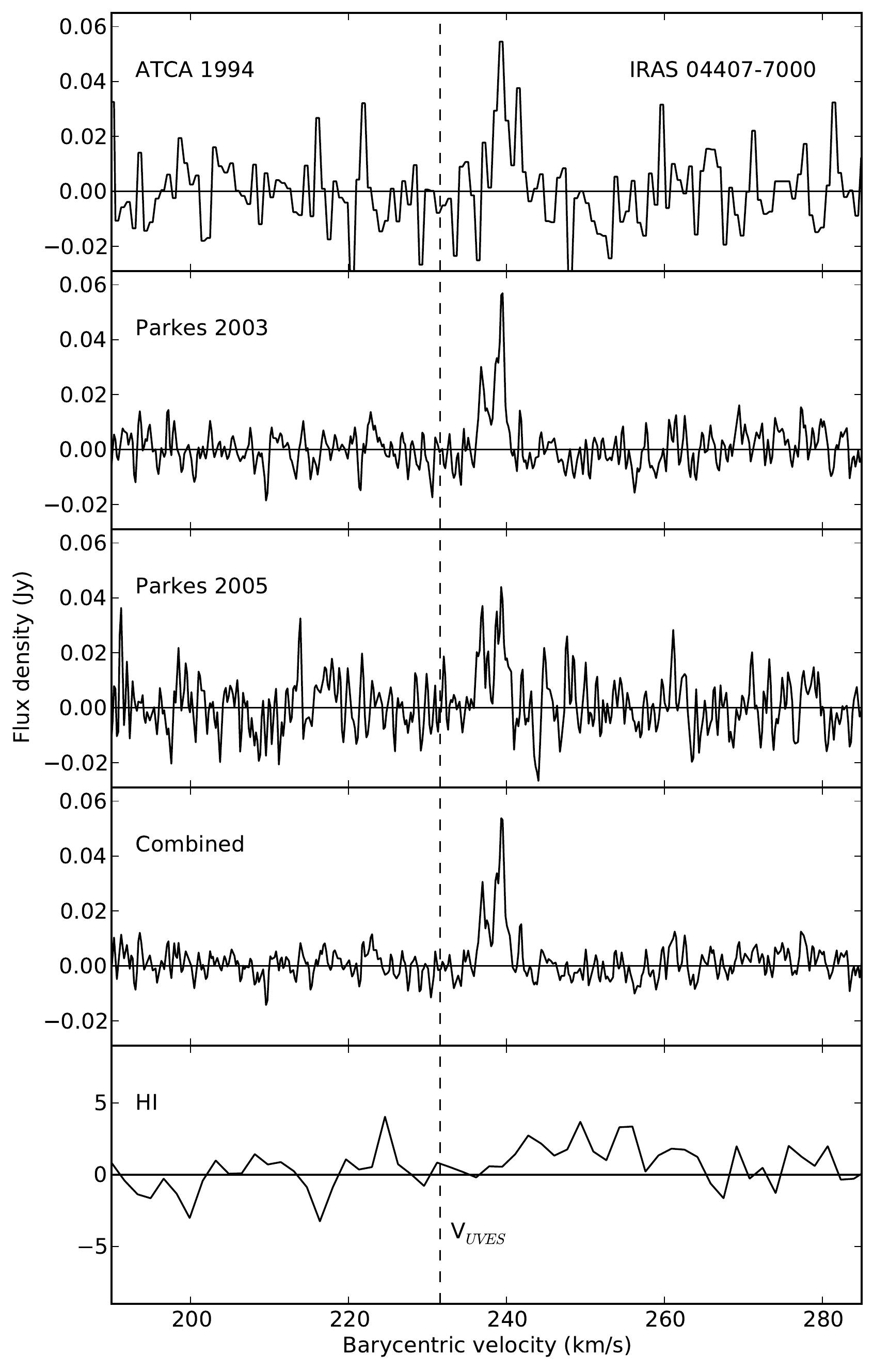}
 \caption{OH 1612-MHz maser emission from IRAS 04407$-$7000 with the velocity derived from cross-spectrum fitting of UVES spectra denoted by V$_{UVES}$.}
 \label{fig:04407}
\end{figure}

\begin{figure}
 \centering
 \includegraphics[width=\columnwidth]{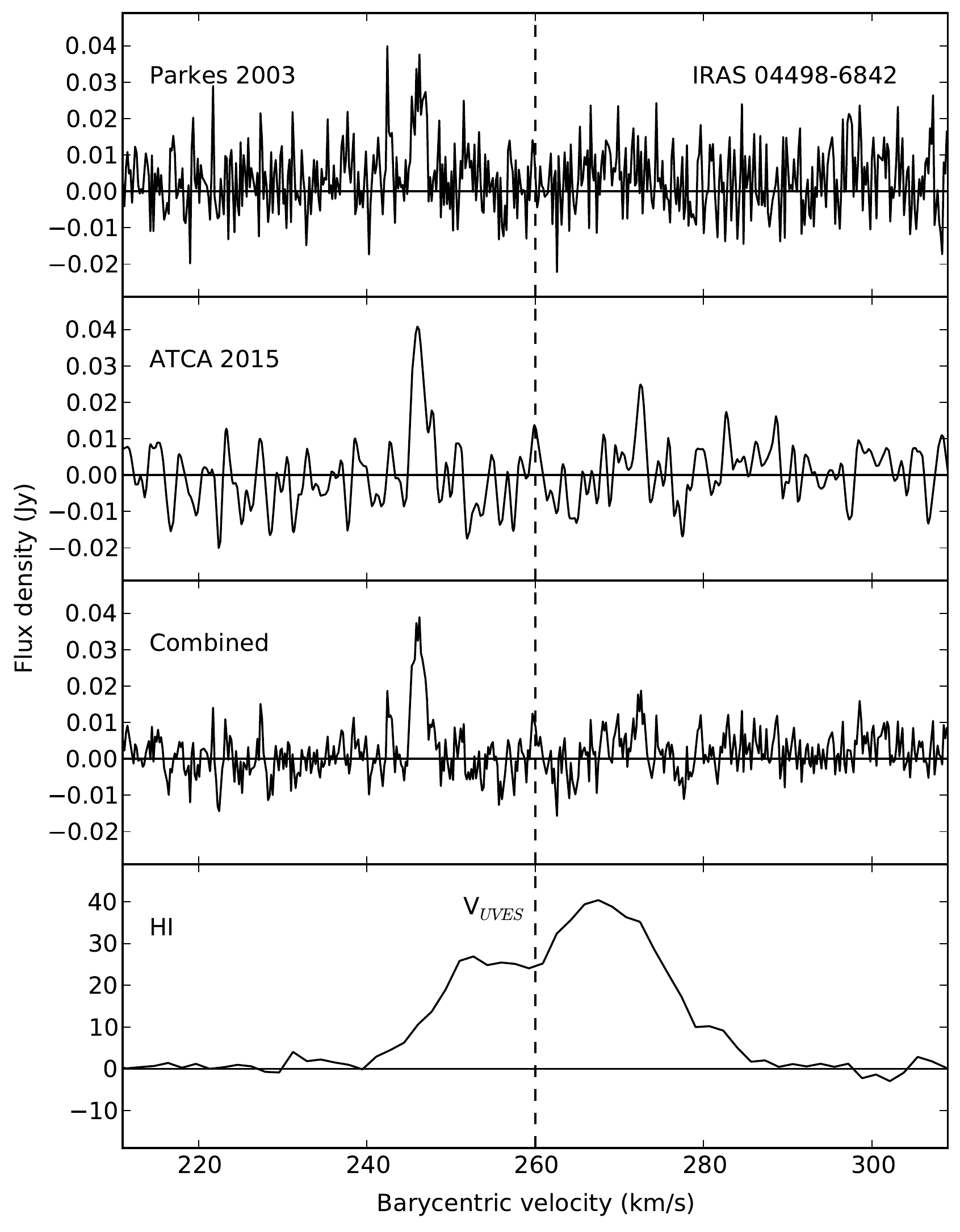}
 \caption{OH 1612-MHz maser emission from IRAS 04498$-$6842 with the velocity used as a model for the cross-spectrum fitting of UVES spectra denoted by V$_{UVES}$.}
 \label{fig:04498}
\vspace{-0.2cm}
\end{figure}

\begin{figure}
 \centering
 \includegraphics[width=\columnwidth]{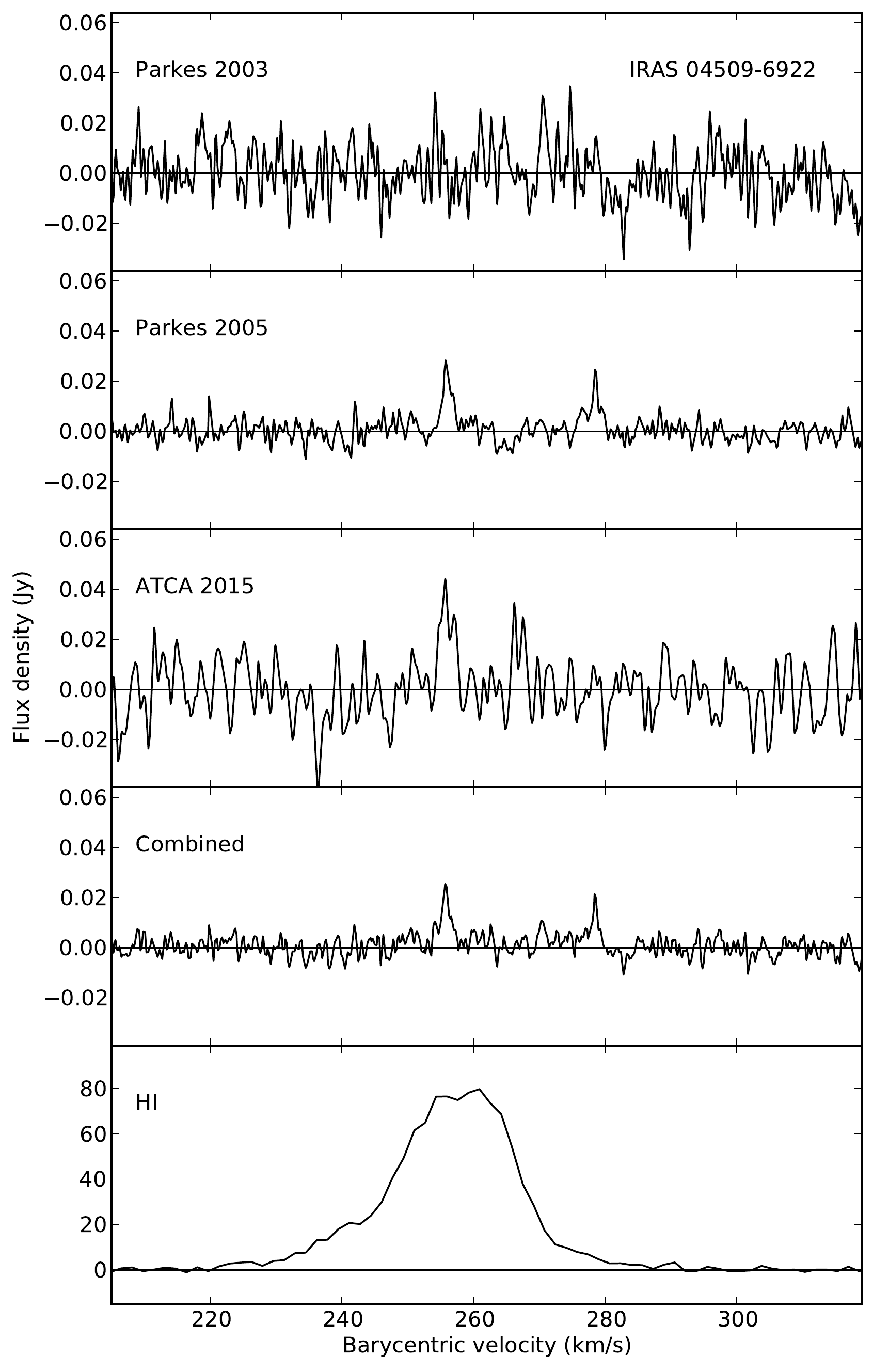}
 \caption{OH 1612-MHz maser emission from IRAS 04509$-$6922.}
 \label{fig:04509}
\end{figure}

\begin{figure}
 \centering
 \includegraphics[width=\columnwidth]{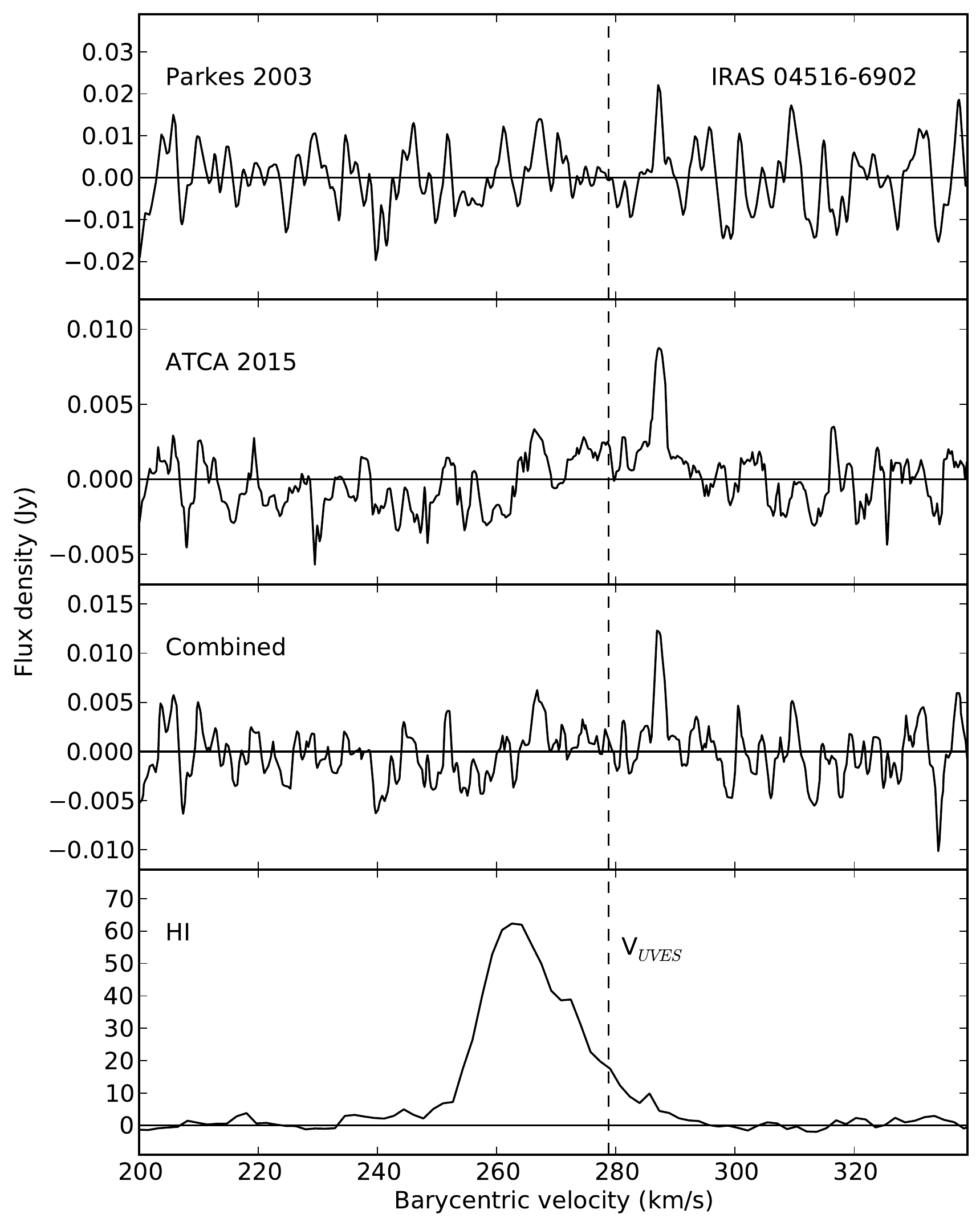}
 \caption{OH 1612-MHz maser emission from IRAS 04516$-$6902 with the velocity derived from cross-spectrum fitting of UVES spectra denoted by V$_{UVES}$.}
 \label{fig:04516}
\end{figure}

\begin{figure}
 \centering
 \includegraphics[width=8.2cm]{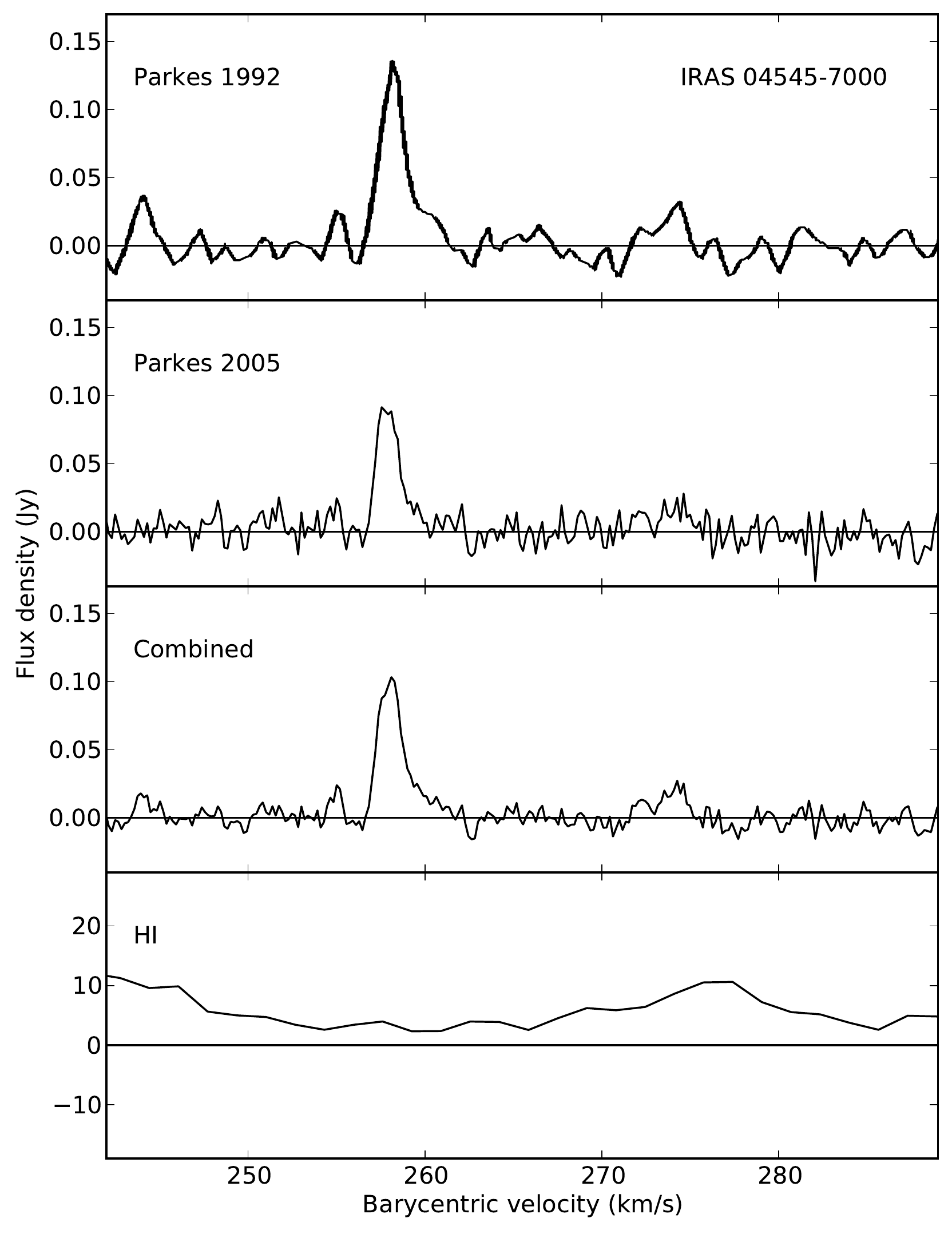}
 \caption{OH 1612-MHz maser emission from IRAS 04545$-$7000.}
 \label{fig:04545}
 \vspace{-0.0cm}
\end{figure}

\begin{figure}
 \centering
 \includegraphics[width=8.2cm]{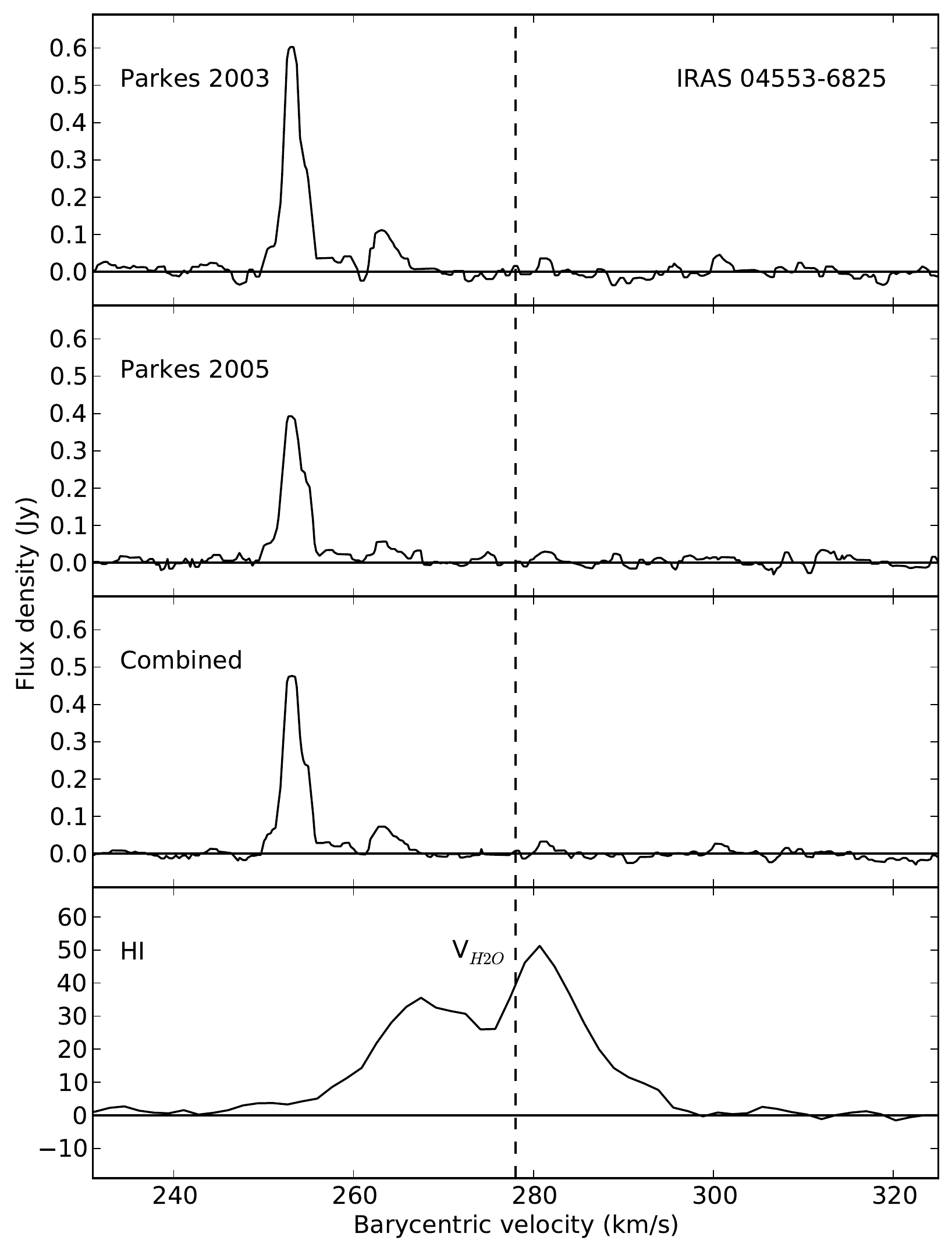}
 \caption{OH 1612-MHz maser emission from IRAS 04553$-$6825 with the H\textsubscript{2}O maser velocity from van Loon et al. (2001b) denoted by V$_{H2O}$.}
 \label{fig:04553}
\end{figure}

\begin{figure}
 \centering
 \includegraphics[width=\columnwidth]{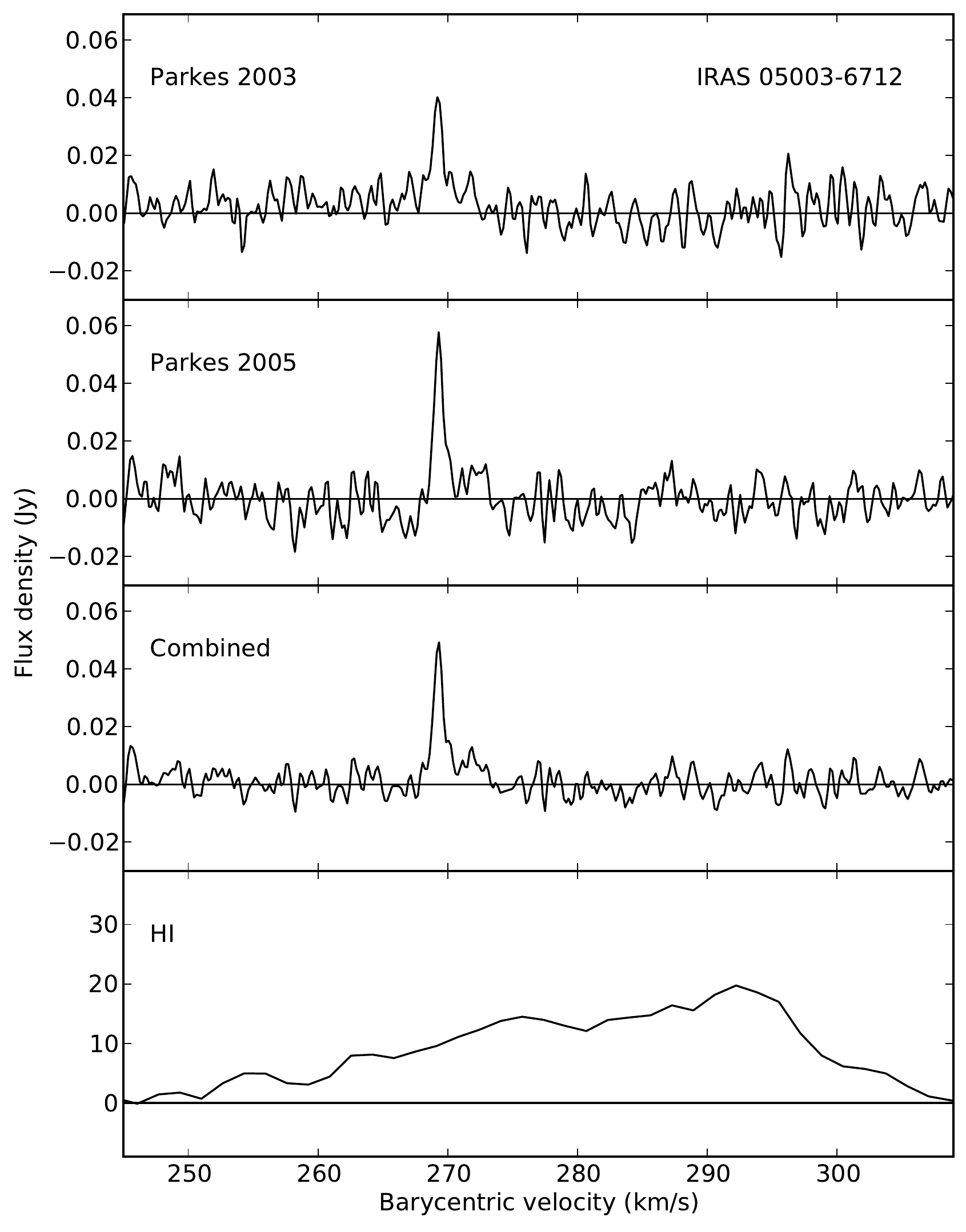}
 \caption{OH 1612-MHz maser emission from IRAS 05003$-$6712.}
 \label{fig:05003}
 \vspace{-0.5cm}
\end{figure}

\begin{figure}
 \centering 
 \includegraphics[width=\columnwidth]{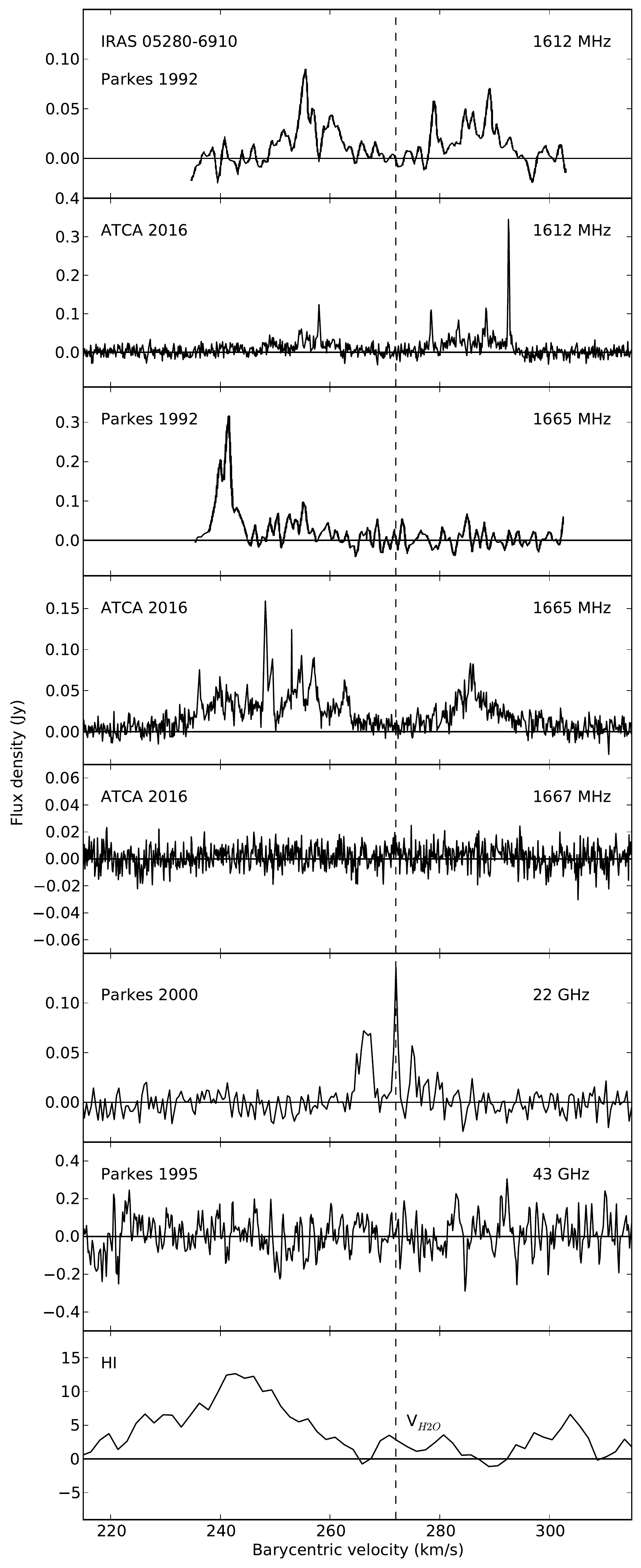}
 \caption{OH 1612-MHz and 1665-MHz maser emission from IRAS 05280$-$6910 with the SiO and H\textsubscript{2}O maser velocity from van Loon et al. (2001b); the water maser velocity has been plotted and denoted by V$_{H2O}$.}
 \label{fig:05280}
\end{figure}

\begin{figure}
 \centering 
 \includegraphics[width=\columnwidth]{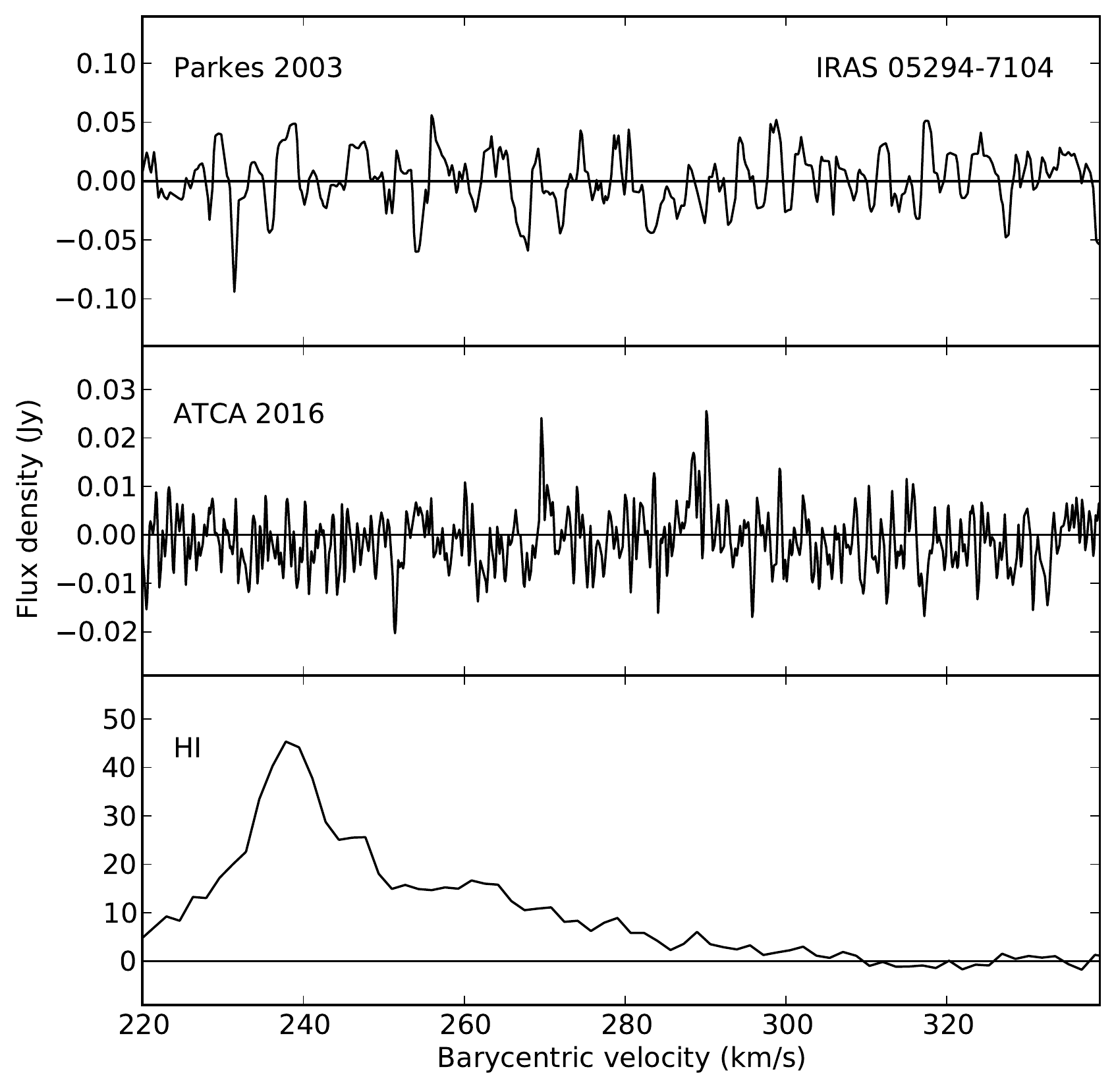}
 \caption{OH 1612-MHz maser emission from IRAS 05294$-$7104.}
 \label{fig:05294}
\end{figure}

\begin{figure}
 \centering
 \includegraphics[width=\columnwidth]{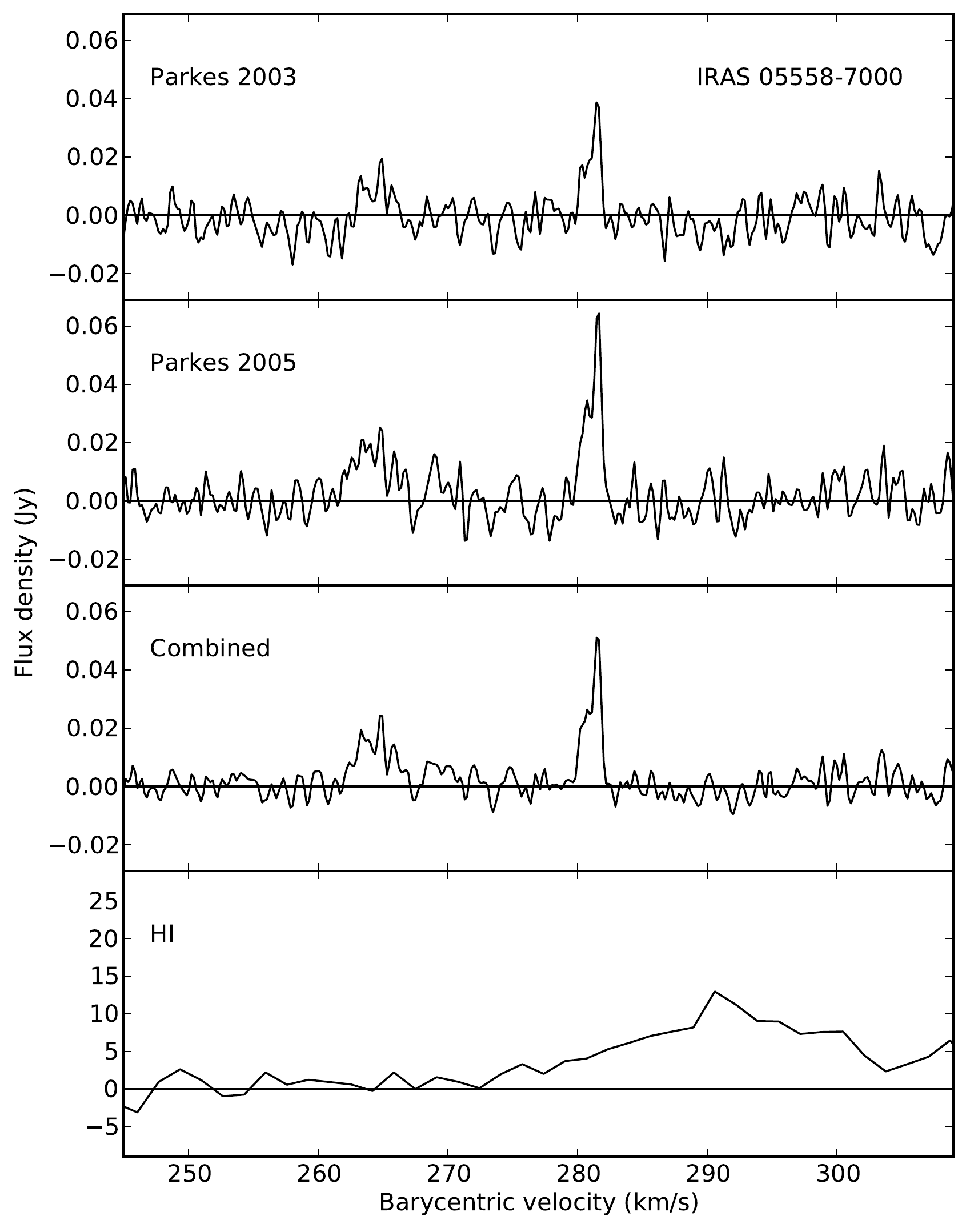}
 \caption{OH 1612-MHz maser emission from IRAS 05558$-$7000.}
 \label{fig:05558}
\end{figure}

\begin{figure}
 \centering
 \includegraphics[width=\columnwidth]{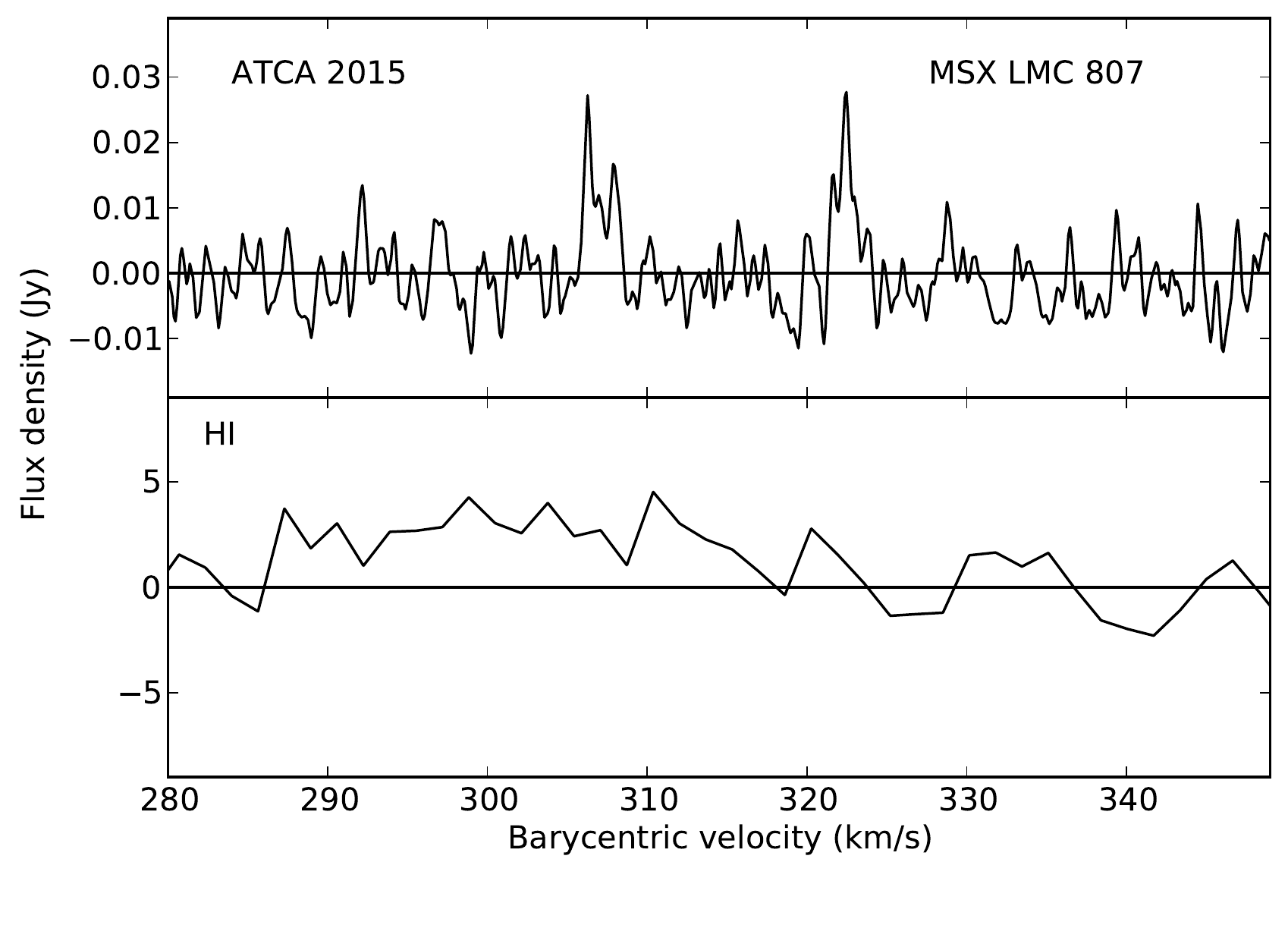}
 \caption{OH 1612-MHz maser emission from MSX LMC 807.}
 \label{fig:msxlmc807}
\end{figure}

\begin{figure}
 \centering
 \includegraphics[width=\columnwidth]{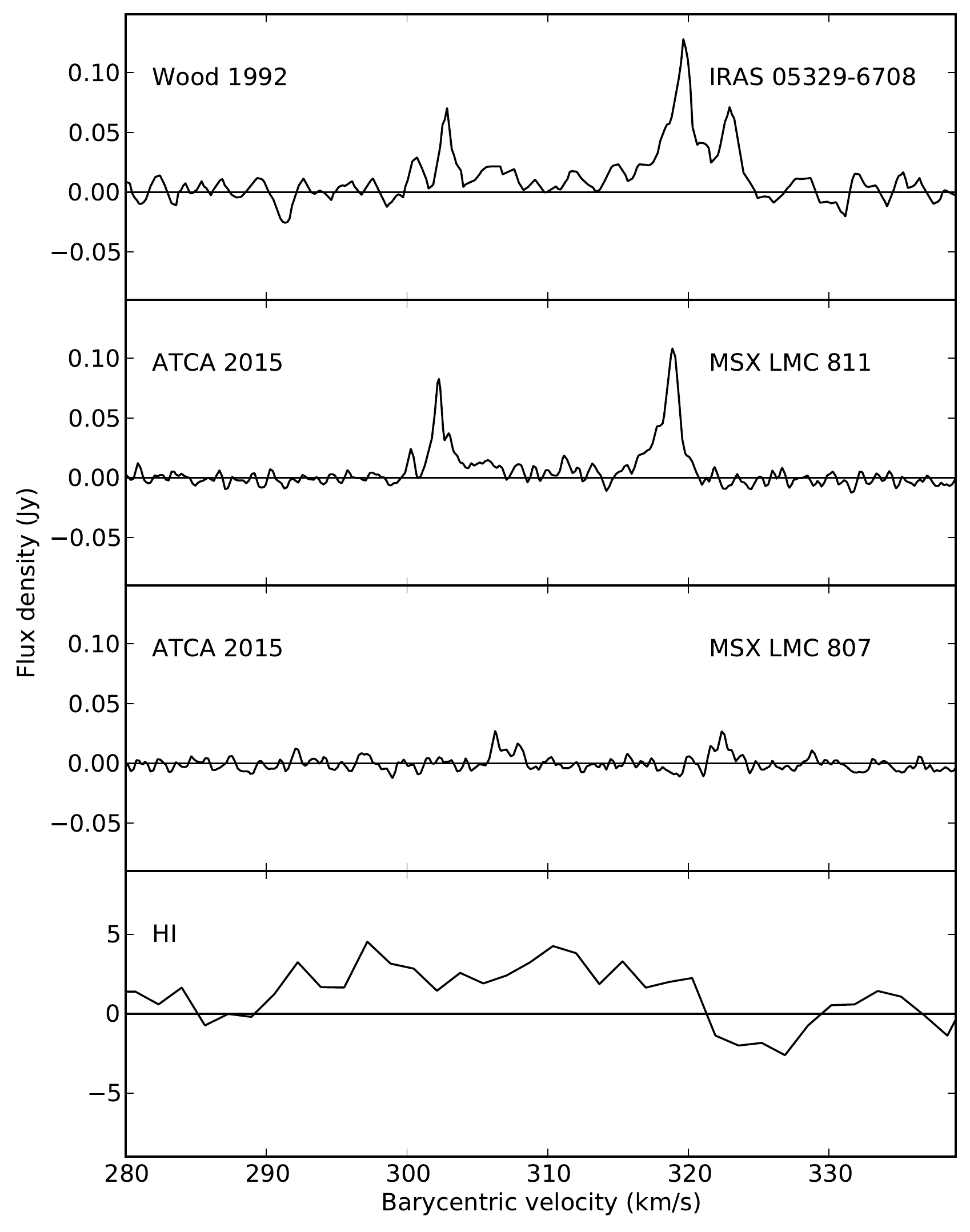}
 \caption{OH 1612-MHz maser emission from MSX LMC 811 and the newly discovered MSX LMC 807. Originally thought to originate from IRAS 05329$-$6708, the true components of the maser emission have now been identified. }
 \label{fig:msxlmc811}
\end{figure}

\begin{figure}
 \centering
 \includegraphics[width=\columnwidth]{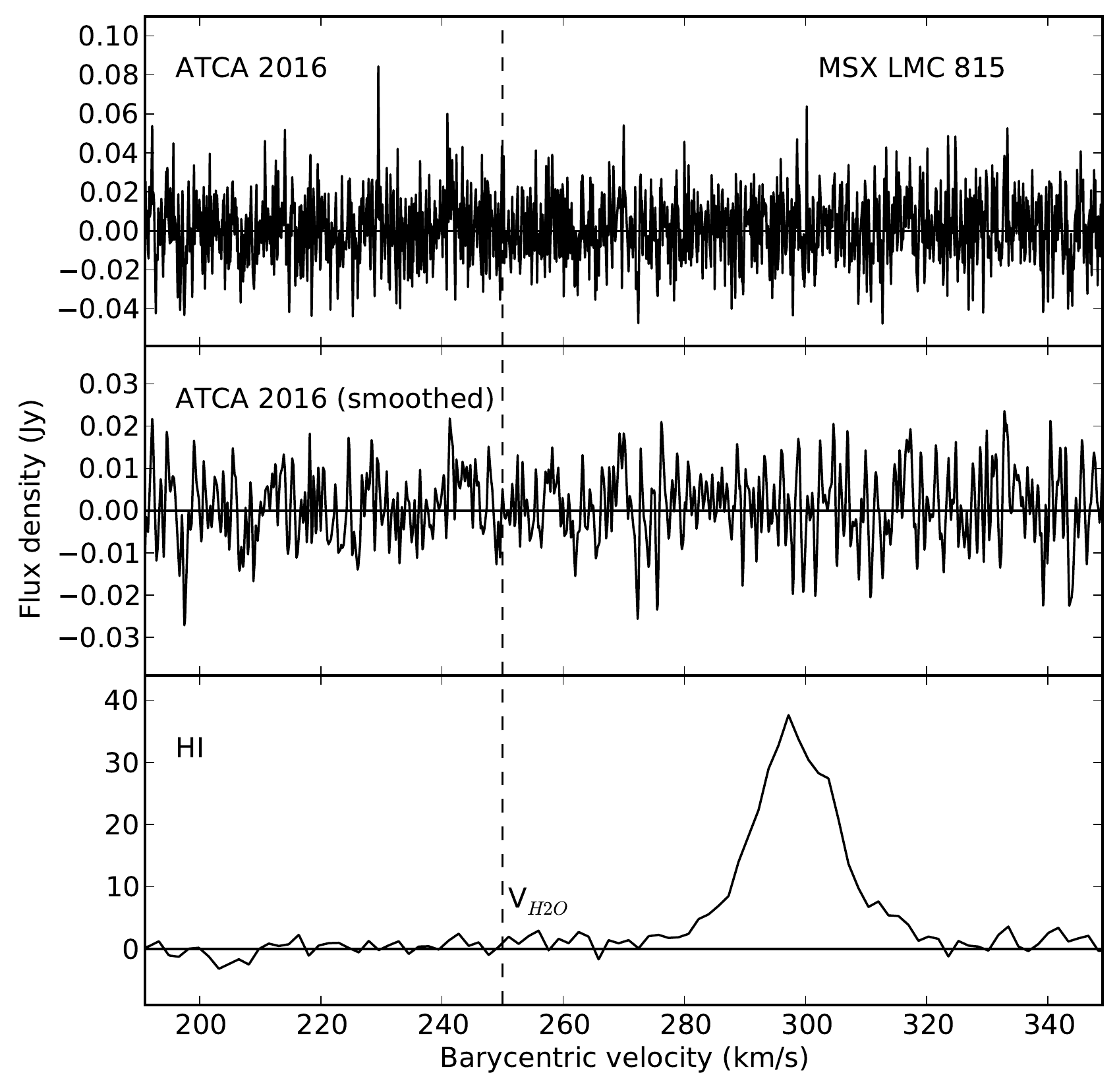}
 \caption{OH 1612-MHz nondetection from MSX LMC 815, with the velocity of the water maser detection made by Imai et al. (2013) denoted by V$_{H2O}$.}
 \label{fig:msxlmc815}
\end{figure}

\subsection{Parkes and ATCA combined H\,{\sc i} Maps}
In addition to the Parkes and ATCA observations, we have also included 1420 MHz H\,{\sc i} data from the Staveley-Smith et al. (2003) combined ATCA and Parkes multibeam H\,{\sc i} maps. Peaks in the H\,{\sc i} spectra indicate the likely systemic velocities of our sources. While some of our sources lie outside these H\,{\sc i} regions, they provide supporting evidence for sources that lie in higher density regions of the LMC. For each source in our LMC sample, a spectrum was extracted using a one arcminute region centred on the sources. The resulting spectra have been plotted below the maser spectra in Section 4 to indicate the likely systemic velocity.

\section{Results}
\subsection{Individual maser sources}
\subsubsection{IRAS 04407$-$7000}
The OH maser emission from the dusty AGB star IRAS 04407$-$7000 was initially detected with the ATCA (van Loon et al. 1998a). The initial detection showed a single peak at 240 km s$^{-1}$; we now present subsequent observations with Parkes (Fig. \ref{fig:04407}). We see a degree of variability between the 2003 and 2005 observations. With the first observation we see two peaks at around 237 and 240 km s$^{-1}$. With a peak separation of 3 km s$^{-1}$, it is unlikely a measure of the full expansion velocity, rather substructure within the circumstellar envelope. We also see a much smaller peak at 223 km s$^{-1}$ in the 2003 observation. For the much shorter 2005 observation, we see the peaks at 237 and 240 km s$^{-1}$, flanked by two smaller tentative peaks at 214 and 261 km s$^{-1}$ but an overall increase in noise. When the spectra are combined and weighted by the noise, the dominant profile is that of the 2003 observation. Using a cross-spectrum fitting technique (see Appendix C) we have fit a UVES spectrum of IRAS 04407$-$7000 from the ESO reduced spectra data archive, to that of IRAS 04498$-$6842 using molecular bandheads around 8440 \AA. We have used a systemic velocity of 260 km s$^{-1}$ for IRAS 04498-6842, assumed from its maser profile. From the phase shift we calculate a systemic velocity for IRAS 04407$-$7000 of 231.6 km s$^{-1}$ (denoted by V$_{UVES}$) confirming the smaller blue maser peak at 223 km s$^{-1}$ and the maser profiles of both sources. 

\subsubsection{IRAS 04498$-$6842}
IRAS 04498$-$6842 was observed in 2003 with a clear peak at 246 km s$^{-1}$ (Fig. \ref{fig:04498}). The new observation with the ATCA has confirmed the peak at 246 km s$^{-1}$ and revealed a red counterpart at 272 km s$^{-1}$. The Parkes multibeam H\,{\sc i} data of our target peaks around 260 km s$^{-1}$. As we expect the H\,{\sc i} to come from the LMC disk, and our source is likely to have a similar stellar velocity, this provides evidence for our detection. Additionally, we have used the systemic velocity derived from this source's maser profile as a reference to fit both IRAS 04407$-$7000 and IRAS 04516$-$6902, and the velocities of all three maser profiles are consistent with our results. IRAS 04498$-$6842 is one of the more extreme stars within our sample (possibly a RSG, or super-AGB star), with an M10 spectral type (van Loon et al. 2005) and high luminosity. Thus it is not surprising that our maser profile yields a high expansion velocity of 13 km s$^{-1}$.

\subsubsection{IRAS 04509$-$6922}
IRAS 04509$-$6922 is a new detection, with a clear double-peaked profile centred around 268 km s$^{-1}$ (Fig. \ref{fig:04509}). The profile expansion velocity of 11.5 km s$^{-1}$ is also high within our sample, but expected given its low temperature and high luminosity. We see a peak around 255 km s$^{-1}$ over all three observations, but the red component is only clear in the much longer integration in the 2005 observation. The peak in H\,{\sc i} at around 260 km s$^{-1}$  supports our detection.  

\subsubsection{IRAS 04516$-$6902}
IRAS 04516$-$6902 was observed in 2003, with no clear maser emission. Given our new ATCA data (Fig. \ref{fig:04516}), it seems that the small peak at around 285 km s$^{-1}$ in our Parkes observation is in fact a maser component. We also see a much fainter peak in the combined data around 267 km s$^{-1}$ yielding an expansion velocity of 10 km s$^{-1}$. Using the same cross-spectrum fitting technique as was done with IRAS 04407$-$7000, we calculate a systemic velocity of 278.8 km s$^{-1}$. This velocity lies directly between our two maser peaks, confirming our maser profile.  

\subsubsection{IRAS 04545$-$7000}
The maser profile of IRAS 04545$-$7000 does not seem to vary drastically over our two epochs (Fig. \ref{fig:04545}). The detection has a clear blue-asymmetry, a phenomenon described in Marshall et al. (2004), which has been suggested as a consequence of the contribution of amplified stellar light. While IRAS 04545$-$7000 is not a very luminous source, it is one of the most highly reddened sources in our sample with a (J$-$K) colour of 5.7 mag. 

\subsubsection{Red supergiant IRAS 04553$-$6825}
IRAS 04553$-$6825 has shown the most prominent OH maser emission in the LMC with a peak over 600 mJy around 253 km s$^{-1}$ (Fig. \ref{fig:04553}). This RSG is one of the largest known stars, with an estimated radius of 1540 R\textsubscript{$\odot$} (Levesque et al. 2009). This object has been studied in depth, and has been observed in OH (Wood, Bessel \& Whiteoak 1986; Marshall et al. 2004), H\textsubscript{2}O (van Loon et al. 1998b) and SiO (van Loon et al. 1996). In addition to its peak at 253 km s$^{-1}$ are three additional peaks at 262, 281 and 300 km s$^{-1}$. Given the systematic velocity of 278 km s$^{-1}$ taken from the SiO maser detection, we can say with some certainty that the maser profile extends from its prominent blue peak at 253 km s$^{-1}$ to its smallest peak at 300 km s$^{-1}$. The existence of the interior two peaks has been suggested by Marshall et al. (2004) to be indicative of a second expanding dust shell, flowing primarily in a direction across the sky.

\subsubsection{IRAS 05003$-$6712}
IRAS 05003$-$6712 is the least luminous source of our sample (Fig. \ref{fig:05003}). We see a clear peak at 269 km s$^{-1}$ in both epochs, but the second peak is unclear. The source fits all other criteria expected of an OH/IR star but the secondary peak may be below the detection limit of past observations. From our SED modeling (explained in Section 4.2), we expect an expansion velocity around 10 km s$^{-1}$ (assuming a metallicity of half that of the Sun). The skewed shape of the peak at 269 km s$^{-1}$ suggests we need to be looking for the missing redshifted peak. Nevertheless, the maser profile remains unclear. 

\subsubsection{Red supergiant IRAS 05280$-$6910}
The maser emission of the RSG IRAS 05280$-$6910 was originally observed at 1612 and 1665 MHz with Parkes in 1992 (Wood et al. 1992). We have since observed the source using the ATCA and revealed five maser peaks at 1612 MHz and some interesting structure at 1665 MHz (Fig. \ref{fig:05280}). The IR source was originally misidentified as the cluster NGC 1984. High resolution near- and mid-IR imaging has successfully identified the stellar counterpart (van Loon, Marshall, \& Zijlstra 2005). IRAS 05280$-$6910 is highly obscured and extremely bright in the mid-IR but shows little variability. This source has also shown 22 GHz H\textsubscript{2}O maser emission (van Loon et al. 2001b). The lack of symmetry within these maser spectra as well as variability of not only the flux but also of the location of maser components remains puzzling. We would also expect the mainline 1665 MHz detection to probe the inner shell of OH and thus lie within the 1612 MHz maser profile. We will discuss this source further in Section 5.
 
\subsubsection{IRAS 05294$-$7104}
IRAS 05294$-$7104 was not initially detected during the 2003 observation but was detected with our more recent longer ATCA observation (Fig. \ref{fig:05294}). It is highly reddened with a (J$-$K) colour of 2.97 mag, with a long pulsation period lasting 1079 days and an expansion velocity of 10.3 km s$^{-1}$. It seems that the initial observation lacked the signal-to-noise ratio required for detection.  

\subsubsection{IRAS 05298$-$6957}
IRAS 05298$-$6957 resides in a cluster and thus has cluster metallicity of 0.4 solar metallicity. This source pulsates with a $\Delta$K of 2 mag, and has an initial mass of 4$\textrm{ M\textsubscript{$\odot$} yr\textsuperscript{$-1$}}$ (van Loon et al. 2001a). The source has a clear double-peaked profile yielding an expansion velocity of 10.5 km s$^{-1}$ (Wood et al. 1992). IRAS 05298$-$6957 has shown silicate in absorption yet lies in a region of the LMC sample colour--magnitude diagram that is dominated by carbon rich chemistry (Sargent et al. 2011). This is a well known but often ignored issue that is seen in a number of OH/IR sources (cf. Zijlstra et al. 1996; van Loon et al 1997; Trams et al. 1999).

\subsubsection{IRAS 05402$-$6956}
OH maser emission from IRAS 05402$-$6956 was detected with Parkes in 1992 (Wood et al. 1992) and is known to be within $3^{\prime}$ of an H\,{\sc ii} region. The expansion velocity from the original observation was 10.5 km s$^{-1}$, but, as the emission was detected with a low signal-to-noise ratio, it is unclear if the profile includes the outermost peaks. 

\subsubsection{IRAS 05558$-$7000}
IRAS 05558$-$7000 lies at the edge of the LMC. The source has two clear peaks at 265 and 281 km s$^{-1}$, in both epochs of observation, yielding an expansion velocity of 8 km s$^{-1}$ (Fig. \ref{fig:05558}).

\subsubsection{MSX LMC 807}
MSX LMC 807 is a dusty O-AGB star with a (J$-$K) colour of 3.7 mag and $M$\textsubscript{bol}=$-6.6$ (Whitelock et al. 2003). This star has shown $10$ $\micron$ silicate emission (Buchanan et al. 2006) but a low IR luminosity comparable to many C-AGB stars in the LMC (Buchanan et al. 2006). This is one of our new OH maser detections from ATCA observations. We see a clear double-peaked profile yielding an expansion velocity of 8 km s$^{-1}$ (Fig. \ref{fig:msxlmc807}). 

\subsubsection{MSX LMC 811 (IRAS 05329$-$6708)}
MSX LMC 811 has been found to be the true counterpart of the maser emission thought to originate from IRAS 05329$-$6708. While IRAS 05329-6708, MSX LMC 807, and MSX LMC 811 were previously resolved in the IR, the original Parkes observation of IRAS 05329$-$6708 did not have the resolution to resolve IRAS 05329$-$6708 into MSX LMC 807 and MSX LMC 811. Thus the observation yielded partial emission from both MSX LMC 807 and MSX LMC 811 (Fig. \ref{fig:msxlmc811}). This can be seen in the four peaks found in the 1992 observation, which show the two similar wind speeds with different systemic velocities. As the MSX survey resolved our IR sources, we have now done the same with the OH maser emission. The star is extremely red with a (J$-$K) colour of 5.2 mag; the expansion velocity of 8.5 km s$^{-1}$ is also reasonable for the source.   

\subsubsection{MSX LMC 815}
MSX LMC 815 is an M4 O-AGB star with a K-band magnitude of 8.14 and has shown a number of spurious spikes at 1612 MHz, but no clear maser profile (Fig. \ref{fig:msxlmc815}). H\textsubscript{2}O maser emission has been detected at 237 km s$^{-1}$ by Imai et al. (2013), yet the H\textsubscript{2}O maser detection was a very weak detection; weaker than many H\textsubscript{2}O maser detections of RSGs in the Galaxy. This source has a short pulsation period of 590 days typical of our galactic sources. We expect the 1612 MHz maser flux should peak at close to $25\%$ of the 35 $\mu$m flux due to the efficiency of the pumping mechanism (Elitzur, Goldreich \& Scoville, 1976; Marshall et al. 2004). As this source has a $F_{24}=0.22$ Jy, the maser emission would likely be below our detection limit.

\subsection{Modeling the spectral energy distribution}
\subsubsection{LMC sources}
We have modeled all our AGB and RSG sources in the LMC with radiatively driven wind models produced with the \textsc{dusty} code (Elitzur \& Ivezi{\'c} 2001). Two important outputs from the \textsc{dusty} 1-D radiation transfer code are the SED of the source modeled, and the expansion velocity of the dust-driven wind. The SEDs for the LMC sources are shown in Figure \ref{fig:lmc_sed} and the results are listed in Table 5. The models use a blackbody for the central star and assume dusty envelopes of warm silicates from Ossenkopf et al. (1992), with varying effective temperatures and inner boundary dust temperatures and the standard MRN (Mathis et al. 1977) grain size distribution.

As these are variable sources, to accurately reproduce the true SED with our models we can only fit one epoch of photometry. We have thus used the \textit{Spitzer} IRAC and MIPS photometry (Fazio et al. 2004; Rieke et al. 2004), plotted in red, and fit them with \textsc{dusty} models using a $\chi^2$ fitting technique. The available 2MASS JHK photometry are not of the same epoch and are more sensitive to the effects of geometry. We have thus ignored them in our source fitting. The fitting technique calculates the best fit of the photometry to 5900 models of varying optical depth, central blackbody temperature, inner dust temperature and normalisation factor. We obtain luminosities for the best fit models (plotted with dashed lines) assuming a distance of 50 kpc to the LMC. We expect these luminosities to be more accurate than for Galactic sources because of the well-determined distance to the LMC. These results constrain the derived parameters of the source at the time that the \textit{Spitzer} photometry data were taken (July$-$November, 2005). The current stellar parameters will have fluctuated since as a result of the sources' variability.

For each of our sources we have compiled all available photometry from 2MASS (Cutri et al. 2003), DENIS (Cioni et al. 2000), AKARI (Yamamura et al. 2010), MSX (Egan et al. 2003), and IRAS (Beichman et al. 1988), and overplotted them in the SED fit with small open circles; we expect our SED to lie between the stellar variations. We have additionally overplotted \textit{Spitzer} IRS spectra (Lebouteiller et al. 2011), shown in solid black. It should be noted that the best \textsc{dusty} model was fit to photometry, independently of the IRS spectra. In most cases the model successfully reproduces the shape of the photometry and spectra well, which lends credence to the fit solution. Our choice of dust optical constants will affect the final result. The dust mineralogy will change as the central star evolves (Verhoelst et al. 2009; Dijkstra et al. 2005) and our inability to compensate for this effect will introduce a degree of uncertainty.

We ran the {\sc dusty} code in its radiatively driven wind mode, whereby it solves the hydrodynamical equations for the radial dependence of the density and for the velocities of the gas and dust separately (their difference is the `drift' speed). It is the velocity of the gas -- which dominates the mass by far -- which we can measure from the OH maser profiles, but the velocity of the dust can be greater by a factor of a few if the wind density is relatively low. If, however, the drift speed is under-estimated then the SED fitting would yield a dust mass loss rate that is too low and (assuming the gas-to-dust ratio is correct) also a total mass loss rate that is too low. Ramstedt et al. (2008) found a discrepancy between the mass loss rates derived using {\sc dusty} and those derived from modeling the CO line emission, with the former always yielding lower values. However, if the drift speed was under-estimated then the heating of the CO molecules would also have been under-estimated and thus CO mass loss rates would be derived that are too high.

The power of the description of the wind incorporated in {\sc dusty} lies in the fact that the results can be scaled for different dust content and luminosity according to Eq. (1) and $\dot{M}\propto r_{gd}^{1/2}L^{3/4}$ (Elitzur \& Ivezi\'c 2001) without changing the optical depth -- it is the latter which determines the shape of the SED. This is, however, a simplification as the shape of the SED (and acceleration of the wind) depends on the distance from the star where the dust forms. This inner radius is set by imposing a fixed dust formation temperature. While it scales with luminosity as $L^{1/2}$ it has a subtle dependency on the temperature of the star as the grain opacity is wavelength dependent (it therefore also depends on the grains' optical properties). Also, the acceleration of the wind will be somewhat different depending on the apparent size of the star whereas it is assumed that all stellar flux arises from one point. These effects mostly affect the emission from the warmest dust at near- to mid-IR wavelengths. An outer limit to the size of the envelope is assumed to be $10^4$ times the inner radius, which is typical given the stars we study are near the end of an extended period of heavy mass loss. The emission from the coldest dust at far-IR wavelengths depends on this choice, but not in a sensitive way.

\begin{figure*}
\centering
\begin{minipage}[c]{\textwidth}
\centering
 \includegraphics[width=6.8cm]{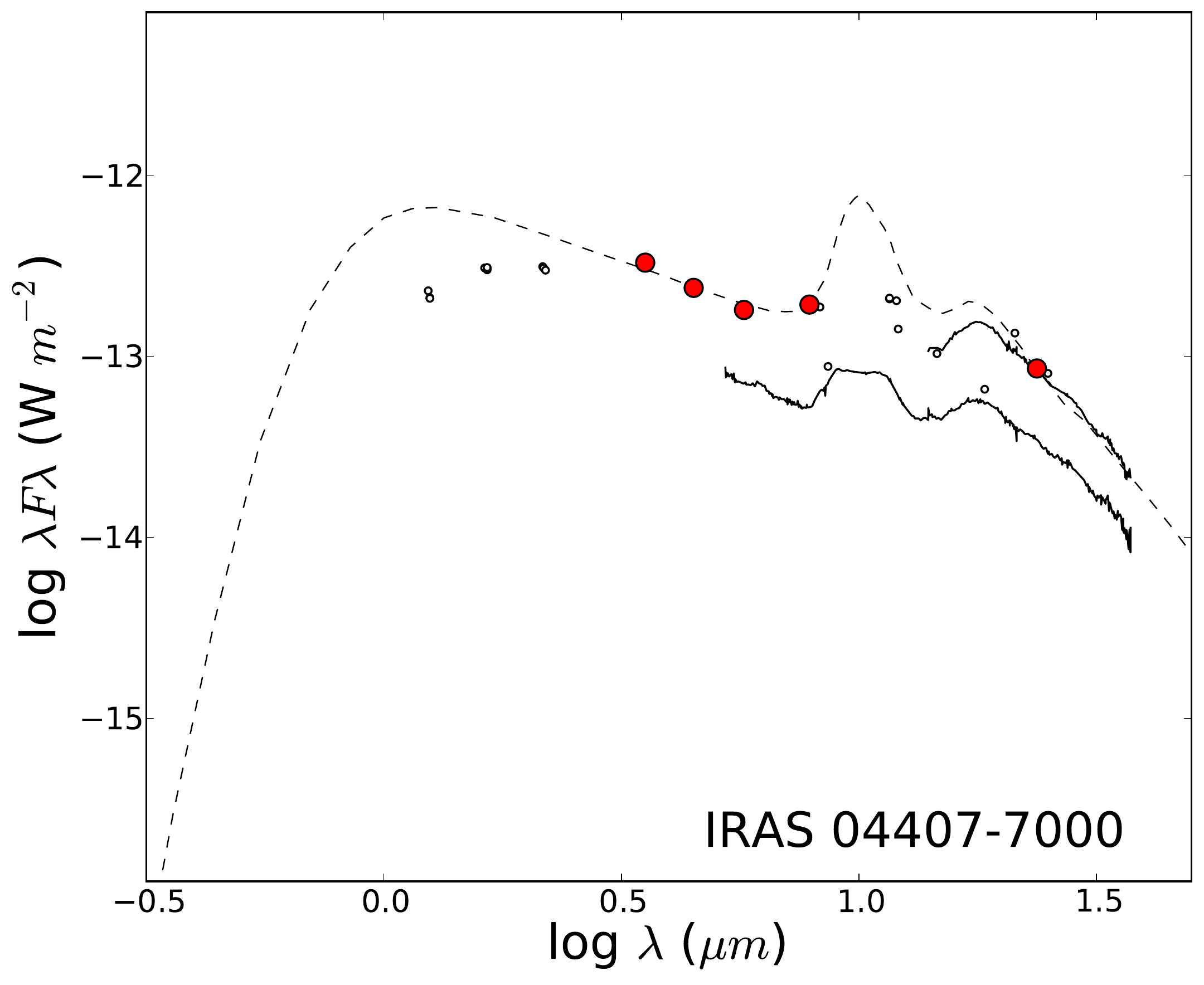}
 \includegraphics[width=6.8cm]{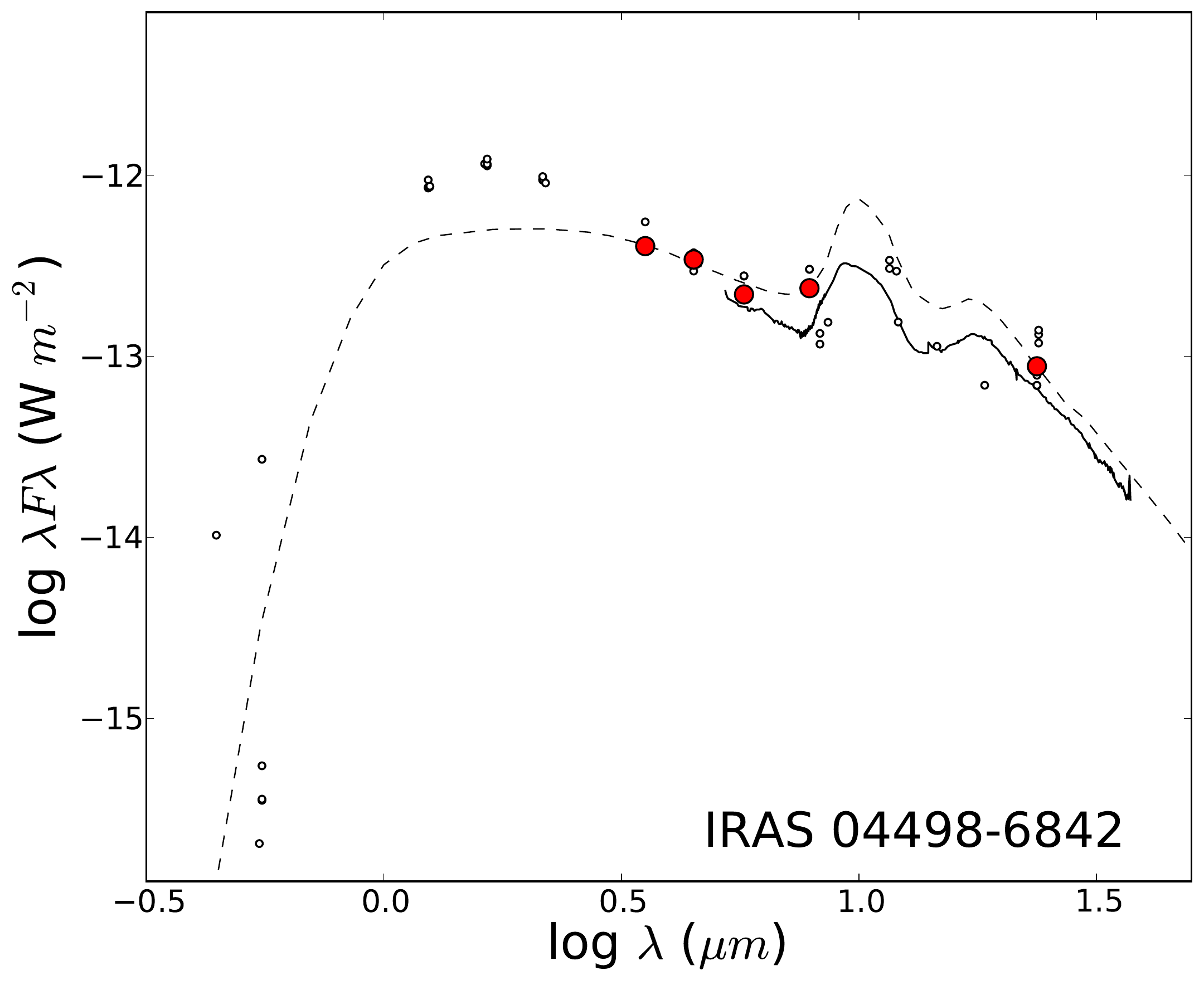}\\
 \includegraphics[width=6.8cm]{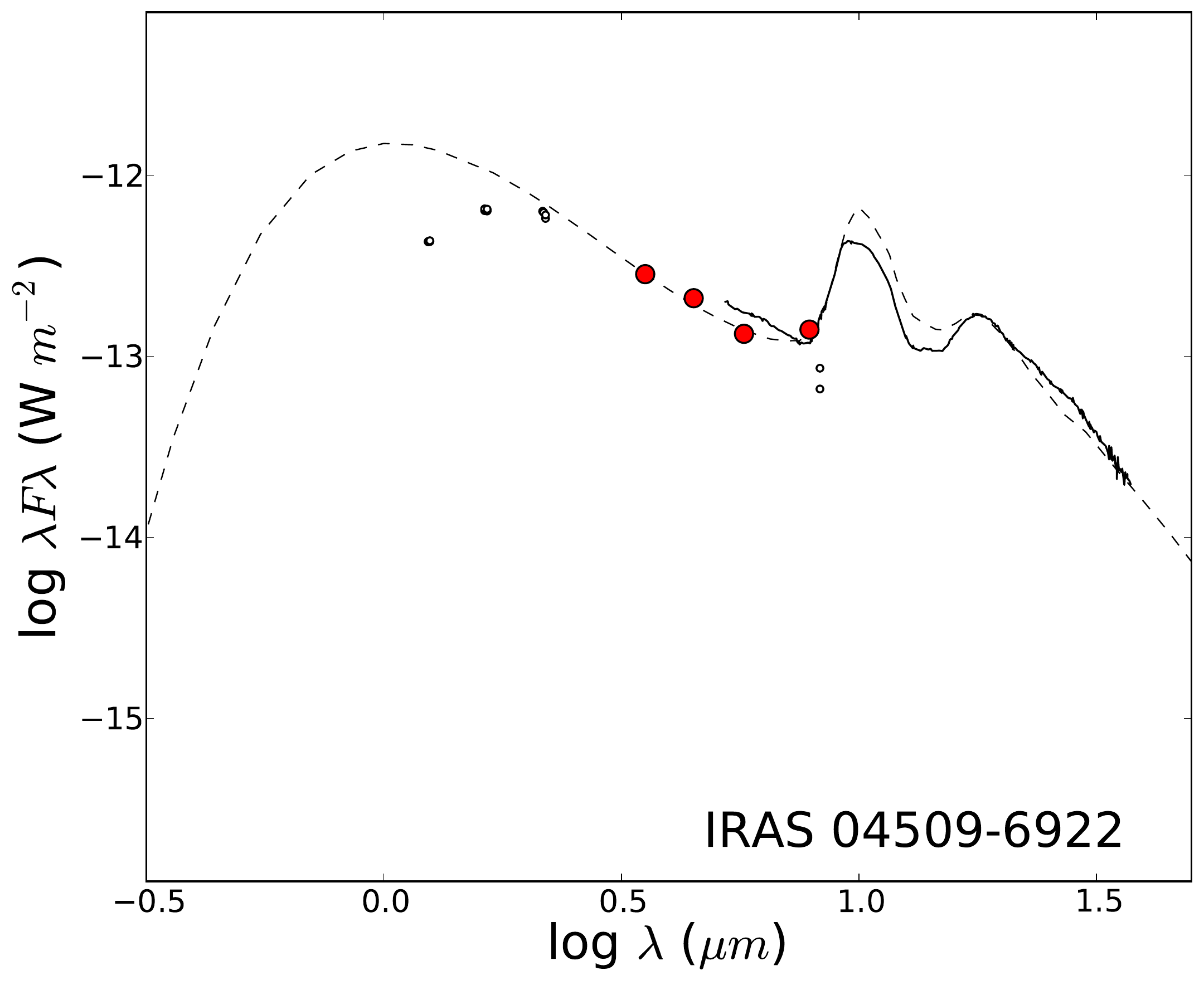}
 \includegraphics[width=6.8cm]{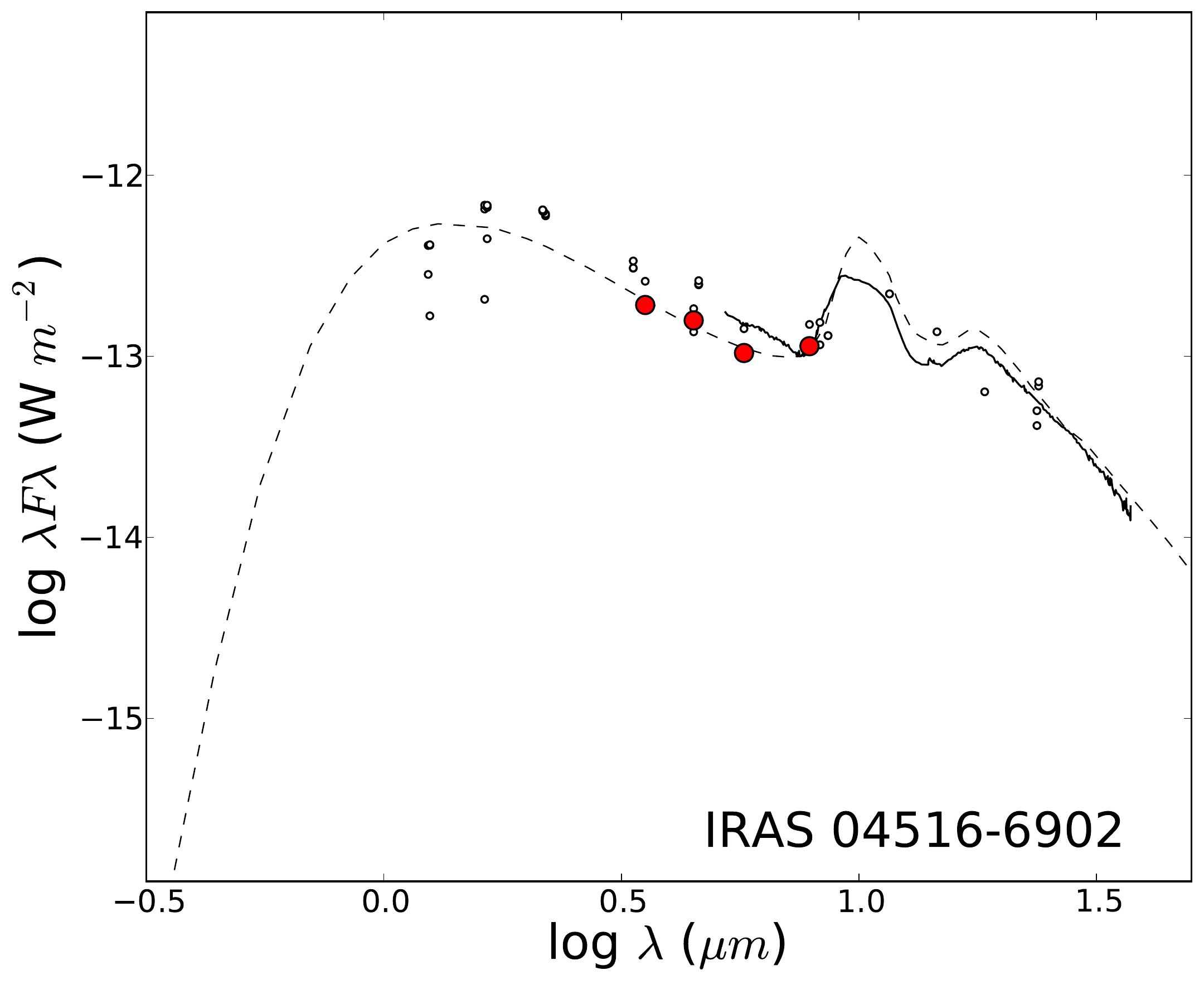}\\
 \includegraphics[width=6.8cm]{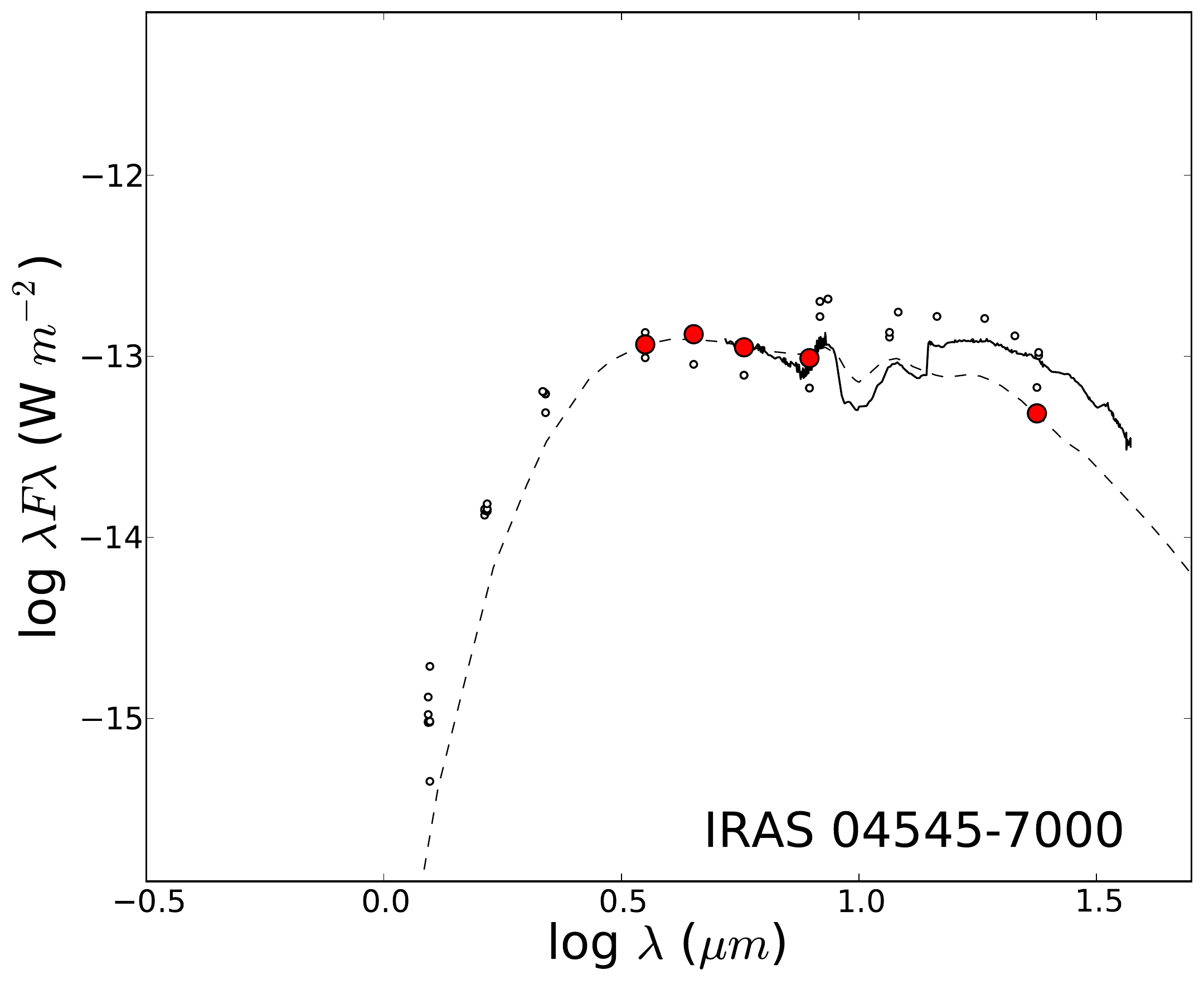}
 \includegraphics[width=6.8cm]{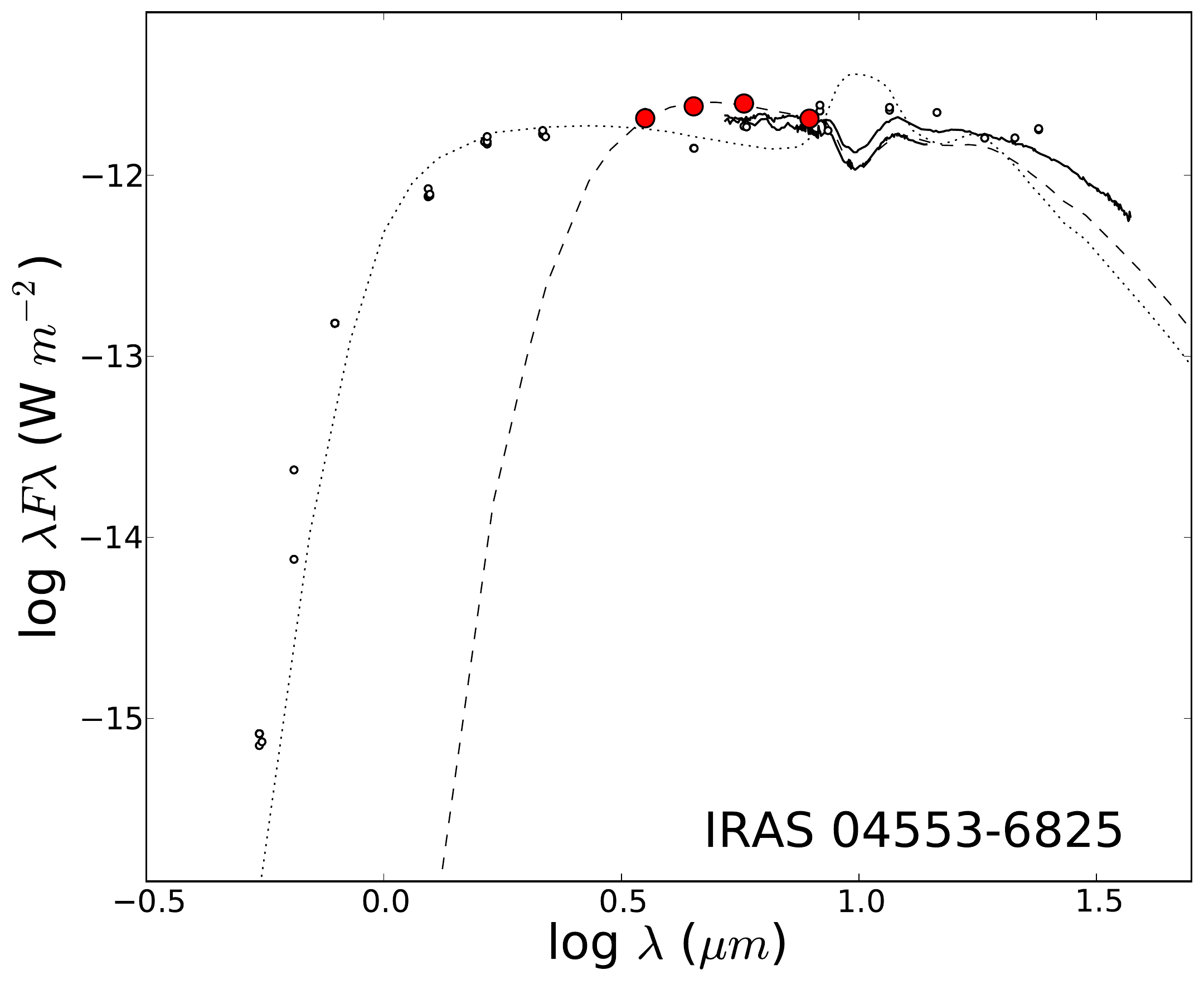}\\
 \includegraphics[width=6.8cm]{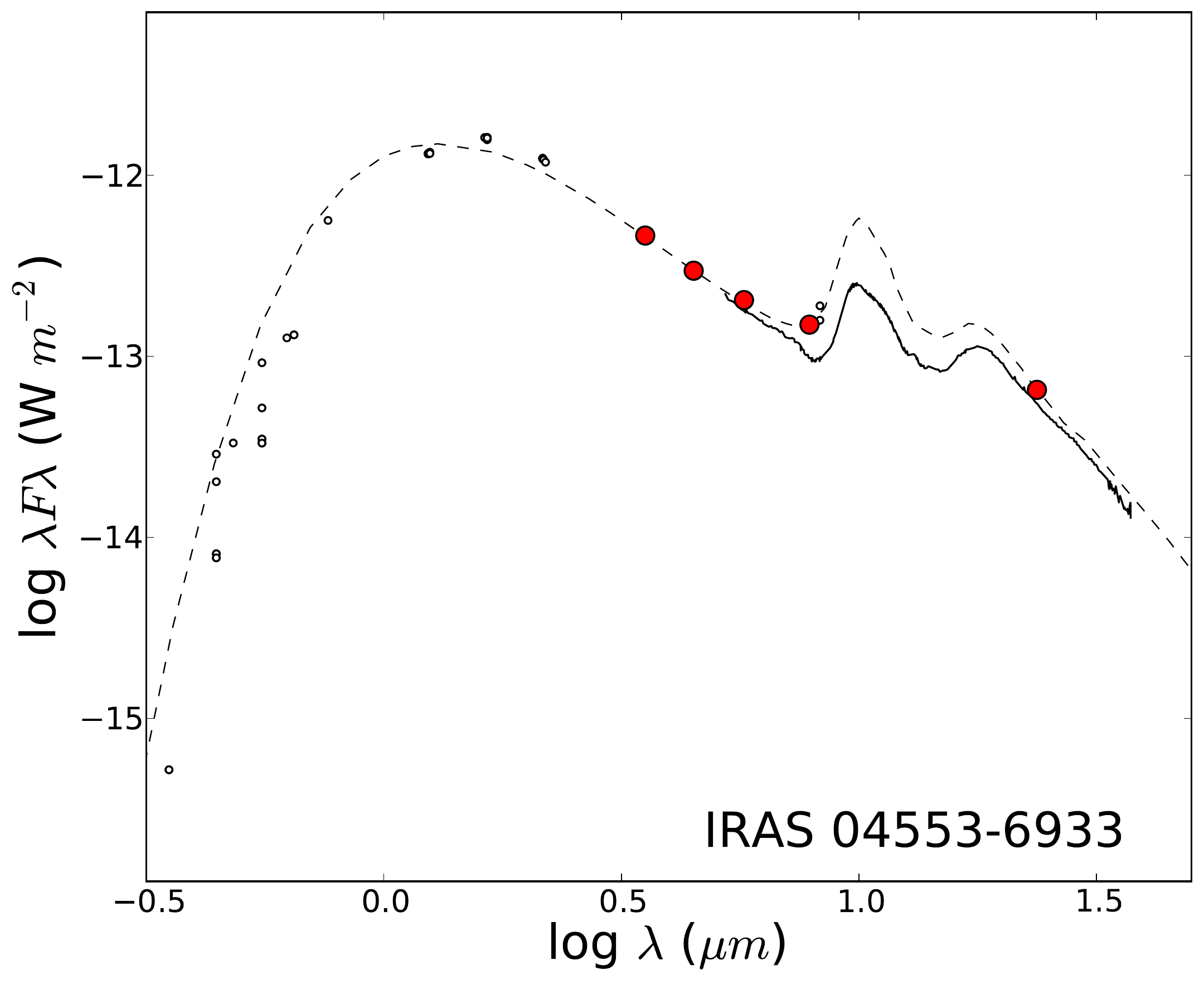}
 \includegraphics[width=6.8cm]{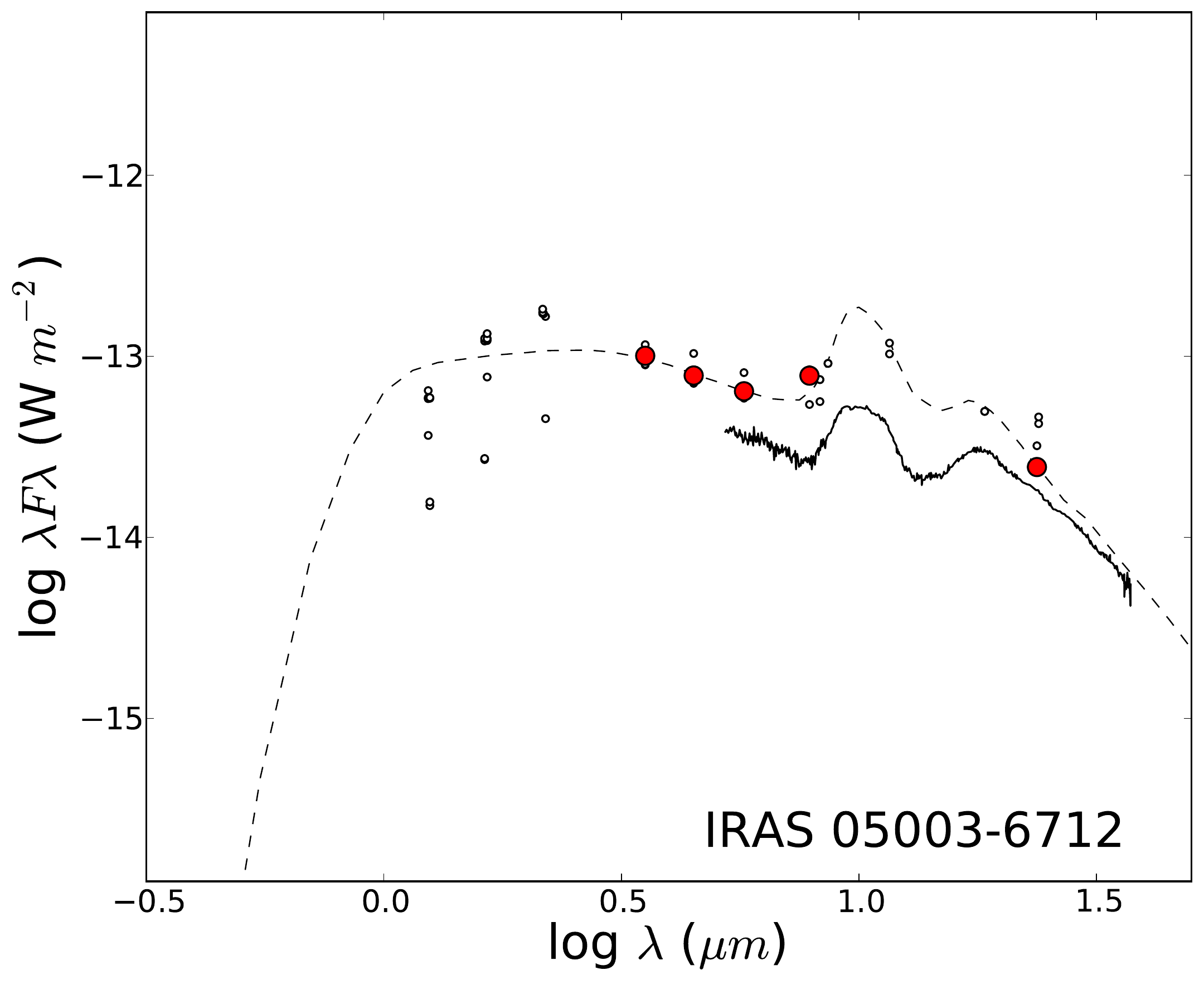}\\
 \caption{The SED fitting of \textsc{dusty} models to our LMC sources, with our best fit model (dashed line), \textit{Spitzer} IRS spectra (in solid black), \textit{Spitzer} IRAC and MIPS photometry (in red), and remaining available photometry (small open circles).}
 \label{fig:lmc_sed}
\end{minipage}
\end{figure*}

\renewcommand{\thefigure}{\arabic{figure}}
\addtocounter{figure}{-1}

\begin{figure*}
\centering
\begin{minipage}[c c]{\textwidth}
\centering
 \includegraphics[width=6.8cm]{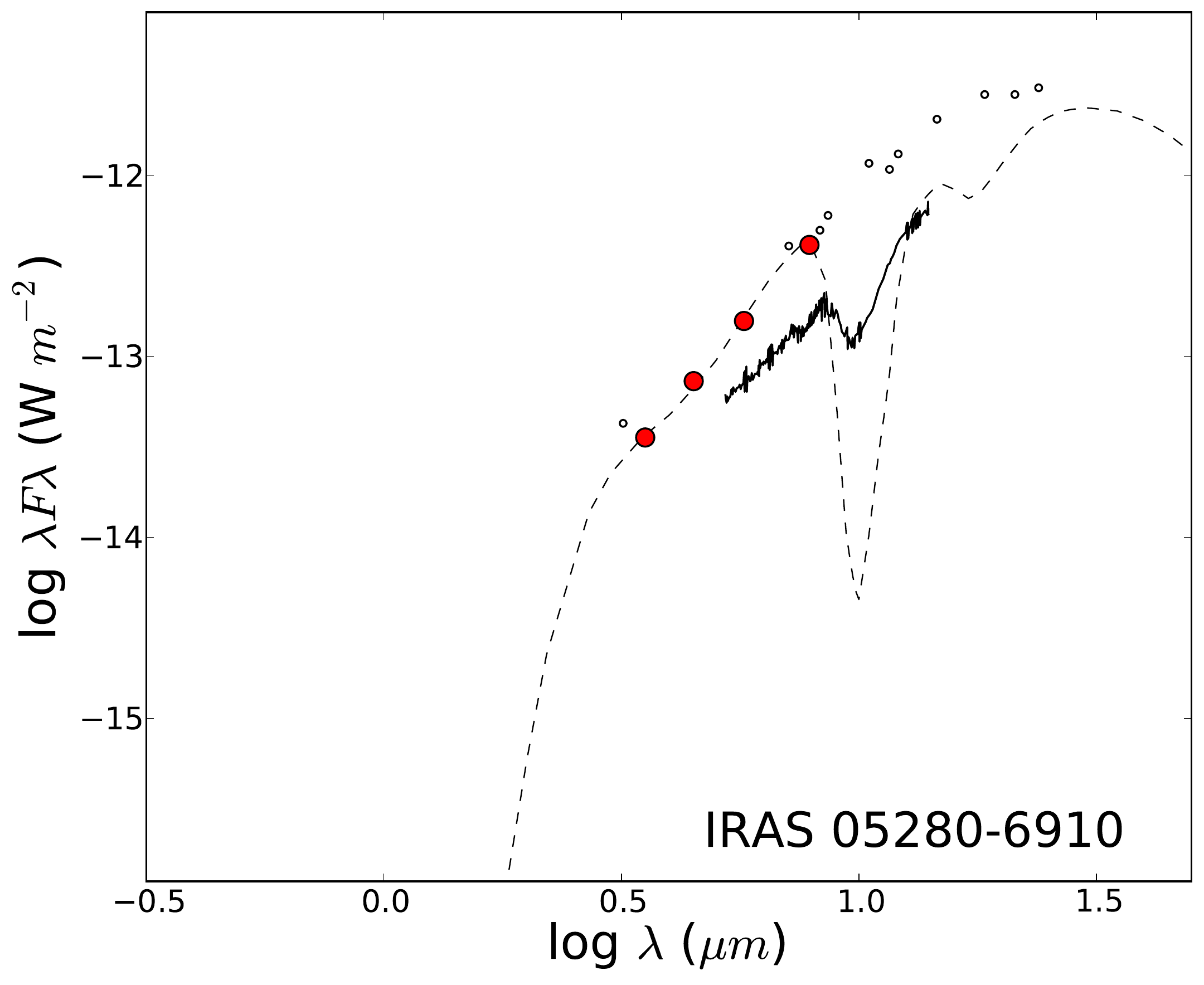}
 \includegraphics[width=6.8cm]{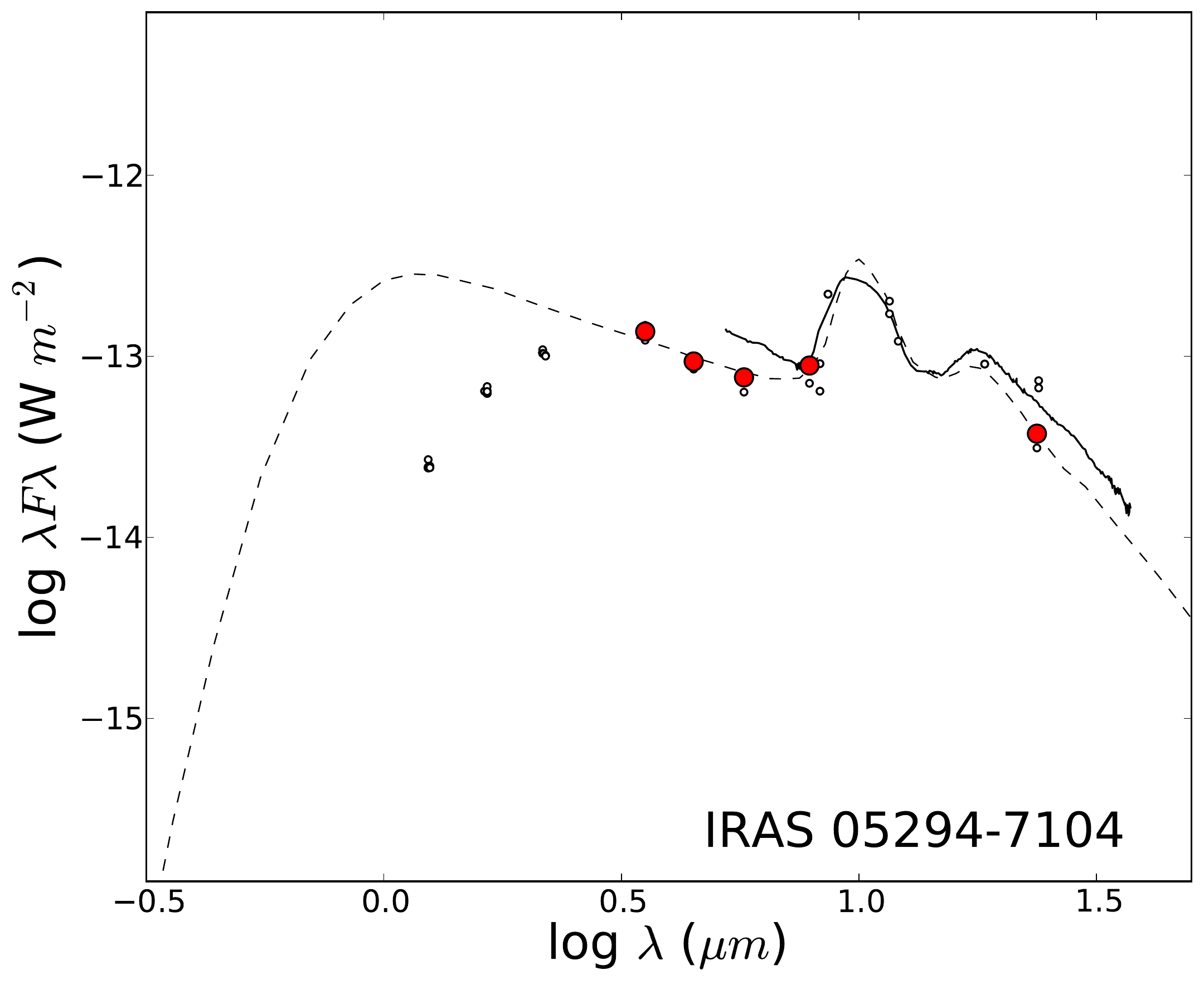}\\
 \includegraphics[width=6.8cm]{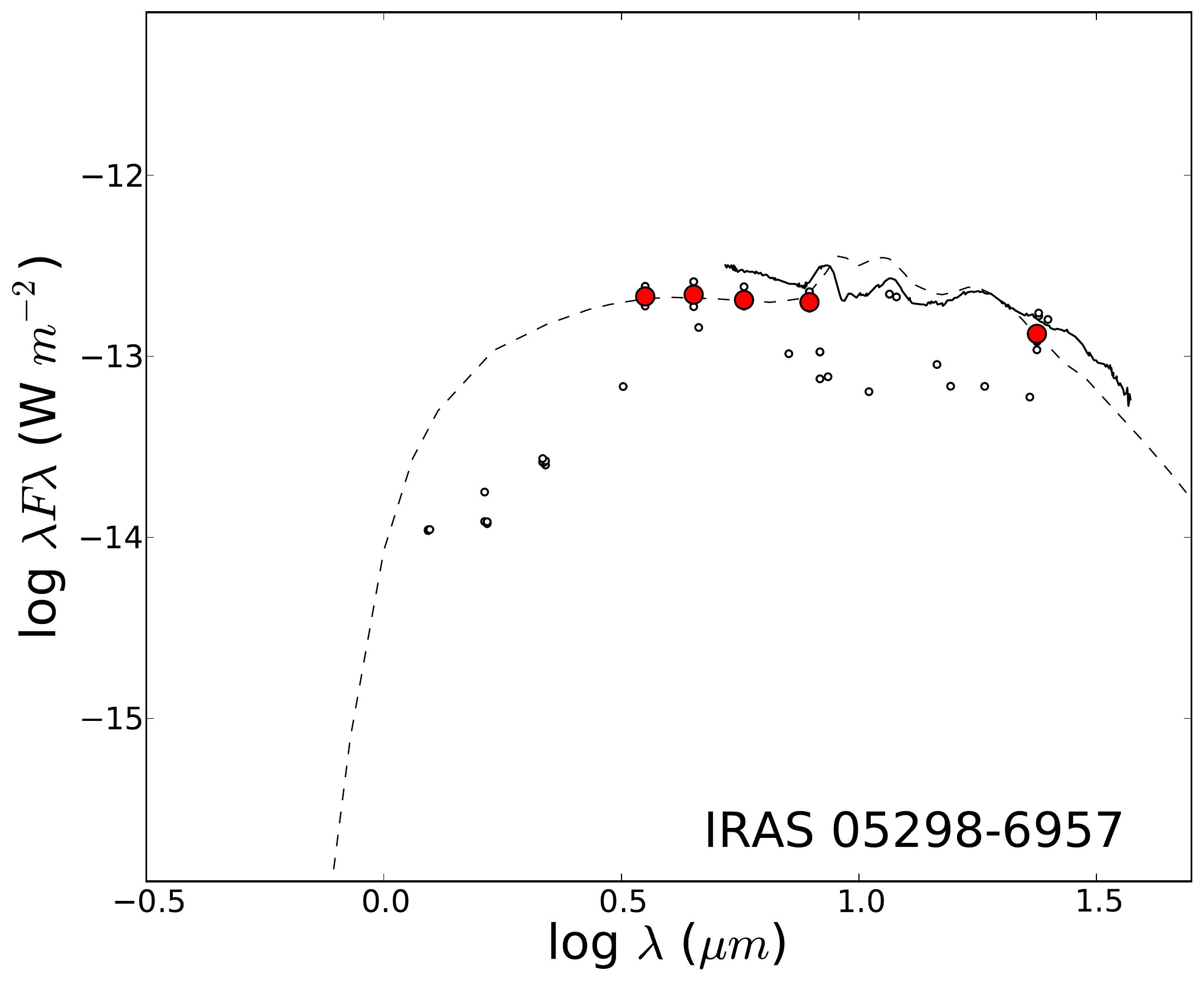} 
 \includegraphics[width=6.8cm]{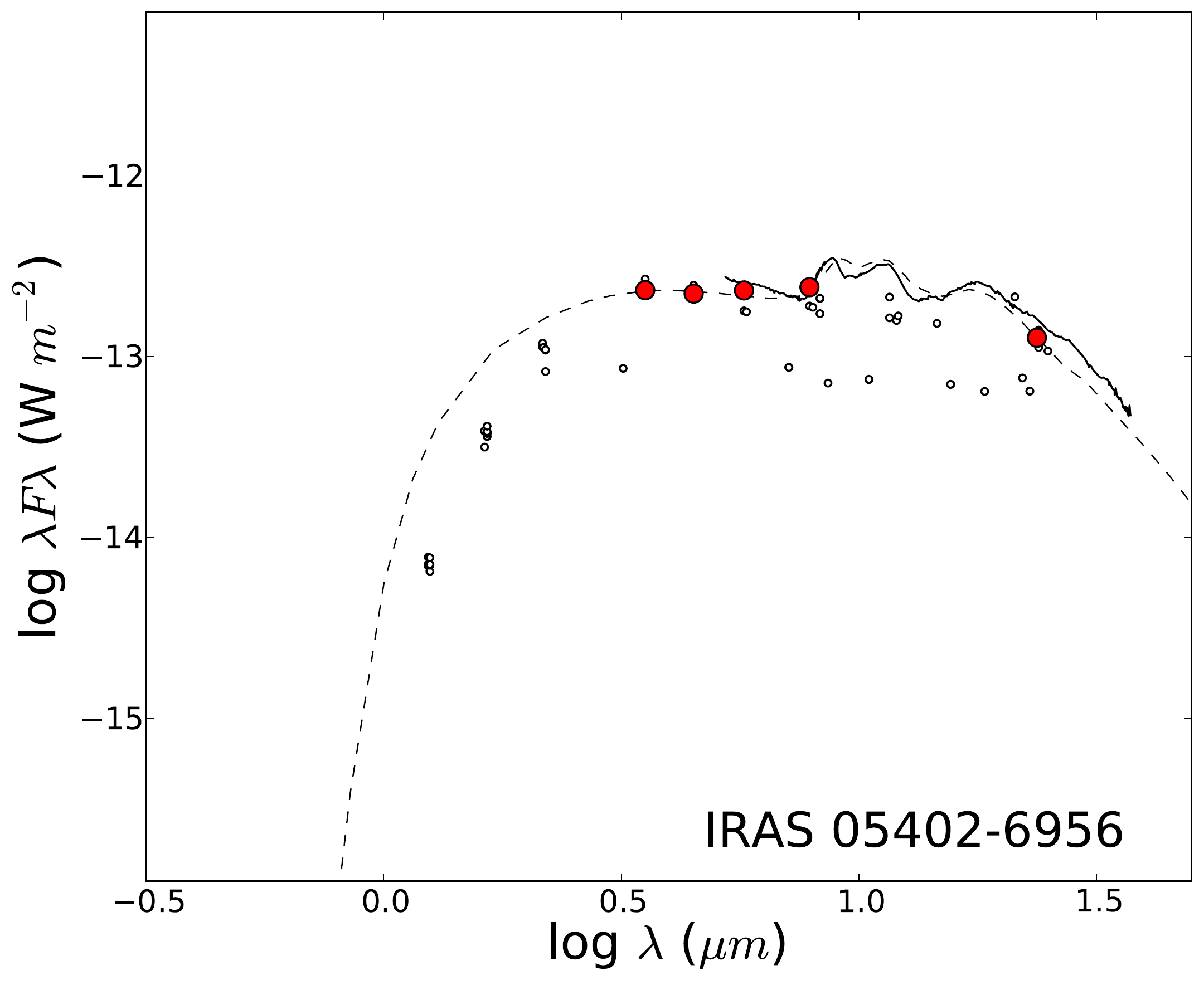}\\
 \includegraphics[width=6.8cm]{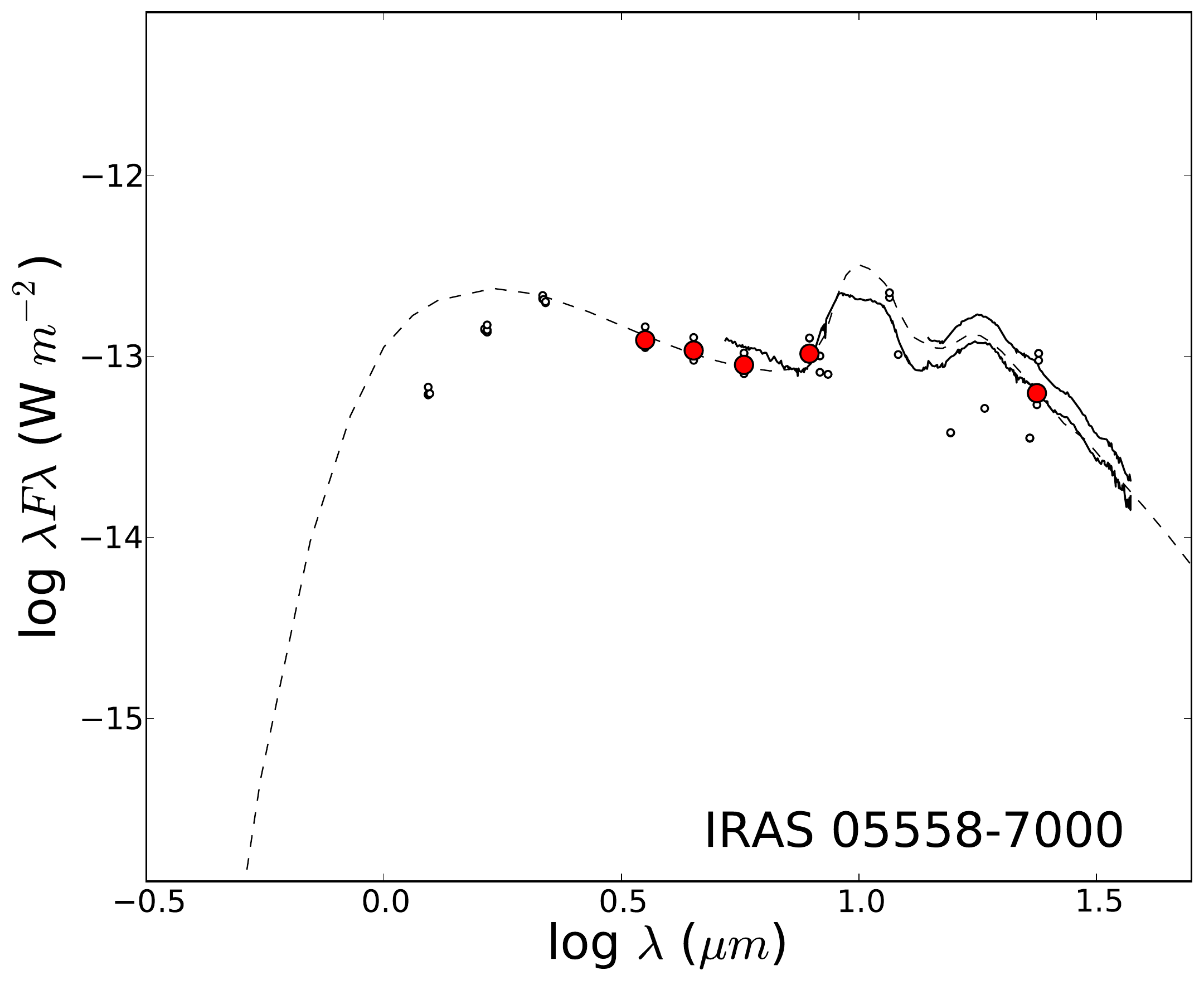}
 \includegraphics[width=6.8cm]{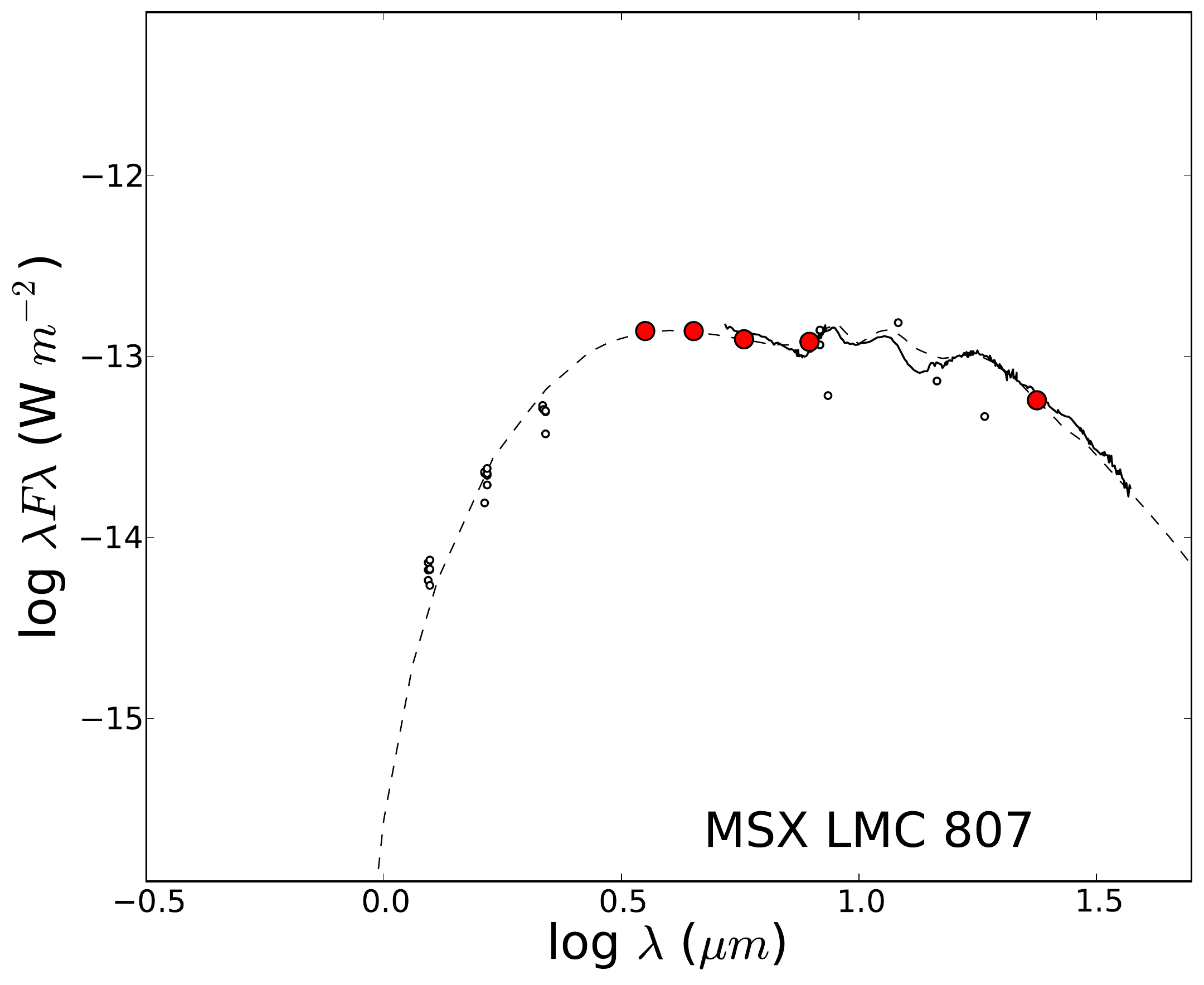}\\
 \includegraphics[width=6.8cm]{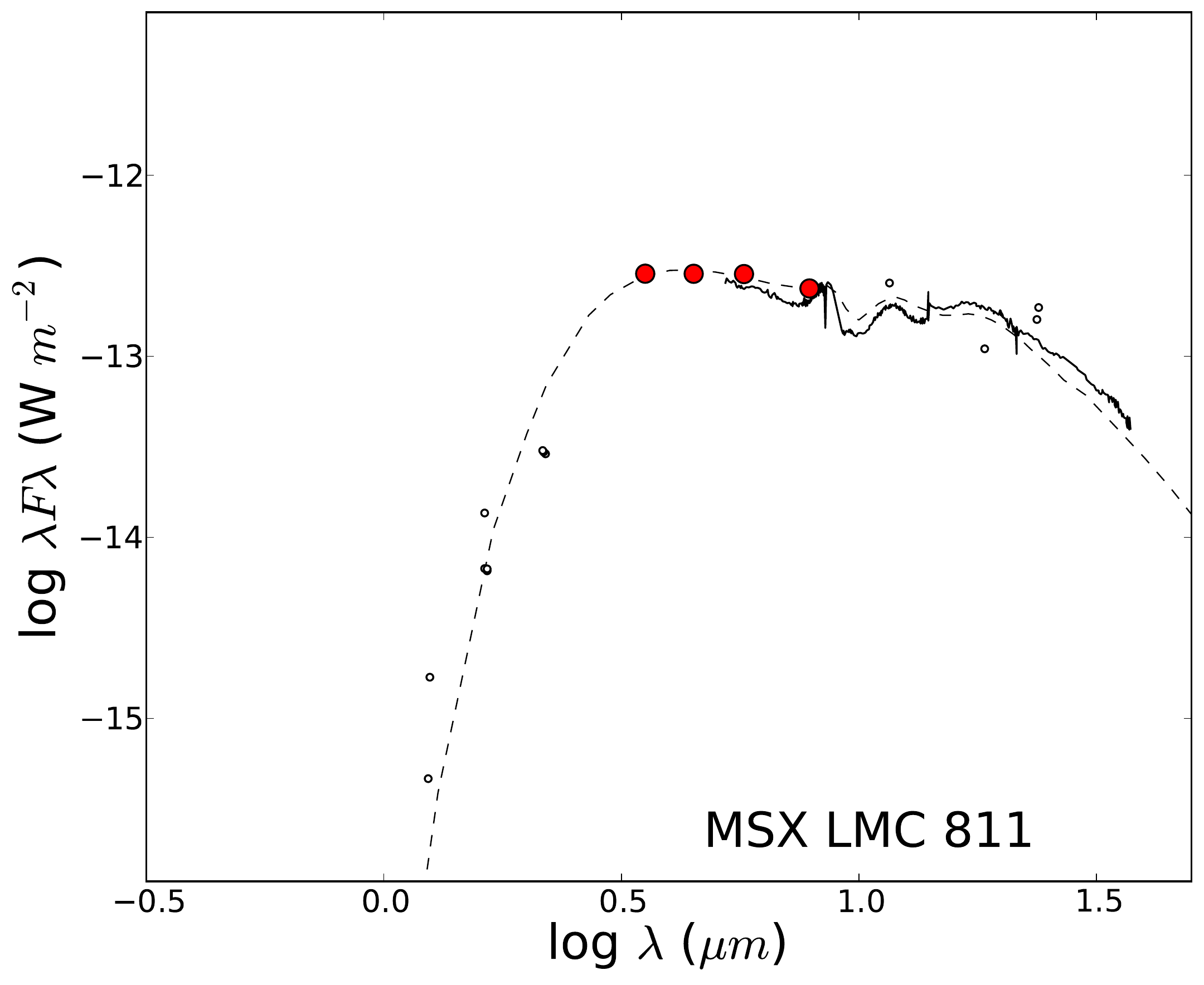}
 \includegraphics[width=6.8cm]{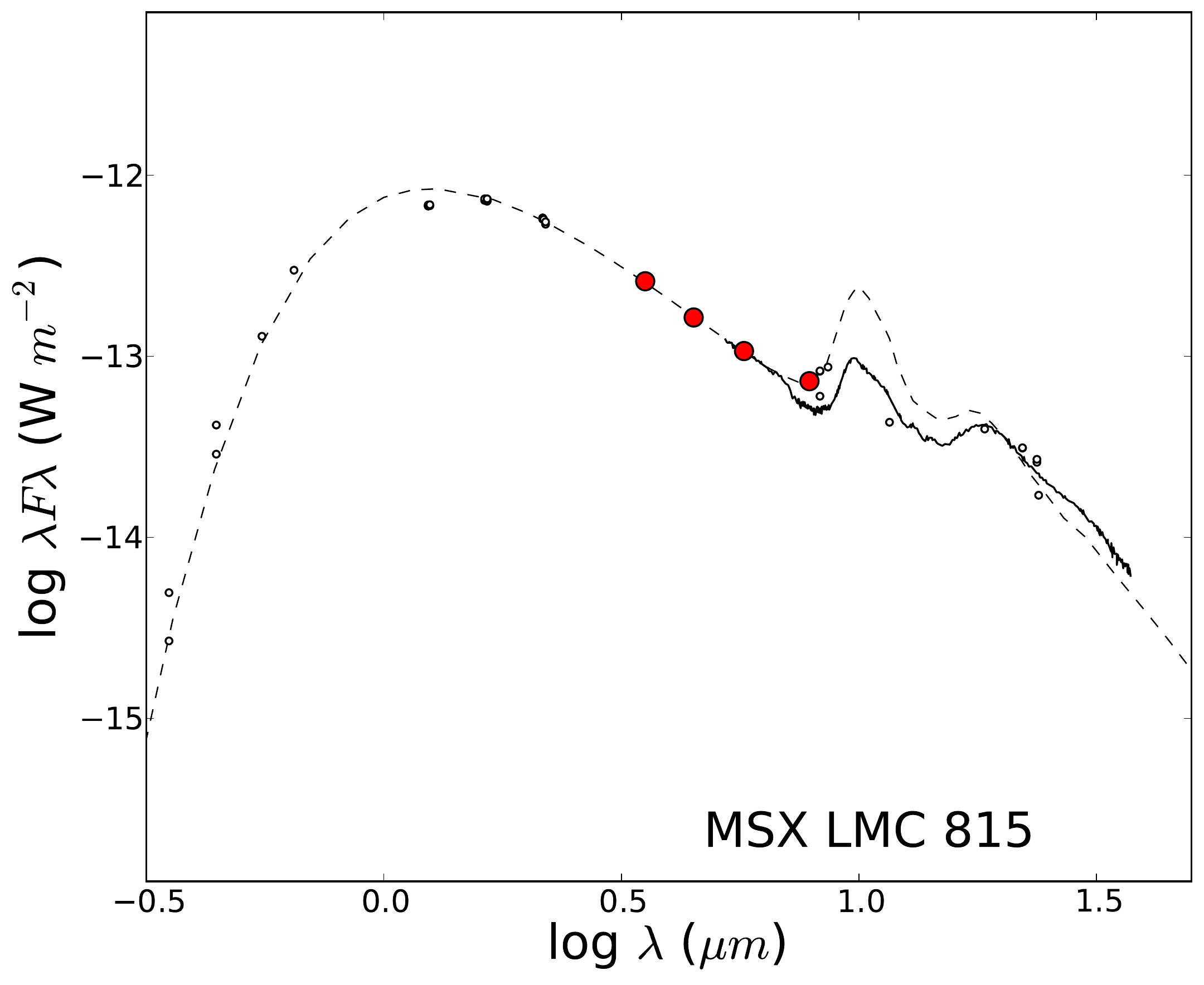}\\
 \caption{continued}
\end{minipage}
\end{figure*}
\renewcommand{\thefigure}{\arabic{figure}}

We have fit the photometry with models of effective temperatures ranging from 2700--3700 K in steps of 100 K, inner dust temperatures ranging from 600--1400 K in steps of 200 K, each with 100 models spanning a logarithmic range of optical depths from 0.1--50. A comprehensive analysis of the inter-dependency of these inputs has been presented by Ramstedt et al. (2008). Given all the above caveats and assumptions (and departures from spherical symmetry, see below) it becomes clear that a precise error analysis is impractical. The formal errors on the fitting of trends we observe between different parameters therefore include both measurement uncertainties and fitting inaccuracies as well as variations in the parameters that we assumed to be fixed.\\

Within our sample we see varying degrees of silicate in  absorption and emission and strong attenuation from dust at shorter wavelengths. We see two sources (IRAS 04553$-$6825 and IRAS 05280$-$6910) where the models struggle to fit the general SED. This is likely due to a non-spherical symmetry of the dust envelope in these sources. These unique sources will be further discussed in Section 5. There are two further sources (IRAS 04553$-$6933 and MSX LMC 815) where the model struggles to fit both the optical and mid-IR data. Both of these sources show no OH maser emission, and extremely high modeled expansion velocities. These sources may not be optically thick enough for the IR radiation to pump the maser or these may be binary systems.

\subsubsection{Galactic Centre and Galactic Bulge samples} 
In addition to our LMC sample, we have also modeled sources from the Lindqvist et al. (1992) Galactic Centre sample of OH/IR stars, and the Jim\'{e}nez-Esteban \& Engels (2015) Galactic Bulge sample of more extreme highly reddened OH/IR stars, to serve as comparison samples at higher metallicity; we have used pulsation periods from Wood et al. (1998) and those we derive (Appendix D) for the Galactic Centre and Bulge respectively. Assuming a distance of 8 kpc to these regions, and using a similar modeling method, \textsc{dusty} models, and sources of photometry, we have successfully modeled 70 Galactic Centre and 21 Galactic Bulge OH/IR stars. The distribution of the resulting luminosities is shown along with the values from the LMC in Figure \ref{fig:lum_hist}. We can see the void in our histogram where at luminosities between those of our lower luminosity Galactic sources and higher luminosity LMC sources, sources are dominated by carbon-rich chemistry (C stars).

As the photometry are more sensitive to extinction, we have fit the SED using all available photometry, expecting the best fit model to fit the median SED. Distances are also less certain for our Galactic sources and we expect this to contribute to the uncertainty in these luminosities. To correct for interstellar extinction, we have used the Schultheis et al. (2014) extinction maps from 8 to 8.5 kpc and applied an extinction curve. Using the closest value (typically within several arc minutes) for the (H$-$K) colour excess we have applied the extinction law from Rieke \& Lebofsky (1985), using interpolated values to fit the wavelength bands associated with our photometry. This was done for all Galactic Bulge and Galactic Centre sources except for IRAS 17030$-$3053 which was not covered by the extinction maps. As a result, we assign it a value of $A_K$ = 0.15 mag, adopted from Jim\'{e}nez-Esteban \& Engels (2015). There has been increasing evidence that the reddening law varies within the Galactic Centre (Nataf et al. 2016; Xue et al. 2016), yet our results are not dramatically affected by this result. This reddening variation is caused by populations of small dust grains and mainly affects shorter wavelength photometry. The results of the \textsc{dusty} modeling of the SEDs for the Galactic Centre sources are shown in Appendix A1, while the modeling results for those in the Galactic Bulge are shown in Appendix A2.

As the Galactic Centre and some of the Galactic Bulge sources lie in crowded fields, we expect a small percentage of photometry points towards shorter wavelengths to come from other sources. We see that our Galactic Centre sources show a variety of silicate in both emission and absorption. This is a stark contrast to the Galactic Bulge sources which all show silicate strongly in absorption, an expectation of this more extreme sample by design. These Bulge sources may also be members of the Galactic Disk, in the foreground of the bulge. Catchpole et al. (2016) have shown that a number of sources assumed to be members of the bulge have had a wide range of distances. This may introduce a large source of uncertainty in our derived luminosities.

\section{Discussion}

\subsection{LMC target population}

Our observing programme targets sources with known but unclear OH maser profiles, as well as sources that are deemed good candidates. In general we target highly luminous and reddened AGB and RSG sources with long pulsation periods and large pulsation amplitude. Within our current sample, IRAS 04407$-$7000 and IRAS 05003$-$6712 only exhibit one maser peak. As these sources and their maser emission are highly variable (Harvey et al. 1974), subsequent observations are much more likely to give us a double-horned maser profile and expansion velocity. 

\begin{figure*}
 \begin{minipage}[c]{\textwidth}
 \centering
 \includegraphics[width=\textwidth]{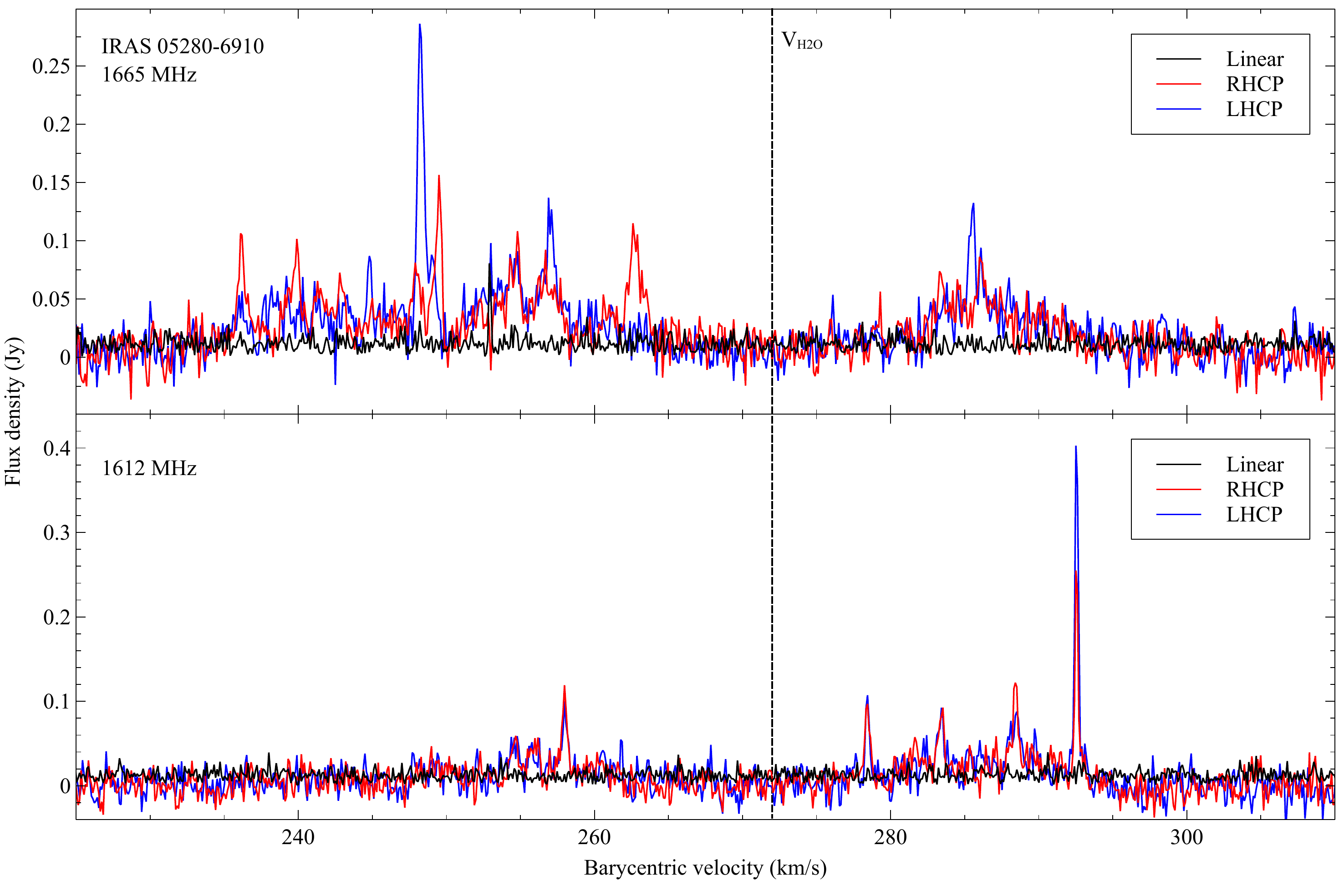}
 \caption{The left and right hand circular polarisation of the 1665 MHz (\textit{Top}) and 1612 MHz (\textit{Bottom}) maser detections of IRAS 05280$-$6910; the systemic velocity from the water maser emission from van Loon (2001) is indicated with V\textsubscript{H2O}.  }
 \label{fig:1665_pol}
 \end{minipage}
\end{figure*}

We expect the LMC to harbour several thousand OH masing sources (Dickey et al. 2013). Our much smaller observed sample likely indicates a sensitivity limit. OH masers are predicted to be excited by line overlapping of far-infrared OH lines (e.g. Elitzur, Goldreich \& Scoville 1976; Bujarrabal et al. 1980). To excite OH masers, these far-infrared OH lines are required to be optically thick. Therefore, only the most luminous, high mass-loss rate AGB stars and RSGs can produce OH masers detectable with our current instruments; the ability to sustain these masers will also depend on metallicity (OH abundance and dust emission). The LMC AGB stars are also dominated by carbon-rich chemistry, decreasing the population of OH masing sources. OH/IR stars of low enough mass not to become carbon stars will also be less luminous, thus restricting the expected detectable sample to the much rarer massive AGB stars ($\gtrsim$ 4 M\textsubscript{$\odot$}). In order to further understand the mass loss mechanism and the effects of varying metallicity and luminosity, we will need to continue to discover more sources in the LMC, and extract what we can from our current sample. The fact that we detected additional sources serendipitously means more OH masers must be awaiting detection at the current sensitivity limits.\newline

\subsubsection{IRAS 04553$-$6825 (WOH G64)}
IRAS 04553$-$6825 is known to be highly non-spherical (Roche, Aitken, \& Smith, 1993; Ohnaka et al. 2008) originally suggested on the basis of optical brightness and strong silicate feature. This was later suggested by van Loon et al. (1999) on the basis of its maser characteristics. The maser source has been observed in OH showing four OH maser peaks. The non-spherical nature of this source makes it quite hard to model. While we were able to successfully model and fit the 10 $\mu$m silicate feature in absorption, something we see in the \textit{Spitzer} and ISO spectra (cf. Trams et al. 1999; van Loon et al. 1999), we are not able to simultaneously fit the general SED. We have therefore overploted a secondary model (plotted with a dotted line) where we do fit the general SED (Fig. \ref{fig:lmc_sed}). The two models differ slightly in both luminosity and mass loss, with a luminosity between 340,000$-$454,000 L\textsubscript{$\odot$} and a mass loss rate between 3.1$-$5.8 x 10$^{-4}$ M\textsubscript{$\odot$} yr$^{-1}$. We expect that our models represent the range of possible luminosities and mass loss rates of this unique source. Ohnaka et al. (2008) suggested that the source is likely to a have a dusty torus that we are observing pole-on. This explanation is consistent with our results.  

\subsubsection{Mass loss of IRAS 05280$-$6910}

The maser emission from IRAS 05280$-$6910 is unique. First, the source shows maser emission at 1612 and 1665 MHz but not at 1667 MHz. Second, the maser profiles are very irregular with a number of peaks that have changed in amplitude and velocity since 1992. And third, the 1665 MHz maser profiles, which should probe a region closer to the star (Richards et al. 1998), extend (at the blue end) to velocities outside of the 1612 MHz profiles. All of these peculiarities come to heel as we look more closely at the polarisation data (Figures \ref{fig:1665_pol}). We see that this source is strongly circularly polarised at 1665 MHz and at 1612 MHz, alluding to magnetic activity close to the star. We suspect Zeeman splitting at two locations in the 1665 MHz spectrum centred at 248.9 and 285.8 km s$^{-1}$. The separation of these bright maser peaks suggests an upper limit of the magnetic fields of around 1.2 mG and 2.2 mG respectively. The peaks at 257 and 262.7 km s$^{-1}$ are unlikely to be a Zeeman pair as the resulting magnetic field would be around 10 mG and would result in a $\sim$1 km s$^{-1}$ split in the 1612 MHz spectra at this velocity, which we do not see. We have used Stokes I + V/2 for RHCP and I $-$ V/2 for LHCP (IAU convention), where RHCP is the rotation of the electric field vector in a counter-clockwise direction as the wave travels from the source to the observer along the line of sight. 

\begin{figure*}
\centering
\begin{minipage}[c c]{\textwidth}
\centering
 \includegraphics[width=8.2cm]{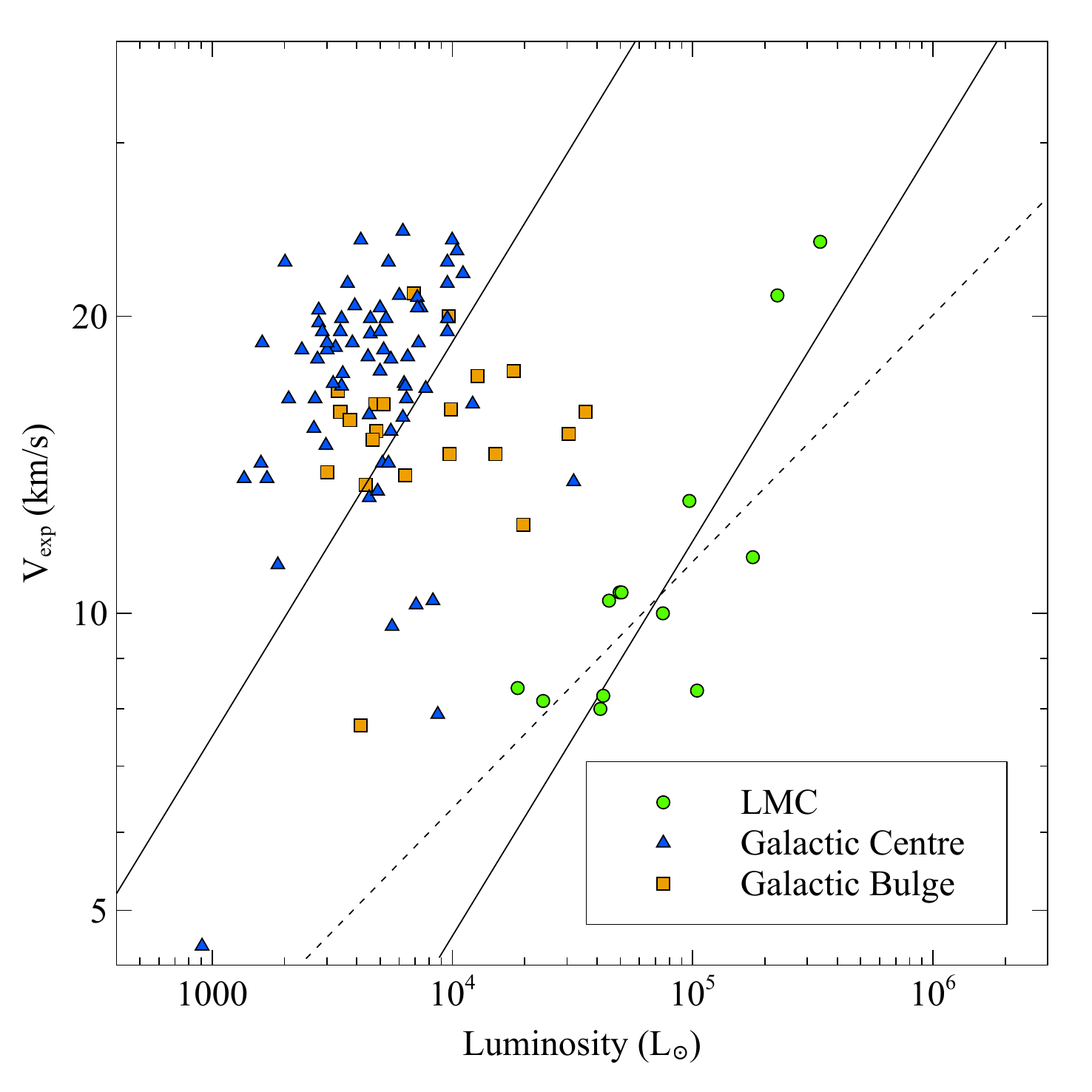} \hspace{0.5cm}
 \includegraphics[width=8.2cm]{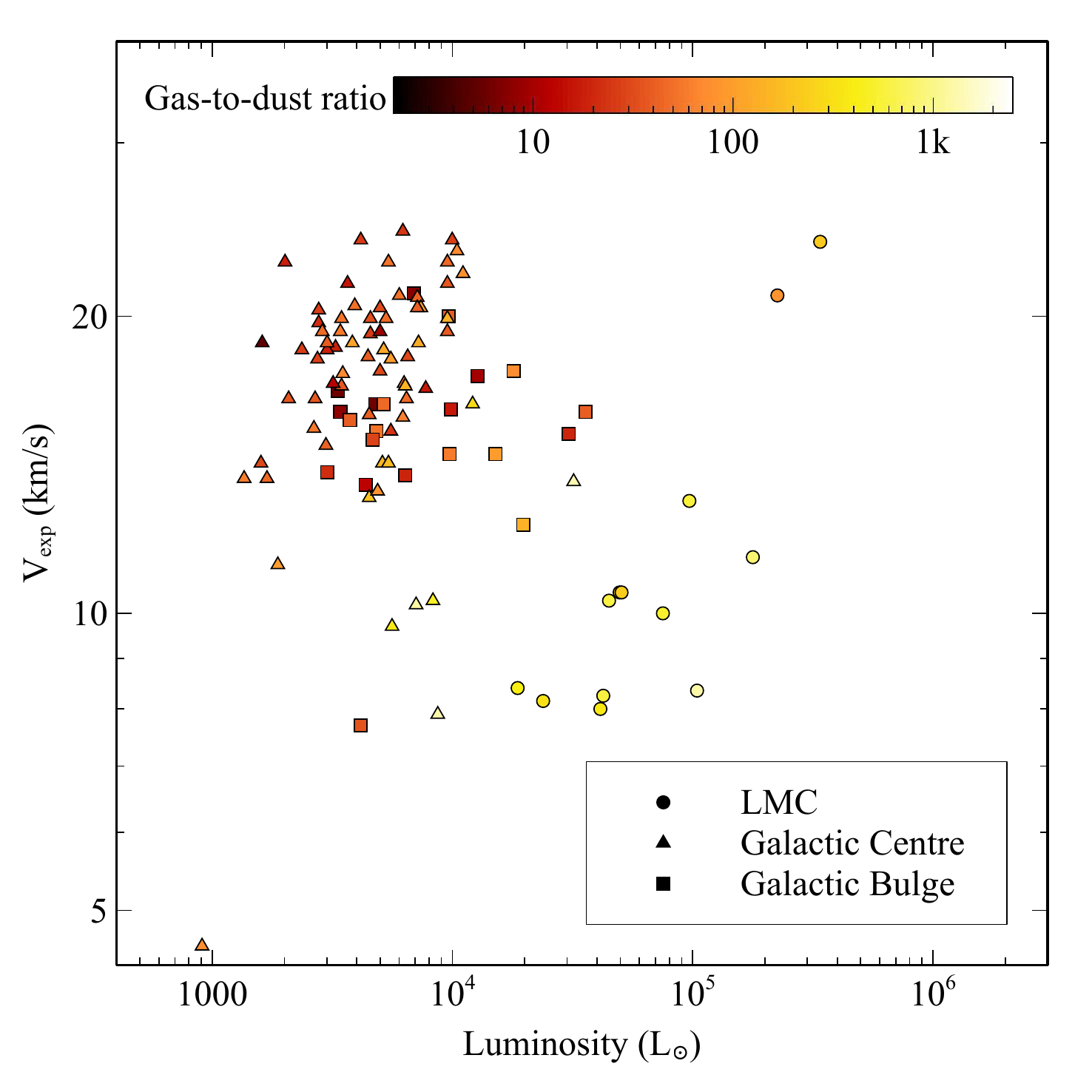}
 \caption{\textit{Left}: The observed wind speed as a function of luminosity for our sample of OH/IR stars in the LMC and Galactic samples. The LMC sample is fit with a power law ($v_{exp} = 0.118 \cdot L^{0.4}$). The power law (solid black) is then scaled to a metallicity characteristic of the Galactic sample (a factor of 2). We have derived a new relation: $v_{exp} \propto Z L^{0.4}$; that fits well with our samples. We compare our relation to that of dust-driven wind theory $v_{exp} \propto r_{gd}^{-1/2}L^{0.25}$ (dotted line)  \textit{Right}: The observed wind speed as a function of luminosity with the gas-to-dust ratio represented in colour, as derived from the scaling of SED models.}
 \label{fig:v_vs_l}
\end{minipage}
\end{figure*}

The magnetic activity coupled with the large velocity range of the maser profile may be suggestive of bipolar outflows, collimated by the magnetic fields. The bluest 1665 MHz maser peaks may either be more highly accelerated, localised ``bullets'' as opposed to the more uniform wind further out that is probed by the 1612 MHz masers, or they may be associated with interstellar masers as they peak at the H\,{\sc i} velocity. The complete absence of 1667 MHz emission must be due to quenching as a result of the strong 1665 MHz masers. Sources like VY CMa and IRAS 04553$-$6825 have shown similar complex profiles with many maser peaks and asymmetric envelopes (Cohen et al. 1987), yet IRAS 05280$-$6910 also has a distinctly unique SED. Studies of VY CMa have shown a number of arcs and knots that may also be present in our source (Humphreys et al. 2005, 2007). 

The \textsc{dusty} modeling of IRAS 05280$-$6910 yields an expected expansion velocity of 13.6 km s$^{-1}$. This is considerably lower than the observed value of 20.6 km s$^{-1}$. We expect this disparity to come from something more than uncertainty or unrealistic modeling. What we may be observing is a near edge-on dusty torus, similar to IRAS 04553$-$6825, that is highly obscured along our line of sight. This result would also support the idea that the source has bipolar outflows. These outflows could explain the models' slow calculated expansion velocity (which could be expected along our line of sight), while not necessarily leading to such a low gas-to-dust ratio or high mass loss rate. 

We see an obvious difference between our \textsc{dusty} models and the IRS spectrum of IRAS 05280$-$6910. The \textit{Spitzer} IRS spectrum of this source shows silicate in absorption, but not to the same degree as our modeled SED. The observed absorption feature may be partially masked by the nearby RSG WOH G347 (van Loon, Marshall \& Zijlstra, 2005), a source that shows silicate in emission. A further investigation of this object in the optical and at high angular resolution in the mid-IR would certainly be worthwhile.

\subsection{Stellar winds at different metallicities}
We now examine the stellar wind expansion velocities of the sources in our LMC, Galactic Centre and Galactic Bulge samples. In OH/IR stars, the mass-loss process and the outflow acceleration are driven by radiation pressure on dust grains. In these dense environments, the gas is dynamically coupled with dust through friction and the outflow behaves like one single fluid. As these stars produce mainly silicate dust, and the production of silicates is limited by initial metallicity\footnote[2]{contrary to C-stars in which the dredge-up events can enhance the amount of carbon initially available in the atmosphere.}, lower metallicity yields a slower wind. We expect that at lower luminosity there is less radiation pressure to force the dust grains outward and drive the wind, and therefore a slower expansion velocity; in general this is observed in both the LMC and Galactic samples as it was shown by Marshall et al. (2004), as well as the following work (Fig. \ref{fig:v_vs_l}).

The LMC and Galactic samples have several differences including very different luminosities and pulsation periods. The LMC sample is biased toward the most luminous sources as they are distant and obscured, and only the most massive and luminous AGB stars remain oxygen-rich in the heavy mass loss phase. The lower luminosity sources are thus dominated by carbon-rich stars. As a result, we expect a selection bias within the LMC. The LMC sample is also biased toward longer pulsation periods as the OH maser emission is brightest in the most evolved and highest mass loss sources. The Galactic Bulge sample is a collection of the most extreme OH/IR sources in the Galactic Bulge and generally lack pulsation periods. We have used data from the Vista Variables in the Via Lactea (VVV) survey (Minniti et al. 2010) to derive periods for seven of our Galactic Bulge sources (Appendix D). The Galactic Centre sources tend to be lower mass and less extreme but still within the superwind phase of mass loss. The fact that this is a older population may be contributing to the sample's lack of higher mass sources, similar to those within our LMC sample. In addition to differences within the sample, we have differences in the quality of our radio observations (Table 4). While we have significantly less angular resolution for our Galactic Bulge sample, we only use resolved and reliable maser profiles; all observations have spectral resolutions around 1 km s$^{-1}$. 

\begin{table}
\caption[]{The resolution of the radio observations of our LMC and Galactic samples. }
\begin{tabular}{ l l l r}
\hline
Sample & Telescope & Spectral & Angular \\
 & & resolution & resolution \\
 & & (km s$^{-1}$) & \\
\hline
Galactic Centre & VLA & \llap{$^1$}1.36 & $6-8^{\prime\prime}$  \\
Galactic Bulge & Parkes & \llap{$^2$}0.9 & $12^{\prime}$ \\
 & Effelsberg & \llap{$^2$}1.1 & $8^{\prime}$  \\
LMC & Parkes & 0.18 & $12^{\prime}$  \\
 & ATCA & 0.4 & $7-8^{\prime\prime}$ \\
\hline
\end{tabular}
\flushleft {References:
$^1$ Lindqvist et al. (1992)
$^2$te Lintel Hekkert et al. (1991)}
\end{table} 
\vspace{1cm}
The majority of stars in the Galactic Centre have metallicities at or above solar, with a small population of low metallicity sources (Do et al. 2015; Wood, Habing, \& McGregor 1998). The Galactic Bulge has a metallicity typically peaking at a near solar value, with a sharp cutoff just above solar, and a tail towards lower metallicity (Zoccali et al. 2003). We expect that our extreme sources likely lie towards the higher end of this metallicity range. 

\begin{figure}
 \includegraphics[width=8.2cm]{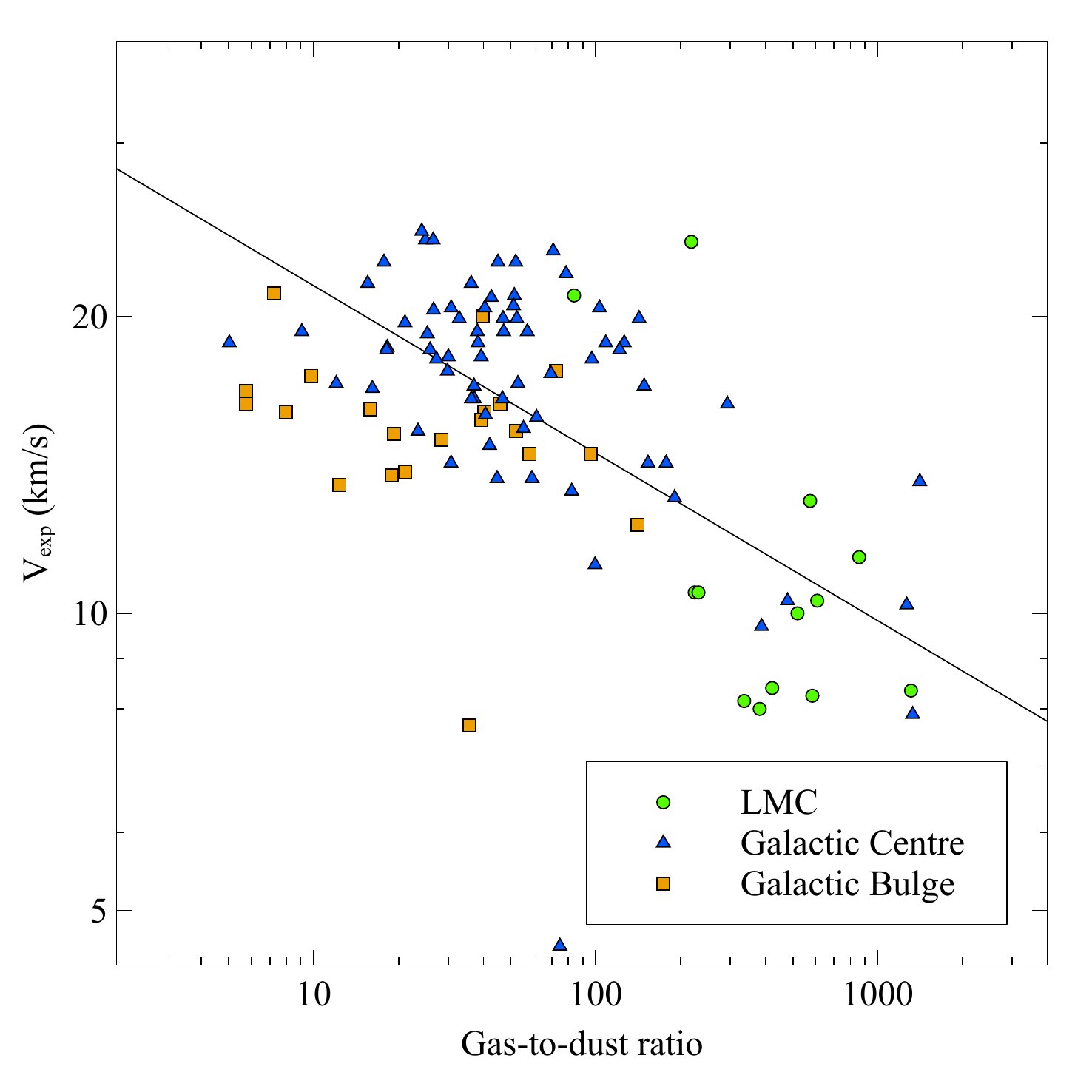}
 \caption{The observed wind speed as a function of gas-to-dust ratios that we have derived from SED fitting. The samples has been fit with a power law that is expected to fit: $v_{exp} \propto r_{gd}^{-0.5}$ (Eq. 1). The actual fit: $v_{exp} \propto r_{gd}^{-0.15}$ is much shallower due to the effects of luminosity on the wind speeds. The higher luminosities of the LMC result in higher LMC wind speeds and a shallower fit.}
 \label{fig:v_rgd}
 \vspace{-0.8cm}
\end{figure} 

\begin{figure*}
\centering
\begin{minipage}[c c]{\textwidth}
\centering
 \includegraphics[width=8.2cm]{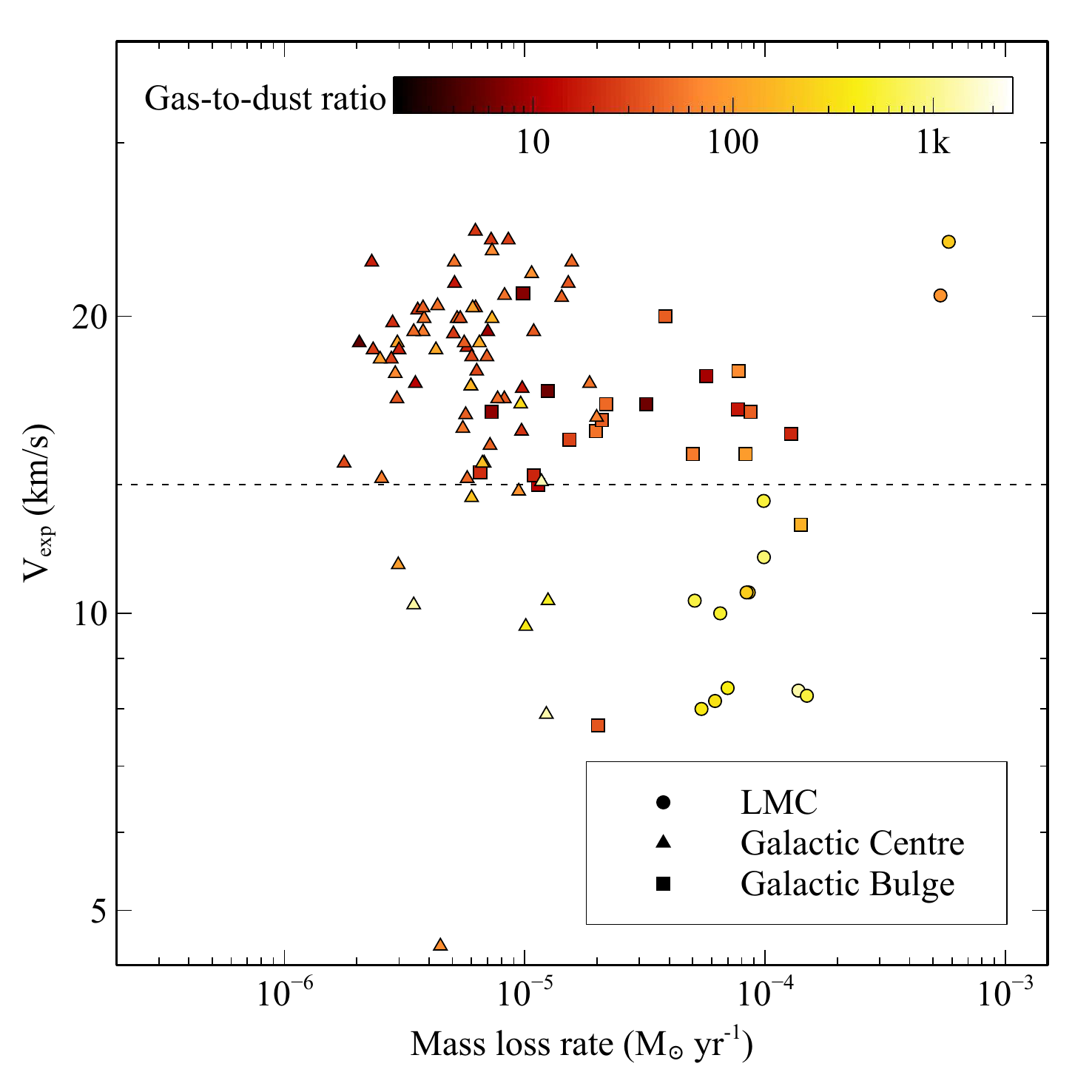} \hspace{0.5cm}
 \includegraphics[width=8.2cm]{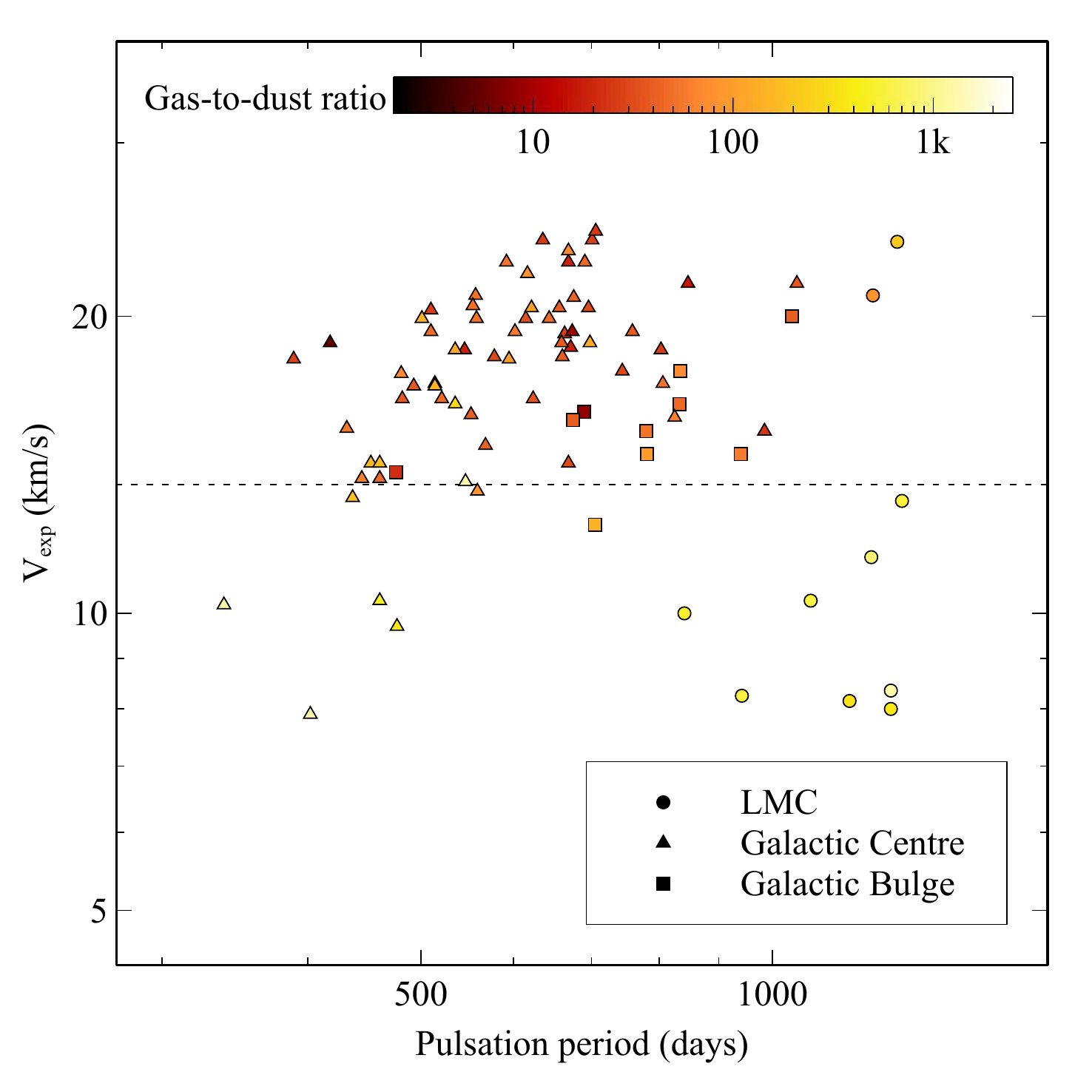}
 \caption{The observed wind speed as a function of mass loss rate (\textit{Left}) and pulsation period (\textit{Right}), with the gas-to-dust ratio represented in colour. Mass loss rates and gas-to-dust ratios are derived from the scaling of SED models (discussed further in section 5.3). The dotted lines indicates the change from the higher to lower metallicity sample, which we have shown will affect the expansion velocities. Looking at the Galactic samples, there does not seem to be any clear correlation between expansion velocity and either mass loss rate or pulsation period.}
 \label{fig:vexp_vs_others}
\end{minipage}
\end{figure*}

We see that at similar luminosity, the Galactic Bulge sources have lower expansion velocities than those of the Galactic Centre sources (Fig. \ref{fig:v_vs_l}). We see this same effect even more so in the LMC sample with a metallicity ($Z$) even lower, at half solar. This is consistent with what we expect from simple dust-driven wind theory (e.g. van Loon 2000). We have fit the expansion velocities from our LMC sample with respect to their luminosities. Our fit ($v_{exp} = 0.118 \cdot L^{0.4}$) is then scaled to a solar metallicity (a factor of 2) typical of the Galactic samples, which fits well within these samples. We have used these results to derive a new relation for expansion velocity: 
\begin{equation}
v_{exp} \propto Z L^{0.4}
\end{equation}

\noindent We have dramatically increased the number of measured wind speeds from previous studies that relied on five accurate wind speeds. Using our larger sample, we find that our updated relation: $v_{exp} = 0.118 \cdot L^{0.4}$, more accurately fits the data (Figure \ref{fig:v_vs_l}). Nevertheless, these results make several assumptions. We assume spherical symmetry within our sources, scattering of photons to affect the sources' collective radiation pressure, and uniformity of dust properties throughout our samples. We also assume a homogeneous metallicity within each of our samples which we know not to be the case. 

We can see the individual effects of different parameters on the expansion velocity more clearly in Figures \ref{fig:v_rgd} and \ref{fig:vexp_vs_others}. In Figure \ref{fig:v_rgd} we compare the expansion velocity against the gas-to-dust ratios we derive through SED fitting (described further in the following section). We see that the expansion velocity is dependent on the gas-to-dust ratio. These samples have been fit with a power law that is expected to follow: $v_{exp} \propto r_{gd}^{-0.5}$. The actual fit: $v_{exp} \propto r_{gd}^{-0.15}$ is much shallower due to the effects of luminosity on the wind speeds. The higher luminosities of the LMC result in higher LMC wind speeds and a shallower fit. When we compare the expansion velocities to mass loss rate and pulsation period (Fig. \ref{fig:vexp_vs_others}) we see no clear correlation. The scatter is mostly due to differences in luminosity and metallicity between and within the samples. This is contrary to the expectations of the DARWIN models (H\"{o}fner et al. 2016), which do predict a correlation of expansion velocity and mass loss rate. Likely, the correlation reflects the evolution along the AGB, whereas our samples are naturally biased towards the endpoints of evolution.

\subsection{Gas-to-dust ratios in the LMC}
Using our SED modeling we are able to estimate the expected expansion velocity of our sources. These estimates, calculated using the \textsc{dusty} code, rely on the assumption that the winds are dust driven and spherically symmetric. We also have measured expansion velocities from OH maser emission, and can scale our gas-to-dust ratio, assumed to be 200, by the square of the ratio in the expected and measured expansion velocity (Eq. 1). Using this method we have derived gas-to-dust ratios for all of our LMC sources shown in Table 5. 

Adding up all of the refractory elements within oxygen-rich dust and scaling to the metallicity of the LMC, we expect a gas-to-dust ratio $\sim$500. The interstellar gas-to-dust ratio ($r_{gd}$) of the LMC has also been shown by Roman-Duval et al. (2014) to be between 380 and 540. Our median $r_{gd}$ of 422 agrees well with previous estimates, but lies toward the lower end of these estimates. This may be due to our bias towards the most metal rich sources. Our higher $r_{gd}$ (lowest metallicity) objects, IRAS 04407$-$7000 and 04509$-$6922, have far lower inferred metallicities than the rest of the sample. All of these sources have quite reliable OH maser detections, with well fit SEDs. These stars are within close proximity to each other, lying within the same south east region of the LMC, towards the Magellanic Bridge. We can speculate that it may be due to either a lower efficiency of dust formation in these stars, or they may have formed from lower metallicity gas, possibly originating from the SMC (see Olsen et al. 2011). These are not likely to be members of the small subset of counter-rotating sources found by Olsen et al. (2011) as extreme oxygen-rich sources are rare in the SMC, the counter-rotating subset is a small fraction of the whole LMC population and the kinematics of the three sources are consistent with the rest of our LMC sample. 

We derive median gas-to-dust ratios of 45 and 28 for the Galactic Centre and Galactic Bulge respectively. We see little difference between the results of our Galactic Bulge fitting technique and the fitting by Jim\'{e}nez-Esteban \& Engels (2015). We derive similar luminosities and mass loss rates, yet find few cases where our best fit models have similar optical depths. It was suggested by Jim\'{e}nez-Esteban \& Engels (2015) that the existence of these oxygen-rich Galactic Bulge sources can be explained if they have metallicities at or above solar metallicity, which is consistent with our derived gas-to-dust ratios. Our Galactic sources may be alpha-enhanced which would increase the abundance of elements like Si, Ti and Mg, and thus increase dust production. Our bias towards the highest metallicity sources may be magnified by this additional alpha-enhancement.

\begin{figure*}
\centering
 \includegraphics[width=7.3cm]{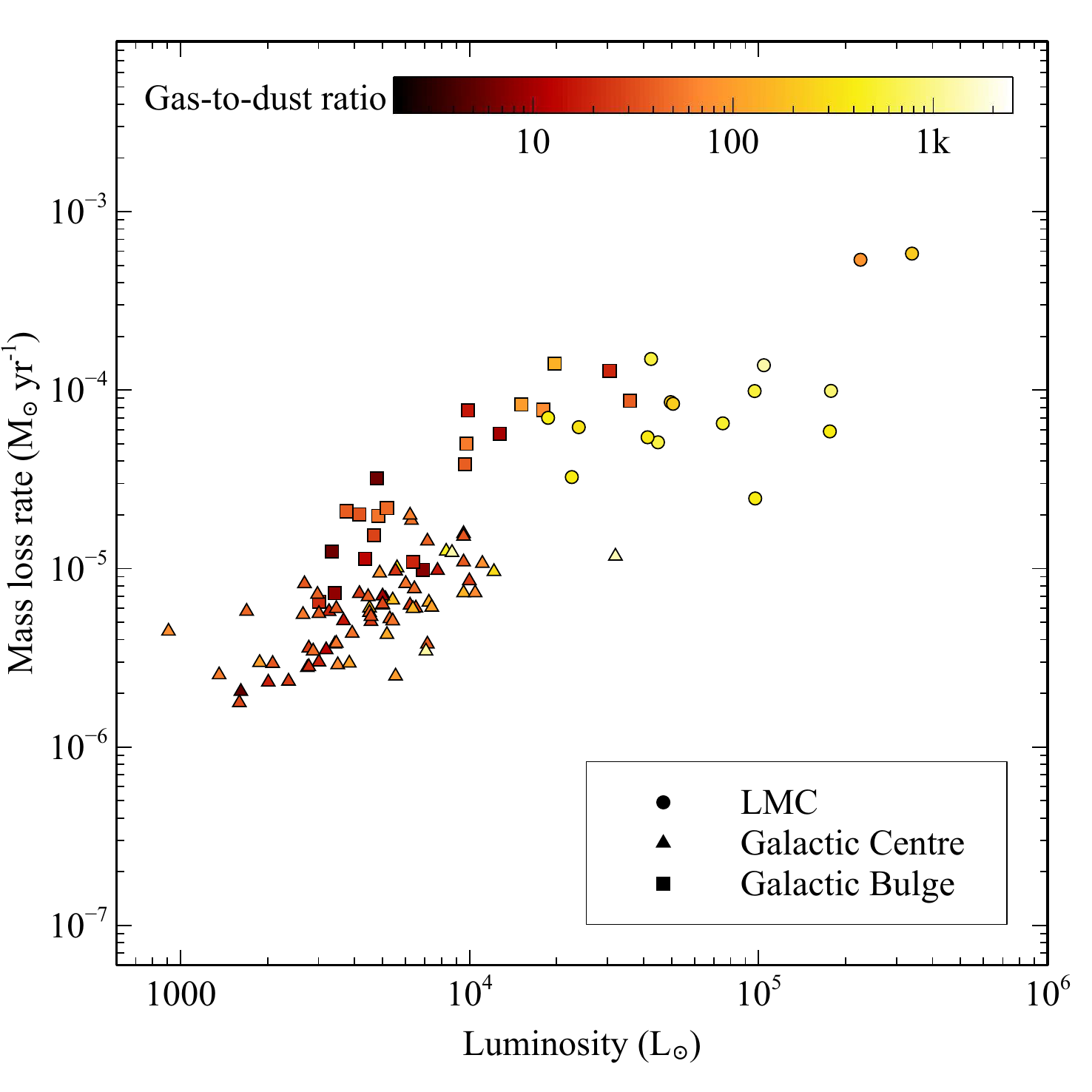} \hspace{0.5cm}
 \includegraphics[width=7.3cm]{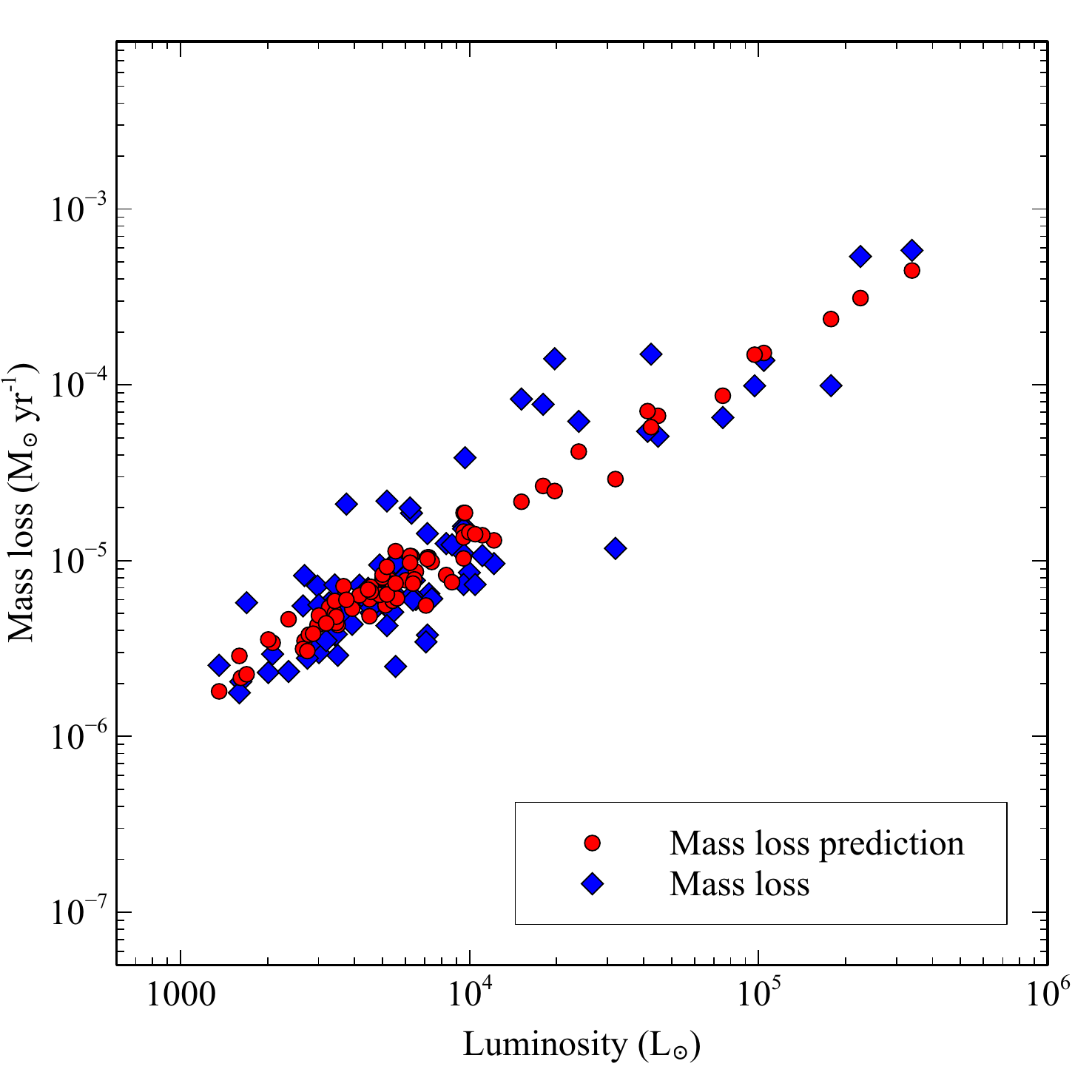}\\ \vspace{-0.2cm}
 \includegraphics[width=7.3cm]{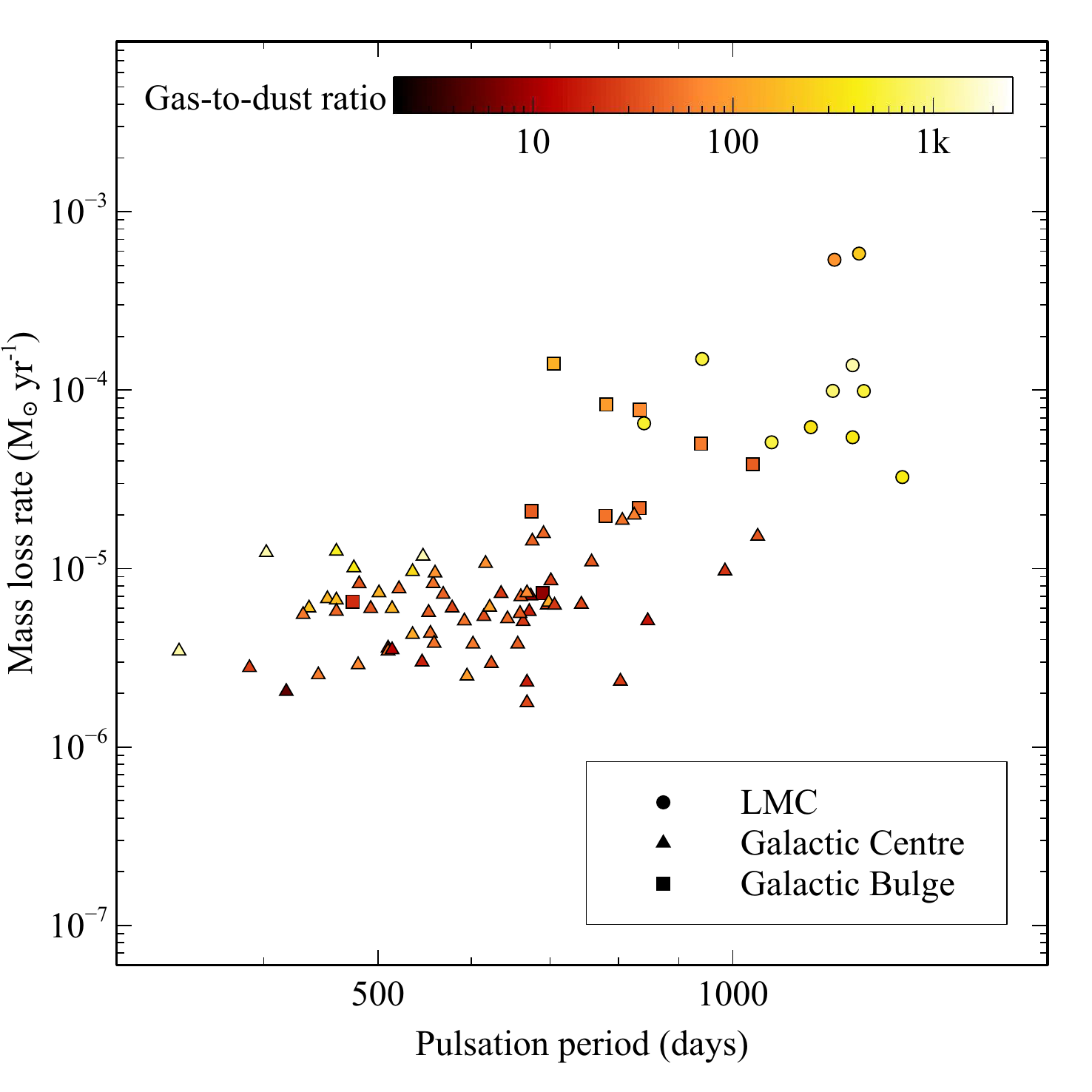} \hspace{0.5cm}
 \includegraphics[width=7.3cm]{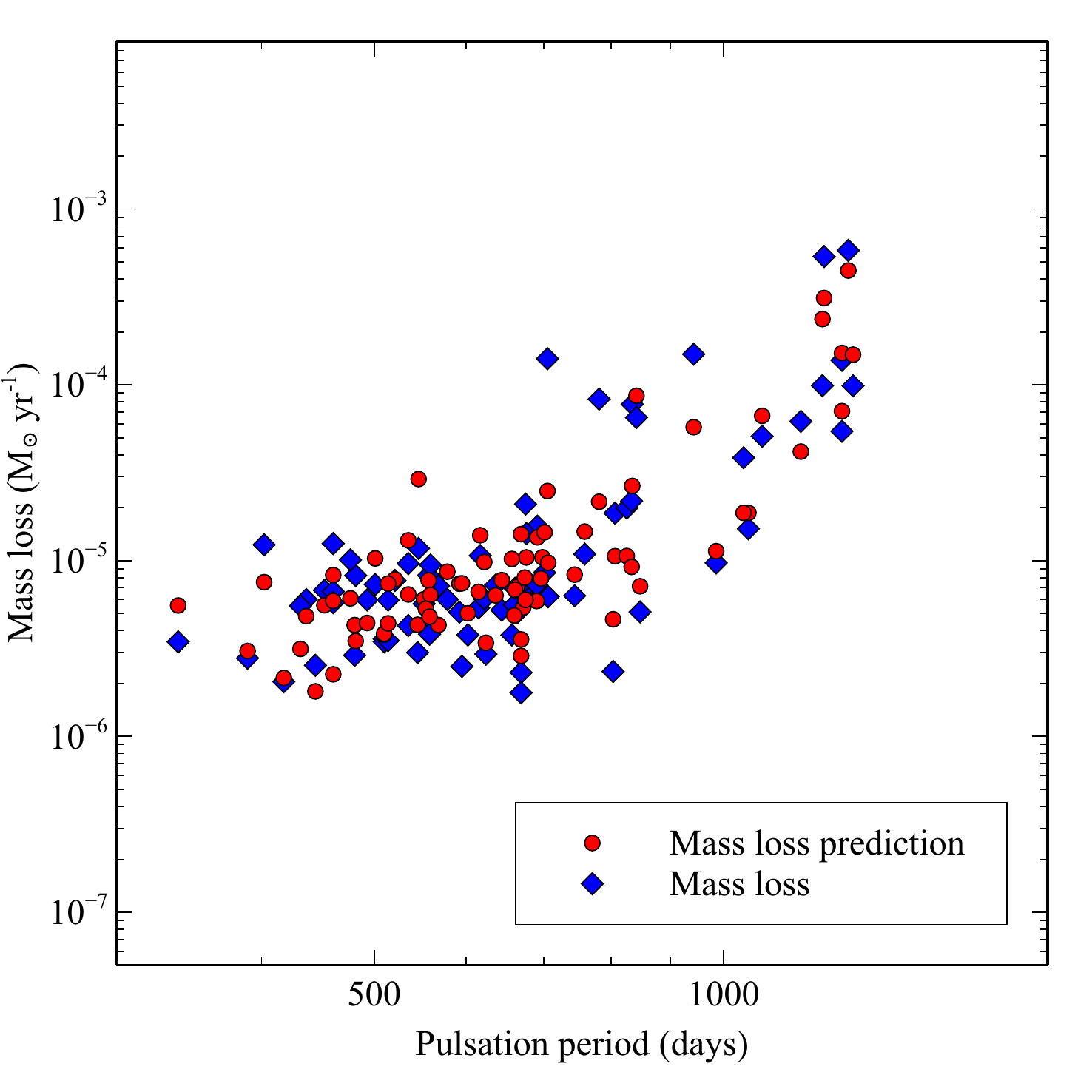}\\ \vspace{-0.2cm}
 \includegraphics[width=7.3cm]{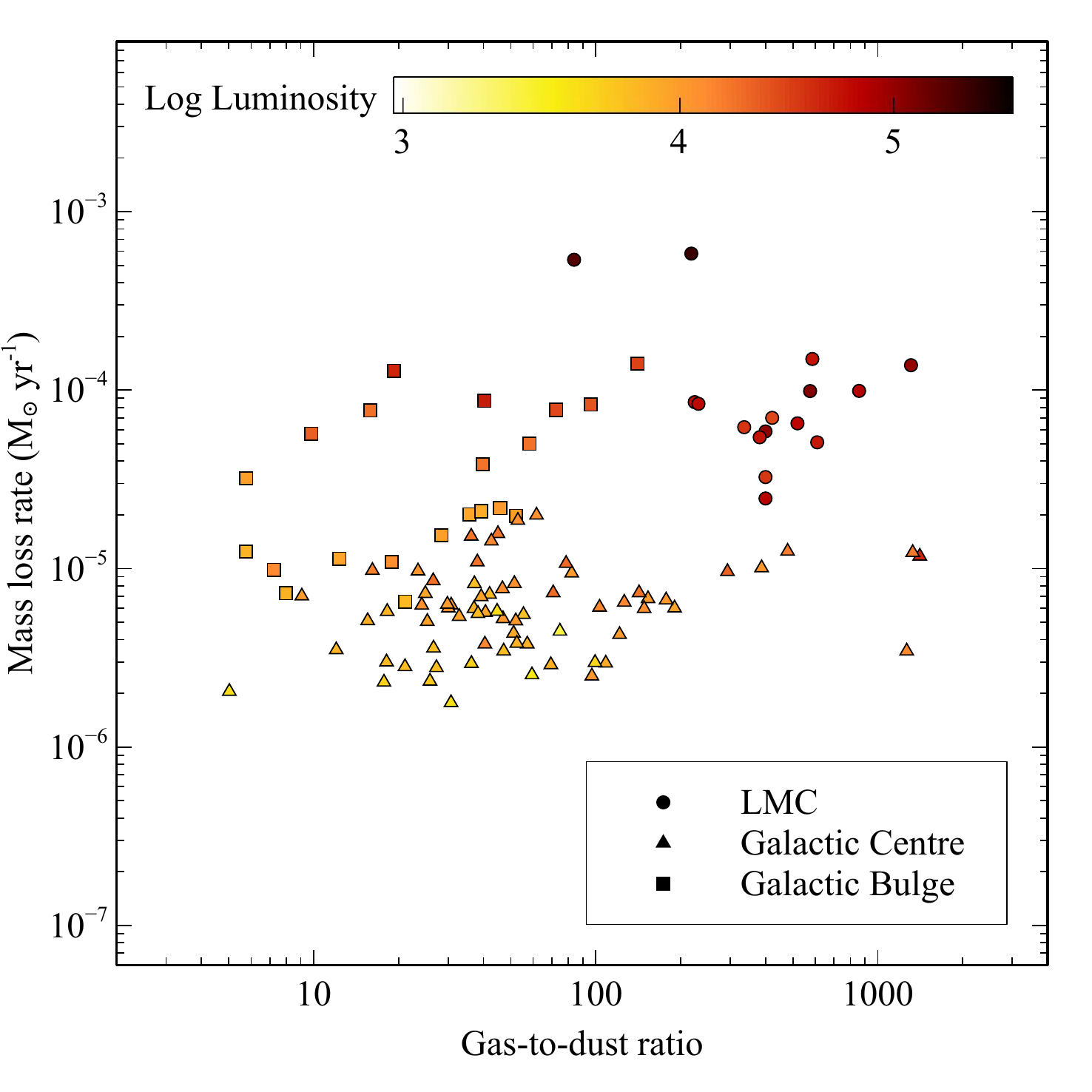} \hspace{0.5cm}
 \includegraphics[width=7.3cm]{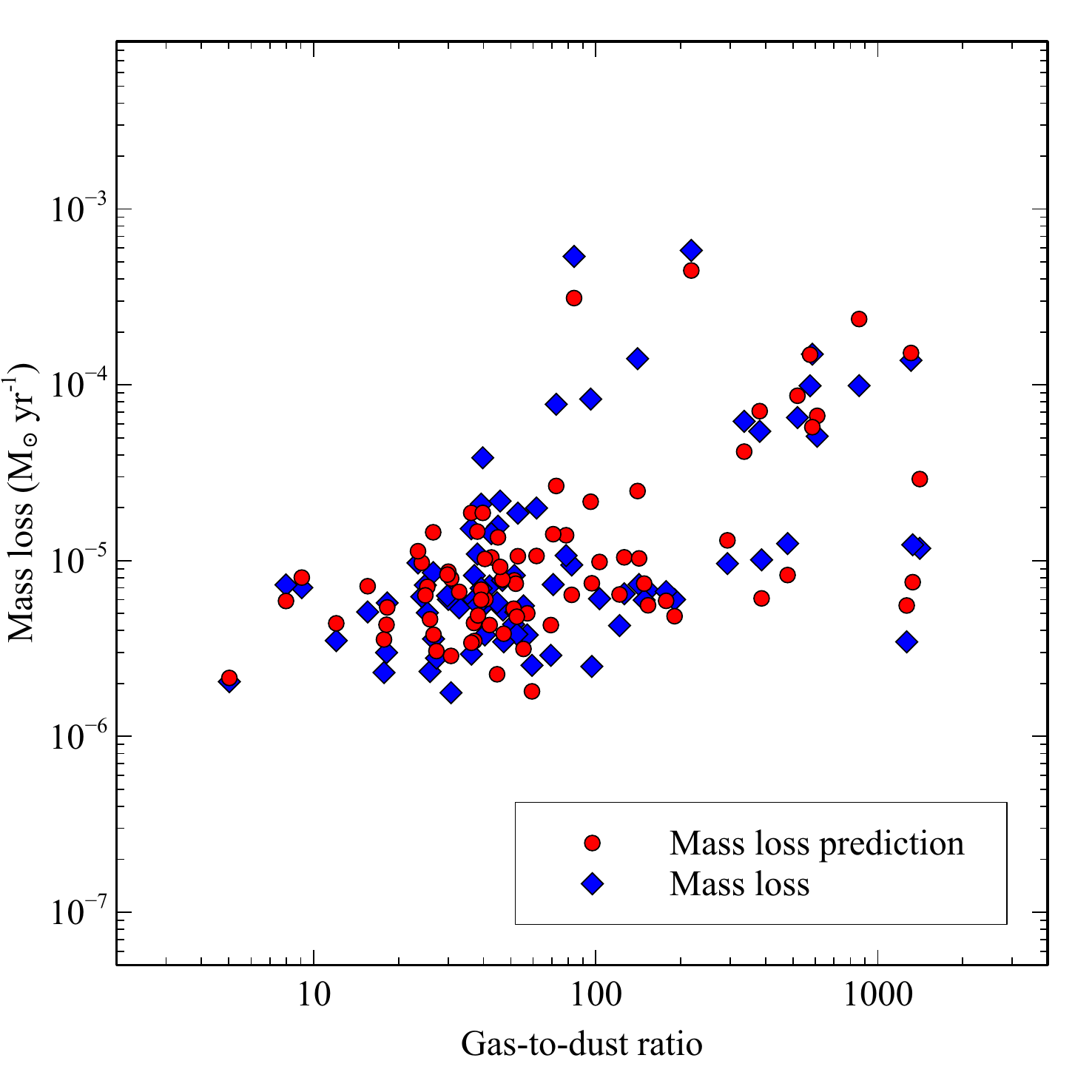} \\
 \caption{Mass loss rate as a function of luminosity (\textit{top}), pulsation period (\textit{middle}), and gas-to-dust ratio (\textit{bottom}). The left-hand column shows the observed mass loss rates. The right-hand column shows the predicted mass loss rates from our new mass loss prescription (Eq. 3) plotted in red, with mass loss rates derived by \textsc{dusty} for our sources with complete data ($L$, $P$, and $r_{gd}$.) in blue. The missing sources in the top right figure (as opposed to the left figure) stem from a lack of pulsation period measurements in the Galactic Bulge sample, which also correspond to the missing sources in the bottom right figure.}
 \label{fig:mdot}
\end{figure*}

There is quite a large range of gas-to-dust ratios in our three samples. When we compare them with respect to luminosity and expansion velocity (Fig. \ref{fig:v_vs_l}) we see a clear difference between our Galactic and LMC samples. When we compare our luminosities and mass loss rates (Fig. \ref{fig:mdot}) we see a clear trend irrespective of gas-to-dust ratio. We also see that the Galactic Bulge sources tend to have higher mass loss rates than the other two samples, expected from this more extreme sample. 

\begin{figure*} 
\centering
 \centering
 \includegraphics[width=5.8cm]{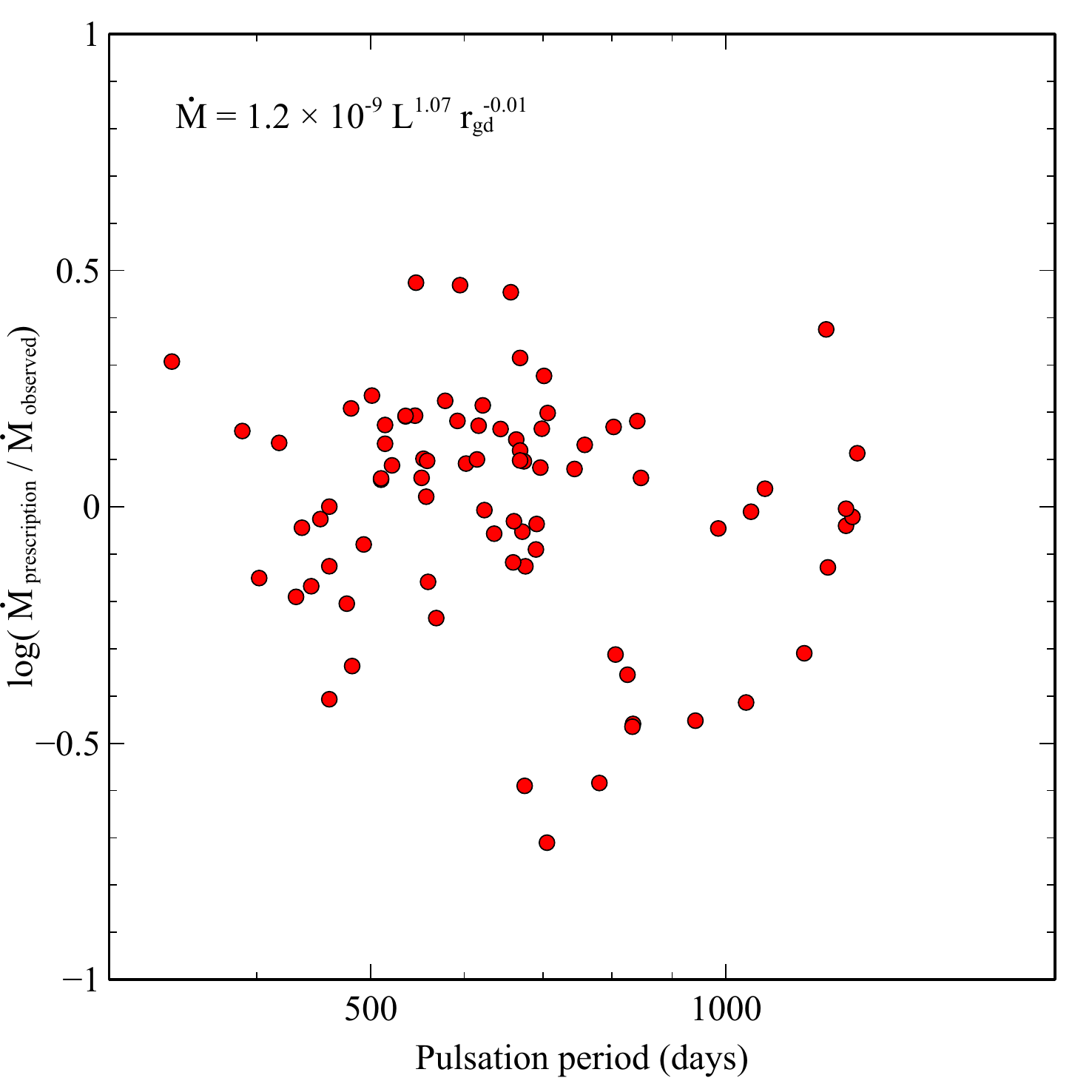}
 \includegraphics[width=5.8cm]{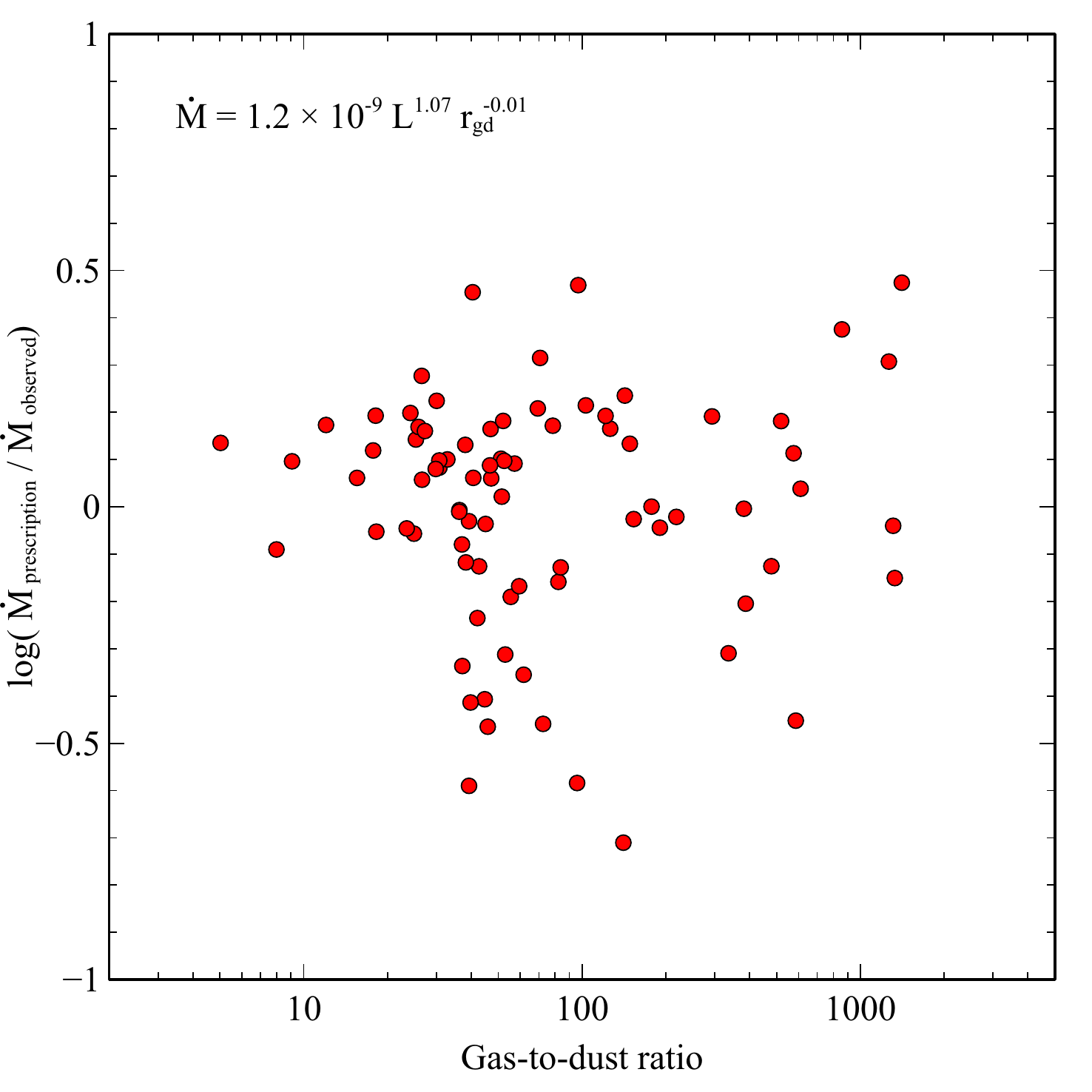}
 \includegraphics[width=5.8cm]{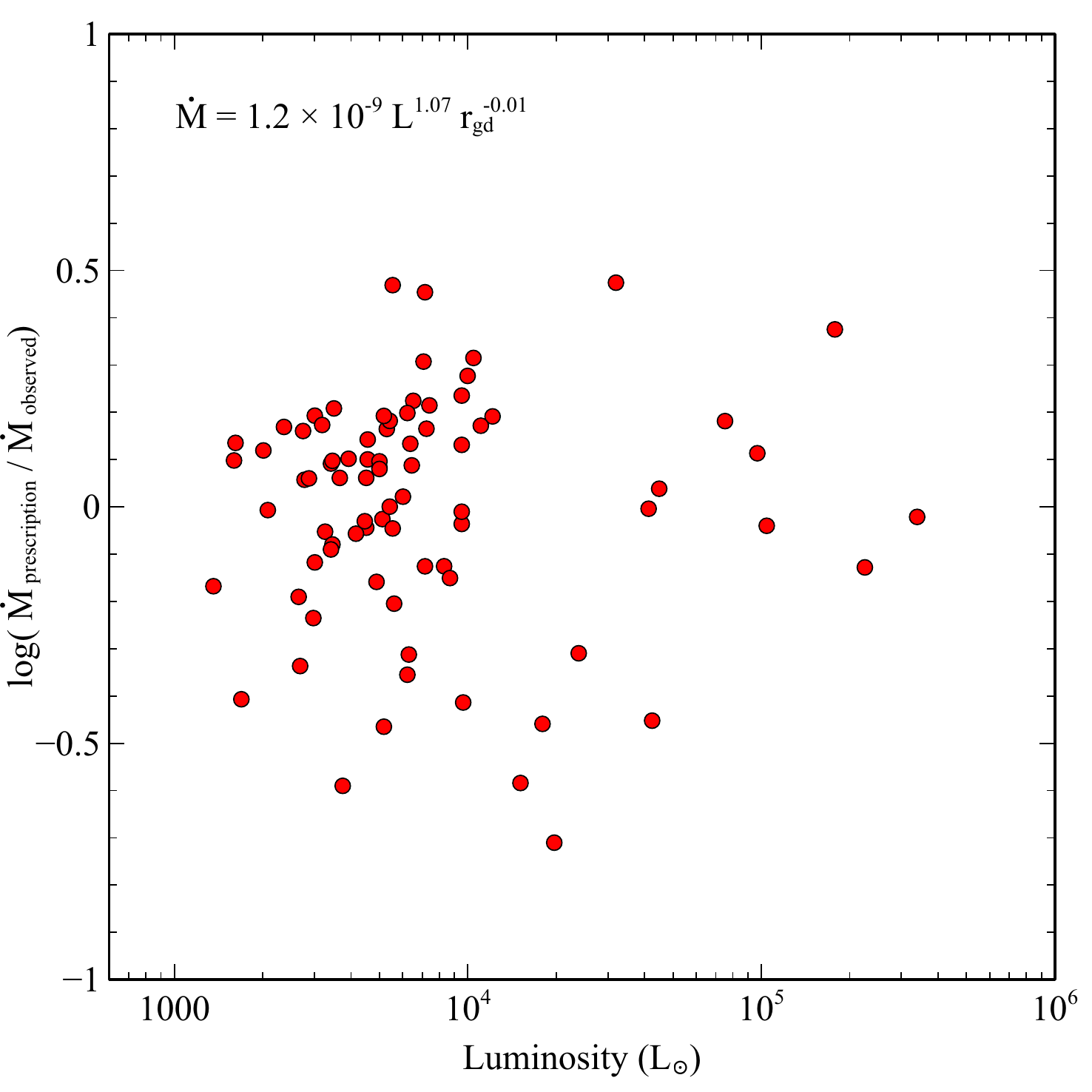} \\
 \includegraphics[width=5.8cm]{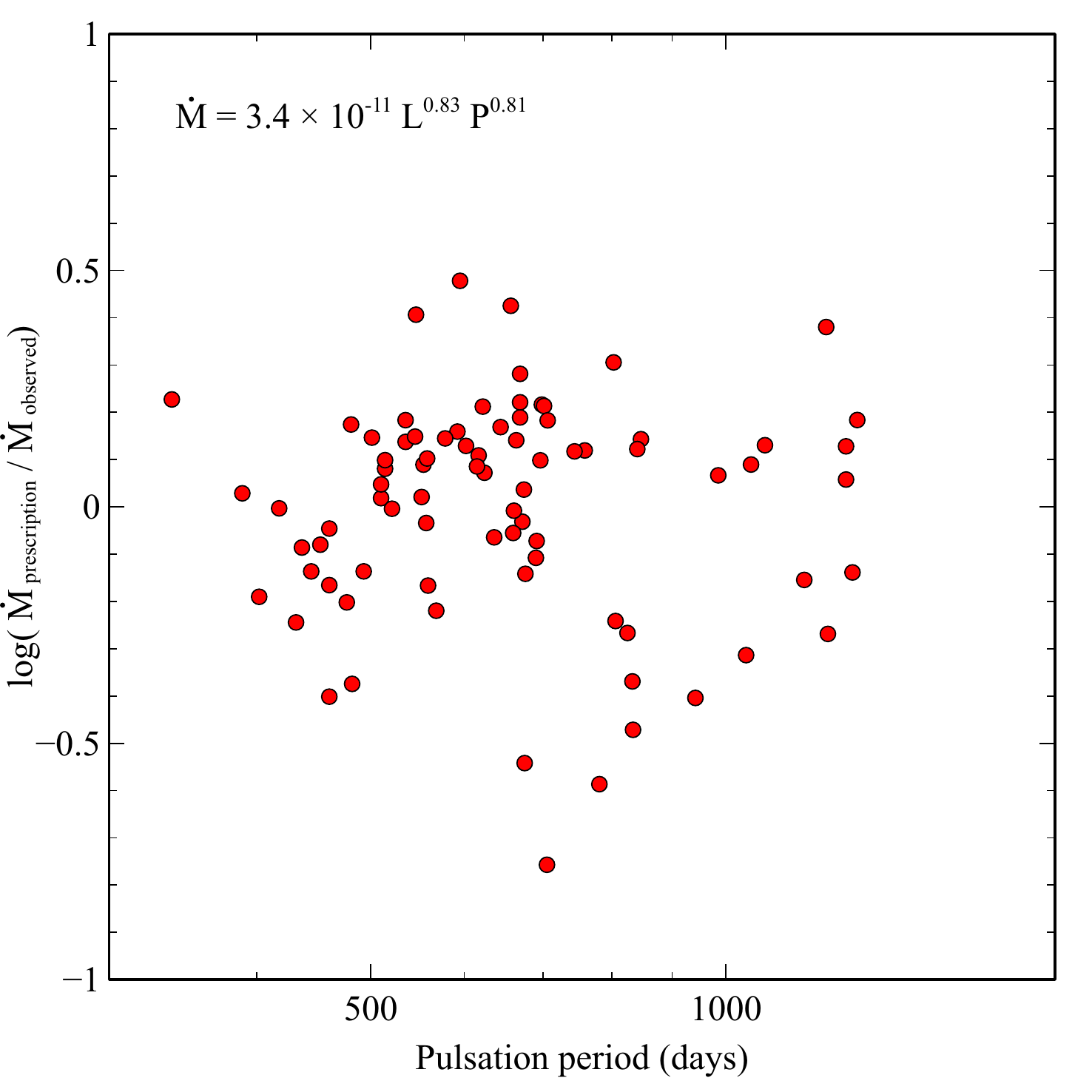}
 \includegraphics[width=5.8cm]{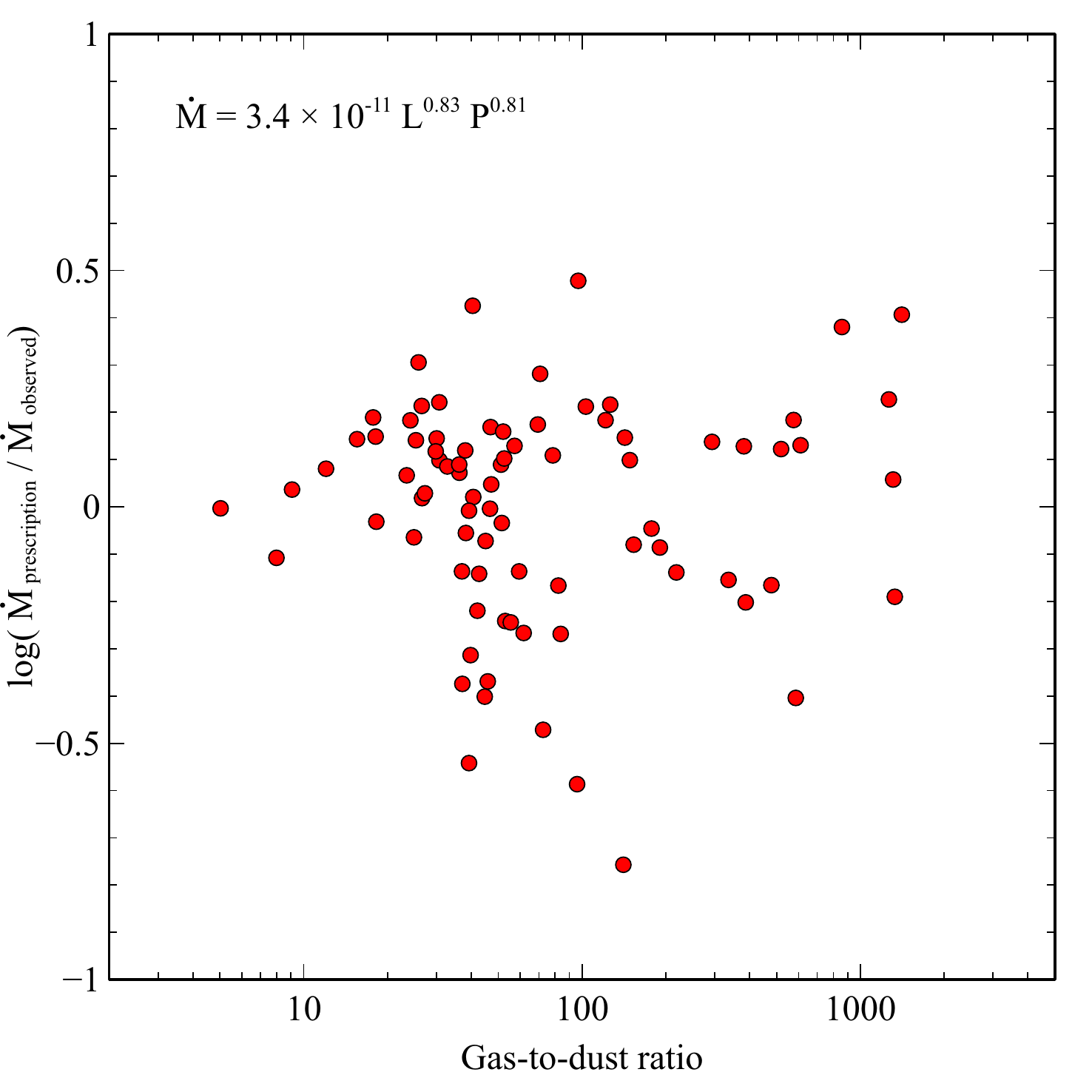}
 \includegraphics[width=5.8cm]{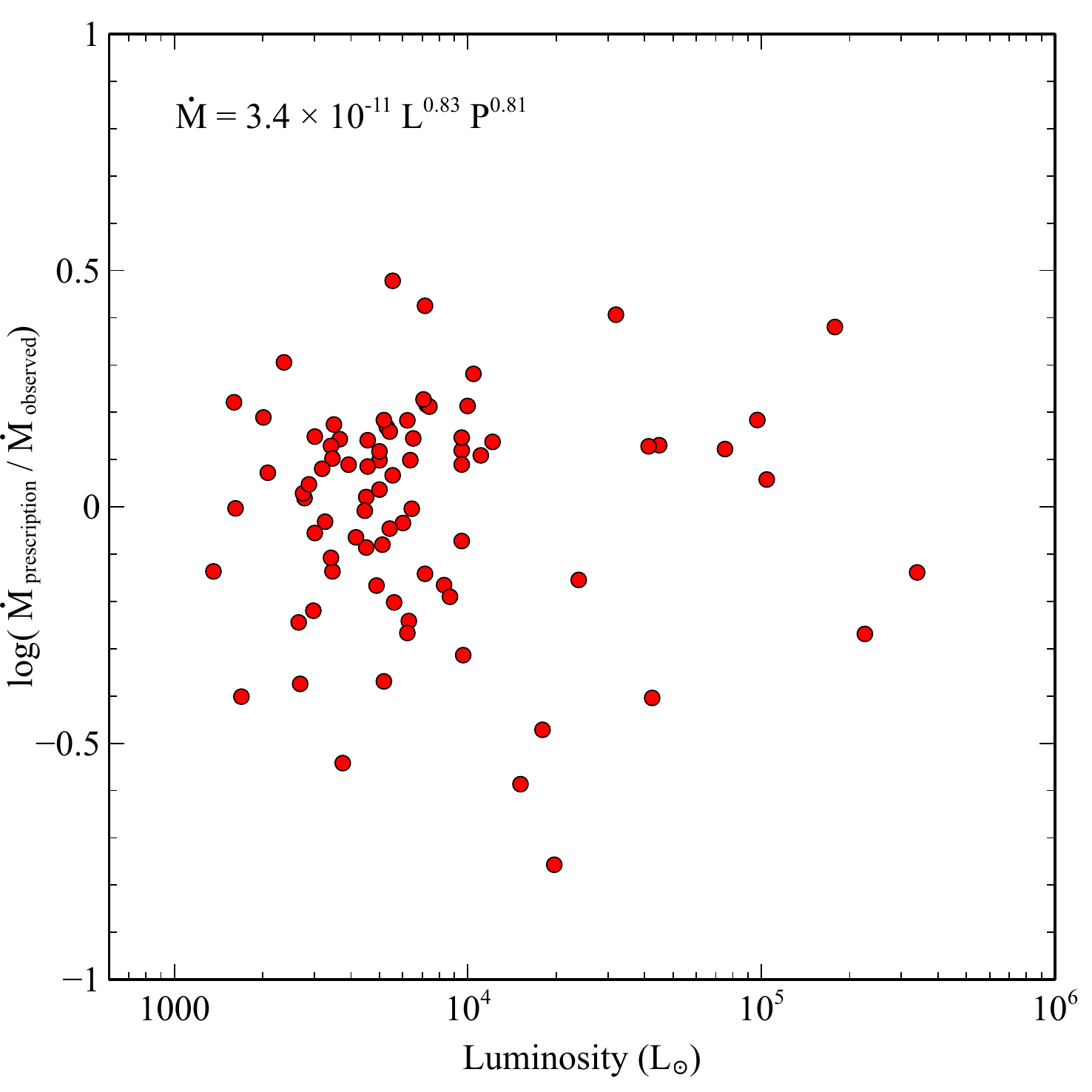} \\
 \includegraphics[width=5.8cm]{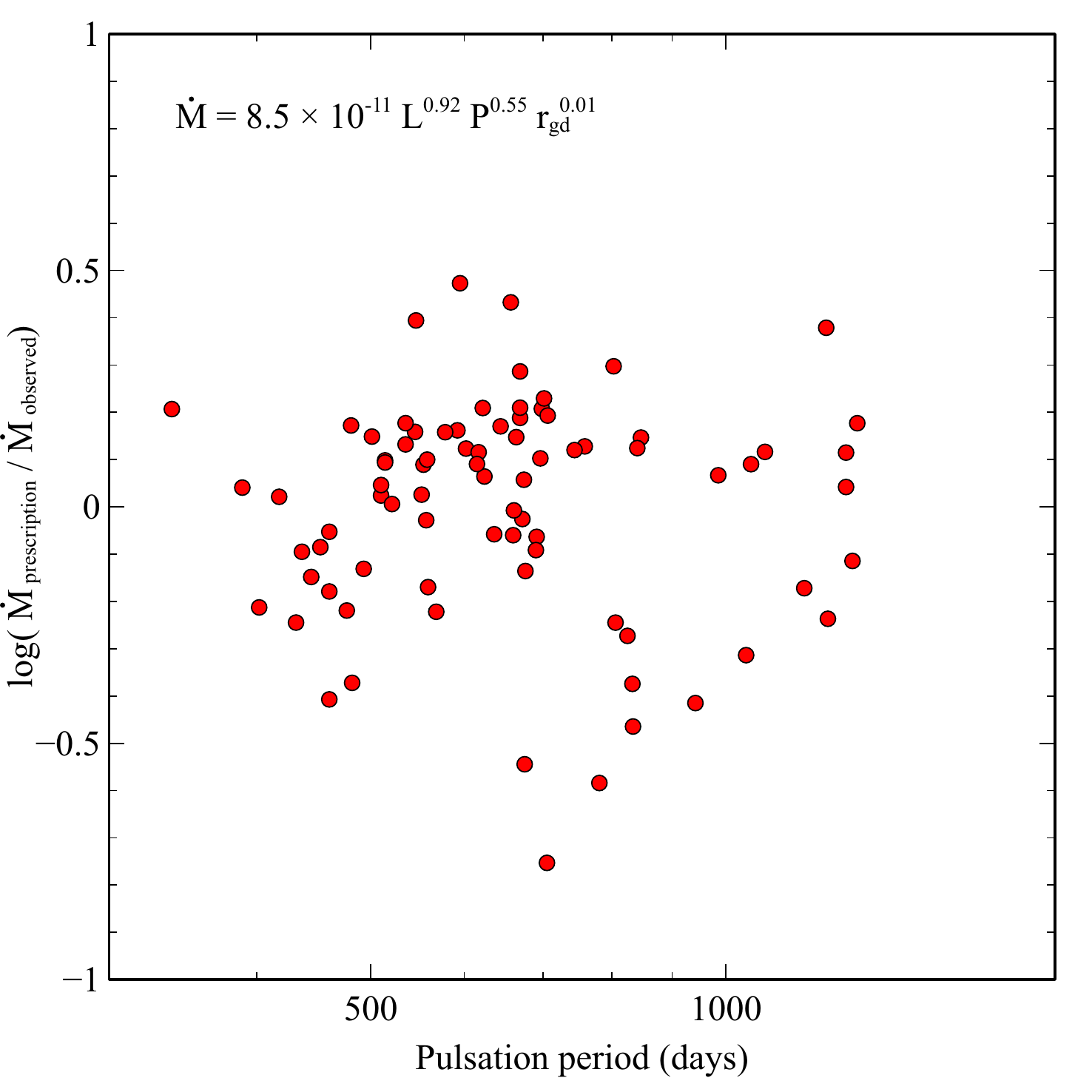}
 \includegraphics[width=5.8cm]{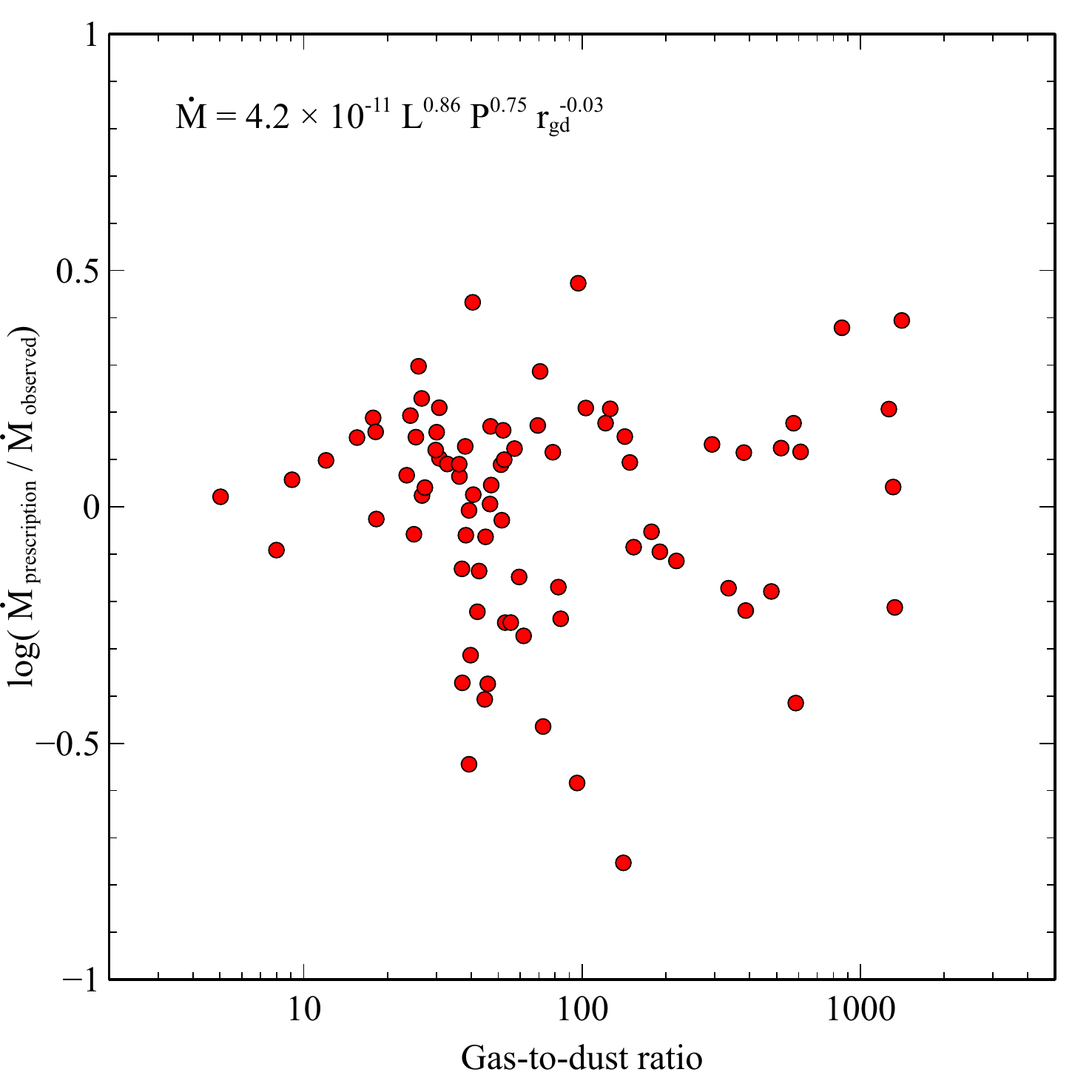}
 \includegraphics[width=5.8cm]{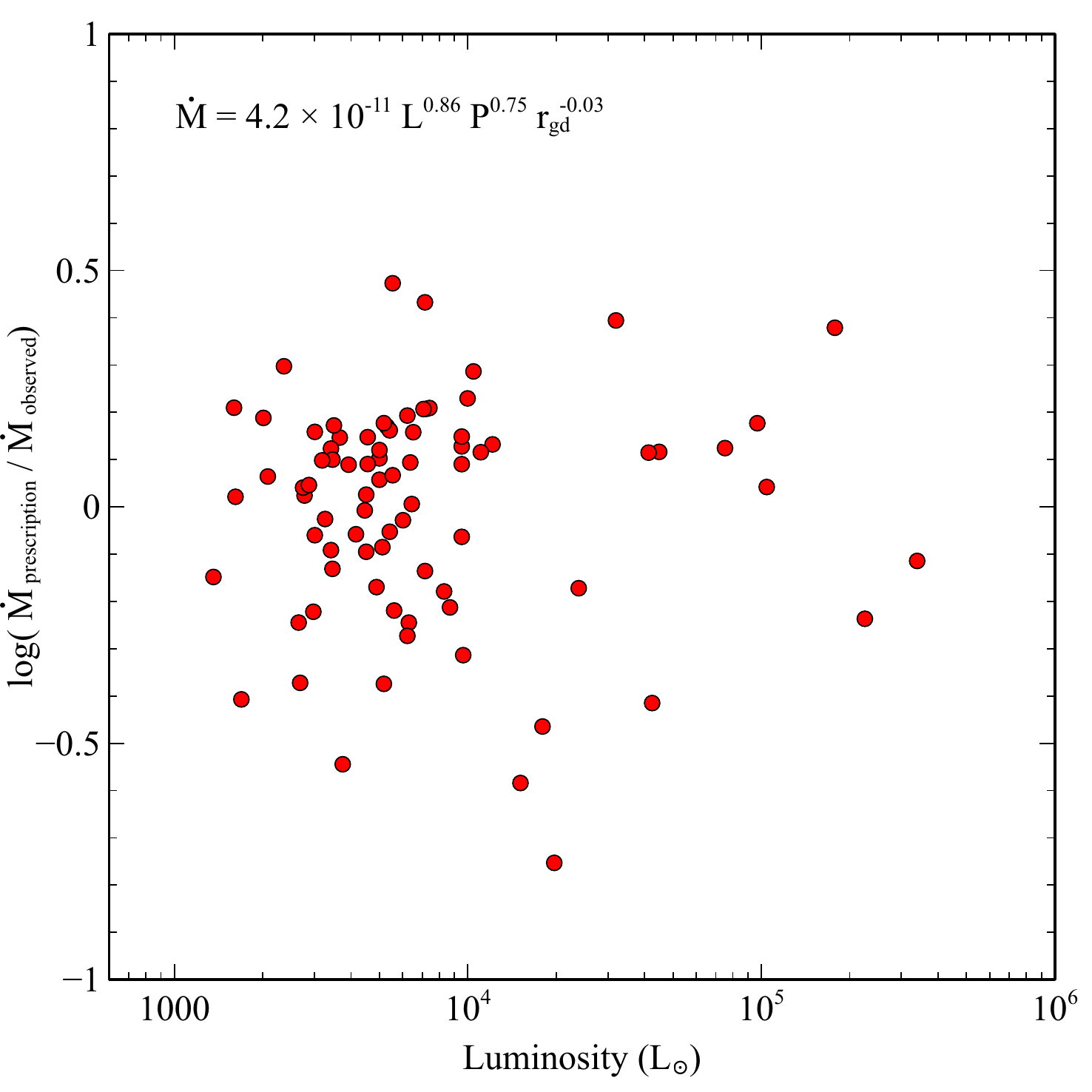}\\
 \caption{A comparison of mass loss prescriptions. The y-axis in each figure is the log ratio of the prescription listed and the ``observed'' mass loss rates derived from \textsc{dusty}. In the top two rows are prescriptions that were only fit with two parameters, in the bottom row is our mass loss prescription that was fit with all three. The two-parameter prescriptions show a only a slightly larger deviation from the \textsc{dusty} mass loss rates than our three-parameter prescription.} 
 \label{fig:logratio}
\end{figure*}
%
%
\begin{table*}
\centering
\caption[]{The LMC inferred gas-to-dust ratios ($r_{gd}$) and mass loss rates ($\dot{M}$) from scaling the resulting \textsc{dusty} expansion velocities (v\textsubscript{exp,DUSTY}) to the expansion velocities obtained from circumstellar OH maser profiles (v\textsubscript{exp,OH}). Luminosities ($L$) and optical depths specified at 10 $\mu$m ($\tau$) are derived from the SED fitting of \textsc{dusty} models that assume the effective temperature ($T$\textsubscript{eff,DUSTY}) and inner dust temperature ($T$\textsubscript{inner}). Mass loss rates scale in proportion to $L^{3/4}(r_{gd}\rho_{s})^{1/2}$, where the dust grain bulk density ($\rho_s$) is 3 g cm\textsuperscript{$-3$}; measured effective temperatures ($T$\textsubscript{eff}) are from van Loon et al. (2005) unless stated otherwise.\\ } 
\begin{minipage}{1\linewidth}
\centering
\begin{tabular}{ l c c c c c c c c c}
  \hline 
Object & 
$L$ & 
$T$\textsubscript{eff} & 
v\textsubscript{exp,OH} &
v\textsubscript{exp,DUSTY} & 
$T$\textsubscript{eff,DUSTY} &
$T$\textsubscript{inner} &
$\tau$ &
$\dot{M}$ & 
$r_{gd}$ \\

name & 
($10^3$ L\textsubscript{$\odot$}) &
(K) &
(km s$^{-1})$ & 
(km s$^{-1})$ & 
(K)&
(K)&
&
(M\textsubscript{$\odot$} yr$^{-1}$) &  
\\
\hline

IRAS\,04407$-$7000 &
104 & 
3008 & 
8.35 &
\llap{2}1.4 &
3400 &
800 &
0.4 &
1.4 $\times 10^{-4}$ &
1312
\\
IRAS\,04498$-$6842 &
97 &
2500 &
\llap{1}3.0 &
\llap{2}2.1 &
3400 &
1000 &
0.7 &
9.8 $\times 10^{-5}$ &
\llap{5}7\rlap{6}
\\
IRAS\,04509$-$6922 &
178 &
2500 &
\llap{1}1.4 &
\llap{2}3.6 &
3700 &
600 &
0.1 &
9.9 $\times 10^{-5}$ &
\llap{8}5\rlap{9}
\\
IRAS\,04516$-$6902 &
75 &
2667\rlap{$^2$} &
\llap{1}0.0 &
\llap{1}6.1 &
2900 &
600 &
0.3 &
6.5 $\times 10^{-5}$ &
\llap{5}1\rlap{9}
\\
IRAS\,04545$-$7000 &
19 &
  &
7.7 &
\llap{1}1.2 &
3200 &
1200 &
4.9 &
7.0 $\times 10^{-5}$ &
\llap{4}2\rlap{2}
\\
IRAS\,04553$-$6825 &
340$-$454 &
3008  &
\llap{2}3.8 &
\llap{2}4.9 &
3300 &
1400 &
8.62 &
5.8 $\times 10^{-4}$ &
\llap{2}1\rlap{8}
\\ 
IRAS\,04553$-$6933 &
176 &
 &
 &
\llap{2}1.9 &
2900 &
600 &
0.1 &
\llap{$^a$}5.9 $\times 10^{-5}$ &

\\
IRAS 05003$-$6712 &
23 &
2667 & 
&
\llap{1}4.5 &
3700 &
1000 &
0.8 &
\llap{$^a$}3.3 $\times 10^{-5}$ &

\\
IRAS\,05280$-$6910 &
225 &
  &
\llap{2}0.6 &
\llap{1}3.6 &
3400 &
350 &
13.4 &
5.4 $\times 10^{-4}$ & 
8\rlap{$4^*$}
\\
IRAS\,05294$-$7104 &
45 & 
2890 & 
\llap{1}0.3 &
\llap{1}7.5 &
3600 &
800 &
0.4 &
5.1 $\times 10^{-5}$ &
610
\\
IRAS\,05298$-$6957 &
50 &
4000\rlap{$^2$} &
\llap{1}0.5 &
\llap{1}1.1 &
3000 &
800 &
2.0 &
8.6 $\times 10^{-5}$ &
\llap{2}2\rlap{4}
\\
IRAS\,05402$-$6956 &
51 &
  &
\llap{1}0.5 &
\llap{1}1.3 &
2700 &
800 &
2.2 &
8.4 $\times 10^{-5}$ &
\llap{2}3\rlap{1}
\\
IRAS\,05558$-$7000 &
42 &
3400\rlap{$^2$} &
8.0 &
\llap{1}1.0 &
2700 &
600 &
0.7 &
5.4 $\times 10^{-5}$ &
\llap{3}8\rlap{1}
\\
MSX LMC 807 &
24 & 
  &
8.2 &
\llap{1}0.6 &
2900 & 
1000 &  
3.2 &  
6.2 $\times 10^{-5}$ &
\llap{3}3\rlap{6}
\\
MSX LMC 811 &
42 &
  &
8.3 &
\llap{1}4.1 &
2700 &
1200 &
5.6 &
1.5 $\times 10^{-4}$ &
\llap{5}8\rlap{6}
\\
MSX LMC 815 &
97 & 
3100\rlap{$^1$} &
&
\llap{2}4.9 &
3000 &
800 &
0.1 &
\llap{$^a$}2.5 $\times 10^{-5}$ &

\\
\\
\multicolumn{2}{l}{Sample median values:}\\
Galactic Centre & 5 &  & \llap{1}8.5 & 8.5 & 3000 & 1200 & 2.8 & 6.0 $\times 10^{-6}$ & 4\rlap{5} \\
Galactic Bulge & 6 &  & \llap{1}5.7 & 4.7 & 3200 & 800 & \llap{1}9.5 & 2.2 $\times 10^{-5}$ & 2\rlap{8} \\
LMC & 63 &  & \llap{1}0.3 & \llap{1}5.3 & 3100 & 800 & 0.8 & 7.7 $\times 10^{-5}$ & \llap{4}2\rlap{2} \\
\hline
\end{tabular}
\flushleft{$^a$ Calculated using assumed gas-to-dust ratio of 400} \vspace{-0.6cm} 
\flushright{References:
$^1$Buchanan et al. (2006)
$^2$Garc\'ia-Hern\'andez et al. (2009)} \vspace{-0.2cm} 
\flushleft{$^*$ Unlikely true gas-to-dust ratio; further explained in section 5.2}
\end{minipage}
\end{table*}

The idea that there may be two different OH/IR populations within our Galaxy was originally suggested by Wood et al. (1998), where sources with expansion velocities lower than 18 km s$^{-1}$ were expected to have a metal content 2.1 times lower than sources with higher expansion velocities; our results suggest that this is not the case. We find a slightly higher median gas-to-dust ratio for our sources with velocities less than 18 km s$^{-1}$.

\subsection{Mass loss as a function of gas-to-dust ratio, luminosity and pulsation period}

By employing our new gas-to-dust ratios and using Eq. 1 we have derived an empirical mass loss prescription that includes the luminosity ($L$), pulsation period ($P$), and gas-to-dust ratio ($r_{gd}$): $\dot{M} = 4.2 \times 10^{-11}\,L^{0.9}\,P^{0.75}\,r_{gd}^{\,\,-0.03}$, (where luminosity is in L$_\odot$, and the pulsation period is in days); normalising this prescription to a typical source, we arrive at the following:

\begin{equation} 
\begin{split}
\dfrac{\dot{M}}{\textrm{ M\textsubscript{$\odot$} yr\textsuperscript{$-1$}}} = 1.06^{+3.5}_{-0.8} \textrm{ $\cdot$ }10^{-5} \left(\dfrac{L}{10^4\,\textrm{L}_\odot}\right)^{0.9 \pm 0.1} \times \\ \left(\dfrac{P}{500\,\textrm{d}}\right)^{0.75 \pm 0.3} \bigg(\dfrac{r_{gd}}{200}\bigg)^{-0.03 \pm 0.07} 
\end{split}
\end{equation} \newline
\noindent The function for the mass loss has been fit using a non-linear least squares fitting technique, to all sources within the three samples and with complete data. The prescription suggests mass loss rates of $1.18 \cdot 10^{-5}$ M\textsubscript{$\odot$} yr\textsuperscript{$-1$} for a typical source with a luminosity of 10,000 L\textsubscript{$\odot$}, pulsation period of 500 days, and gas-to-dust ratio of 200. This is typical for Mira variables and Galactic sources, but not our LMC sources with higher luminosities, longer pulsation periods, and higher gas-to-dust ratios. Figure \ref{fig:mdot} shows a comparison of the parameter dependence of the mass loss rates computed with \textsc{dusty} with the parameter dependence of the mass loss rates from our new formula. There is a general correlation of mass loss with pulsation period, and a much tighter correlation with luminosity. We do not see a strong dependence of gas-to-dust ratio on the mass loss rate. This suggests that the mass loss is insensitive to metallicity.

We have also computed mass loss prescriptions that include only two of the three mass loss parameters. These prescriptions have been compared to our three-parameter prescription on the basis of their ability to correctly predict the mass loss rate given by \textsc{dusty}; the comparison of these prescriptions is shown in Figure \ref{fig:logratio}. There is little difference between our three parameter prescription and our prescription with only luminosity and pulsation period, reflecting the insensitivity of mass loss to gas-to-dust ratio and presumably to metallicity. 

It is possible that no single prescription can accommodate the diversity of all OH/IR stars. We have also fit our individual samples with all three of our mass loss parameters but would urge caution in using these fits. The limited ranges of the parameters within each of the three samples makes these fits much less reliable; the uncertainties on the derived parameters reflect this to some extent, but the fits are overall not nearly as convincing as for the combined sample. The dependence on luminosity (the most reliably fit parameter) is very similar between the three samples, although there is a hint of a trend for a weaker luminosity dependence at higher metallicity. The fits are:

\begin{figure}
\centering
 \vspace{0.62cm}
 \includegraphics[width=\columnwidth]{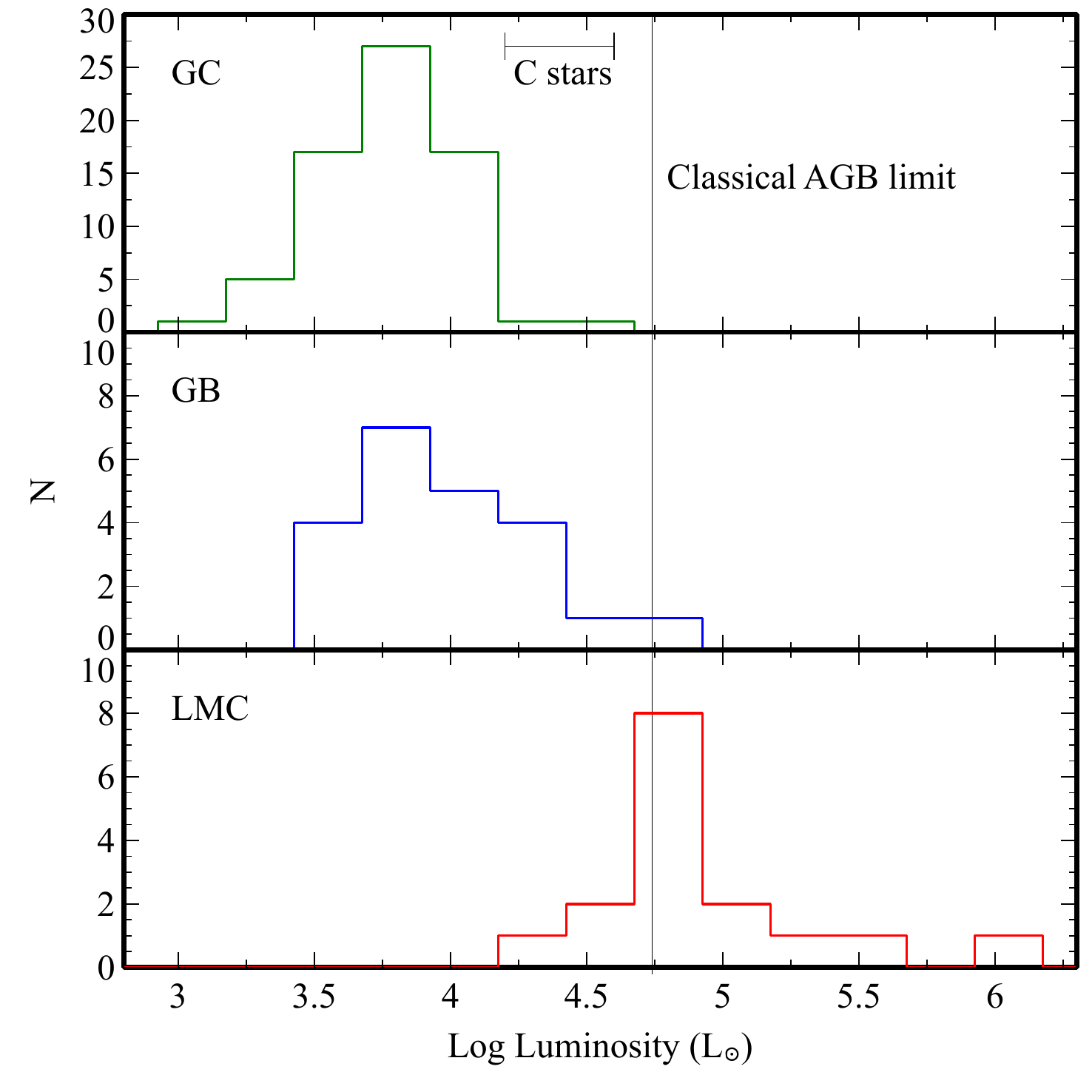}
 \caption{The Galactic Bulge, Galactic Centre and LMC luminosity distributions.}
 \label{fig:lum_hist}
\end{figure}   

\begin{equation} 
\begin{split}
\textrm{GC}: \dfrac{\dot{M}}{\textrm{ M\textsubscript{$\odot$} yr\textsuperscript{$-1$}}} = 8.8^{+40}_{-7.3} \textrm{ $\cdot$ }10^{-6} \left(\dfrac{L}{10^4\,\textrm{L}_\odot}\right)^{0.6 \pm 0.1} \times \\ \left(\dfrac{P}{500\,\textrm{d}}\right)^{0.5 \pm 0.3} \bigg(\dfrac{r_{gd}}{200}\bigg)^{0.05 \pm 0.07} 
\end{split}
\end{equation}

\begin{equation} 
\begin{split}
\textrm{GB}: \dfrac{\dot{M}}{\textrm{ M\textsubscript{$\odot$} yr\textsuperscript{$-1$}}} = 1.1^{+8.8}_{-1.0} \textrm{ $\cdot$ }10^{-4} \left(\dfrac{L}{10^4\,\textrm{L}_\odot}\right)^{0.8 \pm 0.1} \times \\ \left(\dfrac{P}{500\,\textrm{d}}\right)^{-0.3 \pm 0.4} \bigg(\dfrac{r_{gd}}{200}\bigg)^{0.5 \pm 0.1} 
\end{split}
\end{equation}
 
\begin{equation} 
\begin{split}
\textrm{\llap{L}MC}:\dfrac{\dot{M}}{\textrm{ M\textsubscript{$\odot$} yr\textsuperscript{$-1$}}} = 4.4^{+1226}_{-4.3} \textrm{ $\cdot$ }10^{-5} \left(\dfrac{L}{10^4\,\textrm{L}_\odot}\right)^{0.8 \pm 0.2} \times \\ \left(\dfrac{P}{500\,\textrm{d}}\right)^{-0.4 \pm 1.1} \bigg(\dfrac{r_{gd}}{200}\bigg)^{-0.4 \pm 0.2} 
\end{split}
\end{equation}

\begin{figure}
\centering
 \includegraphics[width=\columnwidth]{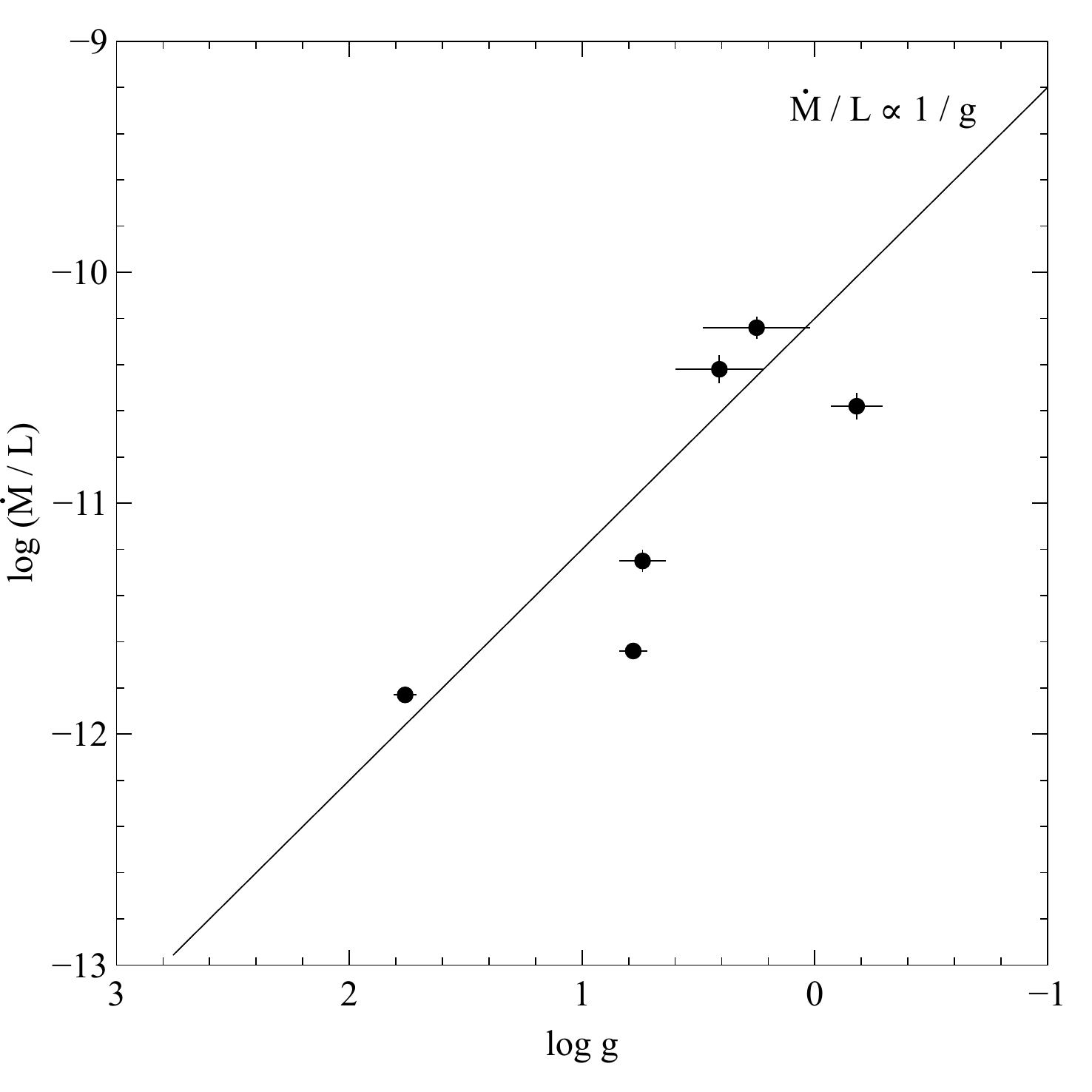}
 \caption{We have compared the values of mass loss and gravity (g) from the RSG sample from Schr\"{o}der and Cuntz (2007). We have divided their mass loss rates by their derived luminosities which, as we have shown (Eq. 3), should be approximately proportional to each other for a given value of gravity. This is done to isolate the dependence of mass loss rate on gravity.}
 \label{fig:1_over_g}
 \vspace{0.6cm}
\end{figure}

\begin{figure}
\centering
 \includegraphics[width=\columnwidth]{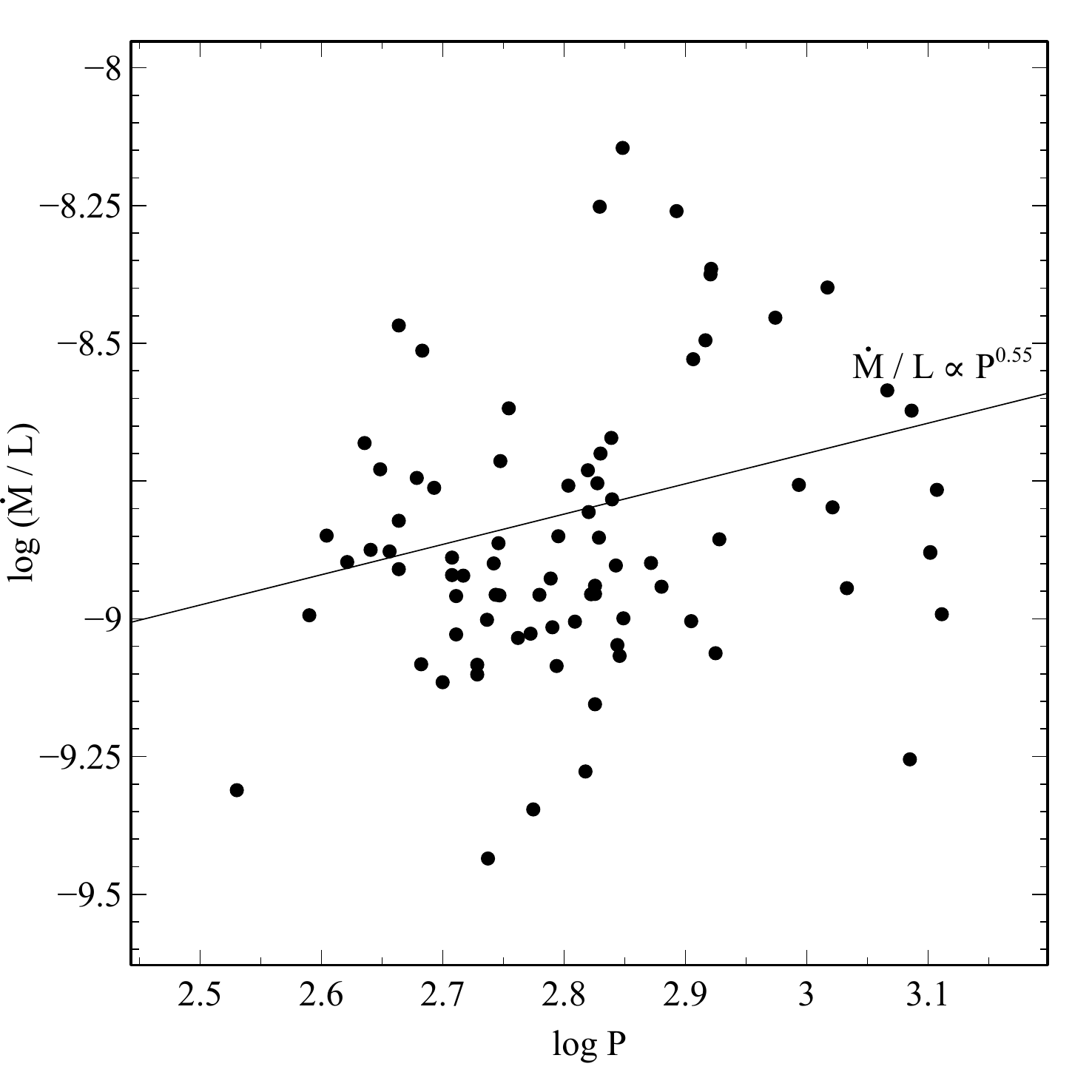}
 \caption{A comparison of mass loss rate against pulsation period. We have again plotted the mass loss rates divided by their derived luminosities to isolate the dependence of mass loss rate on the pulsation period.}
 \label{fig:M_over_lvp}
\end{figure}

\subsection{Interpretation of our prescription}
Our mass loss prescription, based entirely on observational data, matches known physical relationships. From the equation for optical depth (Eq. 1) from van Loon et al. (2000) we know:
\begin{equation} 
\tau(\lambda) \propto \dfrac{\dot{M}}{r_{gd}\,v_{exp}\,\sqrt{L}} 
\end{equation}

\noindent where $\tau$ is the optical depth. From our new relation for the expansion velocity (Eq. 2) shown in Figure \ref{fig:v_vs_l}, we find that: $v_{exp} \propto Z L^{0.4}$. By combining these two we get: $\dot{M} \propto L^{0.9}\,\,\tau(\lambda)\,\, Z\,r_{gd}$. As metallicity has been shown to be inversely proportional to the gas-to-dust ratio (van Loon 2000), and $\tau$ is a measure of the amount of photon scattering, all that remains is $\dot{M} \propto L^{0.9}$, which is exactly what we derive with our mass loss prescription. 

The second term in our prescription is pulsation period. If we describe the pulsation as a harmonic oscillation where gravity, $g$, is the driving force, then we expect that $2\pi/P\sim 1/t_{freefall}\sim \sqrt{g(R)/R}$, where $R$ is the stellar radius. Based on the results from Schr\"oder \& Cuntz (2007), we find that $\dot{M}\propto 1/g$ (Fig. \ref{fig:1_over_g}). This suggests that $\dot{M}\propto P^2/R$. Describing the star as a polytrope, with $MR^3=constant$, we expect $P\propto \sqrt{R/g}\propto M^{-1/2}R^{3/2}=R^3$, and hence $\dot{M}\propto P^{5/3}$. This is a much stronger dependence on pulsation period than our prescription, $\dot{M}\propto P^{0.75}$, however it includes the way mass loss would increase as a result of the star being more luminous, hence larger and thus pulsating with a longer period. At a given stellar effective temperature, $L\propto R^2$, which would suggest $L\propto P^{2/3}$. In our prescription, we find
$\dot{M}\propto L^{0.9}P^{0.75}$. This would then suggest that our prescription has a total dependence on pulsation period according to $\dot{M}\propto P^{1.35}$. This brings it in closer (albeit not perfect) agreement with the above, simplistic expectation.  

Fox \& Wood (1982) and Wood (1990) calculated more sophisticated models for pulsating red giants and found that $P\propto R^{1.8}$ (with some dependence on mass). Replacing the above polytrope by this relationship would yield $\dot{M}\propto P^{1.44}$, which is in remarkable agreement with our empirical relationship.

The last term in our prescription is gas-to-dust ratio. The gas-to-dust ratio has been shown to inversely scale with the metal content. Past studies have shown that metallicity has little to no effect on the mass loss rates within Galactic and Magellanic Cloud AGB and RSG stars (van Loon 2000; van Loon et al. 2005). Our relation of $\dot{M} \propto r_{gd}^{-0.03 \pm 0.07}$ matches this expectation and thus we can conclude that the gas-to-dust ratio has little effect on the mass loss of AGB stars and RSGs within our Galaxy and the LMC. Though the wind is, the mass loss itself does not appear to be driven by dust; the mass loss rate is probably set by pulsation and/or other physical mechanisms originating in the stellar photosphere.

Our new mass loss prescription uses three stellar parameters as gauges of physical differences within the sample. In general, we expect luminosity to give us the initial mass, gas-to-dust ratio to give us the metallicity and pulsation period to give us the precise evolutionary stage within the superwind phase. We expect all three parameters to affect all of these physical differences to some degree.

\begin{figure*}
\centering
\begin{minipage}[c c]{\textwidth}
\centering
\hspace{1.5cm}
 \includegraphics[width=6.75cm]{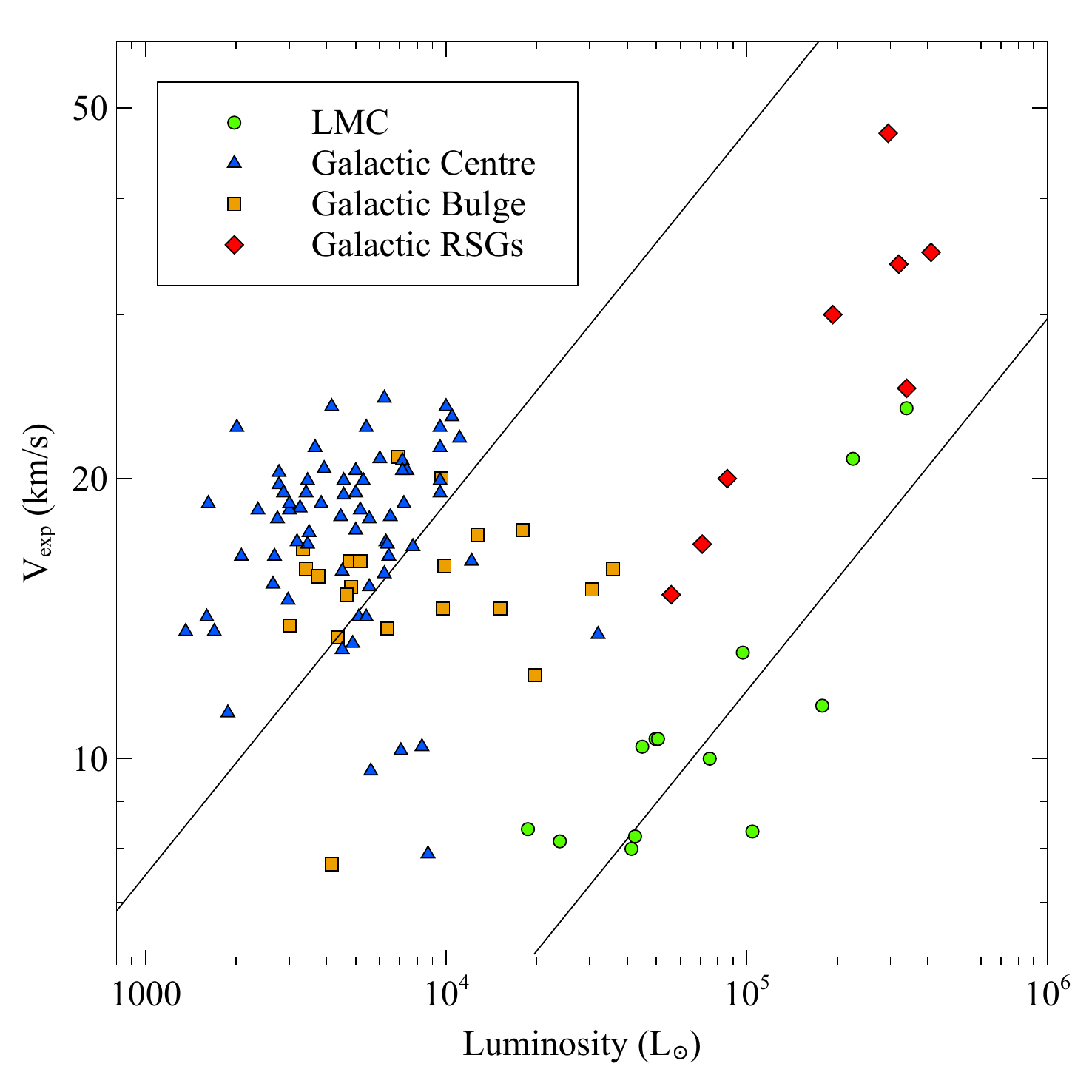} \hspace{-0.0cm}
 \includegraphics[width=9cm]{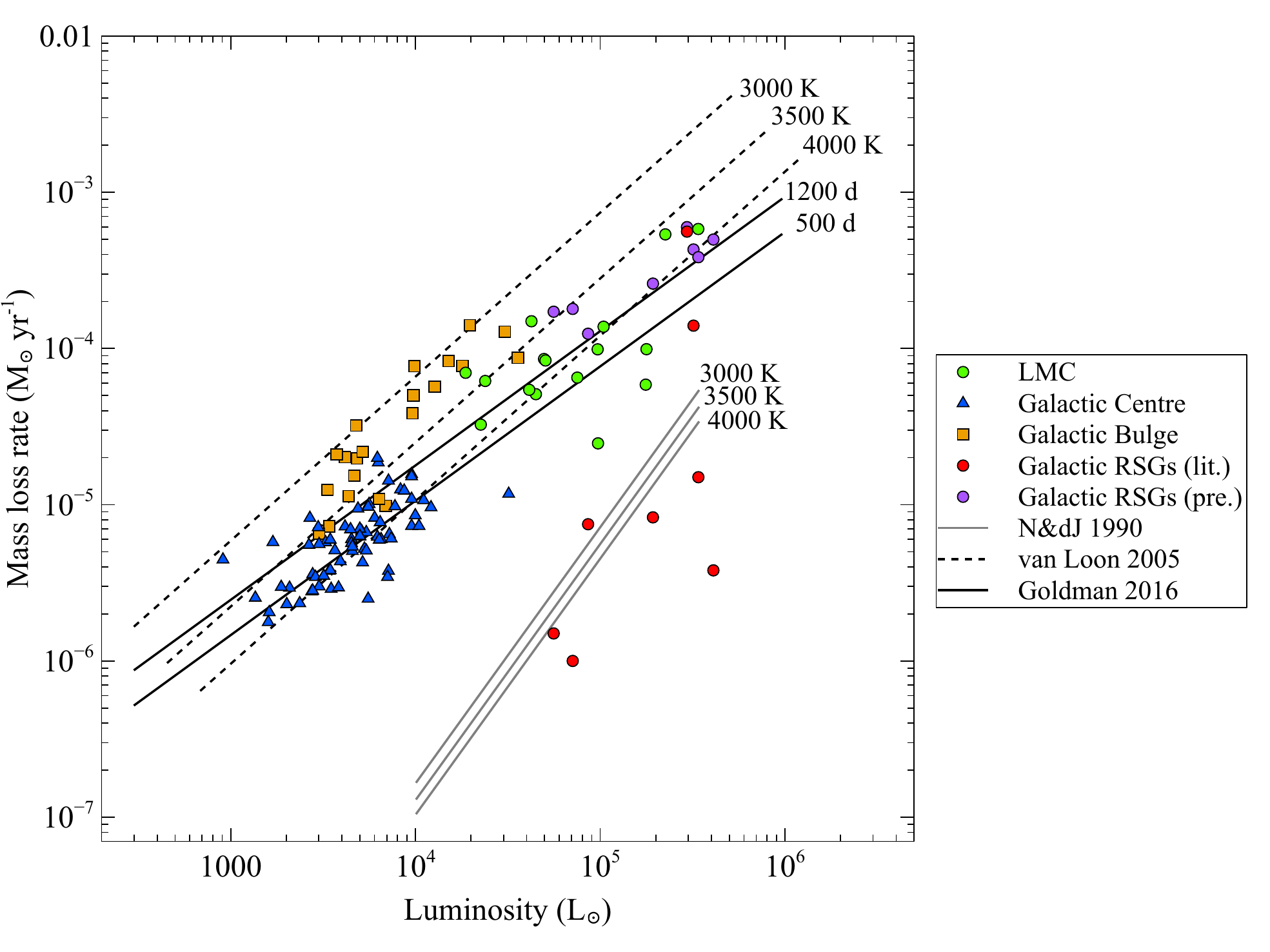}
 \caption{The observed wind speed as a function of luminosity (\textit{Left}), and mass loss rates as a function of luminosity (\textit{Right}) as shown in Figures \ref{fig:v_vs_l} and \ref{fig:mdot}, with the Galactic RSG sample from Mauron \& Josselin (2011); we have used updated literature mass loss rates for VY CMa and $\mu$ Cep from Shenoy et al. (2016). We show literature mass loss values (Galactic RSGs lit.) and the calculated mass loss rates (assuming a $r_{gd}=500$) using our mass loss prescription (Galactic RSGs pre.) and the pulsation periods from Kiss et al. (2006); NML Cyg pulsation period is from Monnier et al. (1997). Also plotted are the mass loss prescriptions by Nieuwenhuijzen \& de Jager (1990), van Loon et al. (2005), as well as our own (Goldman 2016), with a range of values for unknown variables (either effective temperature or pulsation period).}
 \label{fig:mdot_v_vs_l}  
\end{minipage}
\end{figure*}

\subsection{Comparison to previous prescriptions}

Most mass loss prescriptions of AGB and RSG stars typically focus on specific stages of mass loss and are derived using observational data. Past mass loss prescriptions like the Reimers law (Reimers 1975), and the updated Reimers law (Schr\"{o}der \& Cuntz 2005) are used to calculate the mass loss rates in the pre-dust driven wind phase. These are based on the mass loss of red giant branch stars and use luminosity, stellar radius, and mass; the updated prescription also uses effective temperature and gravity. Other prescriptions like that of van Loon et al. (2005) focus on mass loss within the superwind phase and are based on SED modeling of oxygen-rich AGB and RSG stars within the LMC. In-depth comparisons of previous mass loss prescriptions have been done by Mauron \& Josselin (2011) and Rosenfield et al. (2014).

We have compared our values for expansion velocities and mass loss rates, computed by \textsc{dusty}, to the Galactic RSG sample from Mauron \& Josselin (2011), shown in Figure \ref{fig:mdot_v_vs_l}. The Galactic RSGs are expected to have gas-to-dust ratio $\sim 500$ and fit well between our Galactic and LMC samples. Mass loss rates were derived primarily by modeling the wind speeds from CO emission. This excludes the mass loss rates of $\alpha$ Sco, derived by modeling the H\,{\sc ii} region (Reimers et al. 2008), $\alpha$ Ori derived by fitting C\,{\sc i} lines (Huggins et al. 1994), and VY CMa and $\mu$ Cep derived by fitting the SEDs with \textsc{dusty} models (Shenoy et al. 2016). As these were derived using various methods and making different assumptions, we use caution in comparing them individually. We will however use the sample to understand how red supergiants fit within our more general picture of mass loss in evolved stars. We have also displayed the calculated Galactic RSG mass loss rates derived using our prescription (Figure \ref{fig:mdot_v_vs_l}), in which we assume a gas-to-dust ratio of 500 and use pulsation periods from Kiss et al. (2006). 

The literature values for mass loss, which again use different methods of deriving the wind speeds and mass loss rates, are generally much lower than our values for the mass loss rates and we have plotted them in Figure \ref{fig:mdot_v_vs_l}. Along with more accurate distances for our Galactic sources, and thus luminosities and mass loss rates, new data has allowed for a more accurate measure of the optical depth. More accurate \textit{Spitzer} data has allowed for a better measure of infrared excesses as well as a decrease of confusion from contaminating sources within the observing fields. We have also compared our mass loss prescription to previous mass loss prescriptions (Fig \ref{fig:mdot_v_vs_l}) plotting different values for unknown variables (either effective temperature, expansion velocity or pulsation period). The comparison suggests that some of the sources within the Galactic RSG sample may not be within the superwind phase of mass loss. We will discuss the evolutionary stages of these sources further in the following section.
 
\subsection{Fundamental stellar parameters and mass loss} 

Our current analysis covers highly evolved AGB and RSG stars with a luminosity range from 2,000 to over 300,000 L$_{\odot}$ and metallicity range of a half to twice solar. This range allows us to draw conclusions about the mass loss mechanism and how it is affected by fundamental differences within the stars. We expect $L$, $r_{gd}$ and $P$ to relate to mass, metallicity and evolutionary phase. While the links between luminosity and initial mass, and $r_{gd}$ and metallicity are more established (e.g. Bl\"ocker 1995; Vassiliadis \& Wood 1993), evidence for a link between pulsation period and evolutionary stage is still unclear (Groenewegen et al. 2007; Sloan et al. 2012). As our stars progress through the superwind stage, stellar pulsation will become stronger and slower. We expect this pulsation period may be able to give us an indication of the source's stage within the superwind phase. We will discuss each of these relationships in the following subsections.

\paragraph*{Mass (initial and current):} The initial mass will have a dramatic impact on the evolution of these stars. Dredge-up efficiency and hot-bottom burning will vary with initial mass, thus the C/O ratio of these stars will depend on the initial mass (e.g. Boothroyd 1995; Karakas 2010). In terms of mass loss, a higher initial mass will allow the star to get to a higher luminosity before the onset of the superwind phase, thus resulting in higher mass loss rate. As these sources reach more advanced stages within the superwind phase their mass will decrease as their luminosity remains the same. We expect that a higher luminosity will result in a higher mass loss rate. In these stars we know that the dust formation zone will depend on luminosity, being further away for more massive stars. Looking at the ratio of luminosity and current mass (L/M), our higher luminosity RSGs ($\sim 200,000 $ L$_{\odot}$) and lower luminosity sources ($\sim 50,000$ L$_{\odot}$) have masses of around 20 and 5 M$_{\odot}$ respectively and quite similar L/M ratios. Assuming this ratio to be constant within our samples, we expect the ratio of radiation flux and gravity to be independent of radius as both depend on radial distance as $r^{-2}$. The acceleration is determined by the {\it difference} between the radiation pressure and gravity, which at any given radius is larger for more massive (more luminous) stars with the same L/M ratio. However, the wind -- and ultimately the mass that is lost -- is driven by dust grains, which condense at a distance where the dust grain equilibrium temperature matches the condensation temperature. This happens at a certain photon flux, i.e.\ further from a more luminous star by a factor $R^2\equiv L$, thus balancing out the larger difference between radiation pressure and gravity at a given distance. But because the area through which the mass flows also increases in proportion to $R^2$, the total amount of mass flowing through the area increases accordingly, and thus $\dot{M}\propto L$.

As these sources lose mass in the superwind phase and remain at a constant luminosity, L/M will increase making it easier to drive the wind while the decrease in mass will drive a direct increase in P. The source will then evolve to longer pulsation periods at a constant luminosity; we expect this is the case for our extreme Galactic Bulge sample. It is still unclear whether the simultaneous increase in pulsation period affects the mass loss. With longer and larger pulsations, as well as reduced gravity at the surface of a larger star which has lost mass, this should allow stars to levitate more material out to form dust and more easily drive mass loss. We expect L/M to affect the mass loss rate directly, and indirectly through changes in the pulsation period. However, we are not at a stage where we can fully quantify these relationship.

\paragraph*{Metallicity:} 

Gas-to-dust ratios have been shown to scale inversely proportionally with metallicity (van Loon 2000) and thus we can use them to understand the effects of metallicity. Metallicity will have a dramatic impact on the dust production of oxygen-rich AGB and RSG stars as well as the chemical composition of the dust itself (Jones 2013). Unlike carbon stars, these stars can not create the raw materials for dust production. Past studies have shown that while nearby Galactic and Magellanic Cloud oxygen-rich stars have similar total mass loss rates at higher luminosities, the dust mass loss rates show a strong dependence on metallicity (Wood et al. 1998; van Loon 2000; Lagadec et al. 2008); this is consistent with our results. 

Observations of Galactic and Magellanic Cloud carbon AGB stars have been interpreted to show that the quantity of carbon dust produced in both environments is the same. It has also been suggested that LMC carbon stars collectively produce more dust than their oxygen-rich counterparts (Matsuura 2013; Boyer et al. 2011; Sloan et al. 2016). It was suggested by Habing (1996) that carbon-rich LMC AGB stars may have similar gas-to-dust ratios as Galactic samples. If this is the case, it also stands to reason that they will have expansion velocities comparable to those of the Galactic samples. However, van Loon et al. (2008) present evidence suggesting the effect of primary carbon may not be as large, which would mean that also Magellanic carbon star winds are expected to be less dusty, and slower, than solar-metallicity carbon stars. 

It was predicted that AGB mass loss would be dependent on metallicity as an increase in dust optical depth may cause an increase in the number of absorbing and scattering events of photons on dust grains and therefore an increase in the net driving force of the dust-driven wind (Gail \& Sedlmayr 1986). While the dust content and driving of the wind depends on metallicity, we see no evidence to suggest a metallicity dependence on the mass loss. We expect that at lower metallicities the mass is lost in a slower denser outflow with the same net mass loss efficiency.

\paragraph*{Evolutionary phase:} As our sources become more evolved and more extreme, the strength and cycle length of stellar pulsations will increase. In theory we can use the pulsation period at a given luminosity as a gauge of the stage within the superwind phase. As mass decreases due to mass loss, the pulsation period increases and reduced gravity at the surface leads to increased mass loss. We know that as the mass loss ramps up, the gas-to-dust ratio will also change. As the mass loss in our sources has been found to be insensitive to metallicity, this implies that we are looking at sources in the superwind phase where they have reached maximal dust formation efficiency.

Stellar pulsations can be affected by chemical differences in the star. The pulsation amplitude of oxygen-rich AGB stars may be affected by metallicity as TiO absorption is affected by metallicity, but it is unclear if the pulsation period will be affected by changes in metallicity as it is determined deeper inside the star. The efficiency of the dust formation will also vary with evolutionary phase which may leave some evolved stars like Betelgeuse with little dust (van Loon 2013). However, this source has not entered its superwind phase and would be incompatible with our mass loss prescription. 

In this work we have fit a mass loss prescription using several stellar samples selected in the superwind stage of mass loss. We have also included the Galactic RSG sample for comparison as these sources seem to follow a similar mass loss mechanism. However, this sample contains sources not within the superwind phase, and our prescription may not be able to provide a realistic estimation of the mass loss rate of these sources. In Fig. \ref{fig:mdot_v_vs_l} we see that Nieuwenhuijzen \& de Jager prescription fits the bulk of the RSG sample much better. Two of these sources, VY CMa and NML Cyg with mass loss rates of 5.4 and $1.4 \cdot 10^{-5} \textrm{ M\textsubscript{$\odot$} yr\textsuperscript{$-1$}}$ respectively, fit well within our other samples. It is suprising that given VY CMa shows silicate in emission rather than absorption (Harwit et al. 2001) that this source would show such a high mass loss rate compared to the rest of the sample. However, these sources are luminous and have the lowest effective temperatures of the RSG sample, may thus be comparable to our sample while the remaining RSGs may be at an early evolutionary stage. A similar gap in mass loss rates of LMC RSGs on the order of 2 magnitudes has been found in the past by van Loon et al. (1999). It appears that at least within RSGs another parameter is affecting the mass loss. \\

\section{Conclusions}

We present the results of our recent survey of 1612 MHz OH maser emission in the LMC. We have discovered four new sources and increased the number of reliable wind speeds from five to thirteen. The expansion velocities derived from our maser sources fit well with our updated relation for dust-driven winds. We have developed a method of deriving gas-to-dust ratios of OH/IR stars by modeling the spectral energy distribution and scaling the results with the expansion velocities derived from maser observations. Furthermore we have used these new results to develop an empirical mass loss prescription that includes the effects of metallicity and luminosity. The results show a correlation of mass loss with pulsation period, with an even tighter correlation between mass loss and luminosity. We see a clear link between expansion velocity and gas-to-dust ratio, yet the gas-to-dust ratio has little effect on the mass loss of AGB stars and RSGs. This suggests that mass loss is (nearly) independent of metallicity between a half and twice solar. 

\section*{Acknowledgments}

We would like to thank the support staff at CSIRO for their help with the observations, Michele Costa for his help with the SED fitting code, Oliver Turner for his help with the pulsation period analysis, and Malcolm Gray for his helpful discussion on maser pumping mechanisms. We would also like to thank Phillip Edwards for generously allocating directors discretionary time to the project. This research made use of the Duchamp source finder, produced at the Australia Telescope National Facility, CSIRO, by M. Whiting. This paper makes use of the SIMBAD database of the CDS. SRG acknowledges financial support from the Royal Astronomical Society and Keele Postgraduate Association. AN acknowledges the support from the project STARKEY funded by the ERC Consolidator Grant, G.A. No. 615604. PAW acknowledges a research grant from the South African National Research Foundation. JFG is supported by MINECO (Spain) grant AYA2014-57369-C3-3 (co-funded by FEDER). AAZ was supported by the Science and Technology Research Council under grant ST/L000768/1, and JvL under grant ST/M001040/1. HI is supported by JSPS KAKENHI (25610043) and the Strategic Young Researcher Overseas Visits Program for Accelerating Brain Circulation.



\appendix
\renewcommand{\thefigure}{A\arabic{figure}}
\setcounter{figure}{0}

%
%
\begin{table*}
\renewcommand\thetable{B1}
\centering
\caption[]{Results of Galactic maser observations.}
\begin{minipage}{1\linewidth}
\centering
\begin{tabular}{lcccccccrr}
\hline\hline
Object           &
Alternative      &
l                &
b                &
$v_{exp}$        &
$F_{25}$         &
\multicolumn{2}{c}{$F_{int}$} &
\multicolumn{2}{c}{Velocity at peaks}
\\
name             &
name             &
($^{\circ}$)     &
($^{\circ}$)     &
(km s$^{-1}$)    &
(Jy)             &
\multicolumn{2}{c}{(Jy km s$^{-1}$)} &
\multicolumn{2}{c}{(km s$^{-1}$)}             
\\
&
&
&
&
red &
blue &
red &
blue 
\\
\hline 
IRAS\,18115$-$2139 &
OH 9.575 $-$2.032 &
9.57\rlap{4} &
$-$2.03\rlap{1} &
14.2 &
60.8 &
8.8\rlap{5} &
7.4\rlap{7} &
$-$18.4 &
10.0
\\
IRAS\,18139$-$1816 &
OH 12.8 $-$0.9 &
12.819  &
$-$0.90\rlap{1} &
12.1 &
16.9 &
9.2\rlap{1} &
11.22 &
\llap{$-$8}1.1 &
$-56.8$
\\
IRAS\,18257$-$1000 &
OH 21.5 +0.5 &
21.451  &
+0.50\rlap{0} &
18.4 &
\llap{1}20.5 &
\llap{3}9.7 &
\llap{3}0.3 &
\llap{8}2.4 &
119.1
\\
IRAS\,18268$-$1117 &
OH 20.4 $-$0.3 &
20.433  &
$-$0.34\rlap{4} &
17.4 &
30.5 &
6.15 &
9.2\rlap{2} &
9.6 &
44.4
\\
IRAS\,18268$-$1117 B &
 &
 &
 &
16.4 &
 &
0.8\rlap{2} &
0.5\rlap{8} &
\llap{7}8.6 &
111.4
\\
IRAS\,18432$-$0149 &
OH 30.7 +0.4 &
30.715  &
+0.42\rlap{3} &
17.5 &
52.3 &
\llap{1}5.7 &
\llap{1}2.5 &
\llap{3}2.3 &
67.3
\\
IRAS\,18460$-$0254 &
OH 30.09 $-$0.69 &
30.086  & 
$-$0.68\rlap{1} &
20.3 &
\llap{2}79.9 &
\llap{11}6.4 &
\llap{13}1.2 &
\llap{6}2.1 &
102.7
\\
IRAS\,19059$-$2219 &
V\textsuperscript{*} V3880  Sgr &
14.662  &
$-$13.610 &
13.0 &
\llap{2}11.6 &
8.2\rlap{3} &
4.5 &
\llap{$-$}3.0 &
23.0
\\
IRAS\,20077$-$0625 &
V\textsuperscript{*} V1300 Aql &
36.356  &
$-$20.415 &
12.6 &
\llap{10}61.0 &
\llap{1}8.9 &
8.4\rlap{6} &
\llap{$-4$}4.9 &
$-19.8$
\\
\hline
\end{tabular}
\end{minipage}
\end{table*}

\newpage

\section{SED modeling of Galactic sources}
We have modeled a number of Galactic sources from the Lindqvist et al. (1992) Galactic Centre sample of OH/IR stars and the Jim\'{e}nez-Esteban \& Engels (2015) Galactic Bulge sample of more extreme OH/IR stars. Using the same approach as in our LMC sample, we have fitted 69 Galactic Centre and 21 Galactic Bulge OH/IR stars with \textsc{dusty} models. These samples serve as a comparison sample at higher metallicity. 

\begin{figure*}
  \begin{minipage}[c]{\textwidth}
  \begin{center}
     \begin{tabular}{c}
     \vspace{-0.15cm}
     \includegraphics[width=0.243\textwidth]{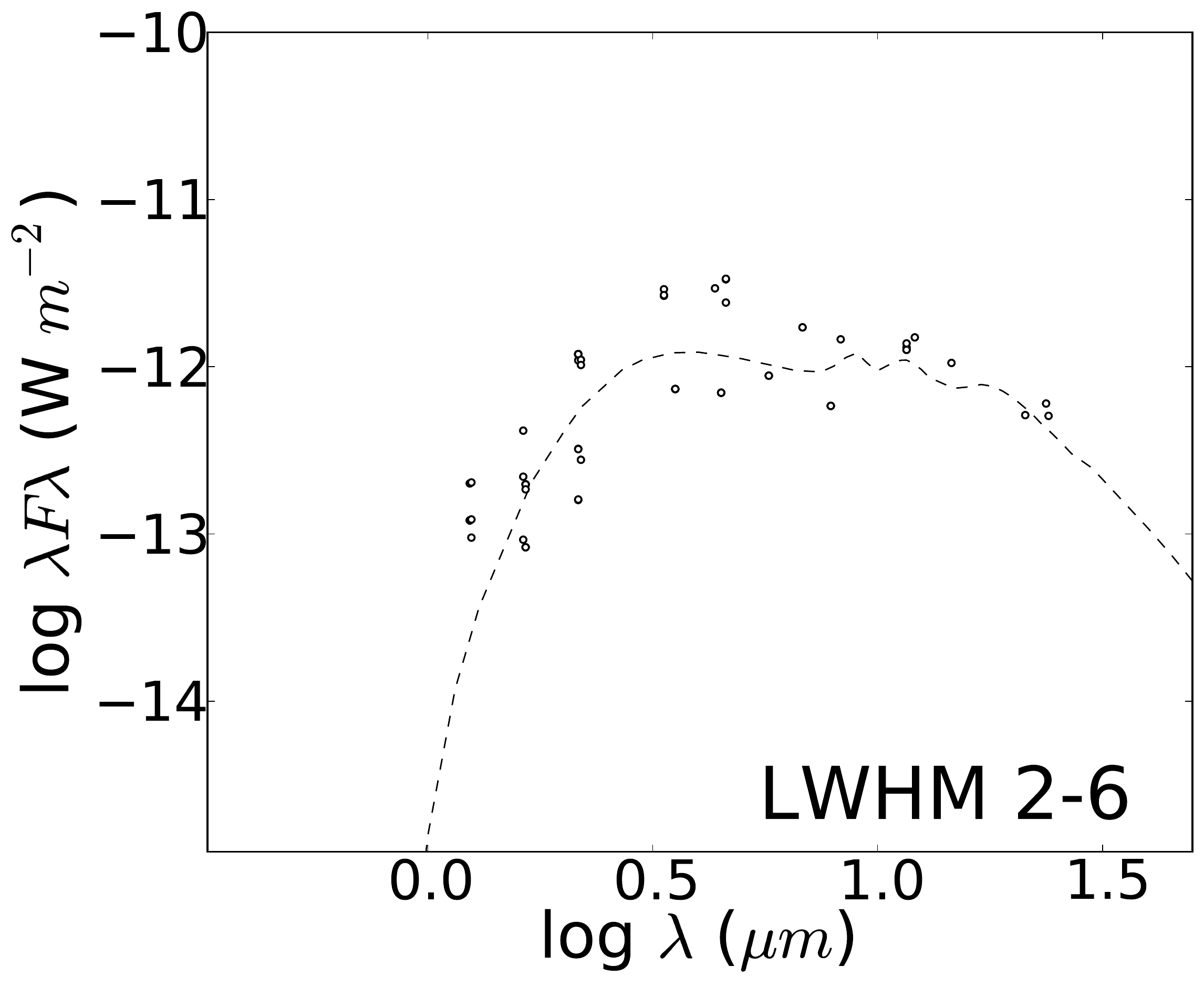} 
     \includegraphics[width=0.243\textwidth]{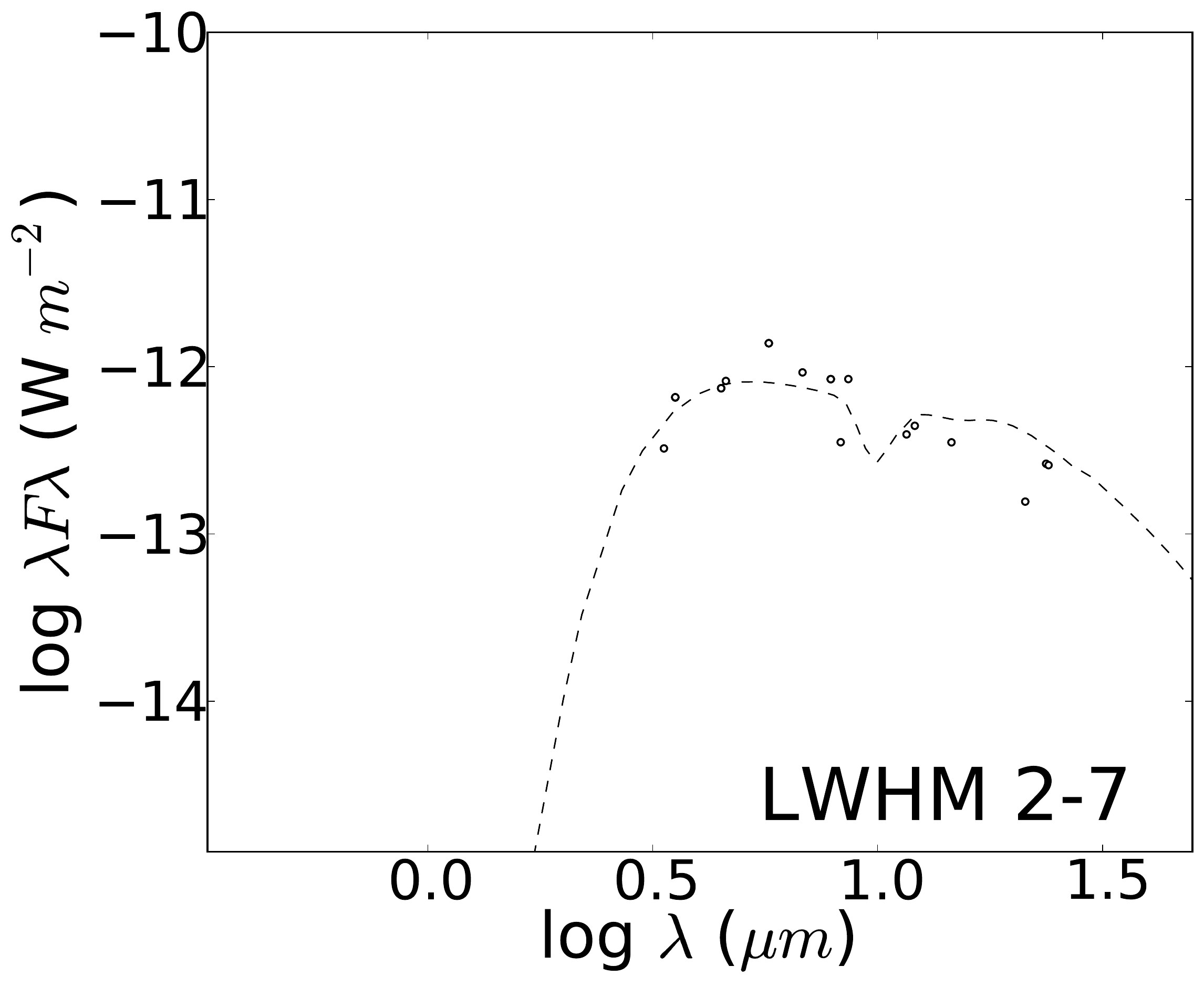} 
     \includegraphics[width=0.243\textwidth]{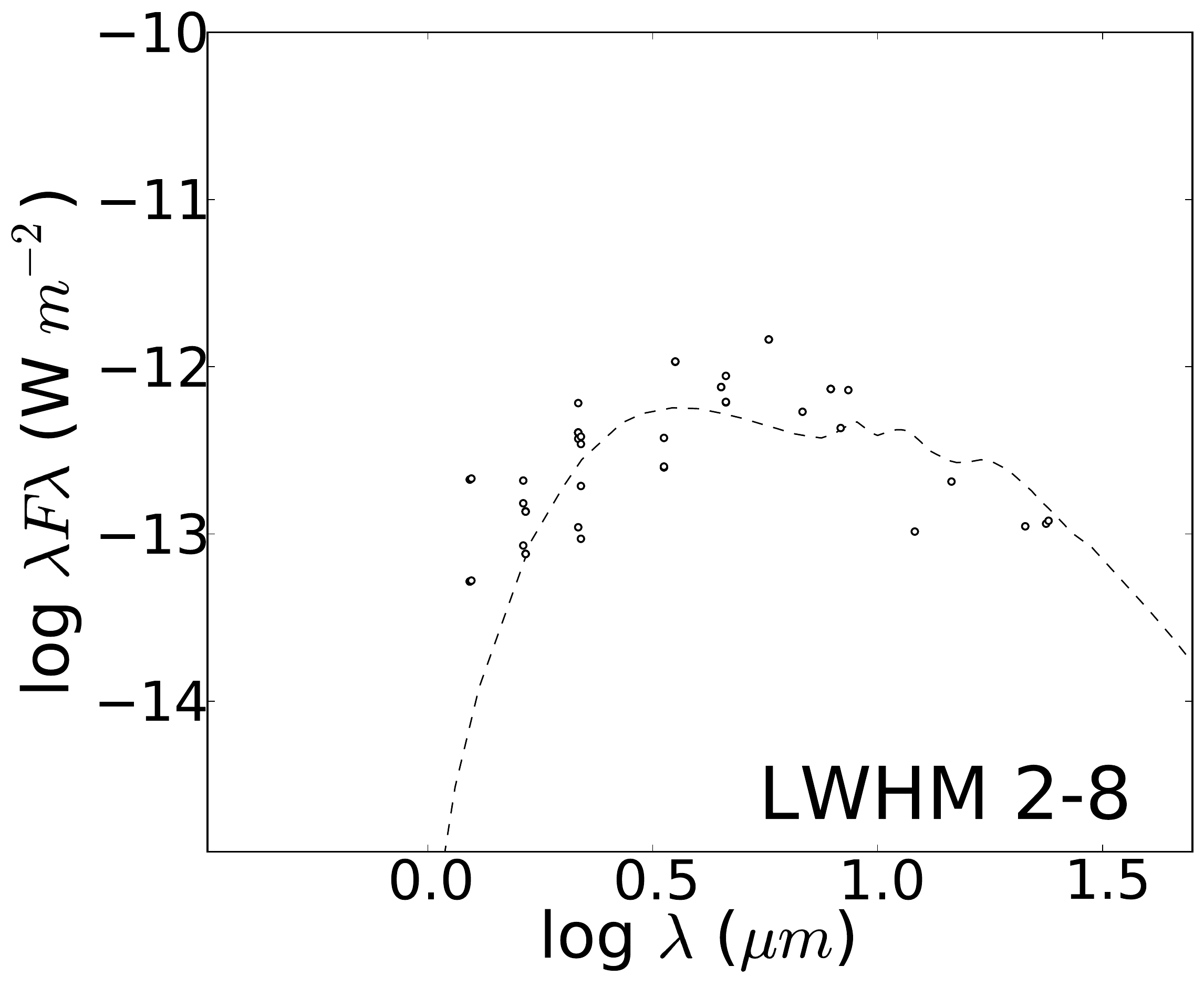} 
     \includegraphics[width=0.243\textwidth]{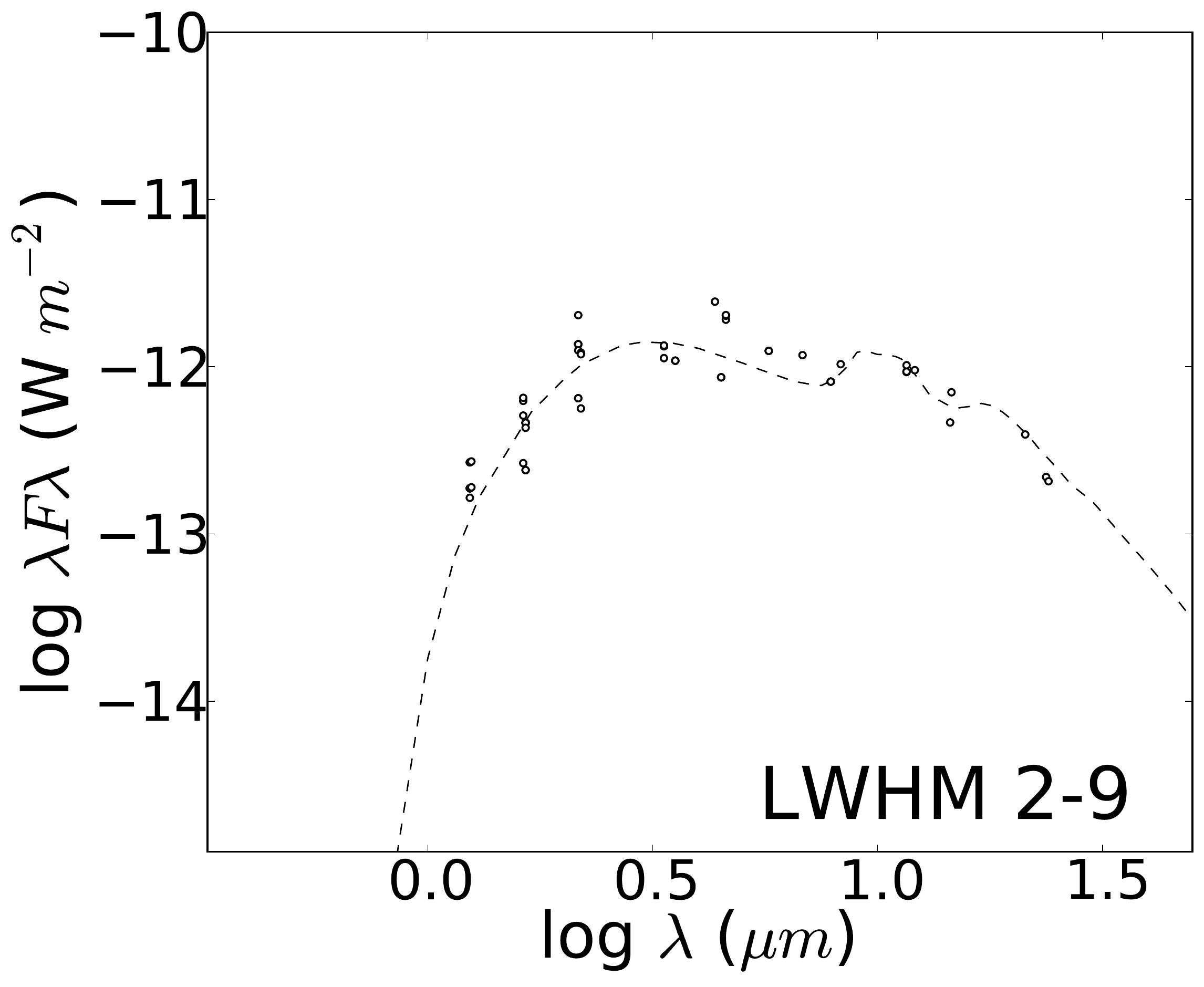} \\
     \vspace{-0.15cm}
     \includegraphics[width=0.243\textwidth]{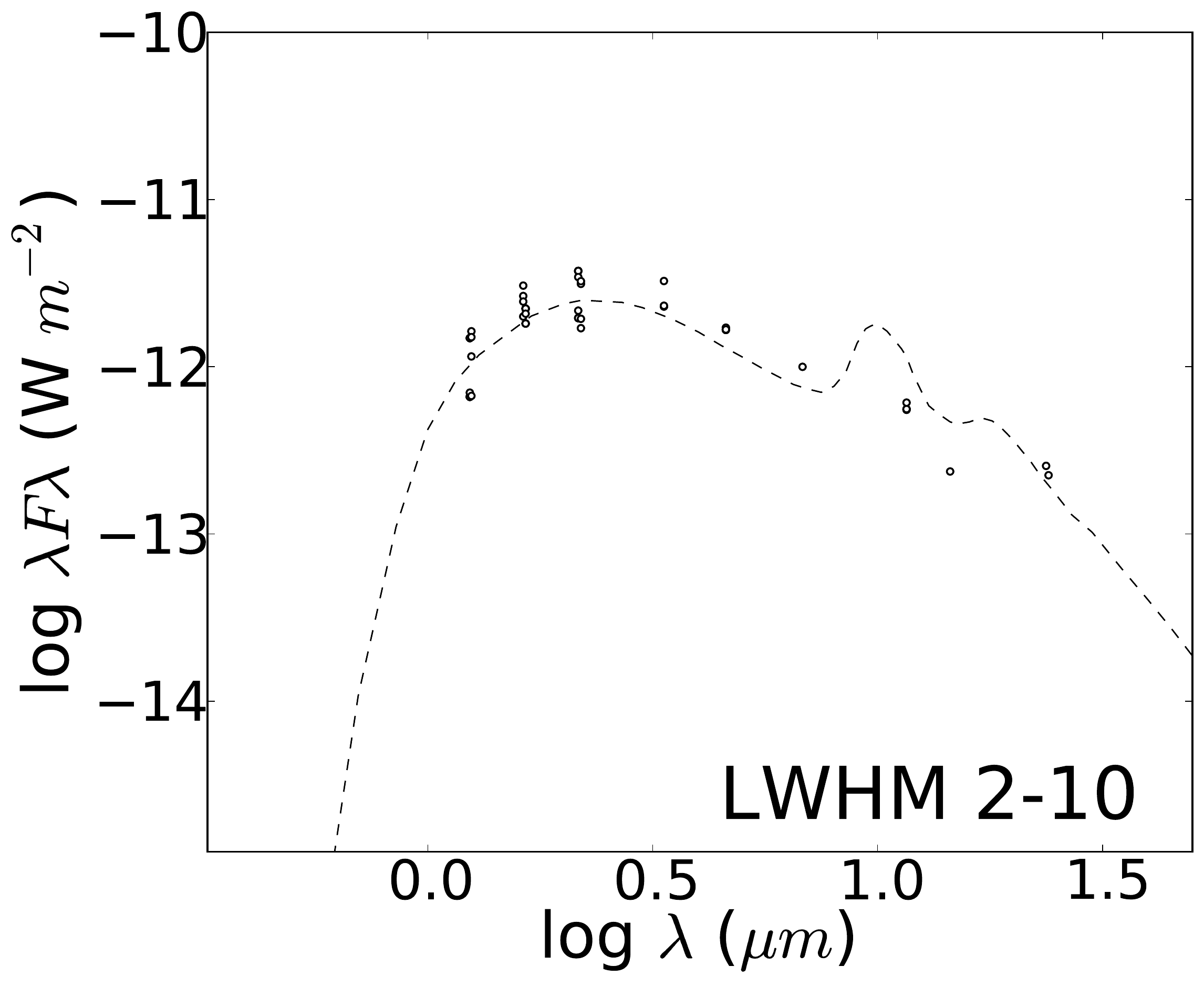}  
     \includegraphics[width=0.243\textwidth]{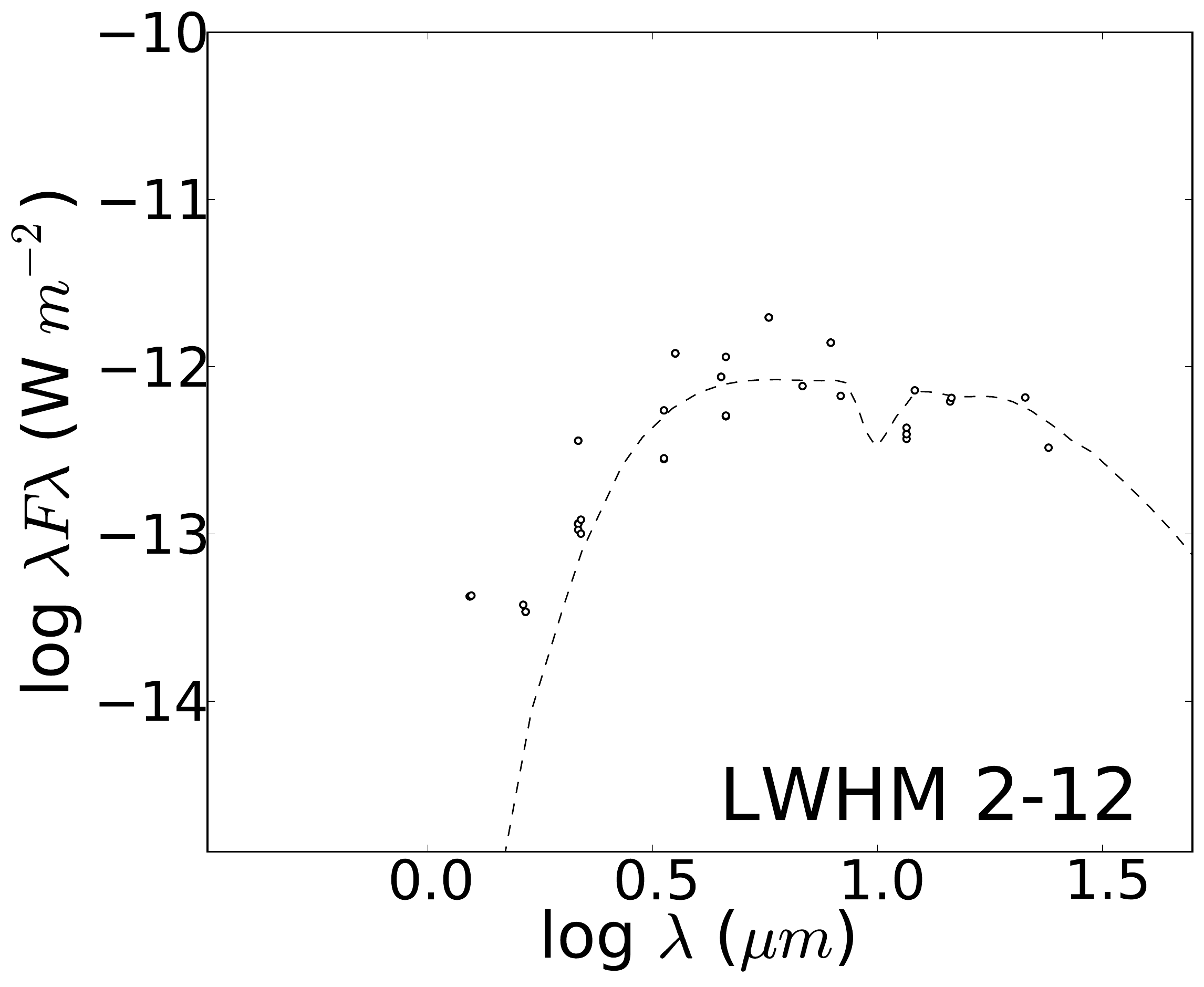}  
     \includegraphics[width=0.243\textwidth]{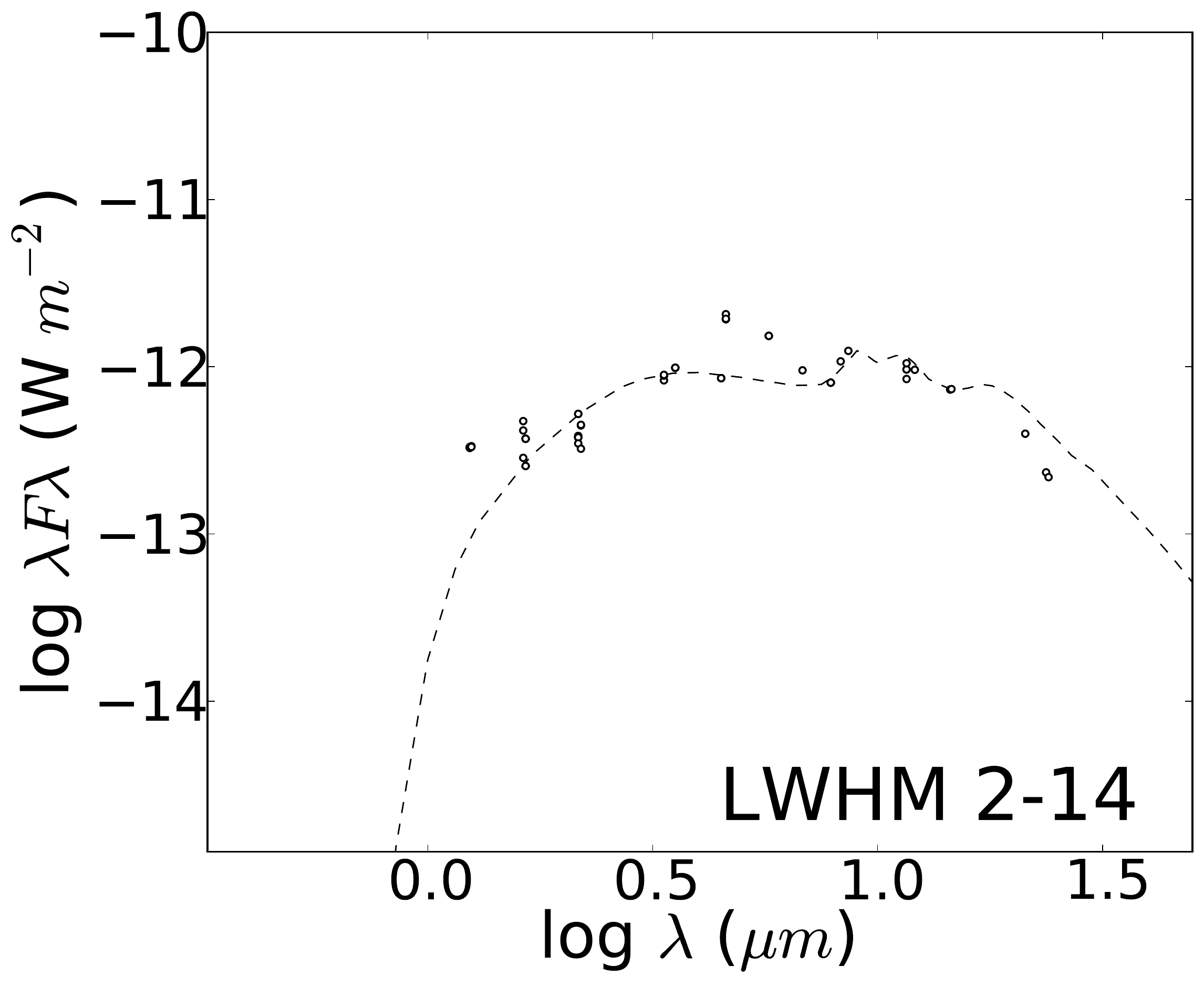}
     \includegraphics[width=0.243\textwidth]{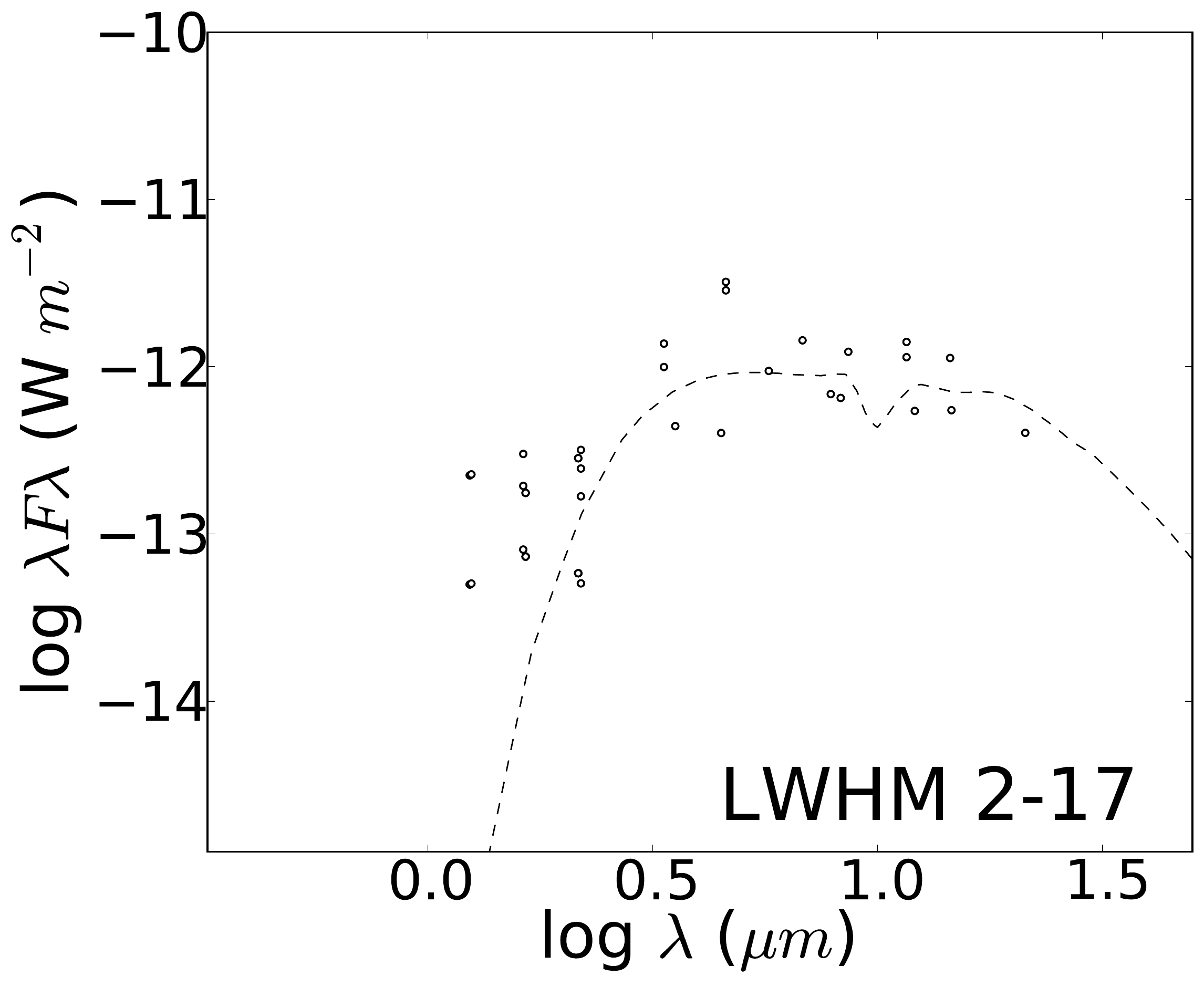}  \\
     \vspace{-0.15cm}
     \includegraphics[width=0.243\textwidth]{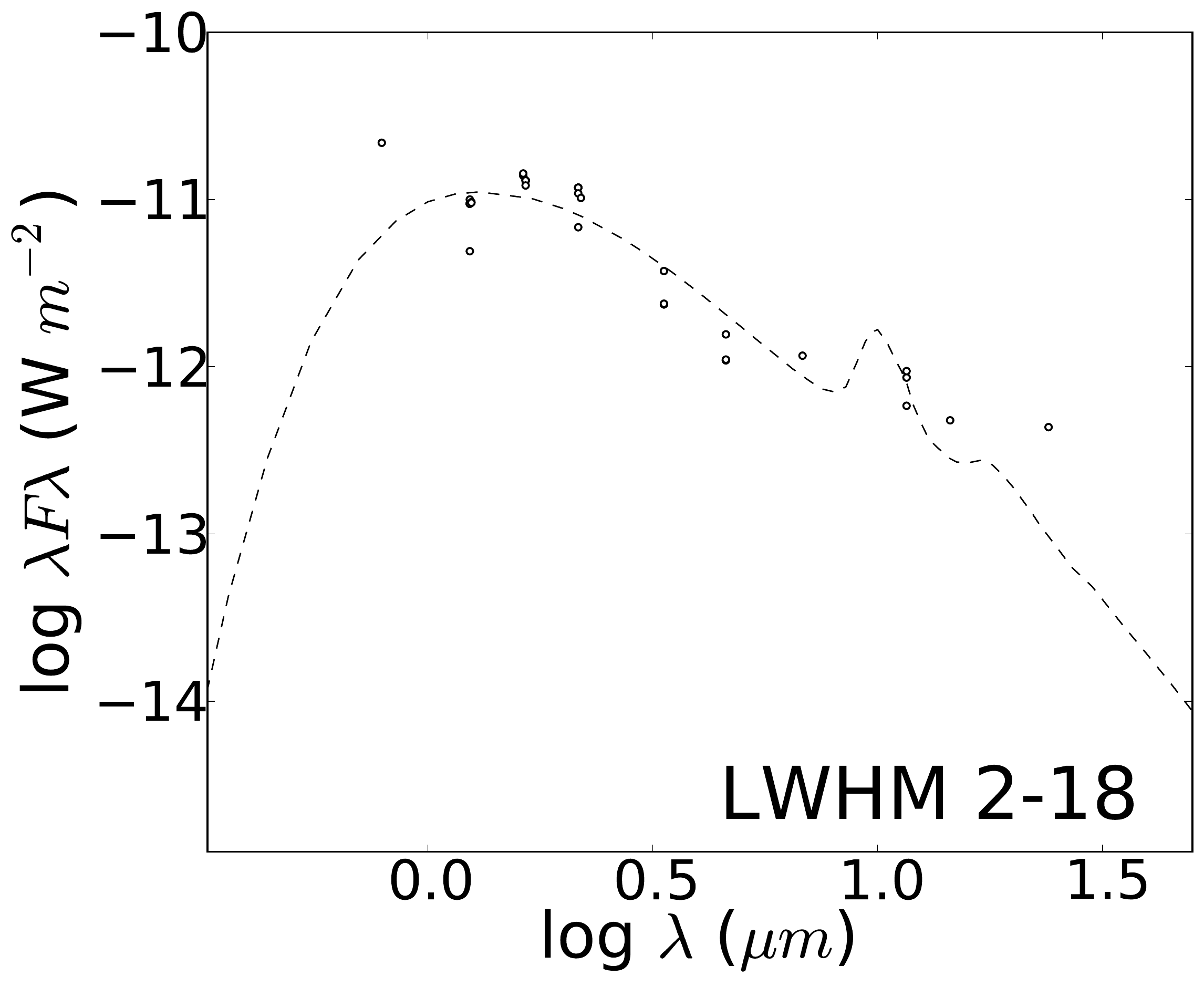}  
     \includegraphics[width=0.243\textwidth]{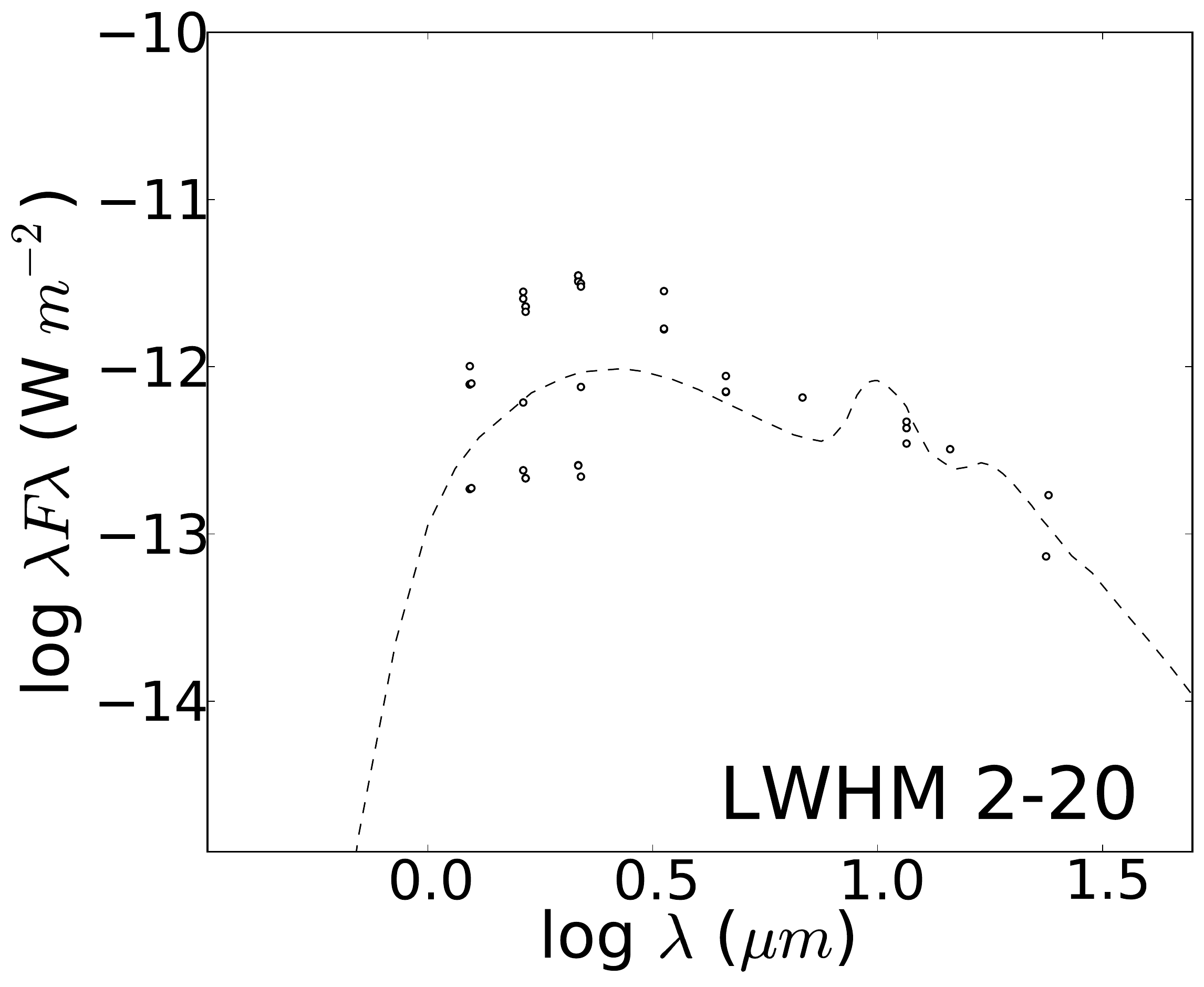} 
     \includegraphics[width=0.243\textwidth]{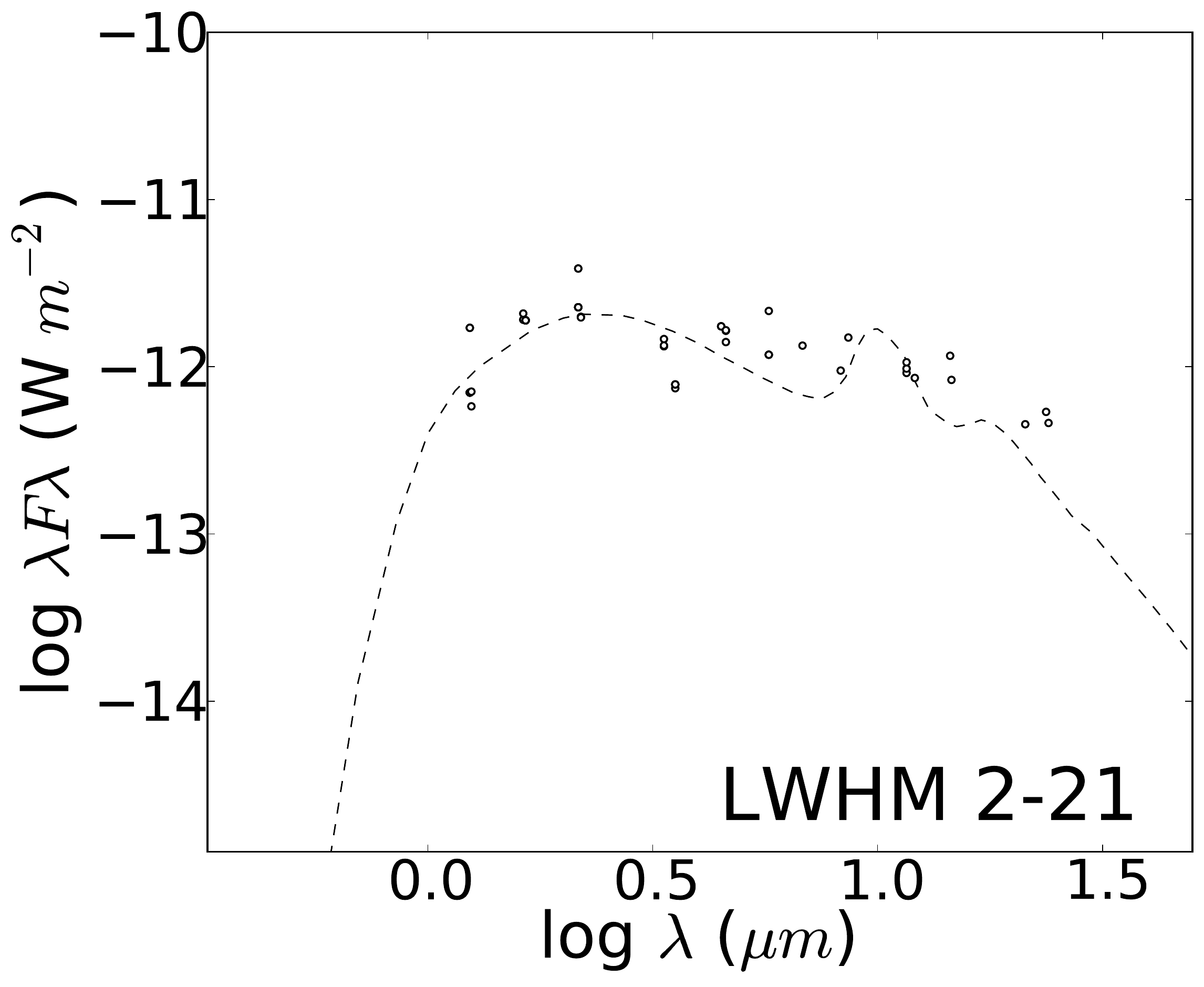}
     \includegraphics[width=0.243\textwidth]{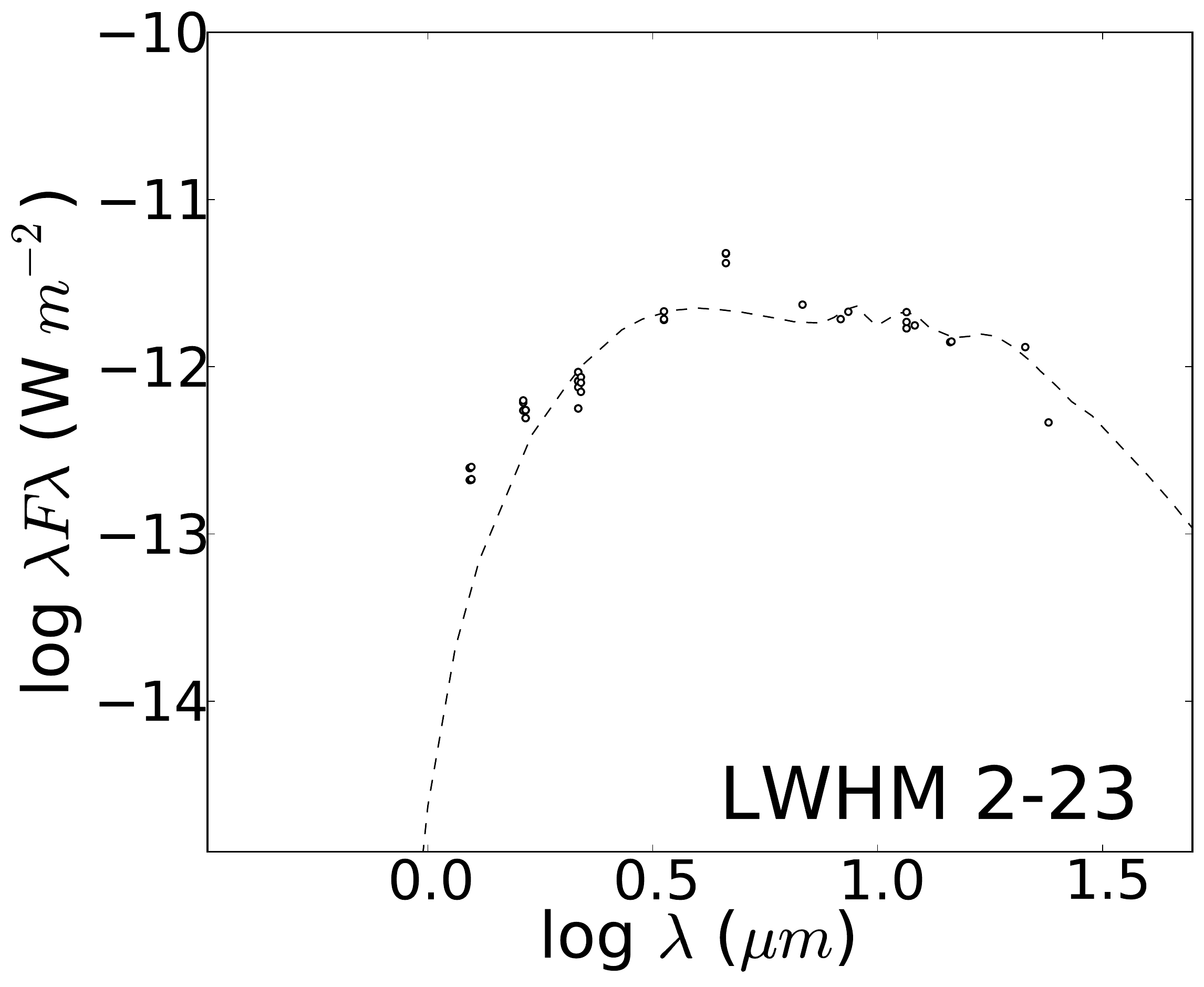}  \\
     \vspace{-0.15cm}
     \includegraphics[width=0.243\textwidth]{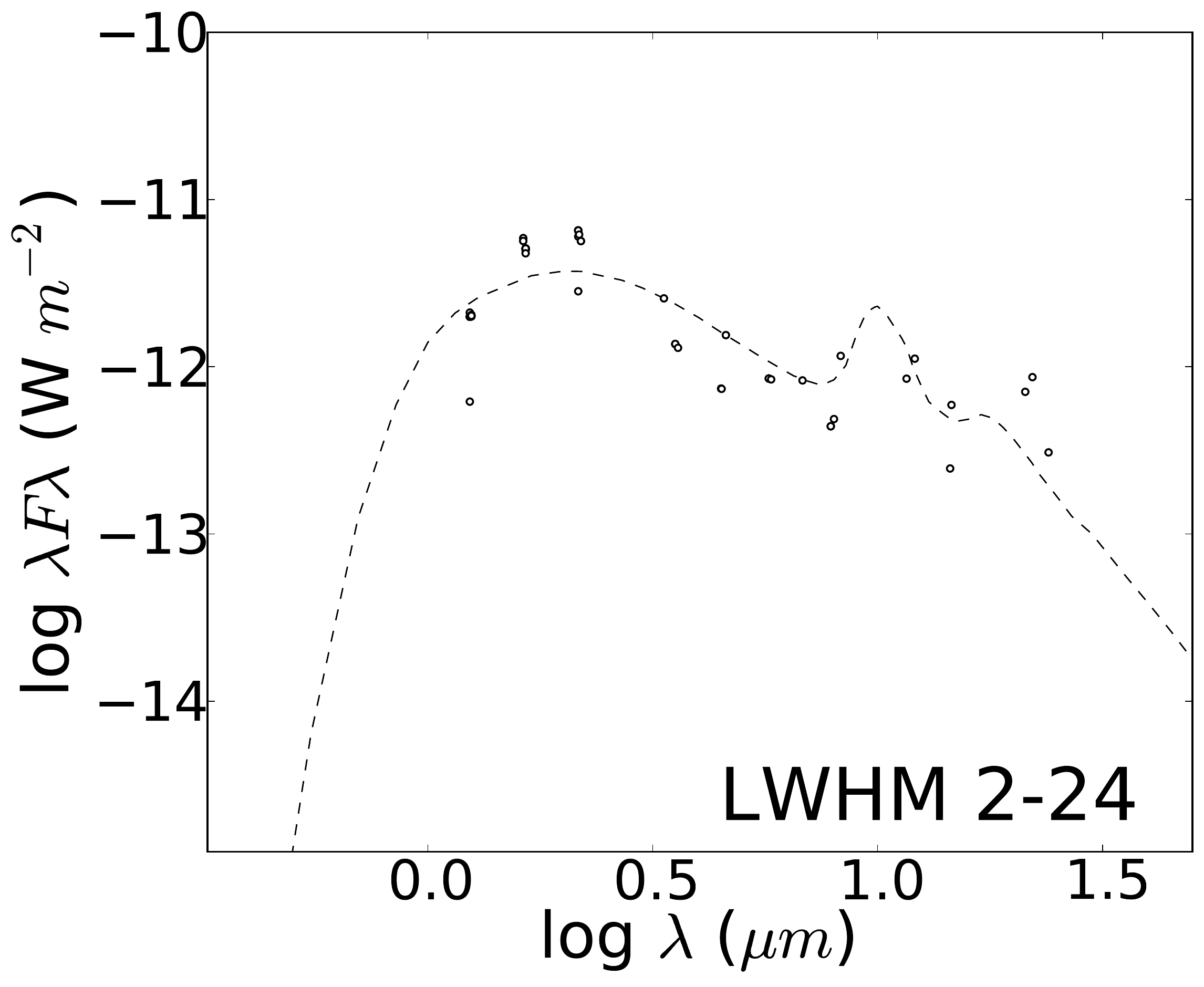}  
     \includegraphics[width=0.243\textwidth]{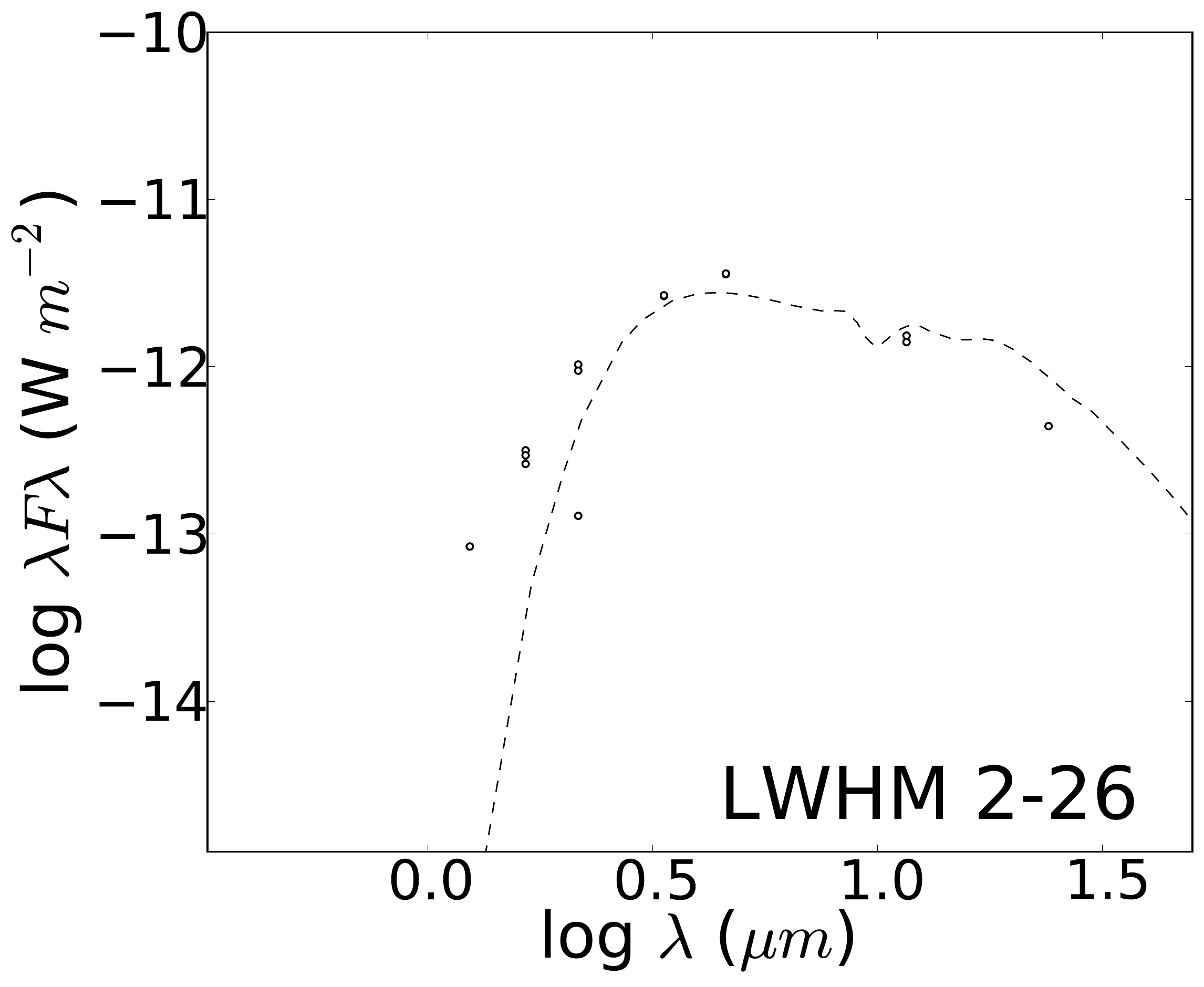}  
     \includegraphics[width=0.243\textwidth]{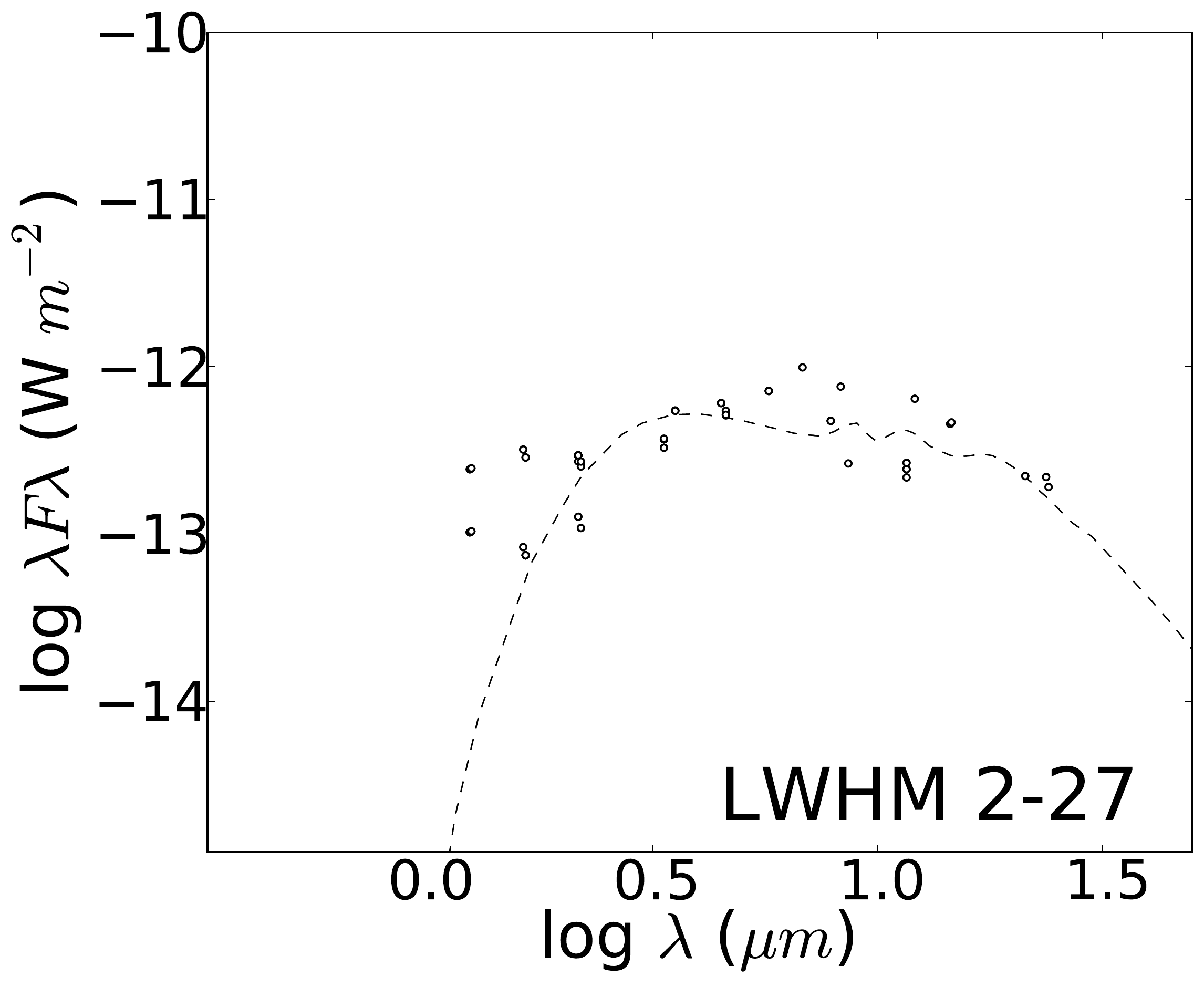}
     \includegraphics[width=0.243\textwidth]{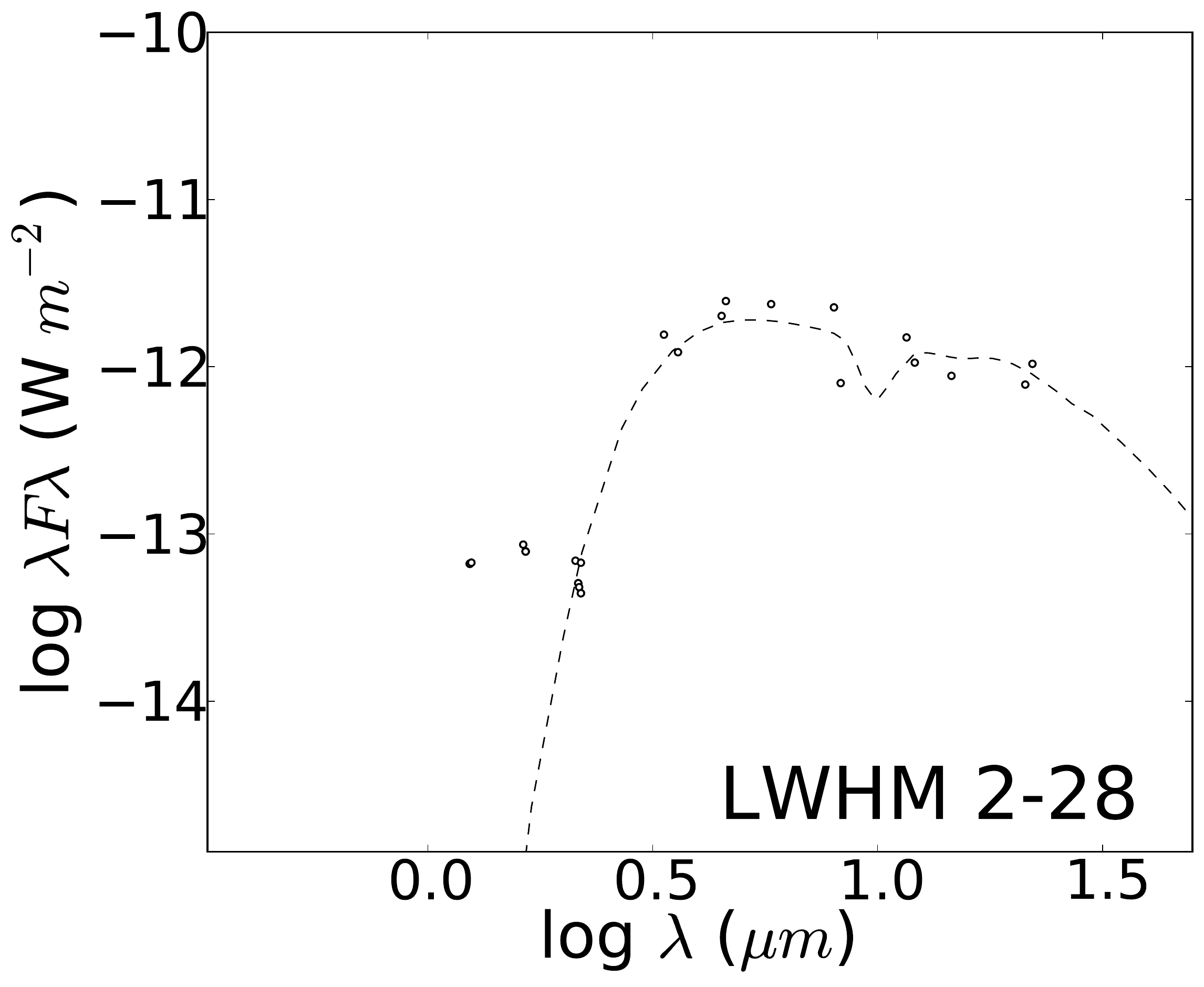}  \\
     \vspace{-0.15cm}
     \includegraphics[width=0.243\textwidth]{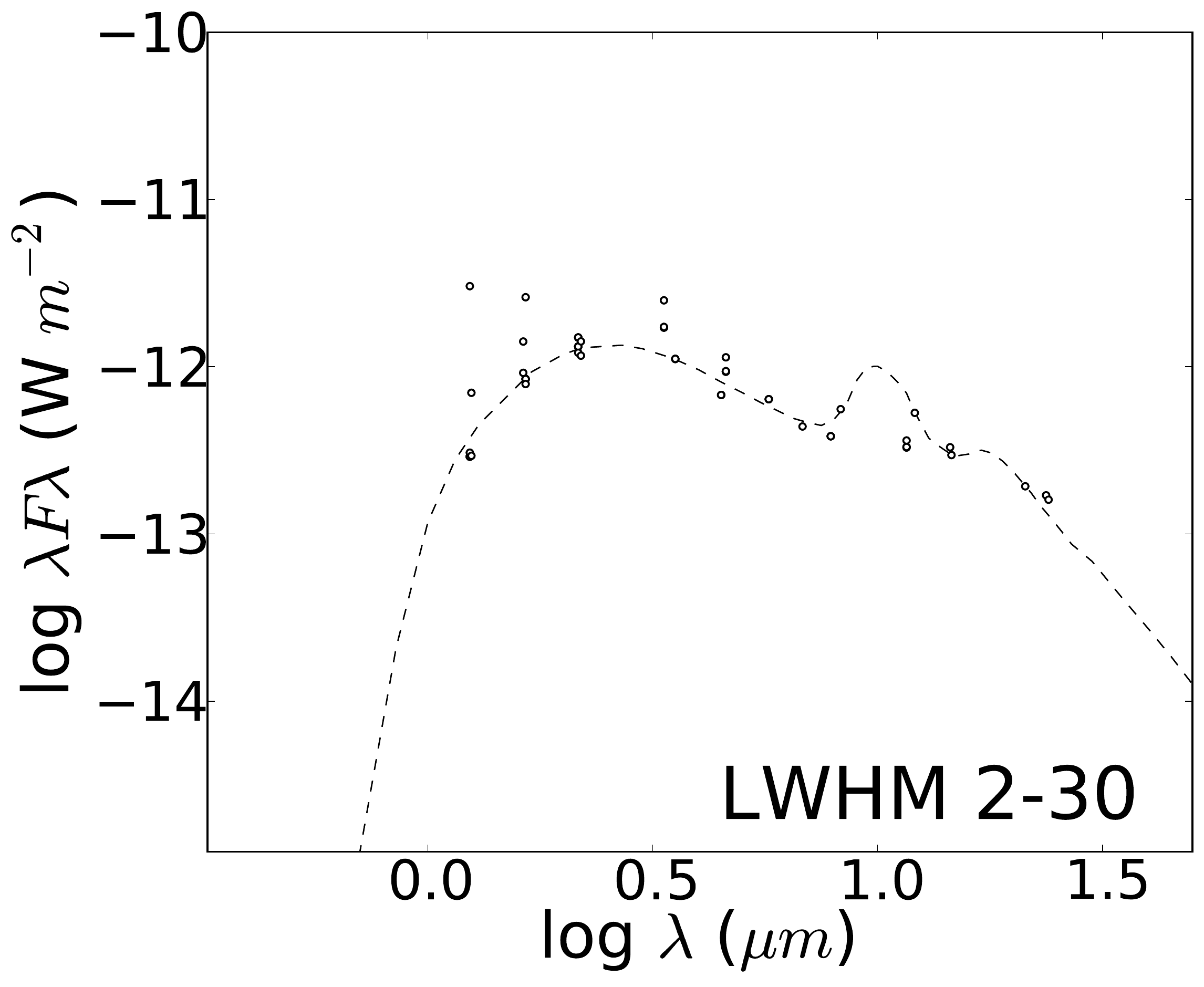}  
     \includegraphics[width=0.243\textwidth]{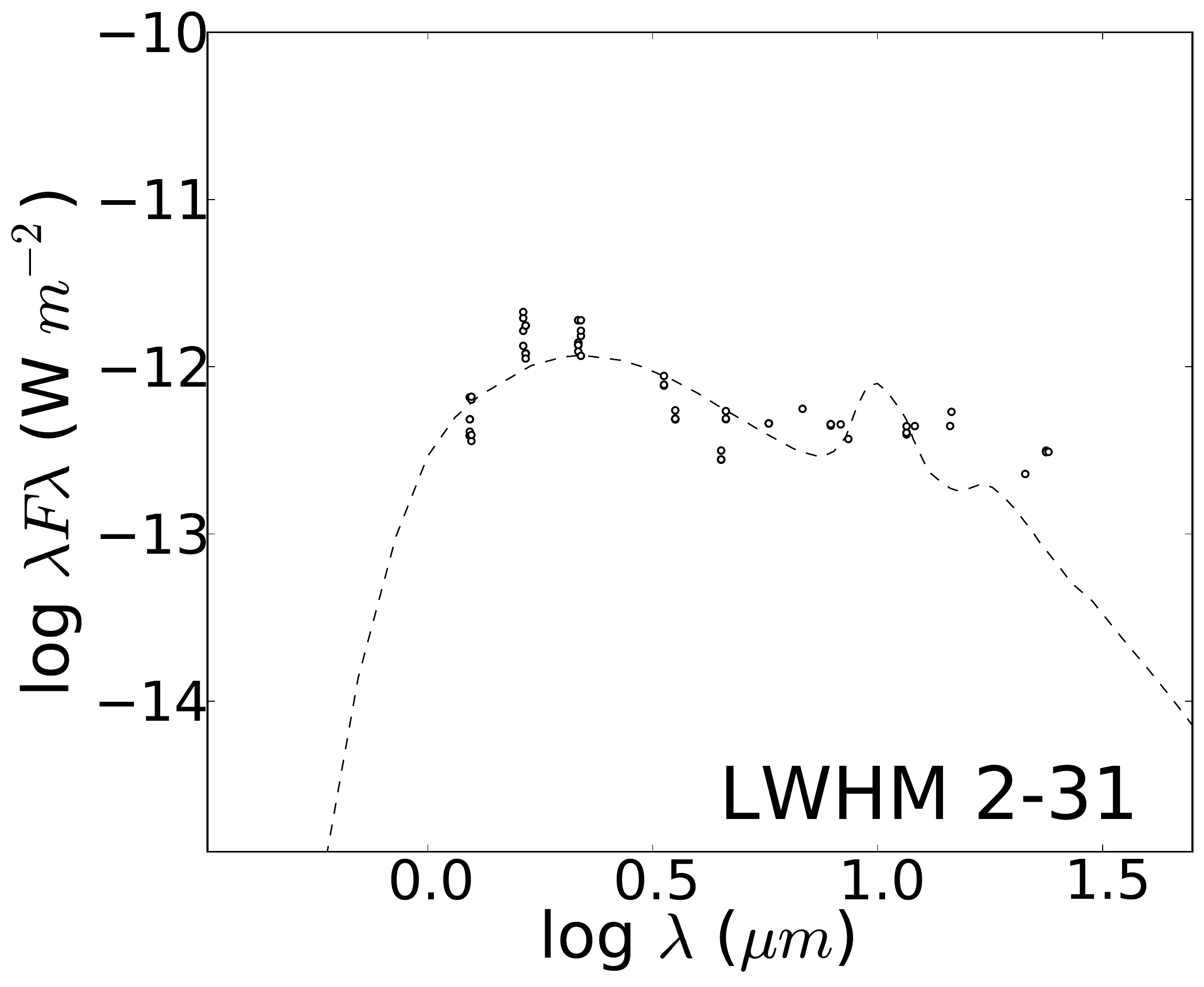}  
     \includegraphics[width=0.243\textwidth]{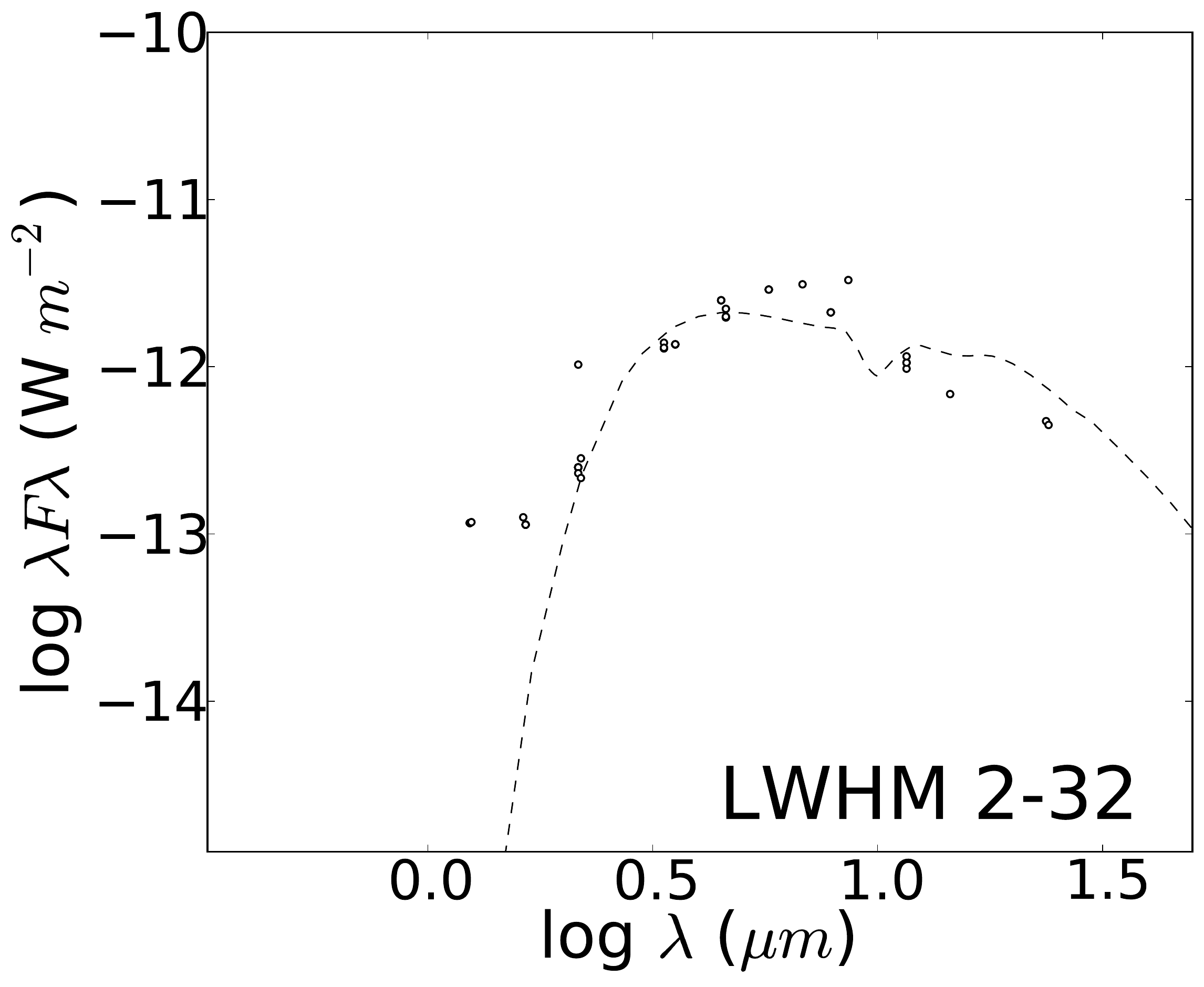}
     \includegraphics[width=0.243\textwidth]{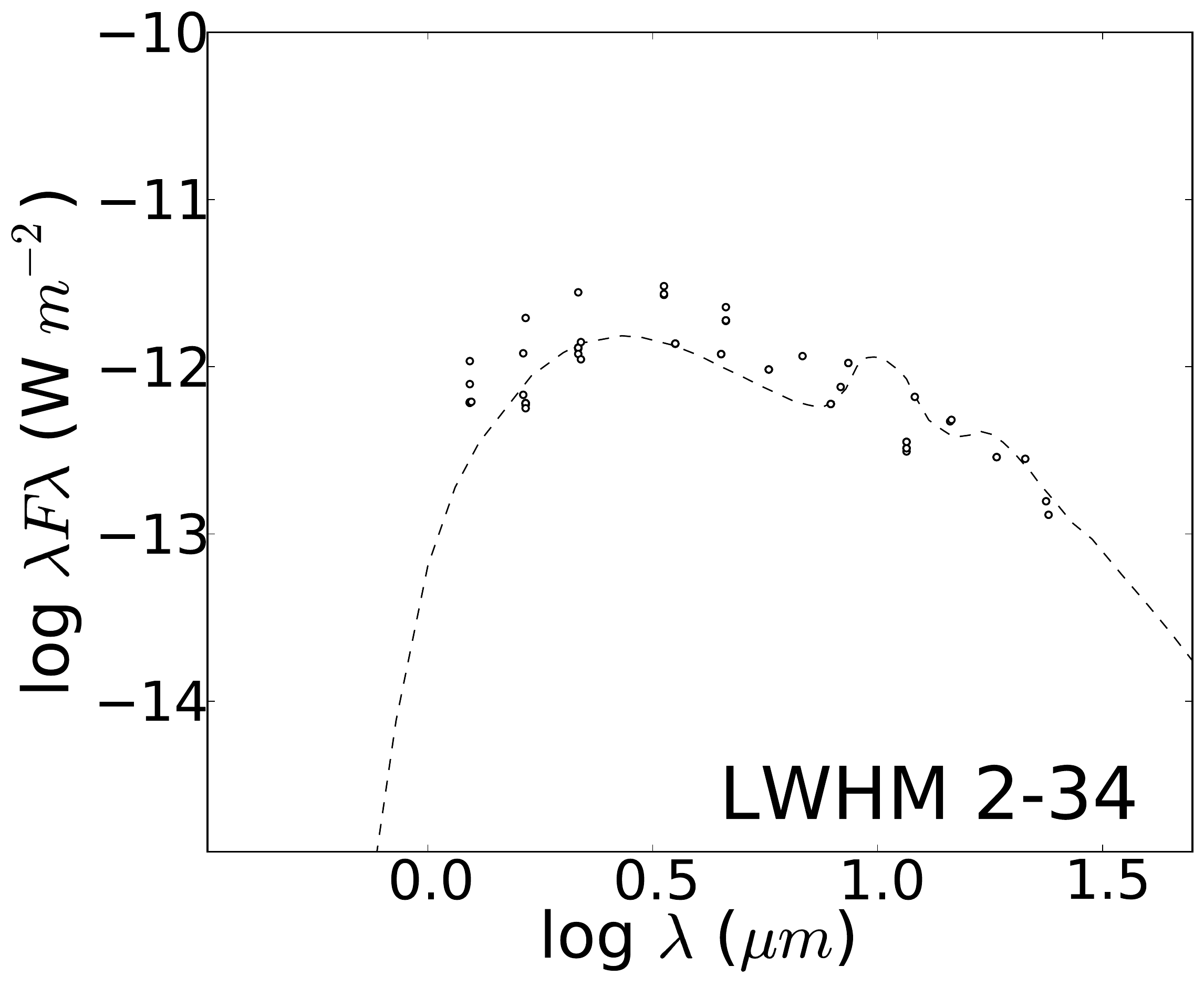}  \\

     \includegraphics[width=0.243\textwidth]{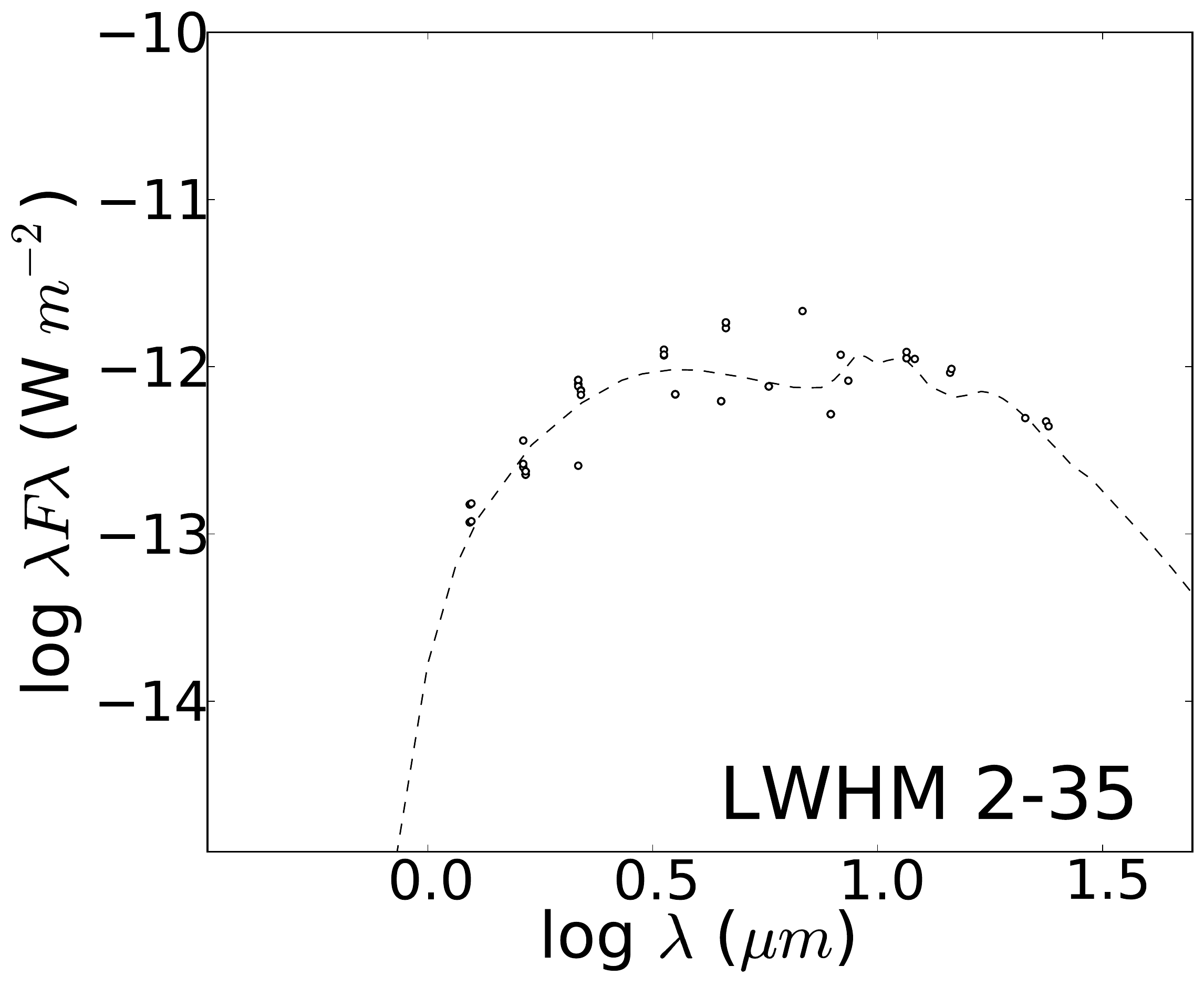}  
     \includegraphics[width=0.243\textwidth]{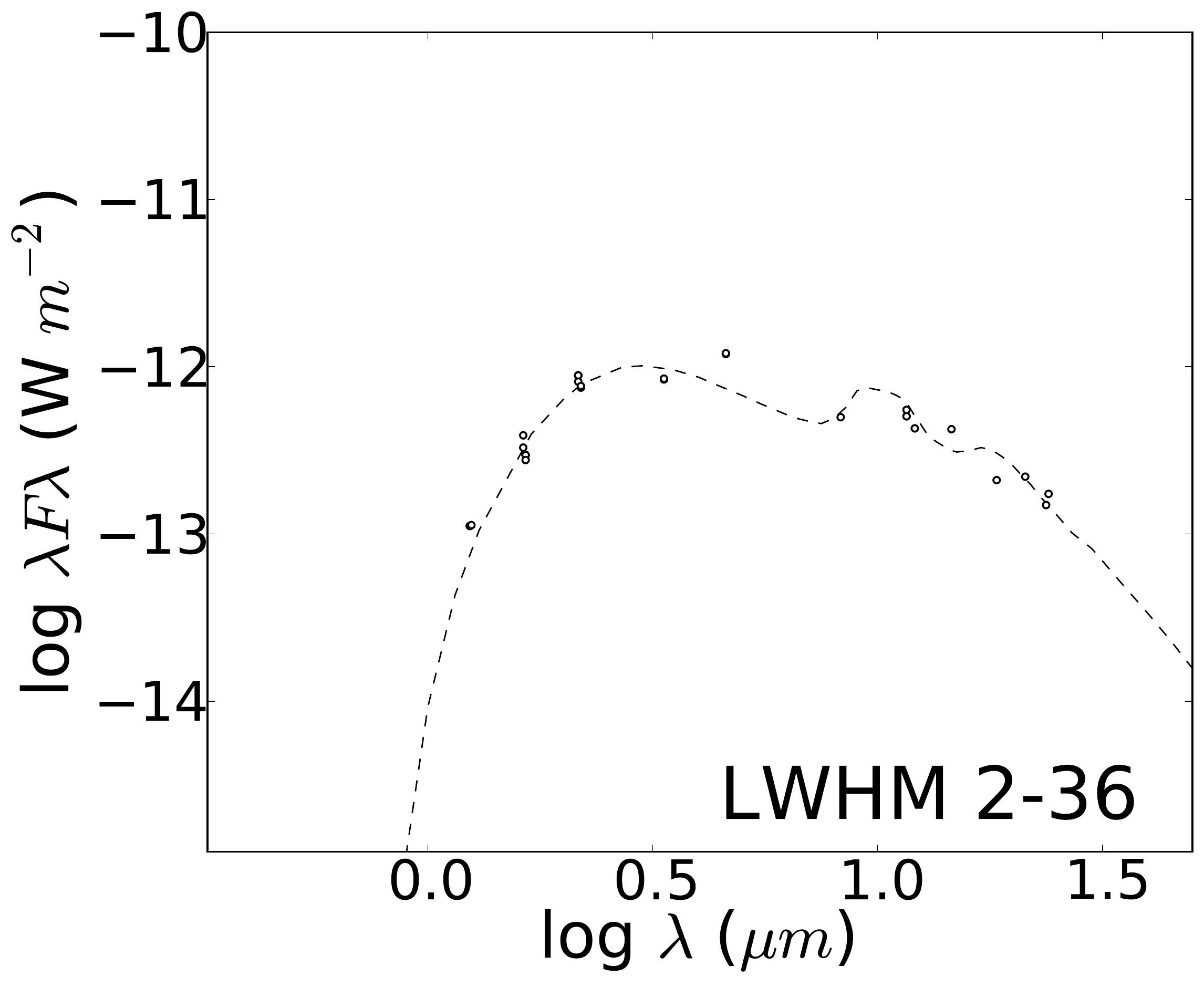} 
     \includegraphics[width=0.243\textwidth]{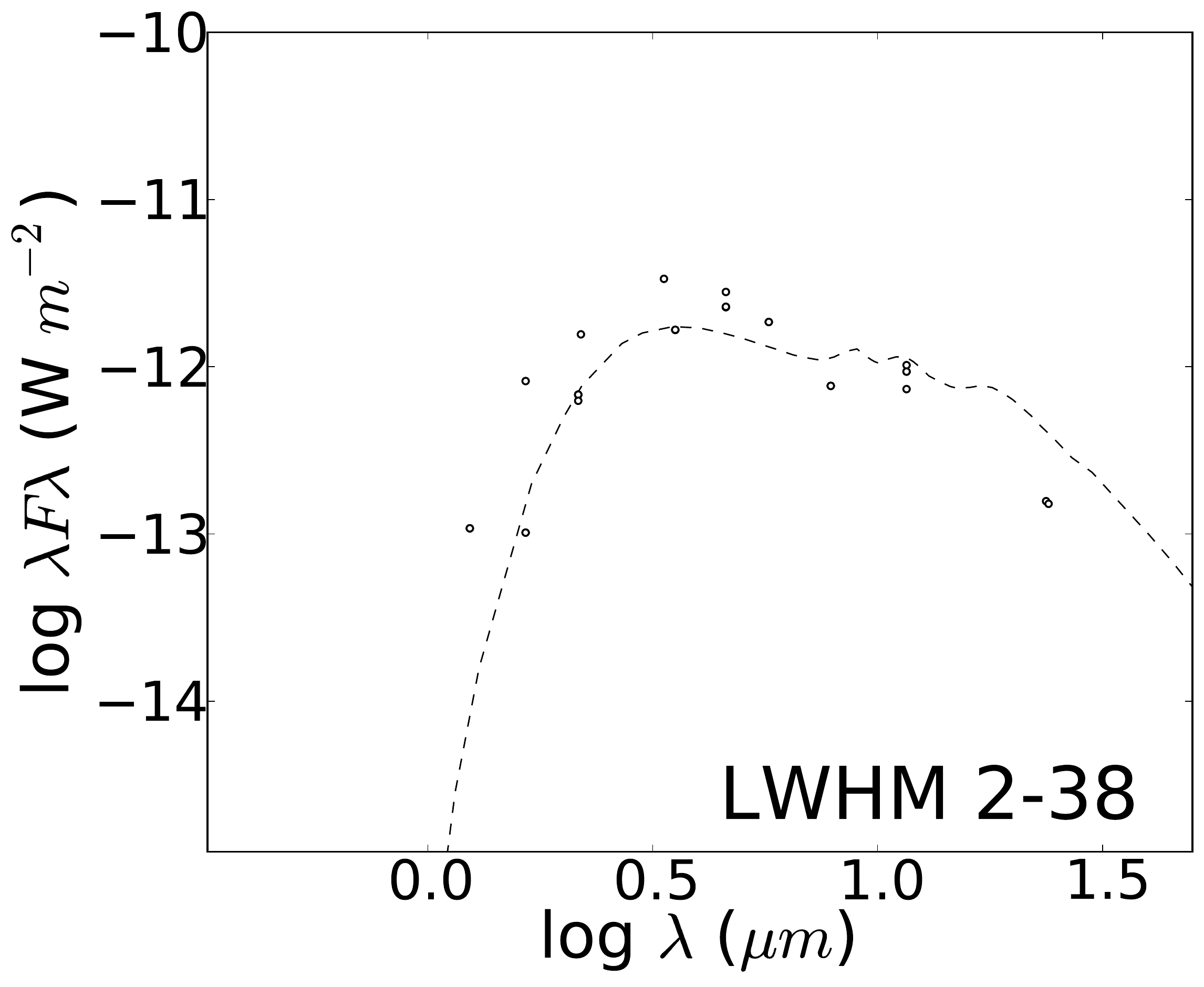}
     \includegraphics[width=0.243\textwidth]{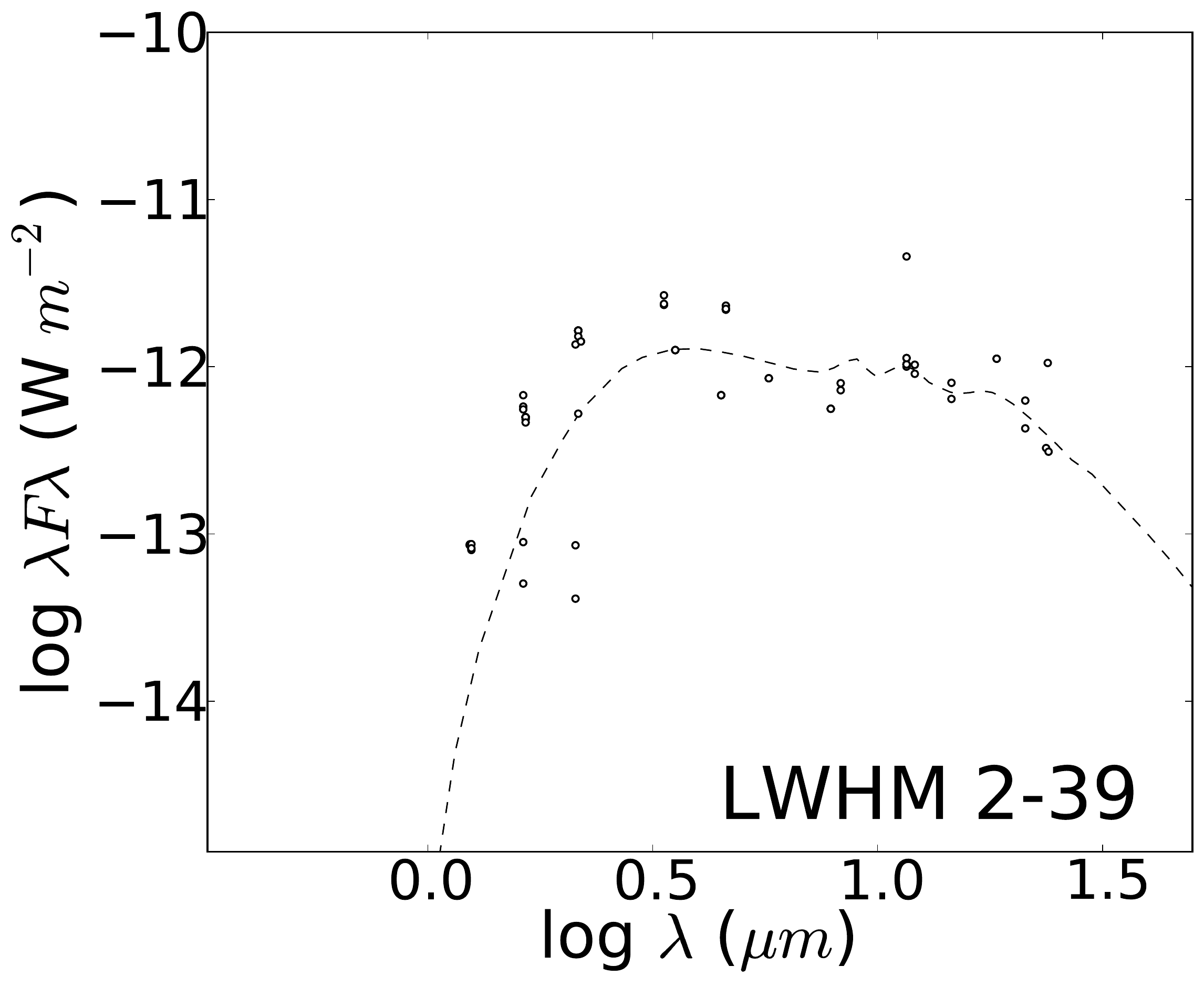}  \\ \\
     \end{tabular}
   \end{center}
   \vspace{-0.4cm}
   \caption{The SED fitting of \textsc{dusty} models to the dereddened photometry from our Galactic Centre sources.}
   \end{minipage}%
\end{figure*}

\renewcommand{\thefigure}{A\arabic{figure}}
\addtocounter{figure}{-1}

\begin{figure*}
  \begin{minipage}[c]{\textwidth}
  \begin{center}
     \begin{tabular}{c}
     \vspace{-0.15cm}
     \includegraphics[width=0.243\textwidth]{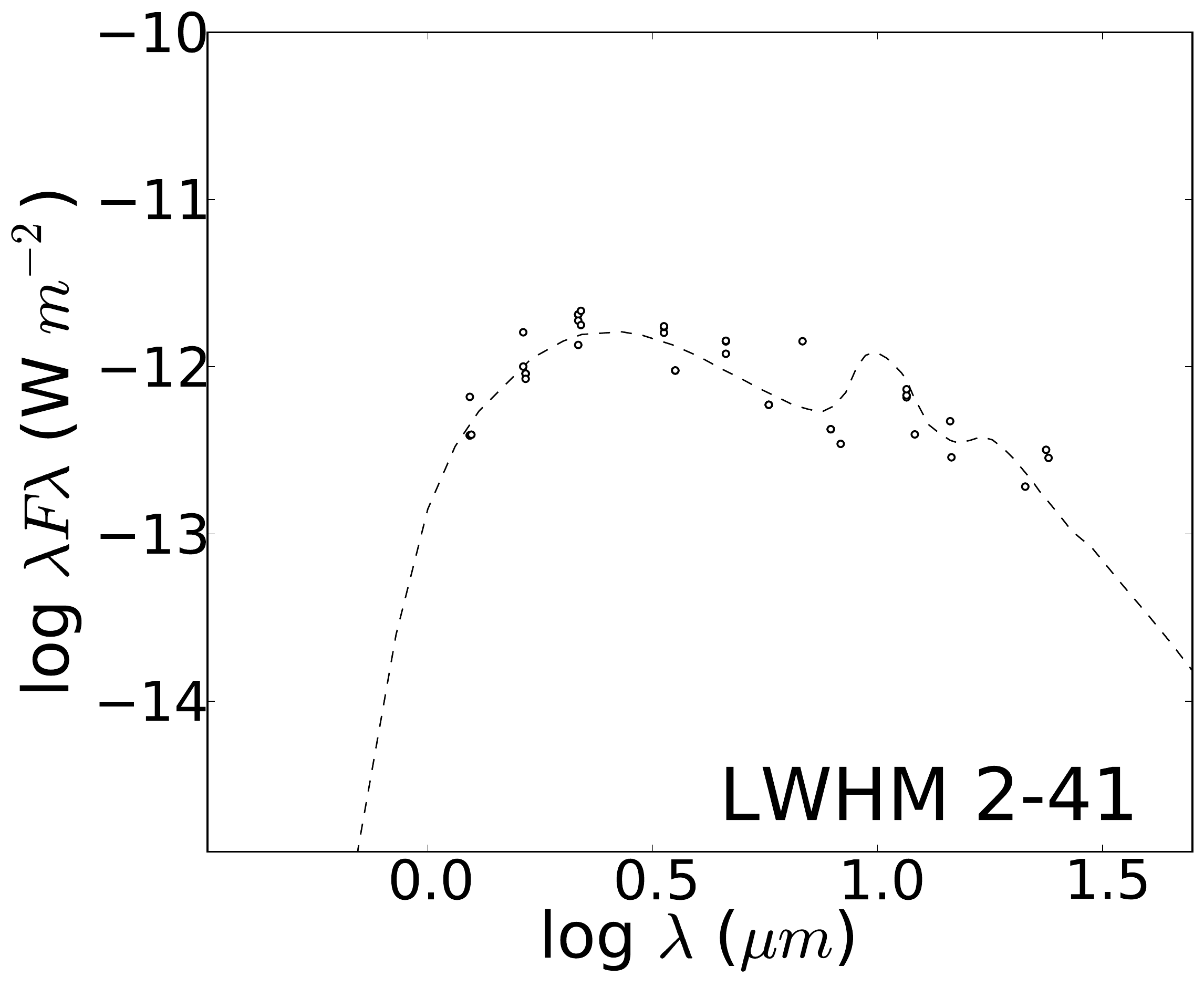}  
     \includegraphics[width=0.243\textwidth]{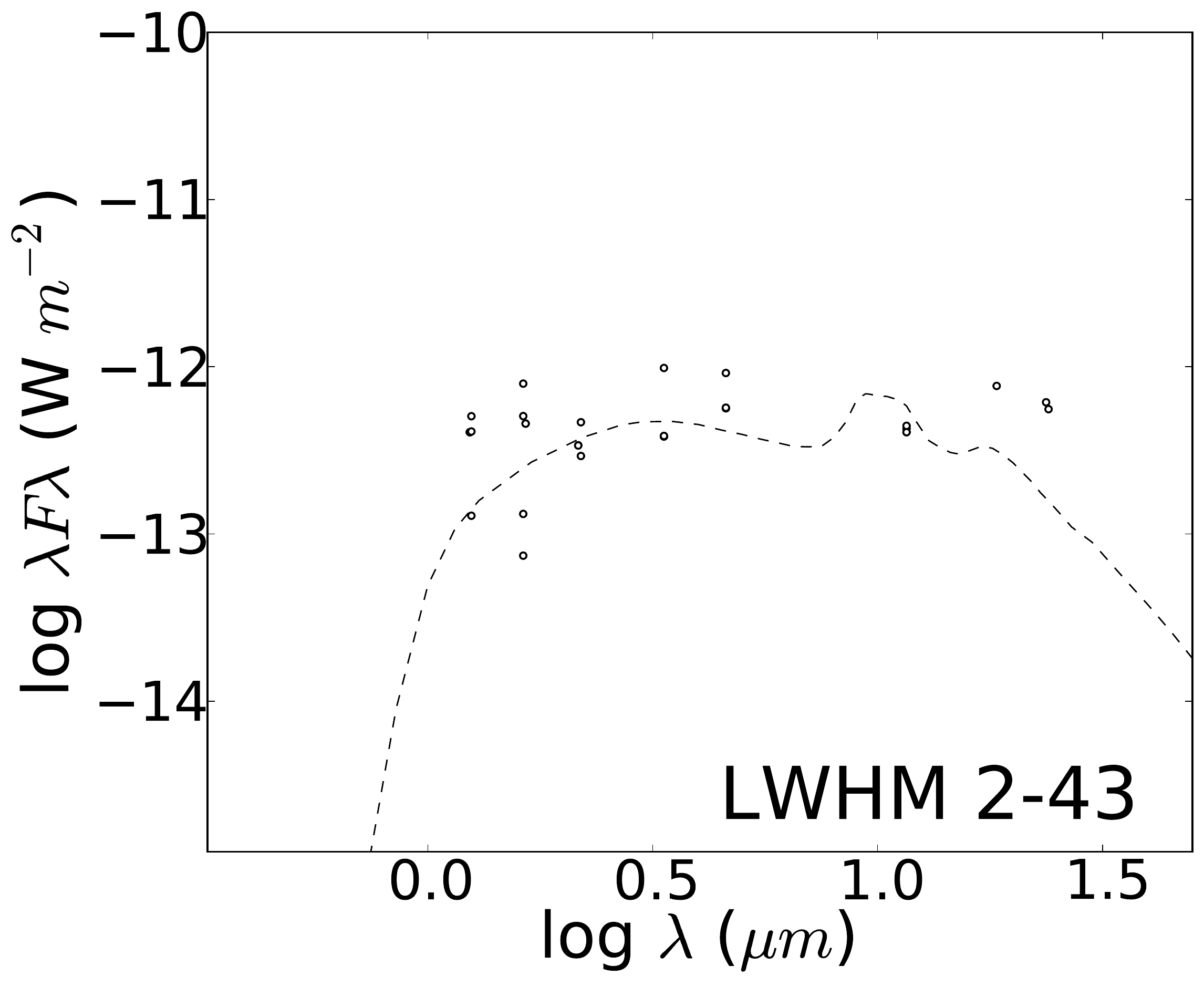}  
     \includegraphics[width=0.243\textwidth]{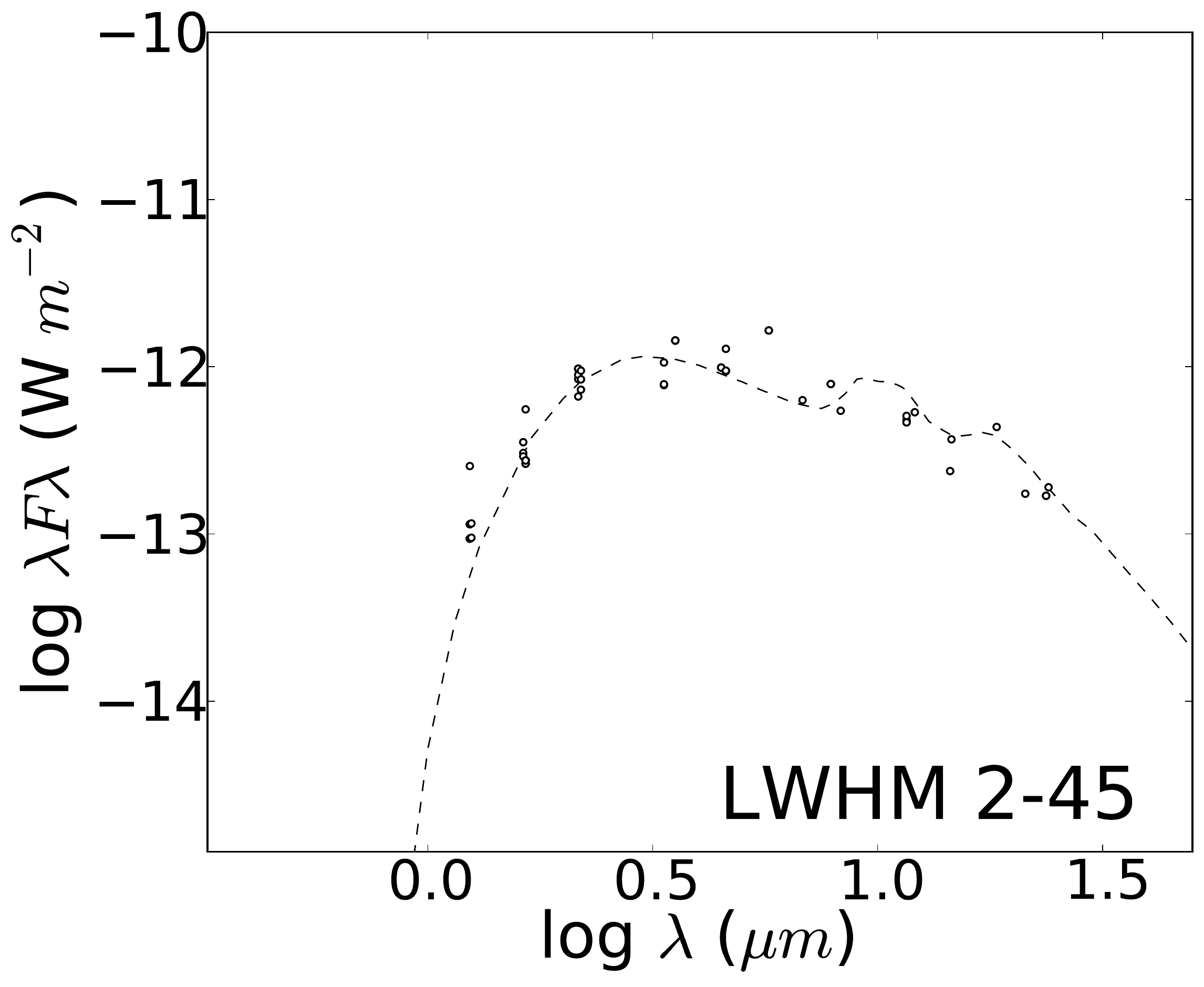}
     \includegraphics[width=0.243\textwidth]{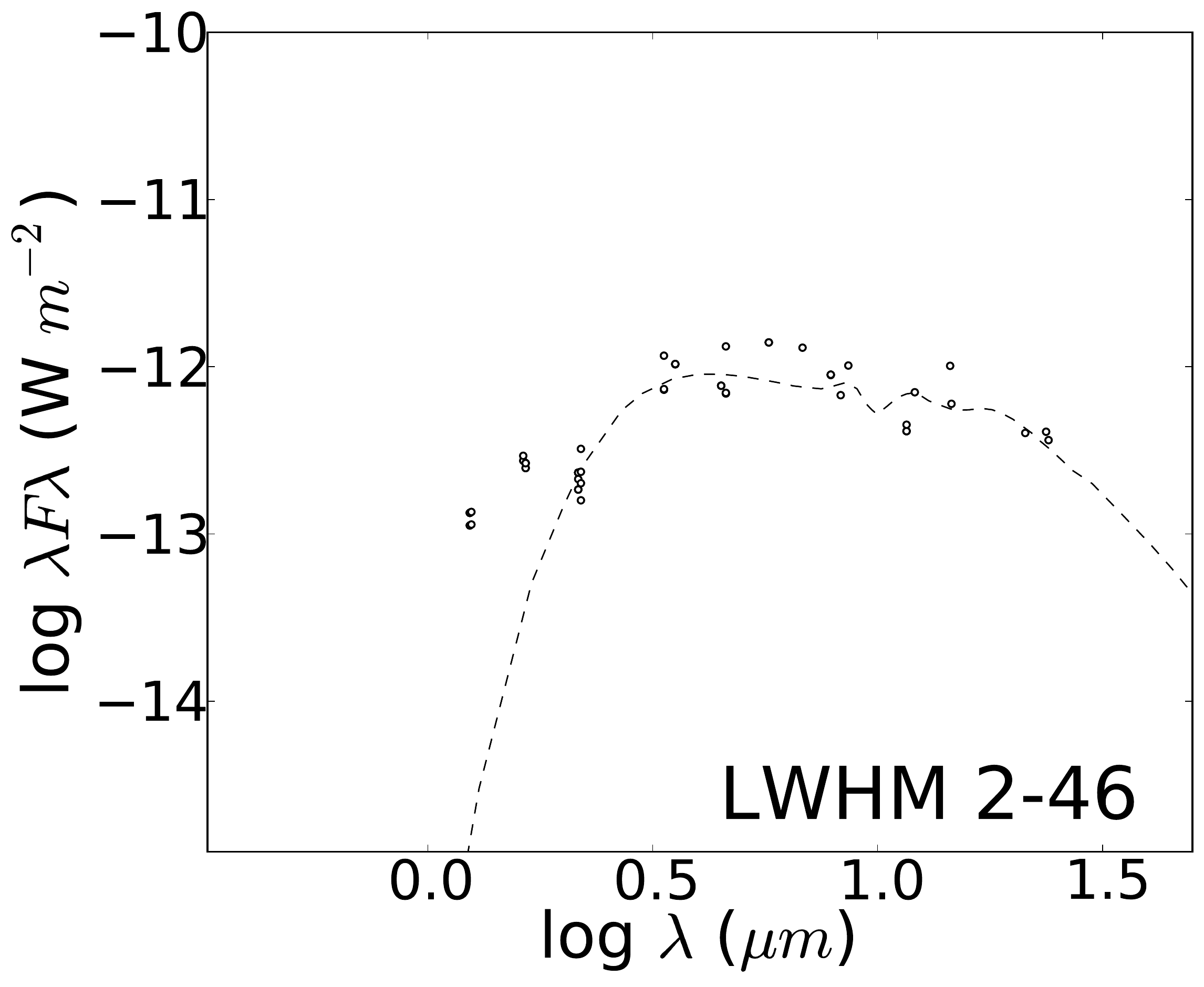}  \\
     \vspace{-0.15cm}
     \includegraphics[width=0.243\textwidth]{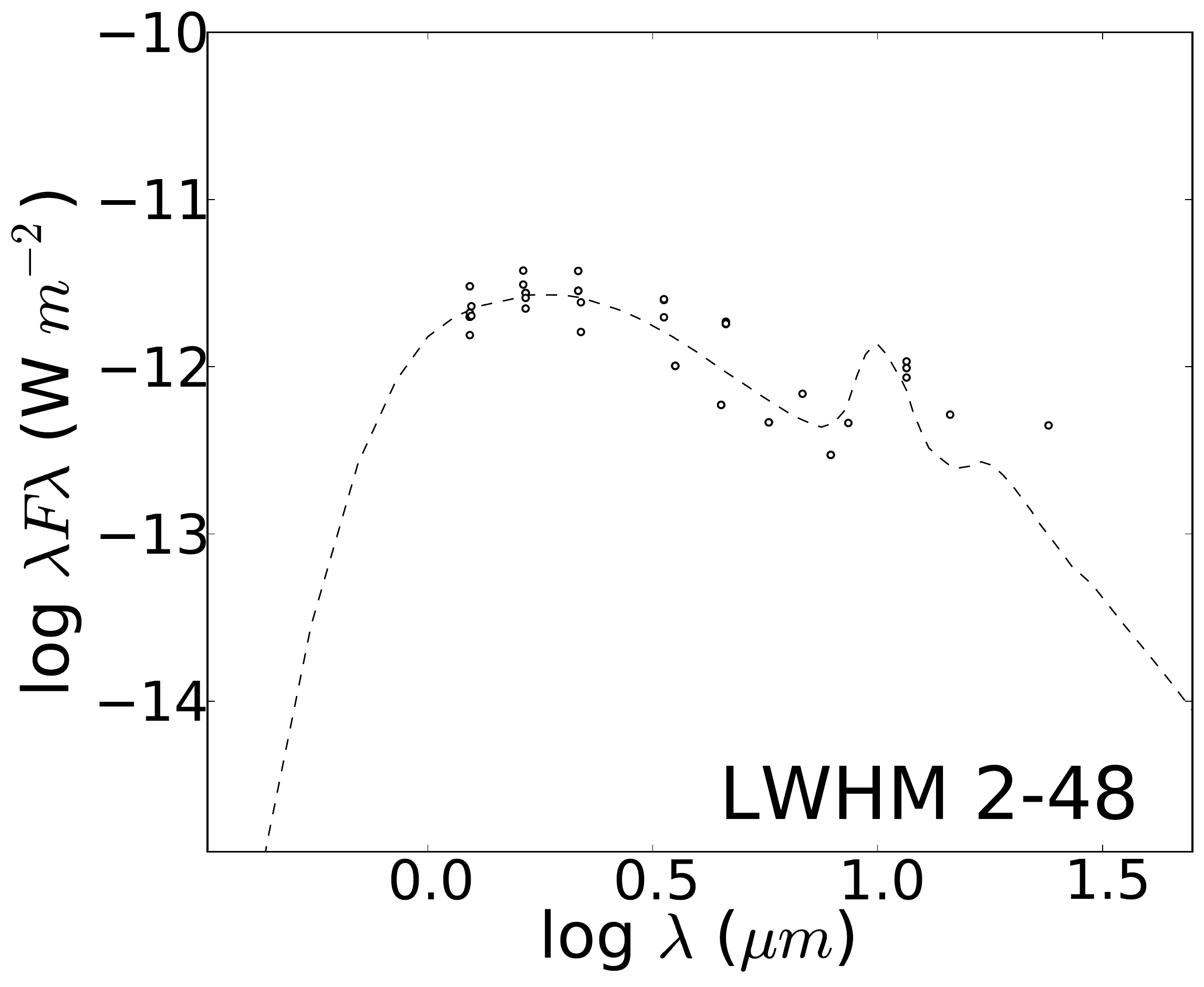}  
     \includegraphics[width=0.243\textwidth]{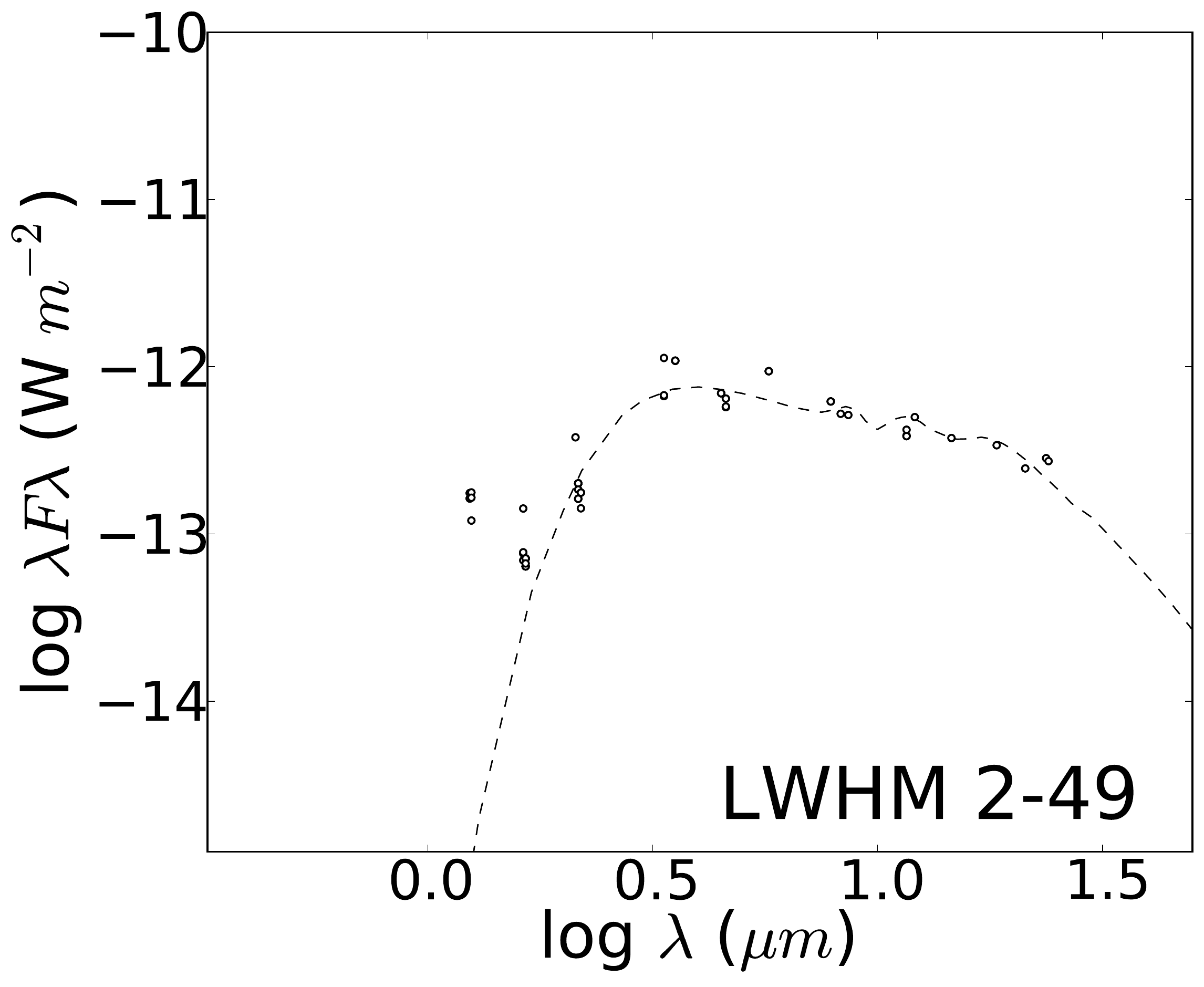}  
     \includegraphics[width=0.243\textwidth]{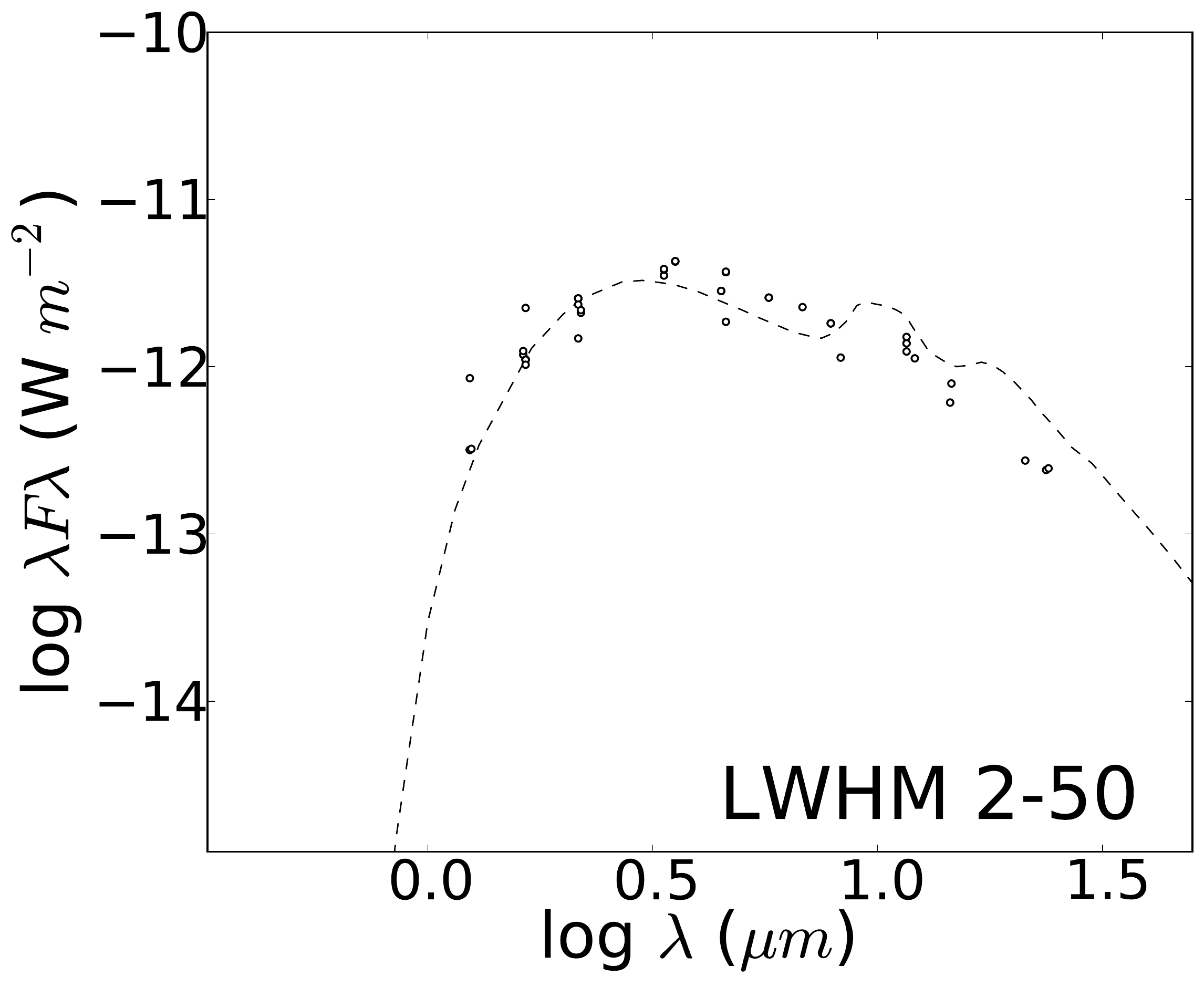}
     \includegraphics[width=0.243\textwidth]{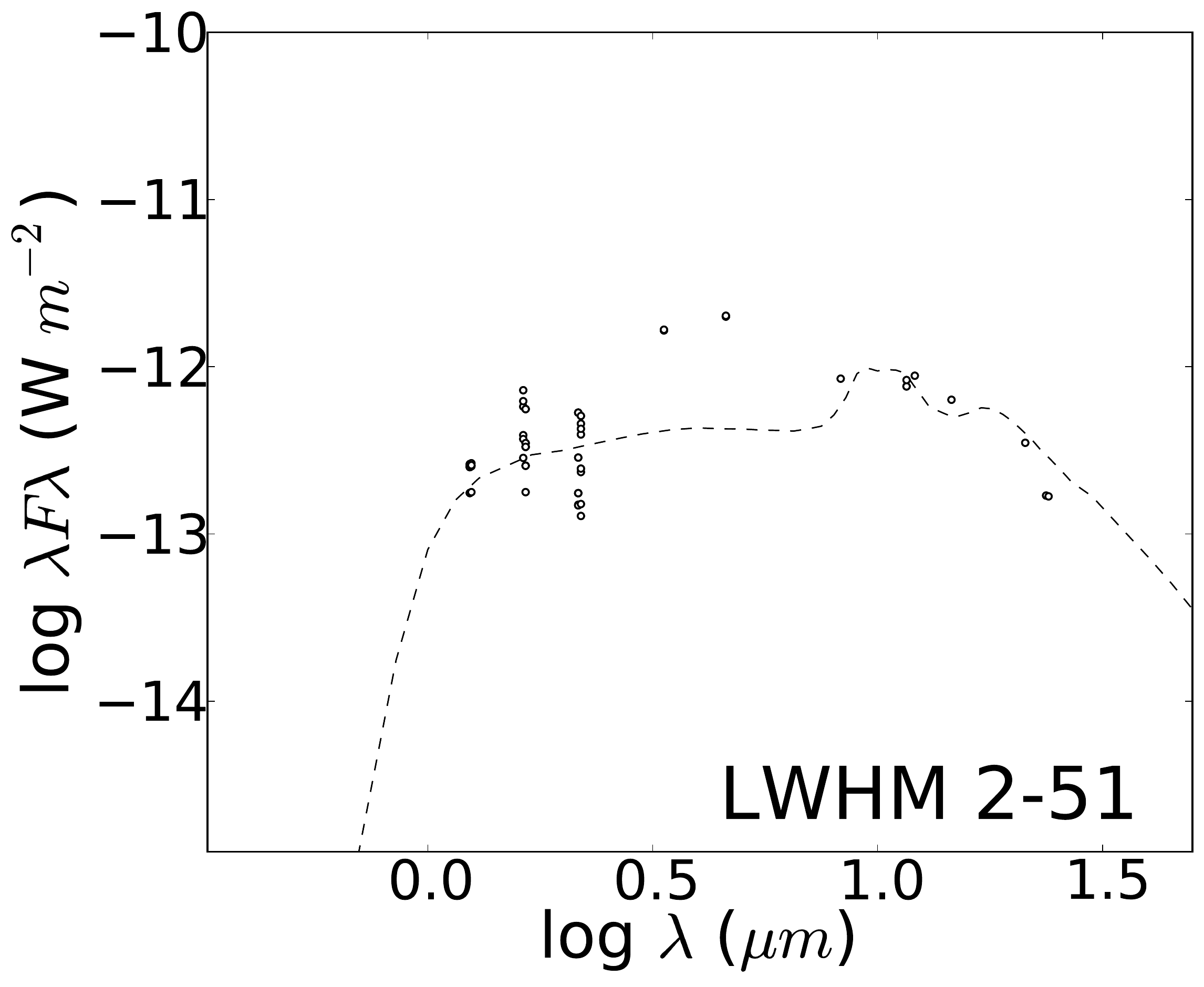}  \\
     \vspace{-0.15cm}
     \includegraphics[width=0.243\textwidth]{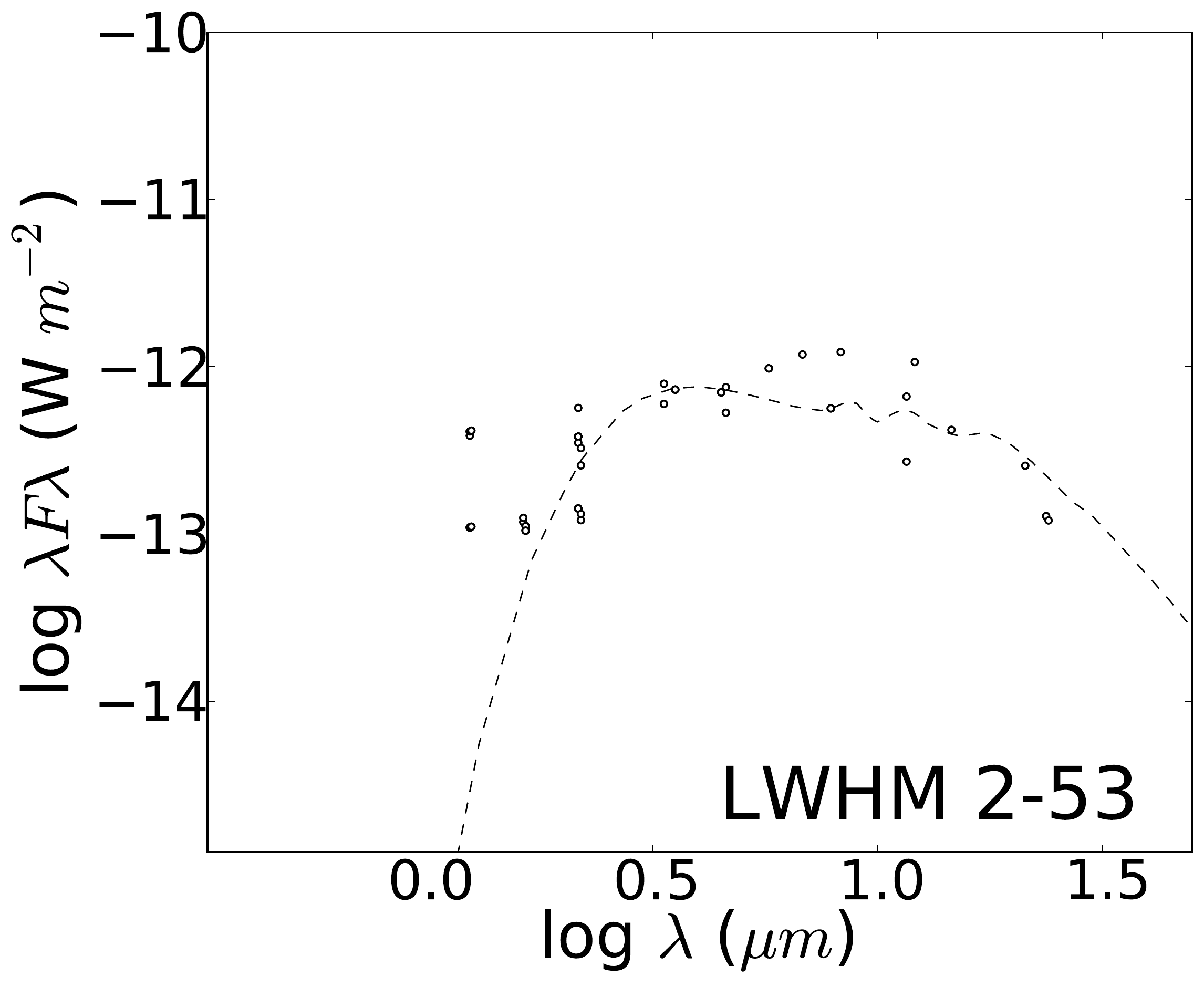}  
     \includegraphics[width=0.243\textwidth]{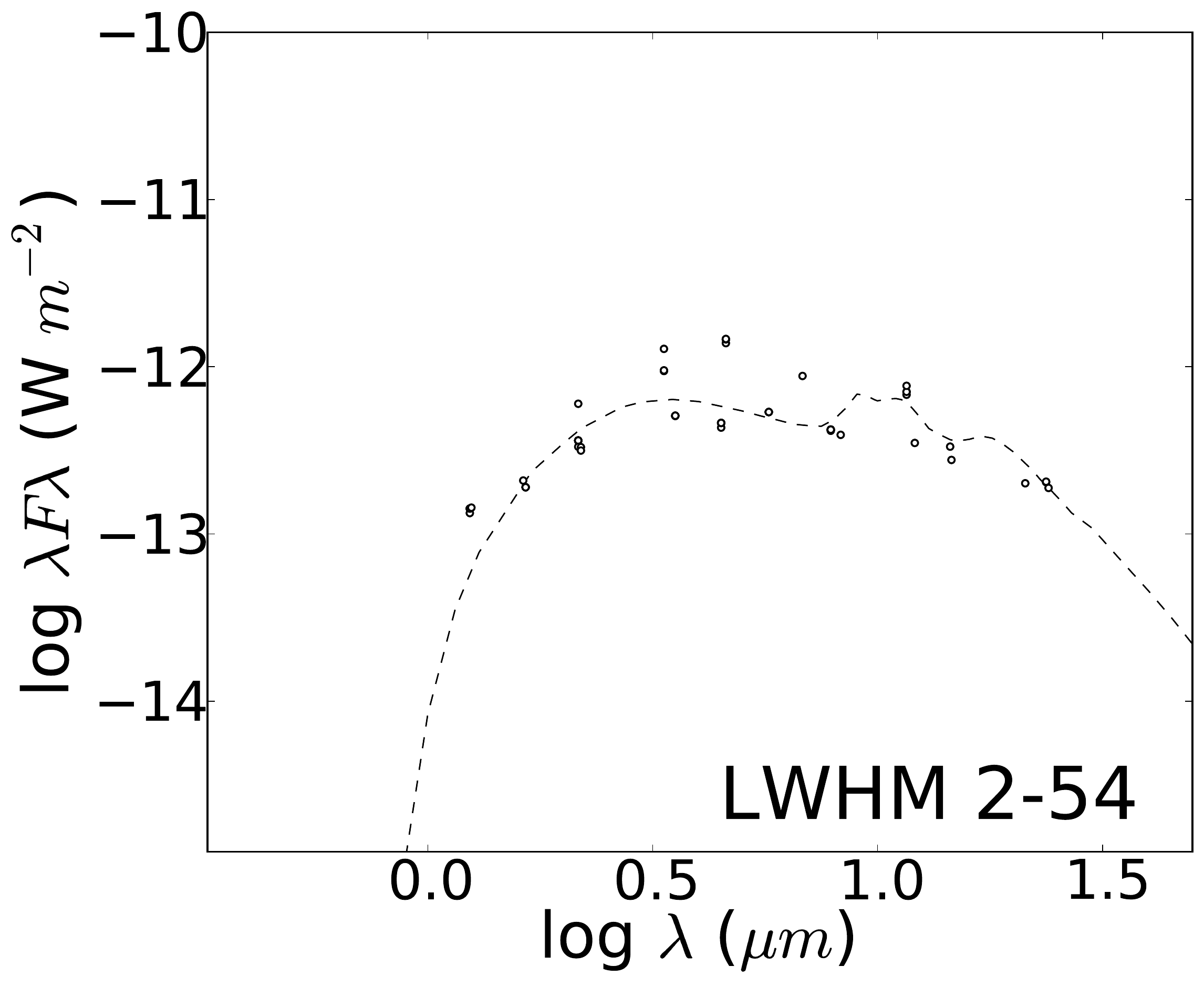}  
     \includegraphics[width=0.243\textwidth]{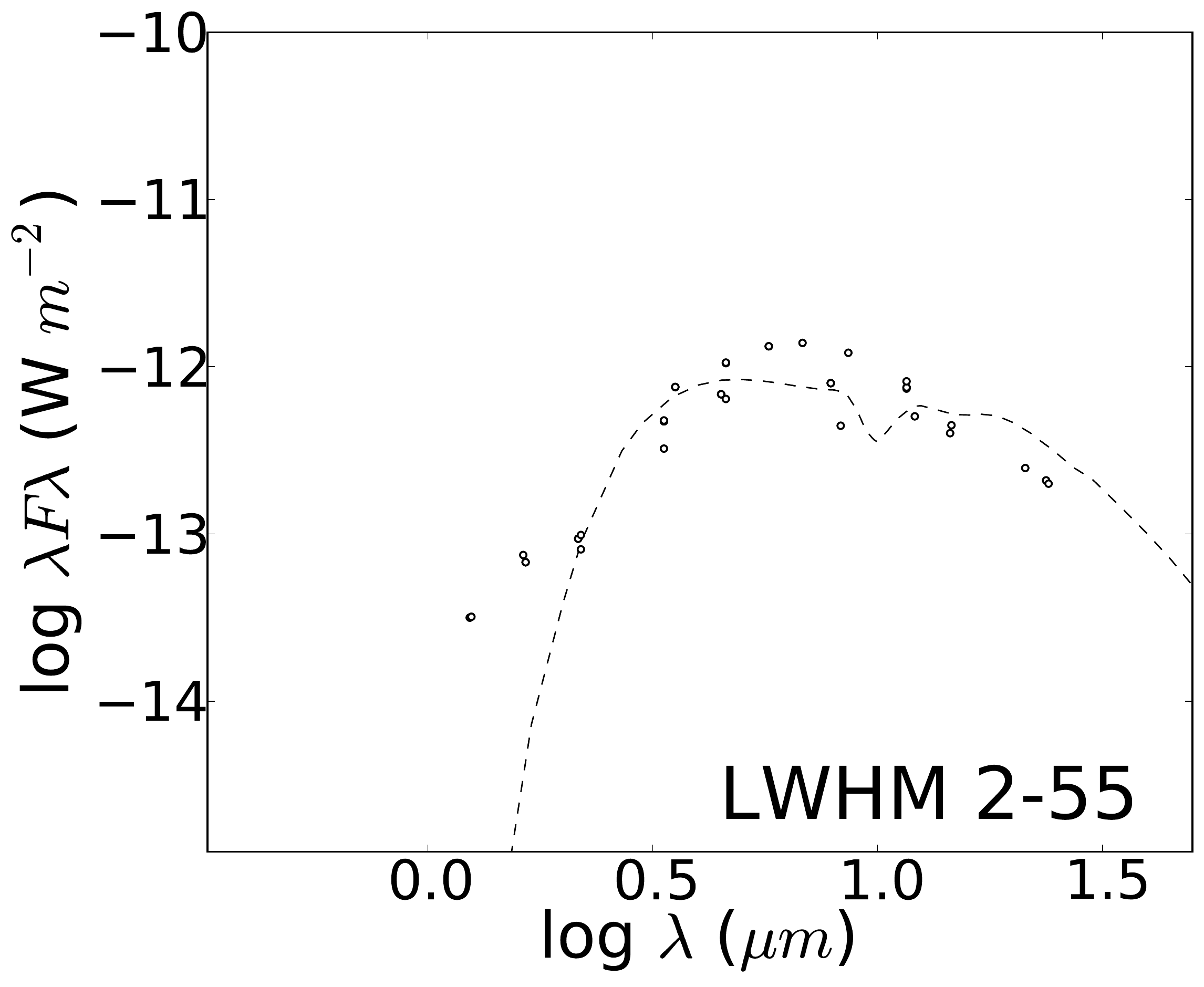}
     \includegraphics[width=0.243\textwidth]{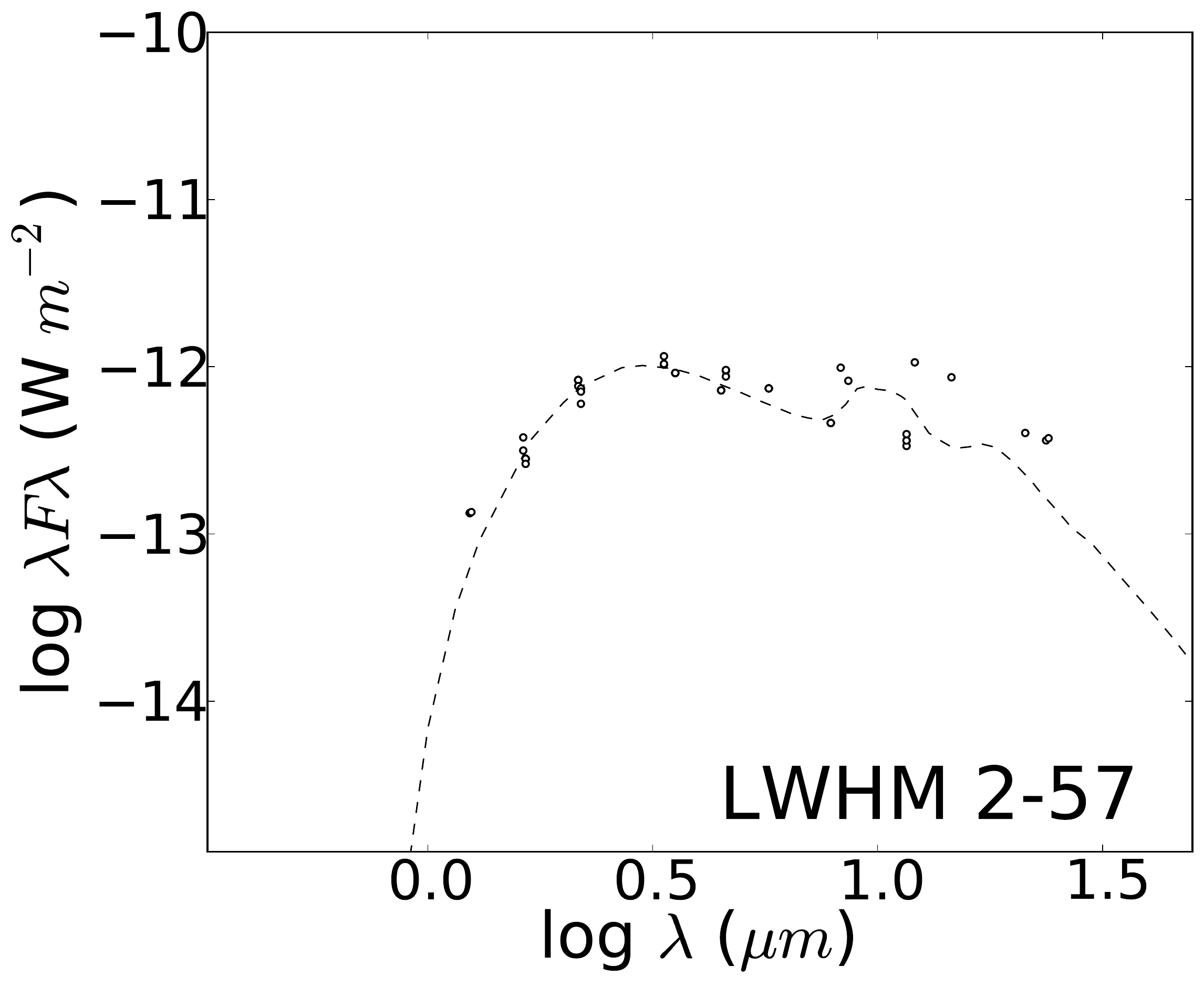} \\
     \vspace{-0.15cm}
     \includegraphics[width=0.243\textwidth]{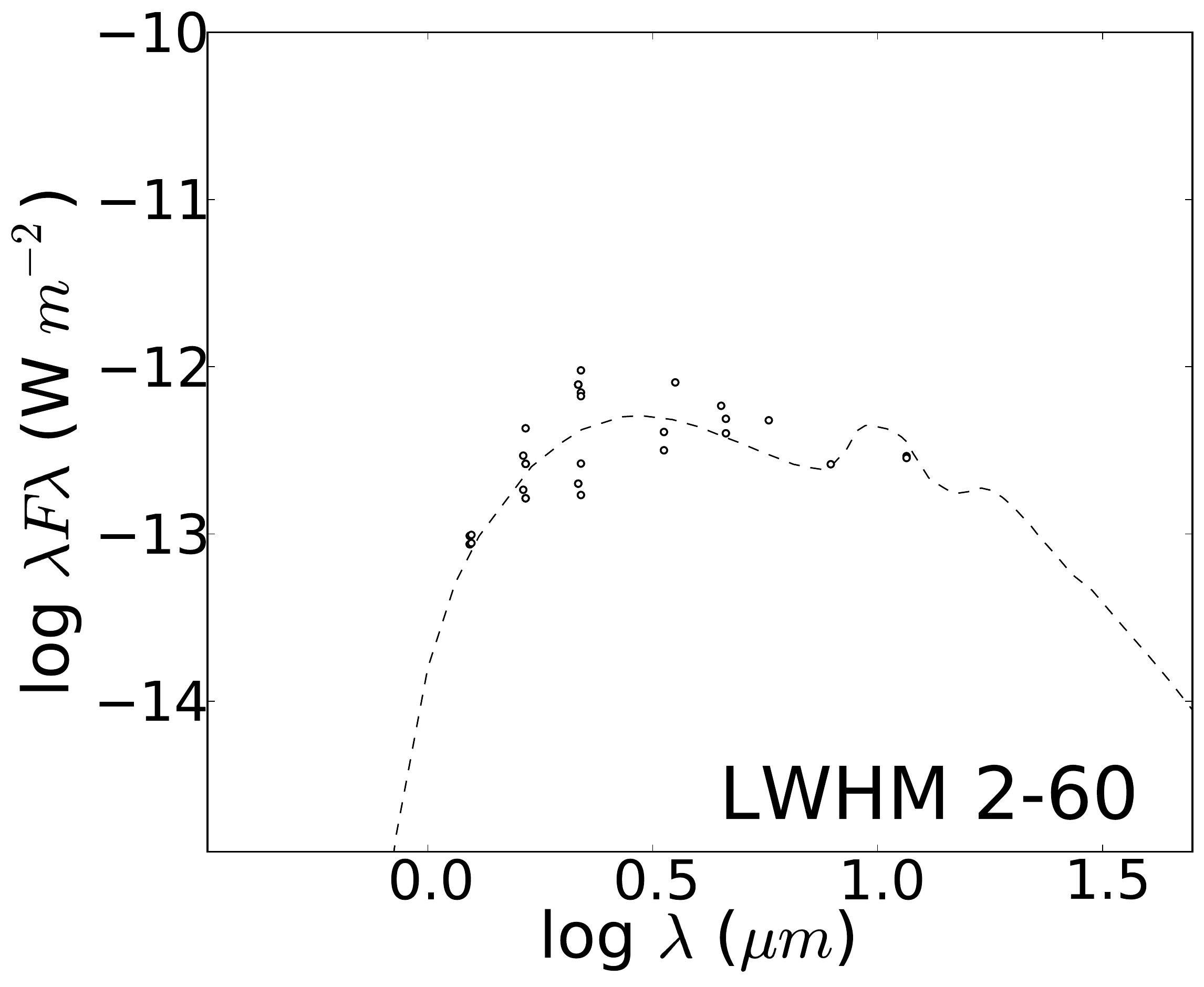}  
     \includegraphics[width=0.243\textwidth]{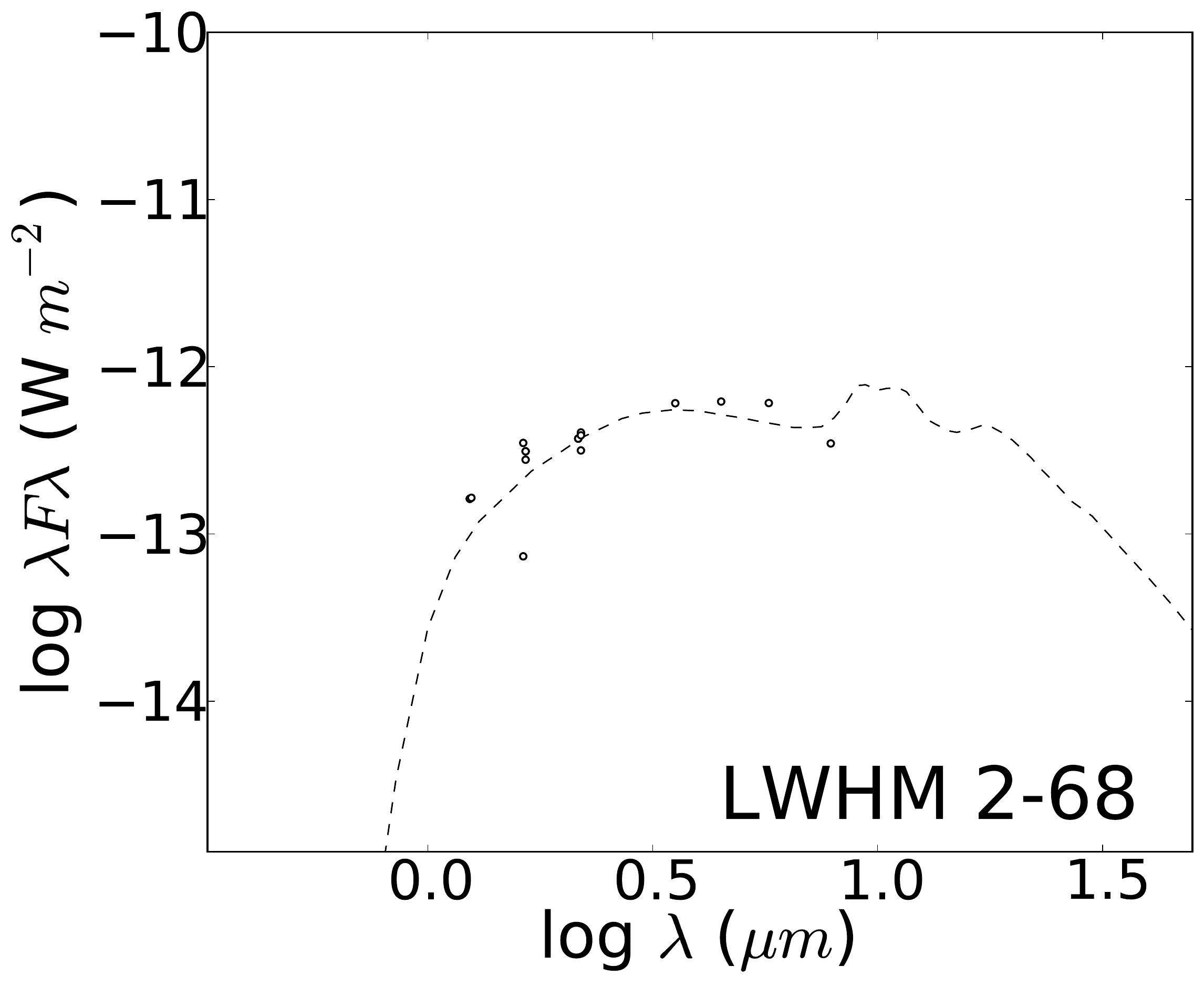}  
     \includegraphics[width=0.243\textwidth]{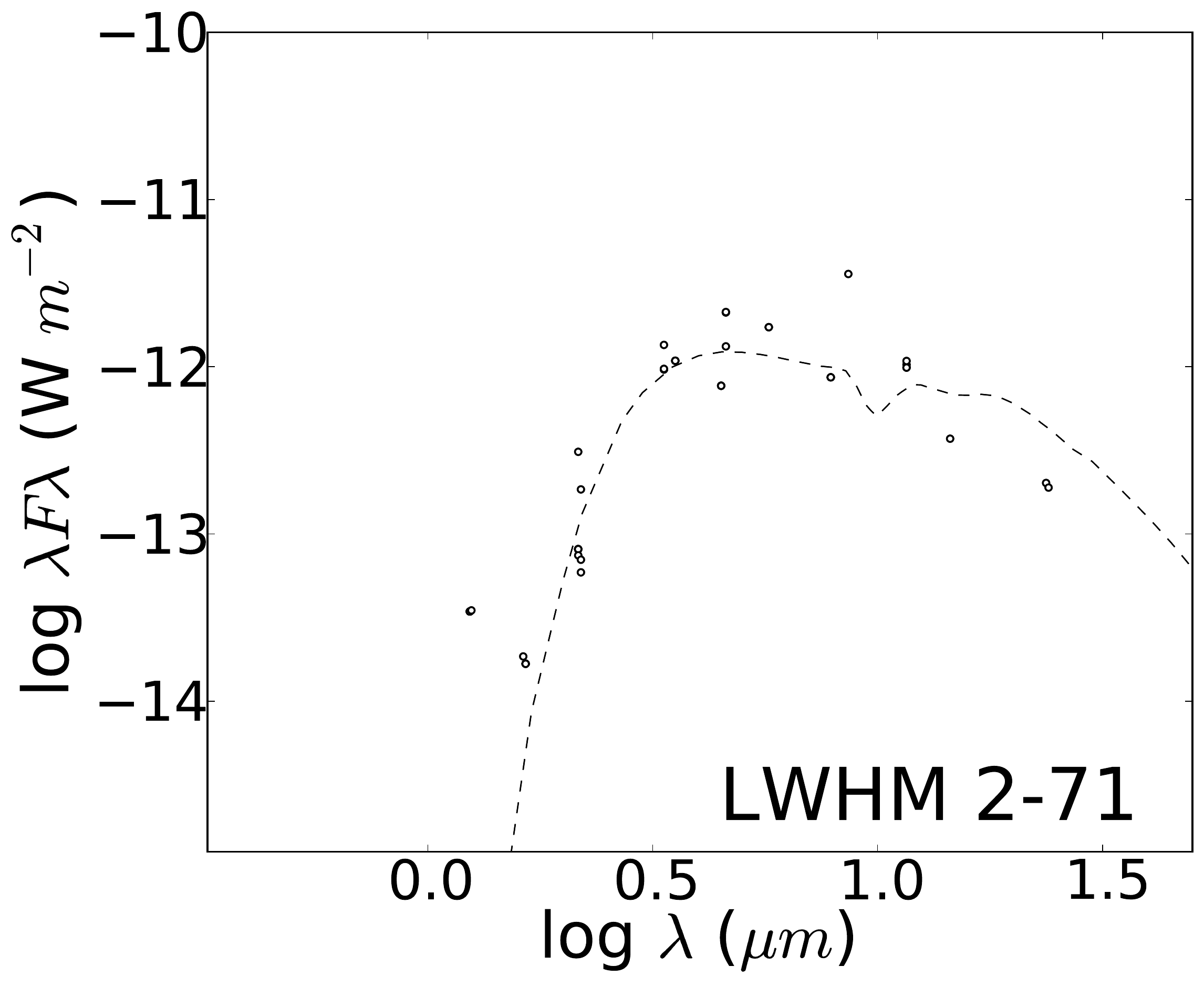}
     \includegraphics[width=0.243\textwidth]{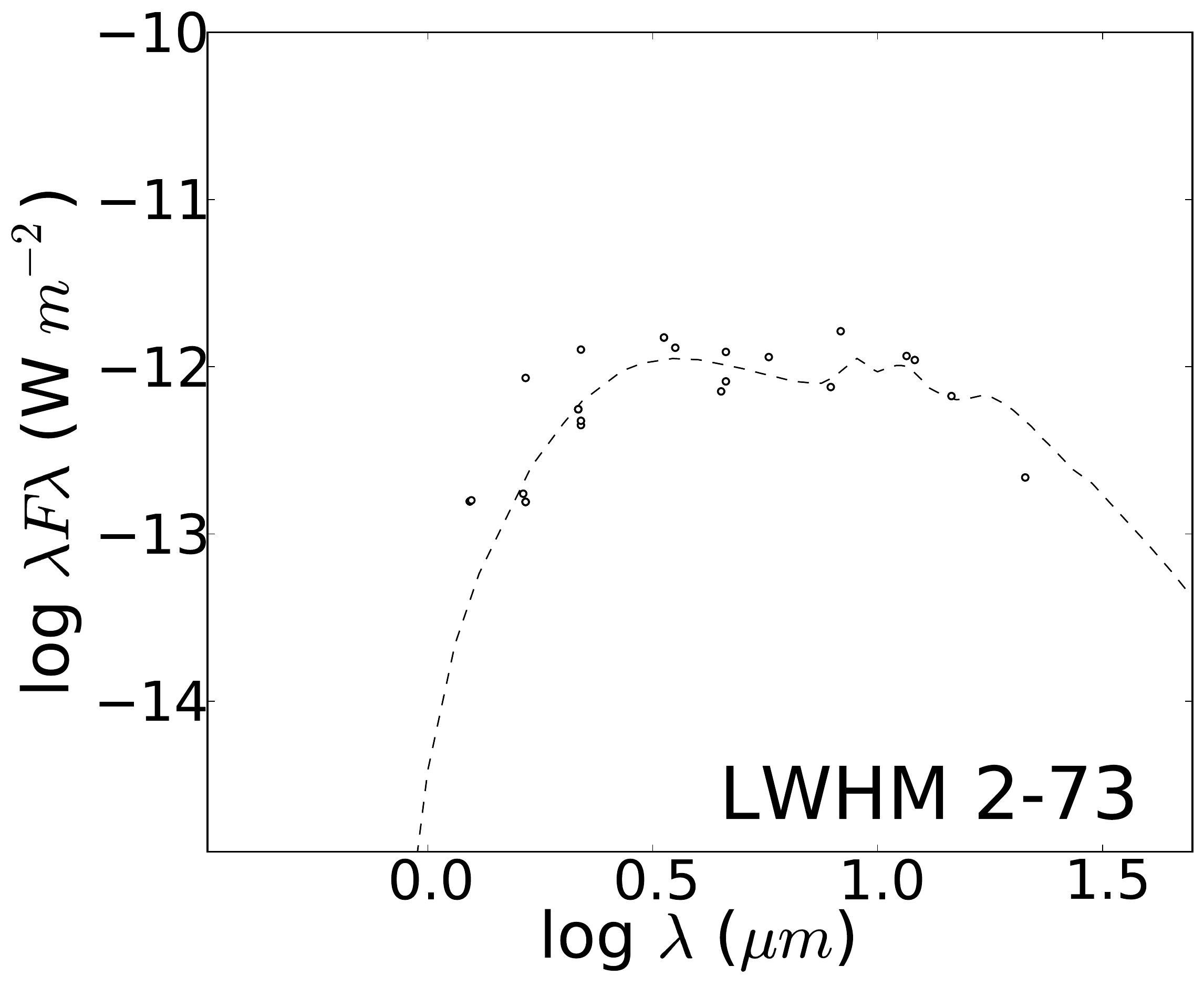}  \\
     \vspace{-0.15cm}
     \includegraphics[width=0.243\textwidth]{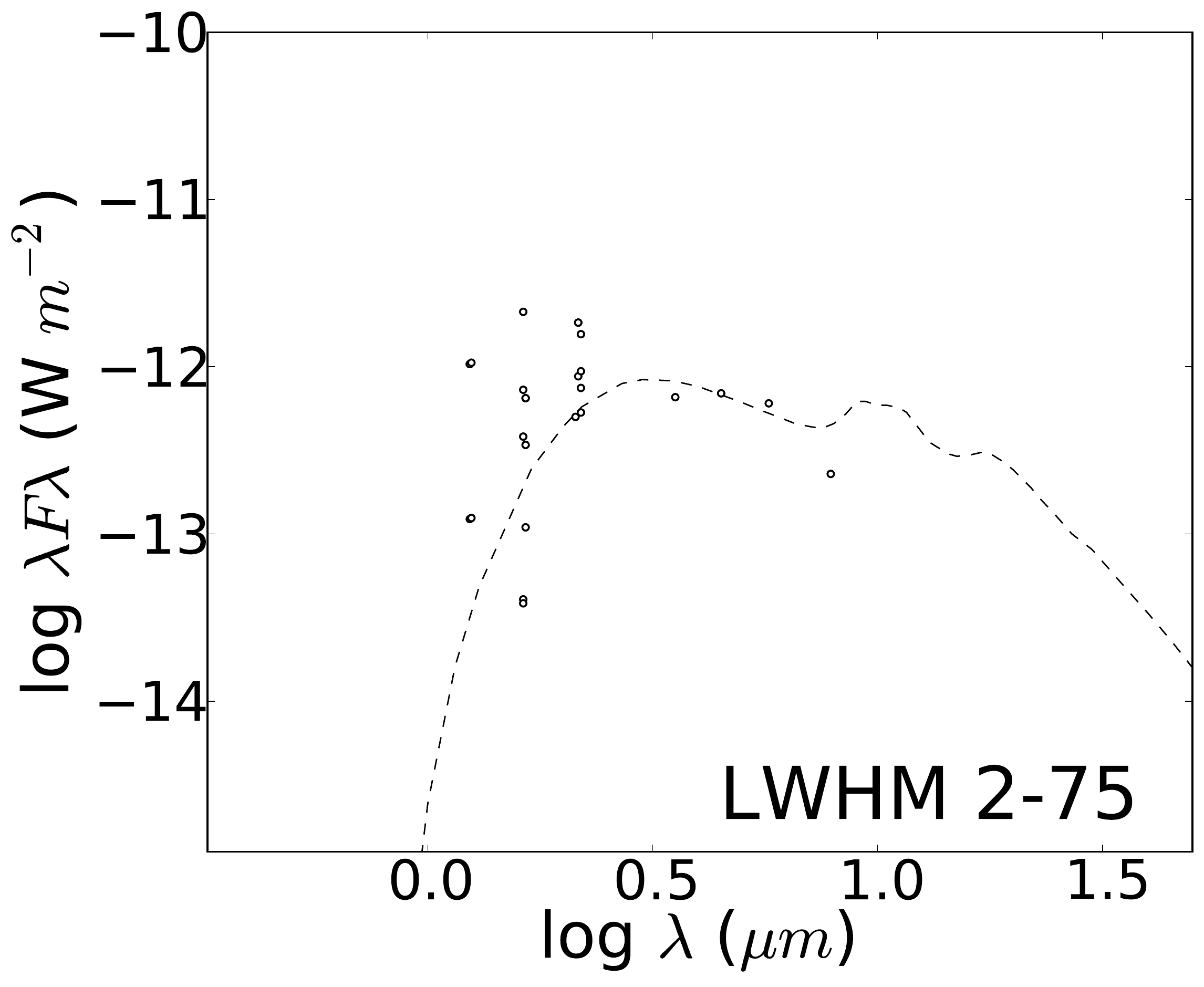}  
     \includegraphics[width=0.243\textwidth]{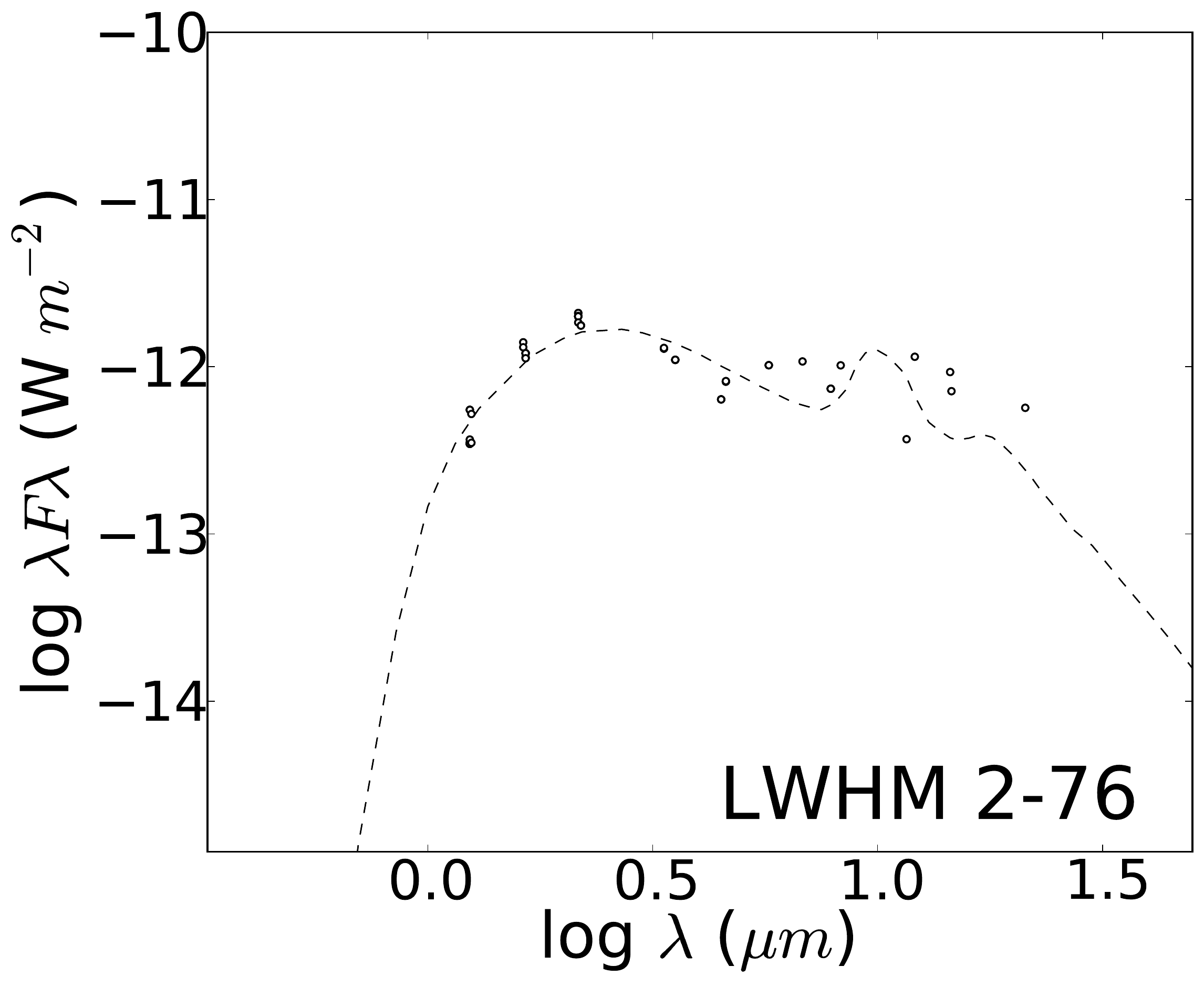}  
     \includegraphics[width=0.243\textwidth]{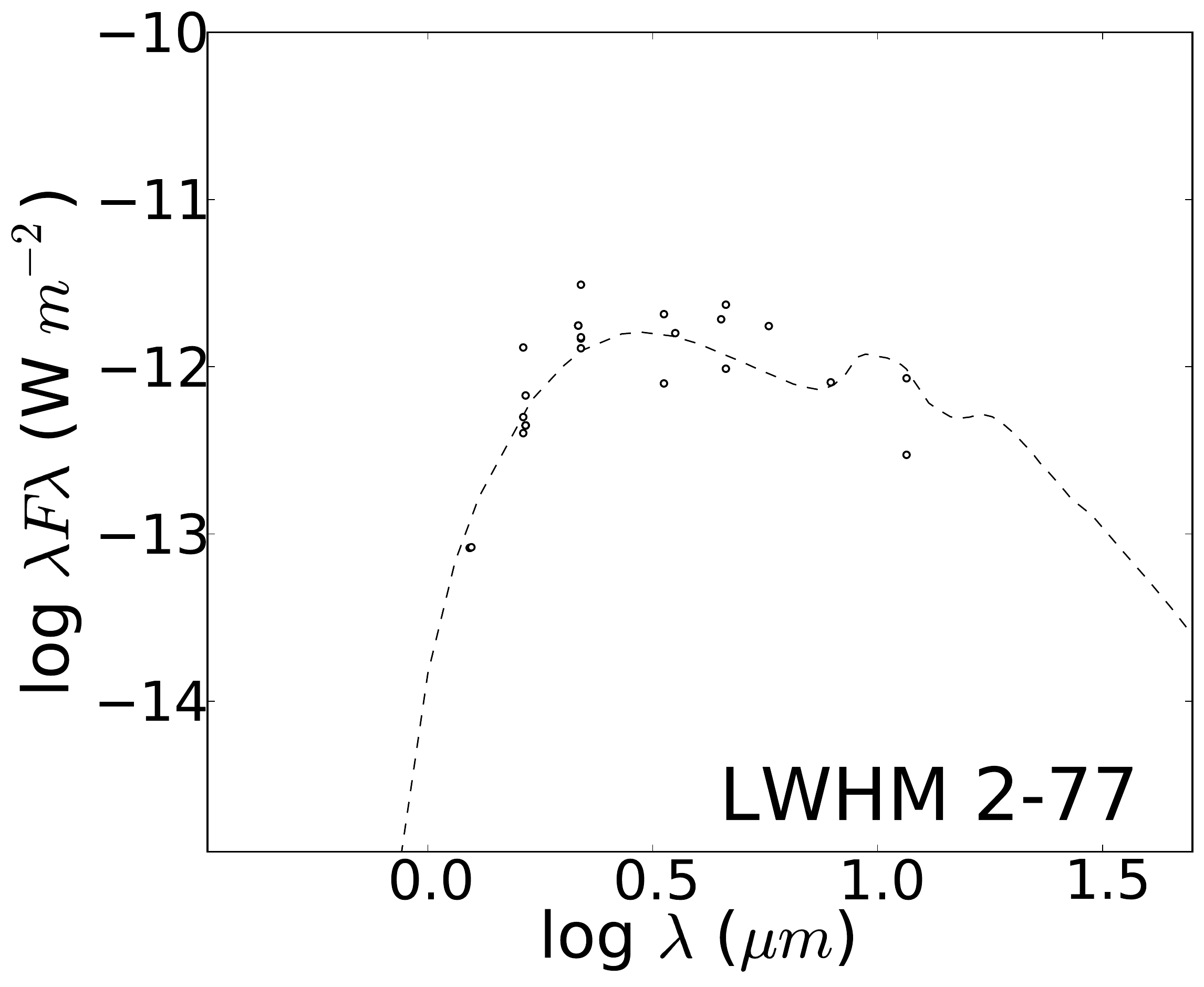}
     \includegraphics[width=0.243\textwidth]{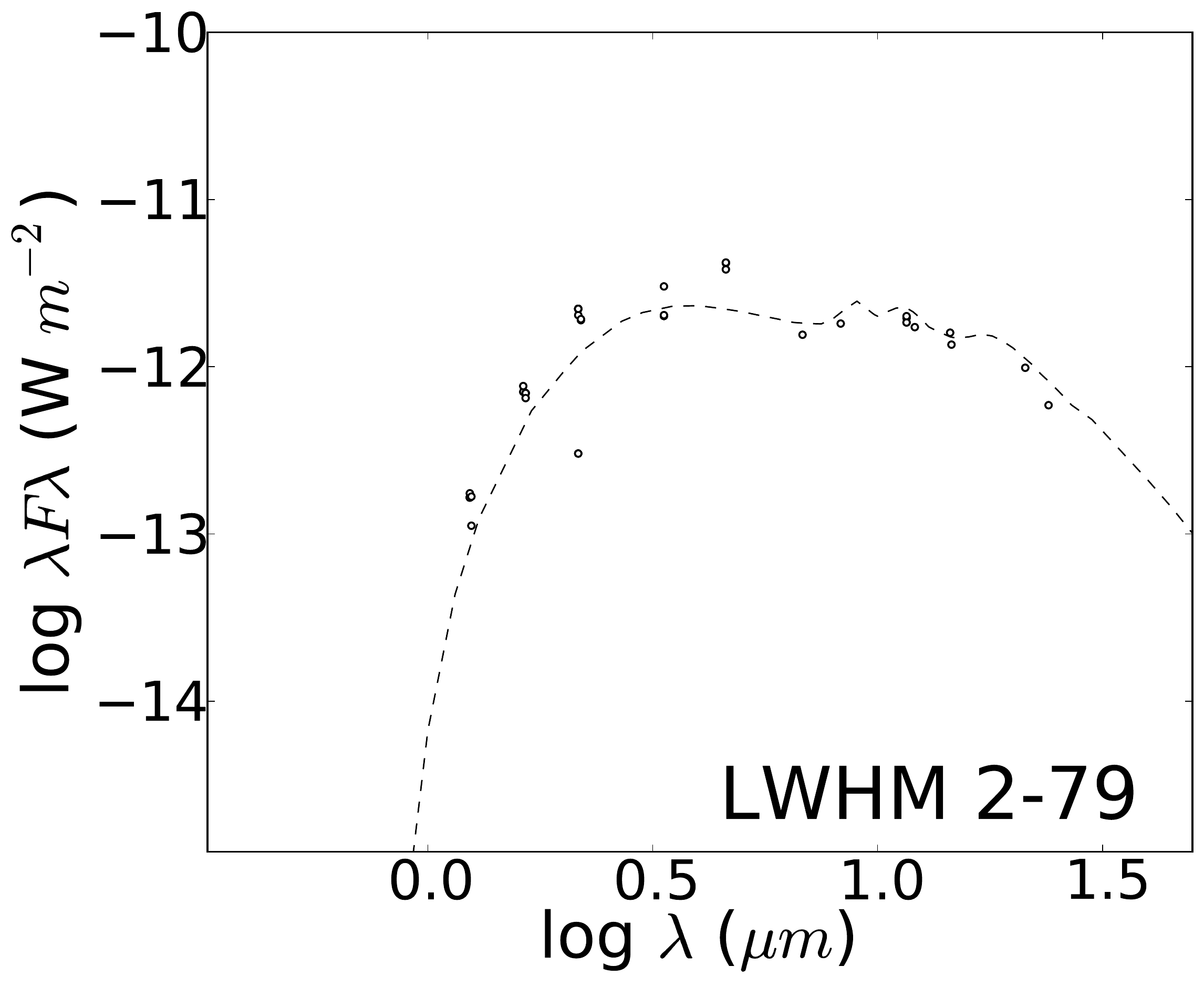}  \\

     \includegraphics[width=0.243\textwidth]{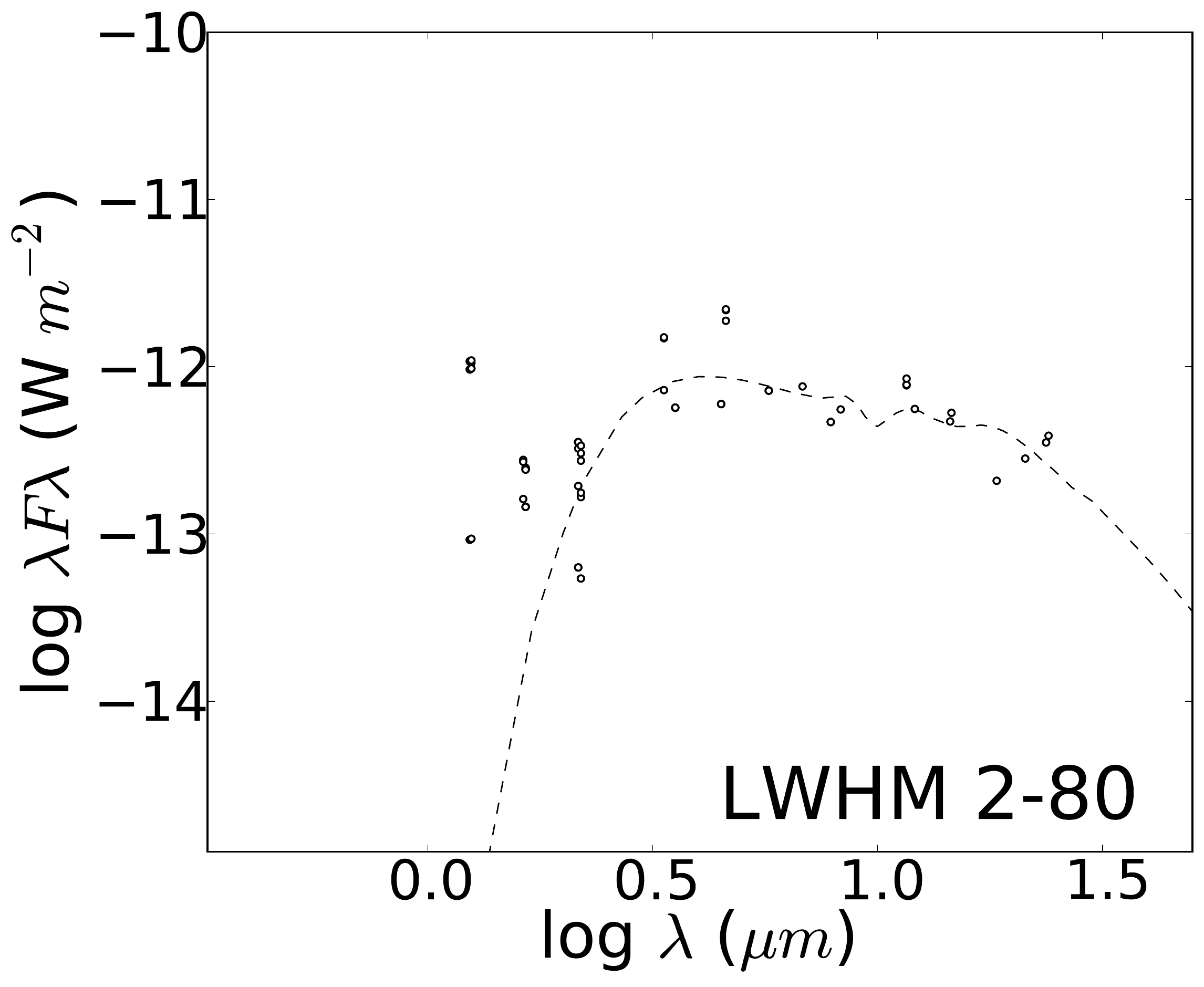}  
     \includegraphics[width=0.243\textwidth]{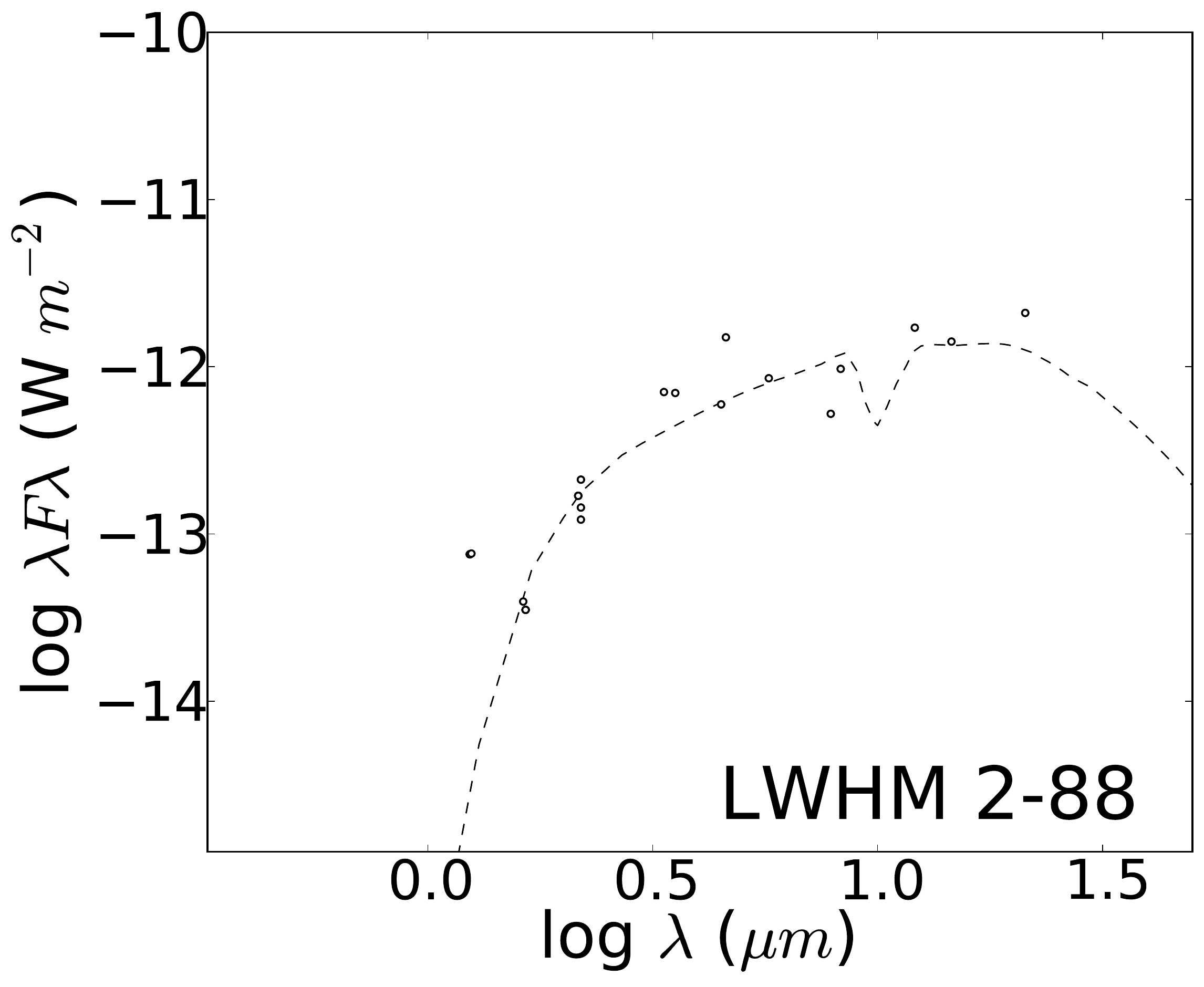}  
     \includegraphics[width=0.243\textwidth]{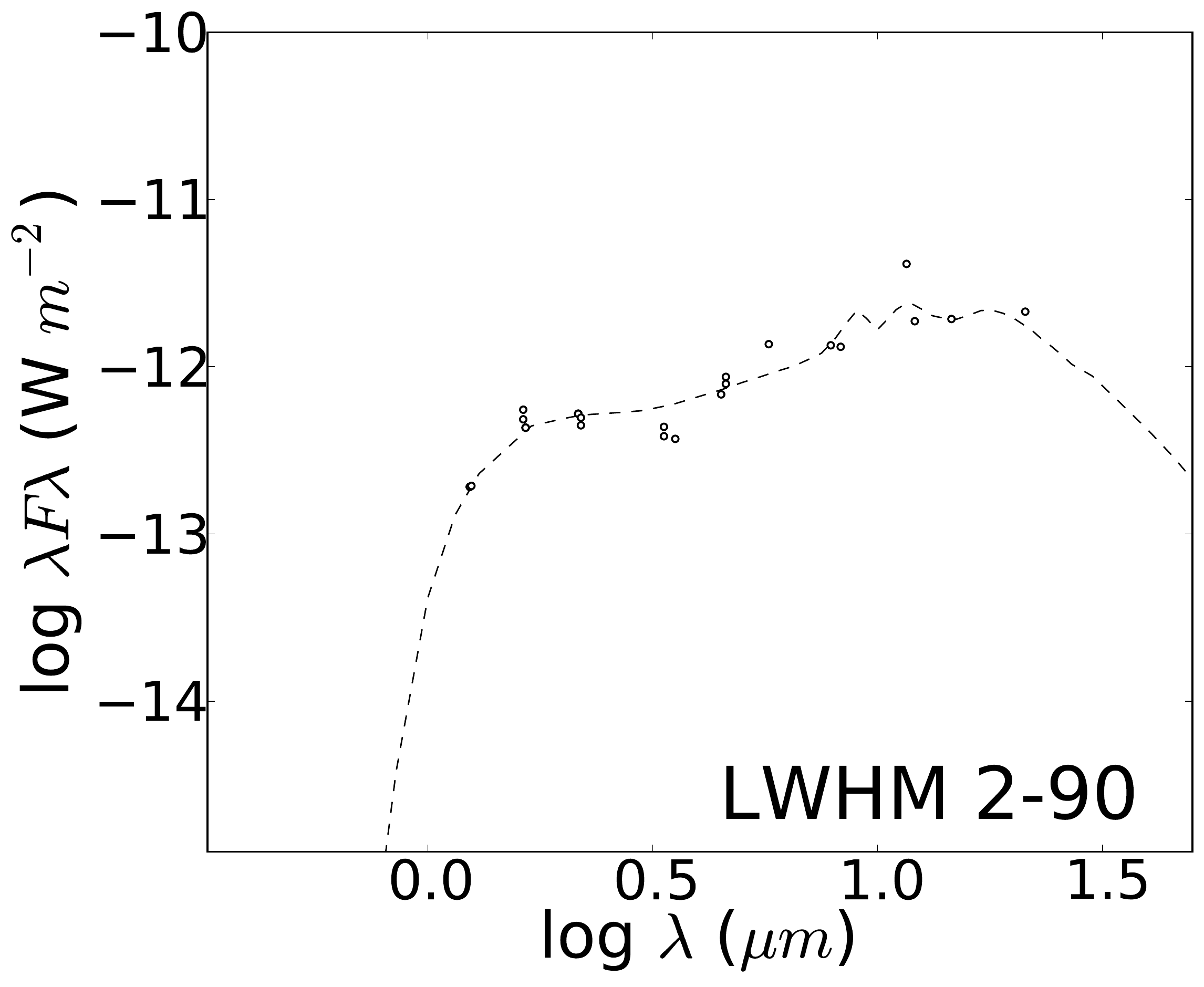}
     \includegraphics[width=0.243\textwidth]{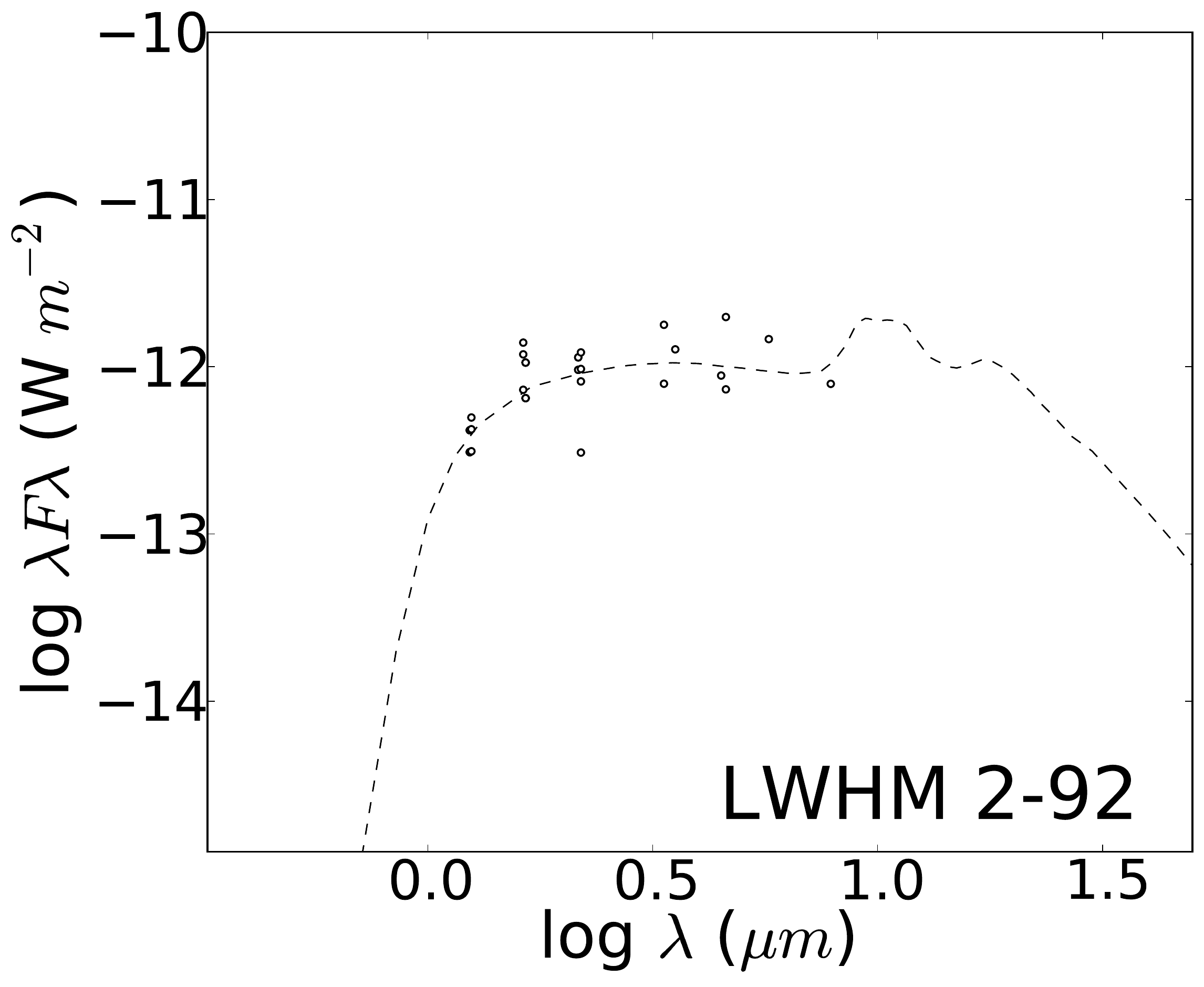}  \\
     \end{tabular}
   \end{center}
   \vspace{-0.4cm}
   \caption{continued}
   \label{sdf}
   \end{minipage}%
\end{figure*} 
\addtocounter{figure}{-1}

\begin{figure*}
  \begin{minipage}[c]{\textwidth}
  \vspace{1.2cm}
  \begin{center}
     \begin{tabular}{c}
     \vspace{-0.15cm}
     \includegraphics[width=0.243\textwidth]{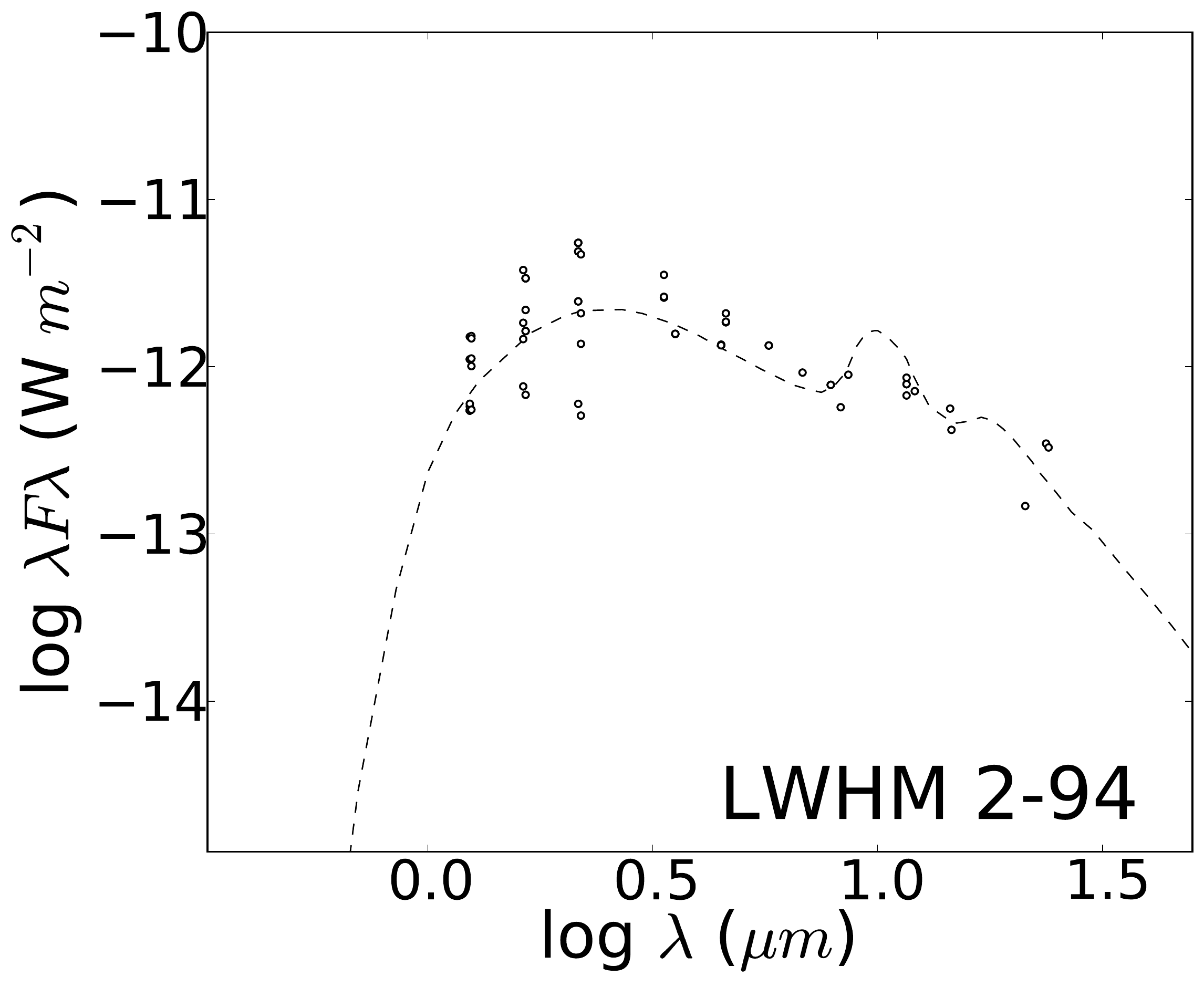}  
     \includegraphics[width=0.243\textwidth]{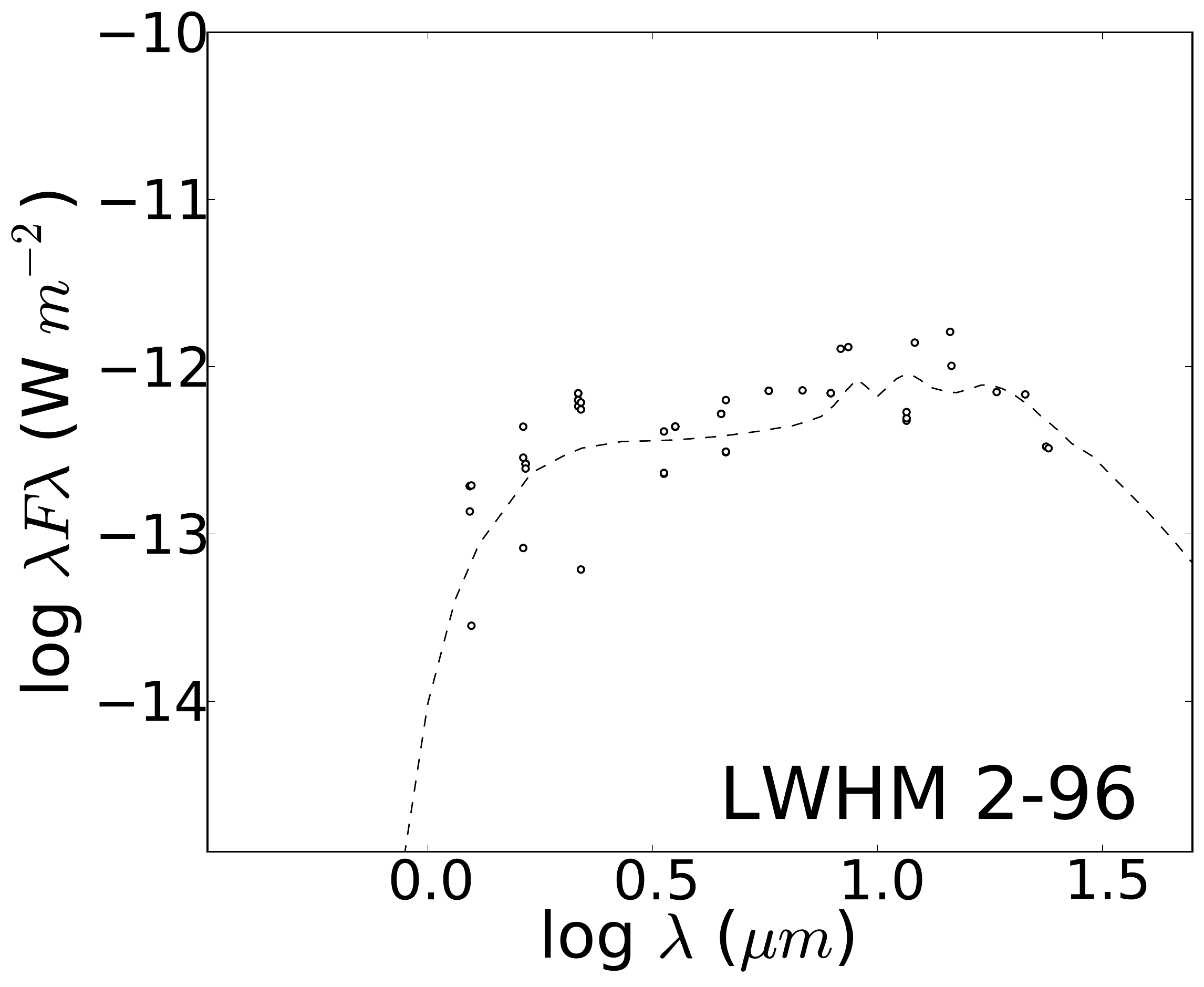}  
     \includegraphics[width=0.243\textwidth]{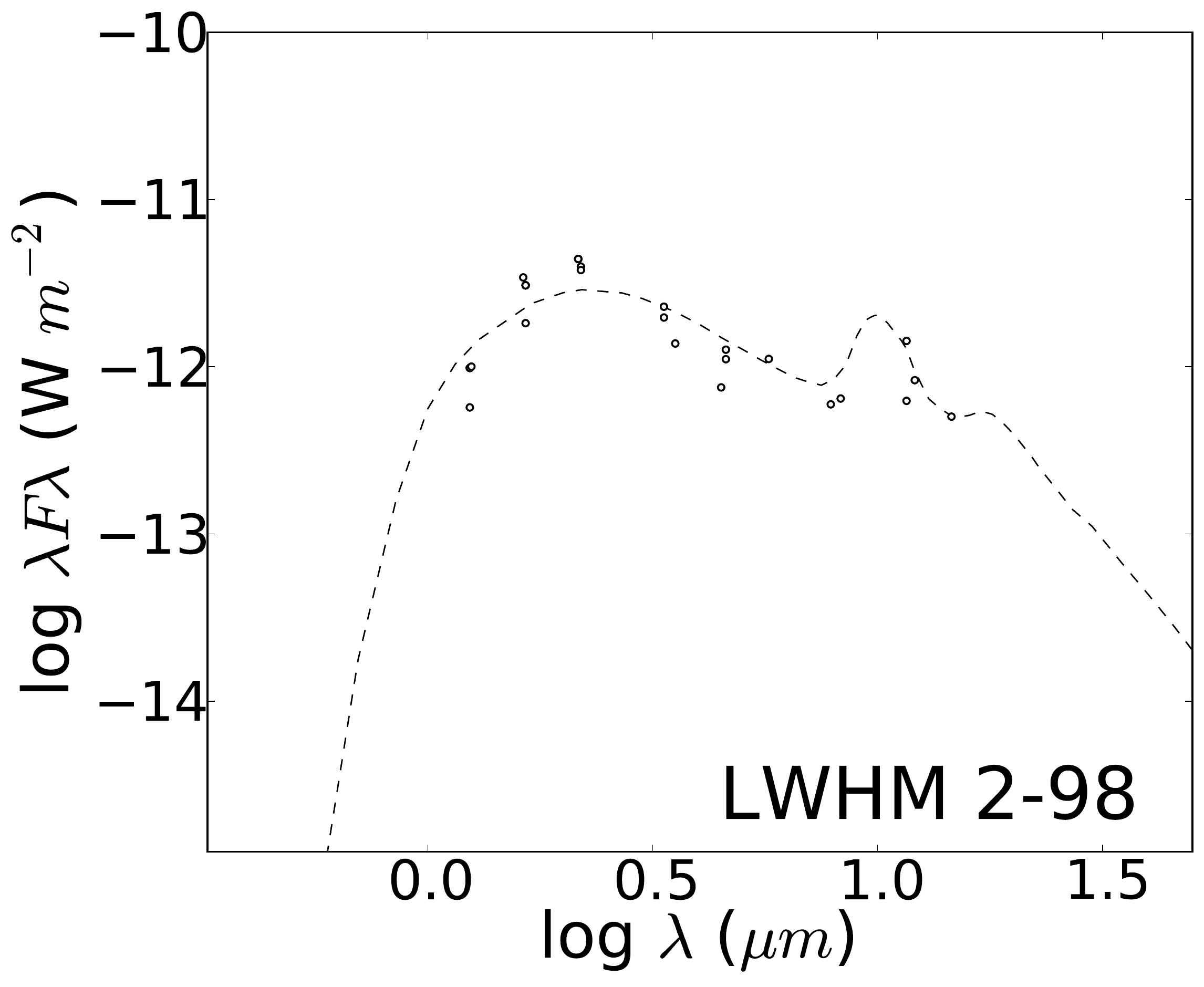}
     \includegraphics[width=0.243\textwidth]{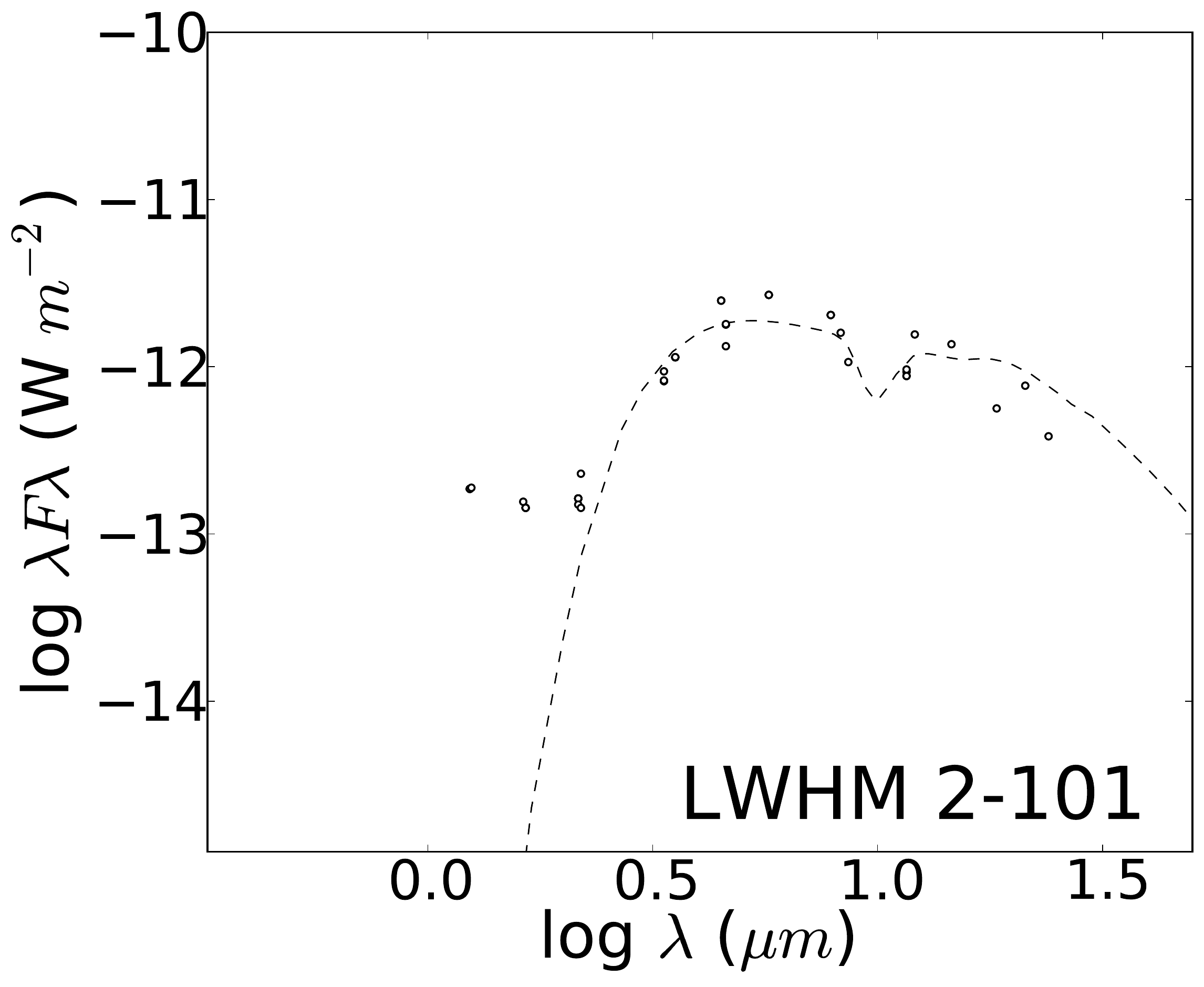}  \\
     \vspace{-0.15cm}
     \includegraphics[width=0.243\textwidth]{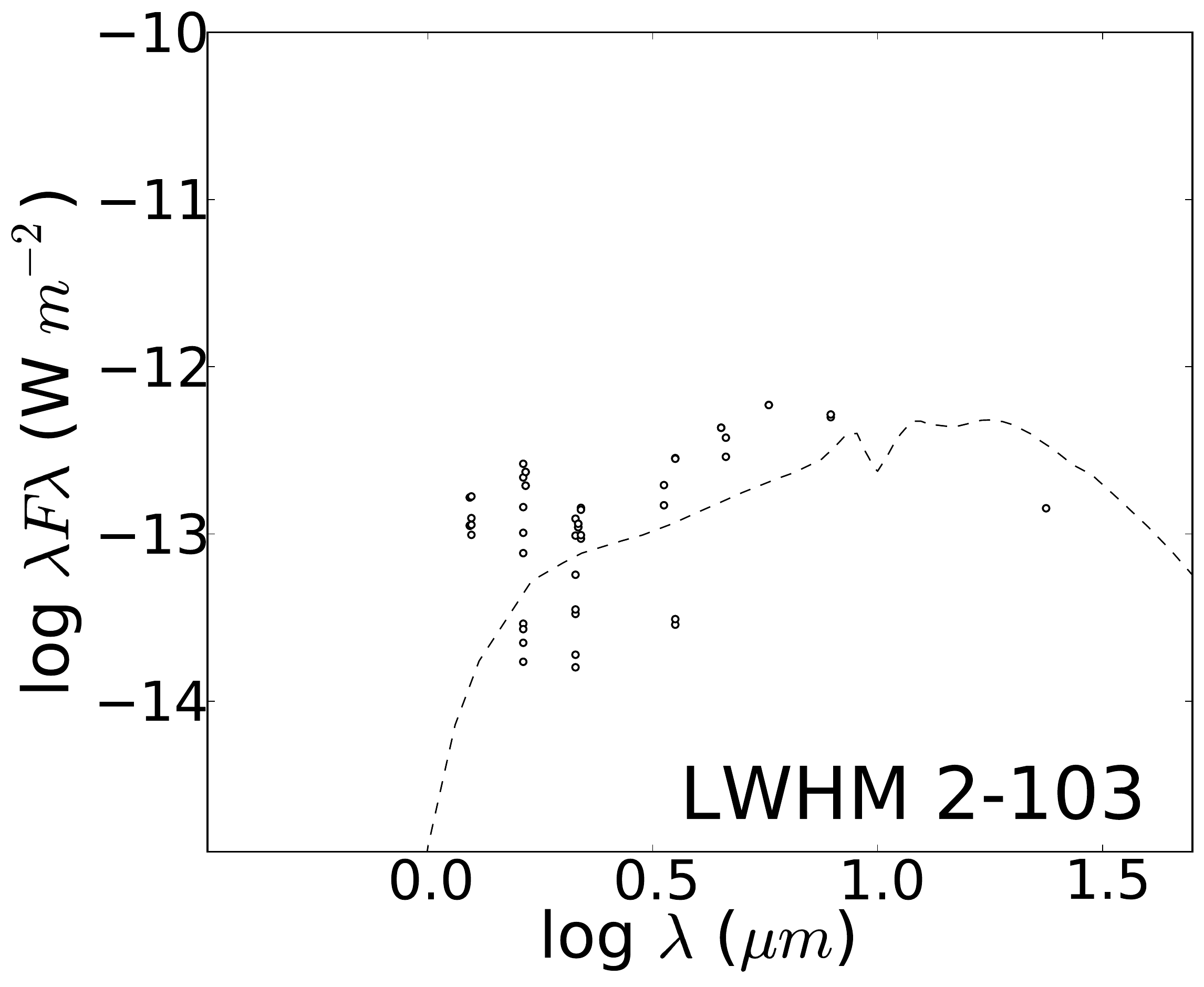}  
     \includegraphics[width=0.243\textwidth]{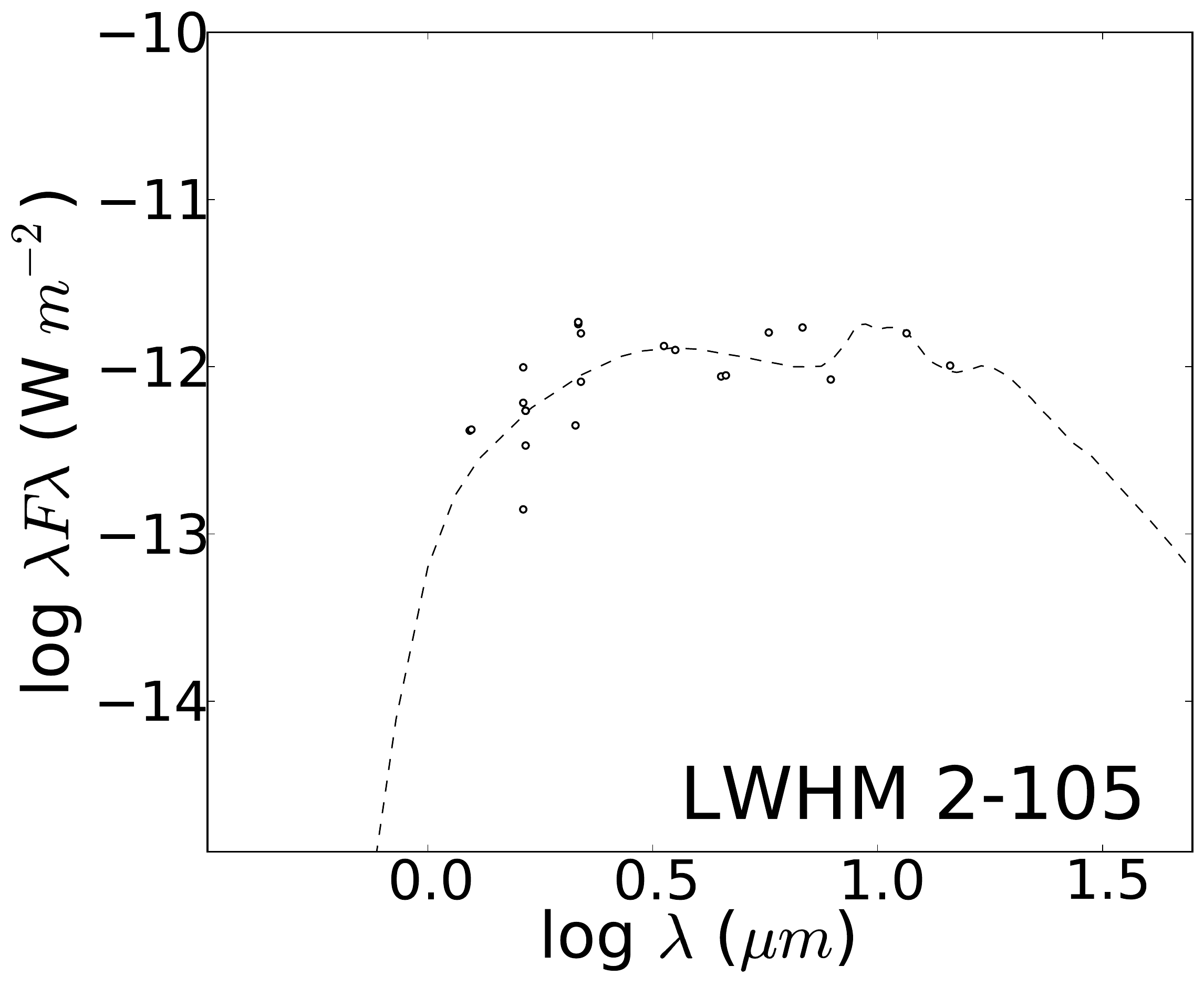}  
     \includegraphics[width=0.243\textwidth]{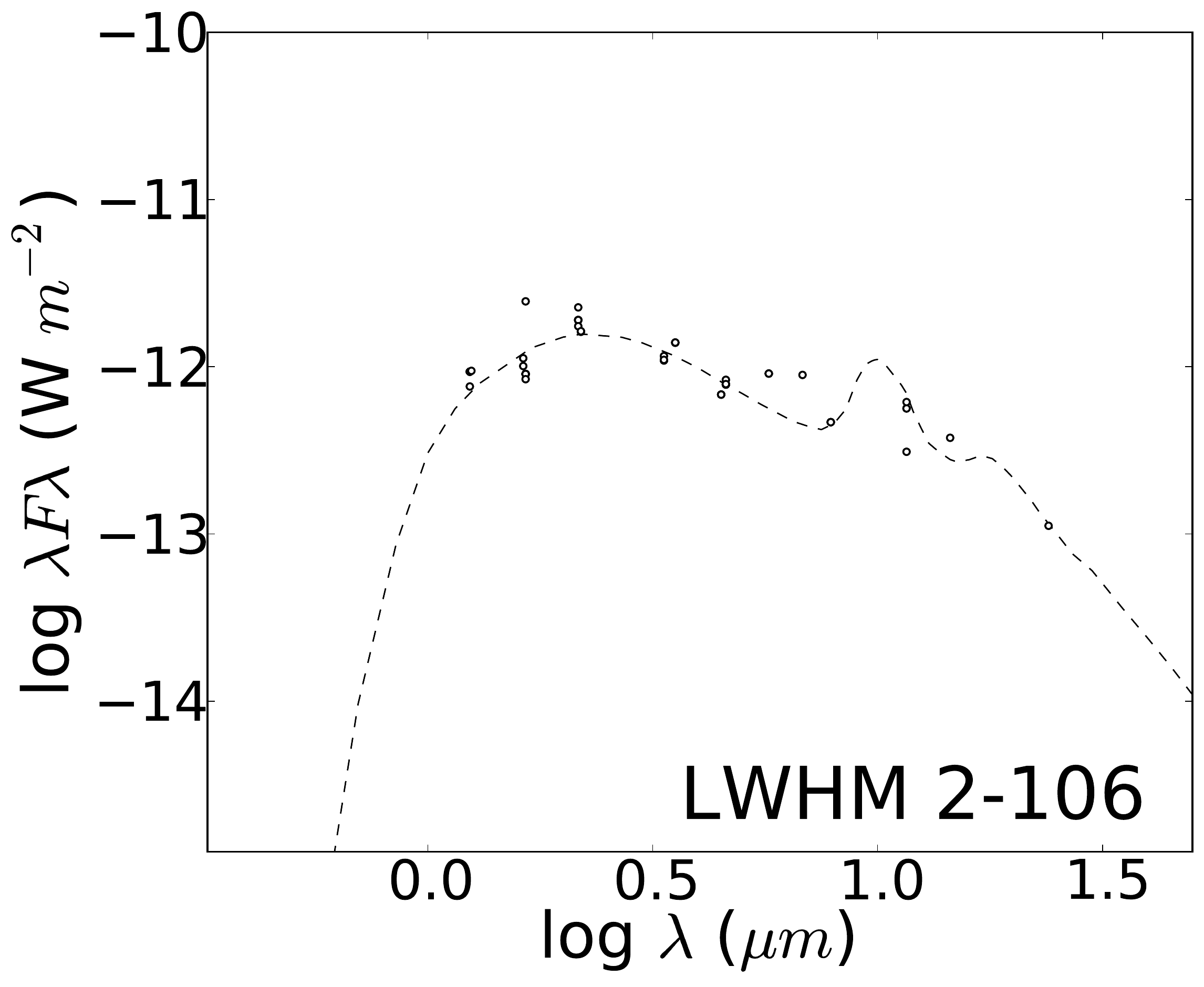}
     \includegraphics[width=0.243\textwidth]{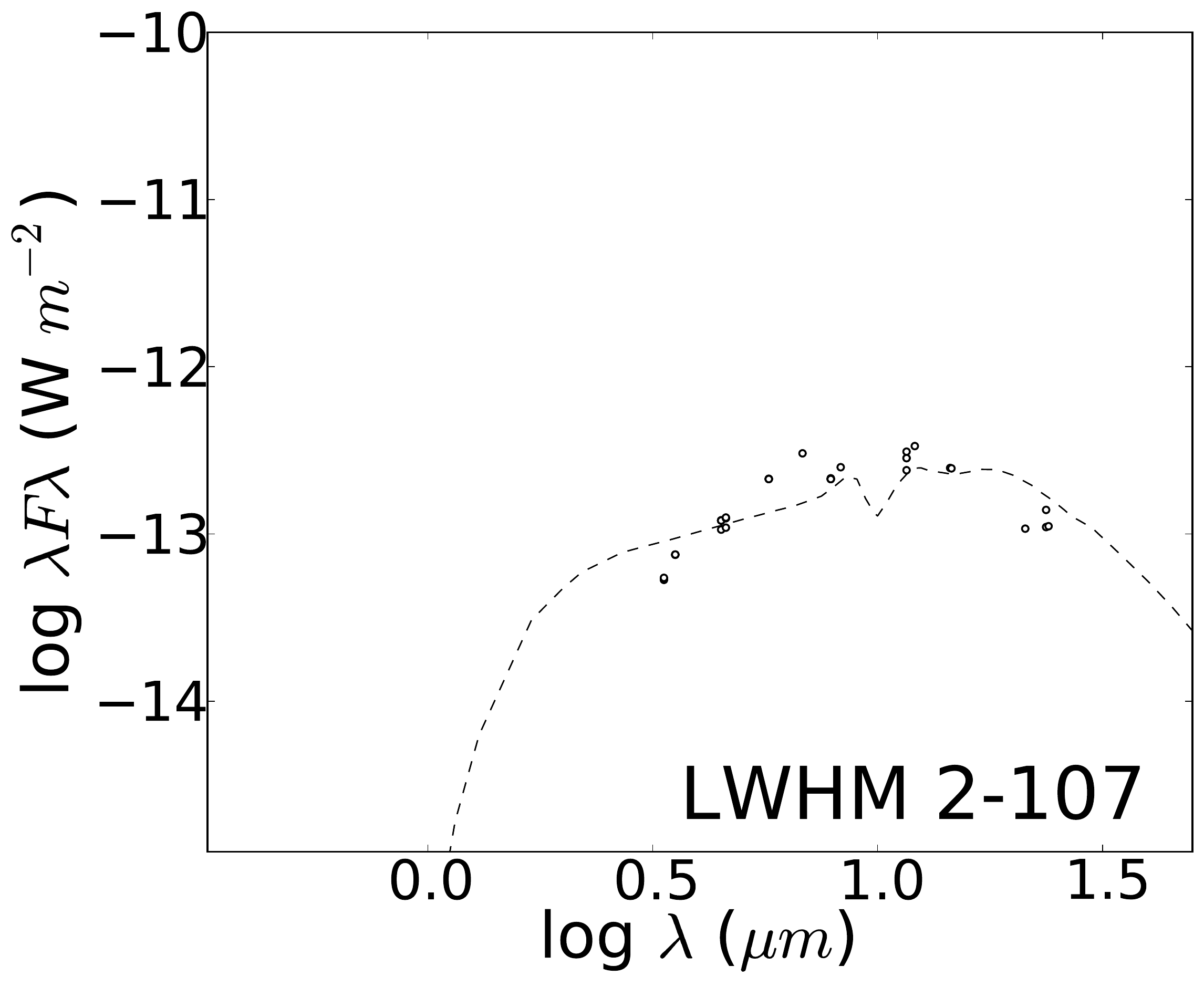}  \\
     \vspace{-0.15cm}
     \includegraphics[width=0.243\textwidth]{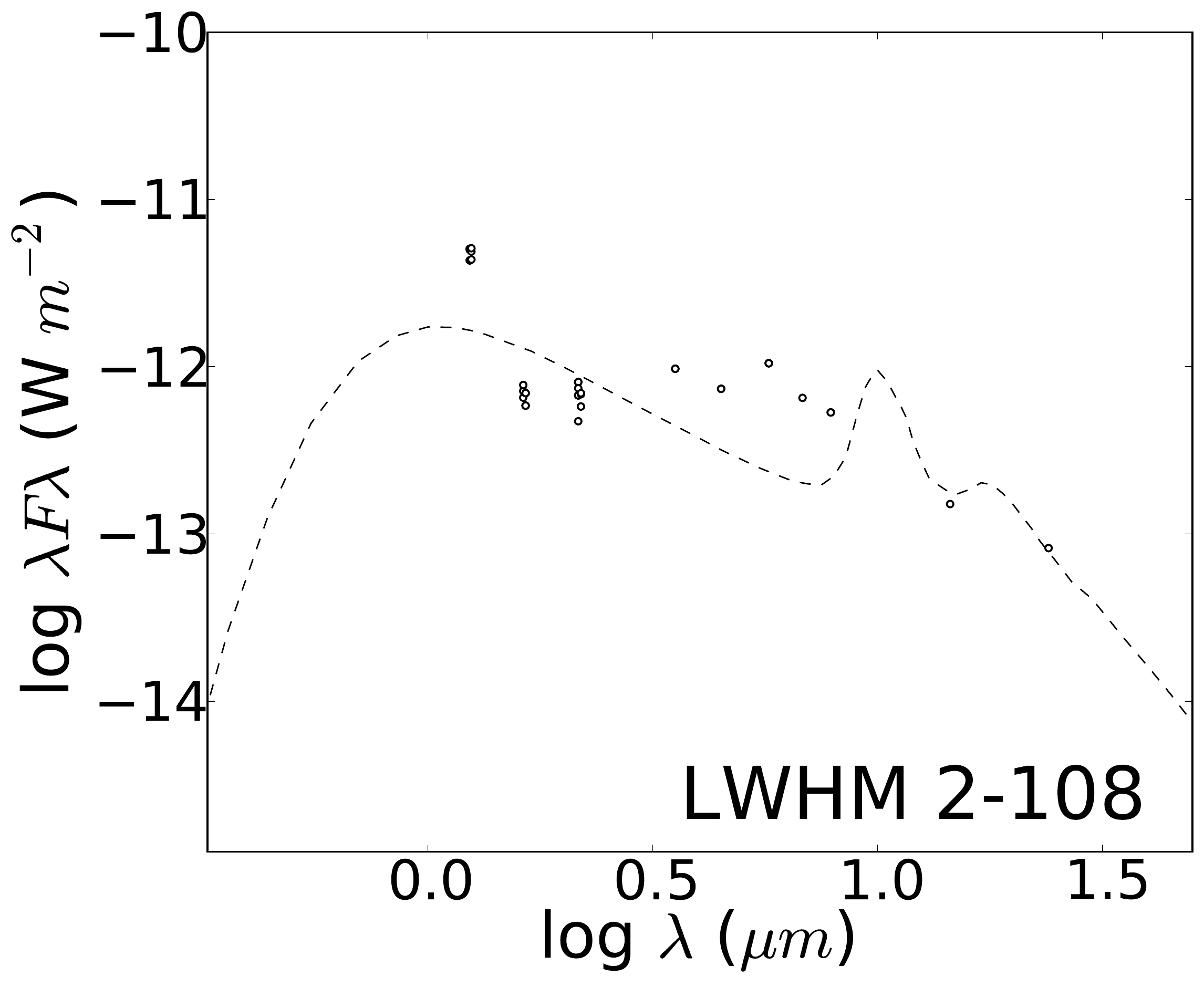}   
     \includegraphics[width=0.243\textwidth]{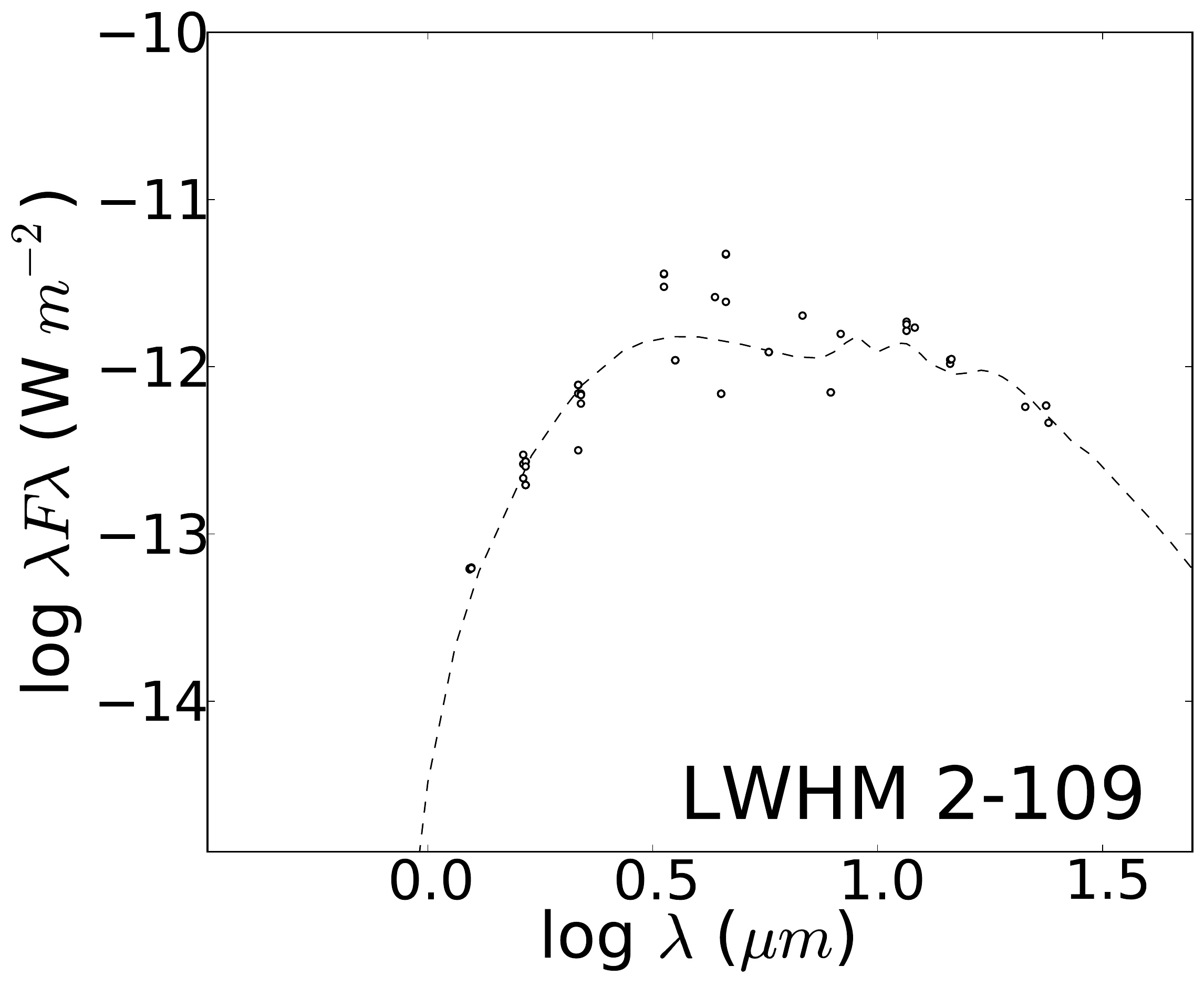}  
     \includegraphics[width=0.243\textwidth]{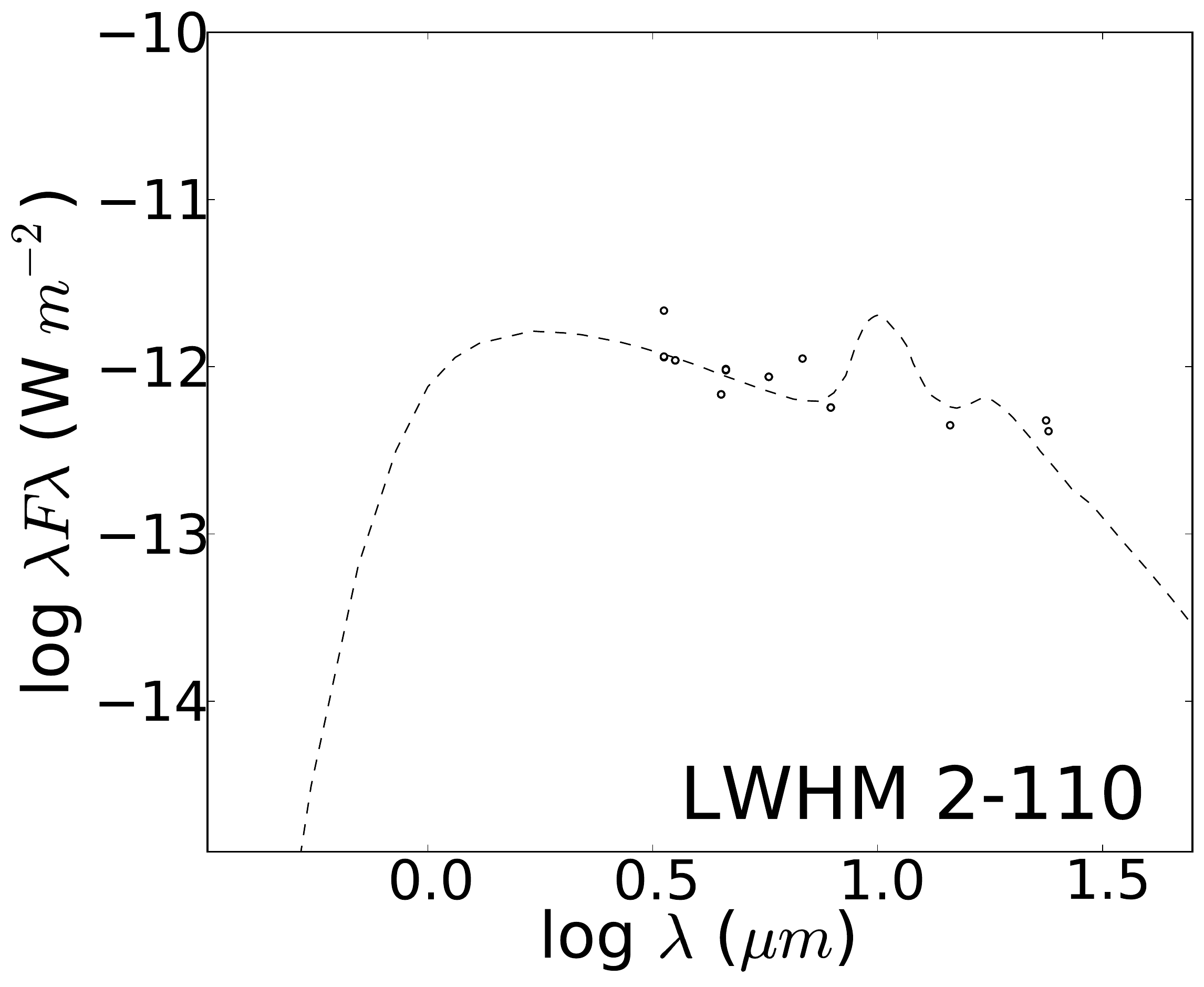}
     \includegraphics[width=0.243\textwidth]{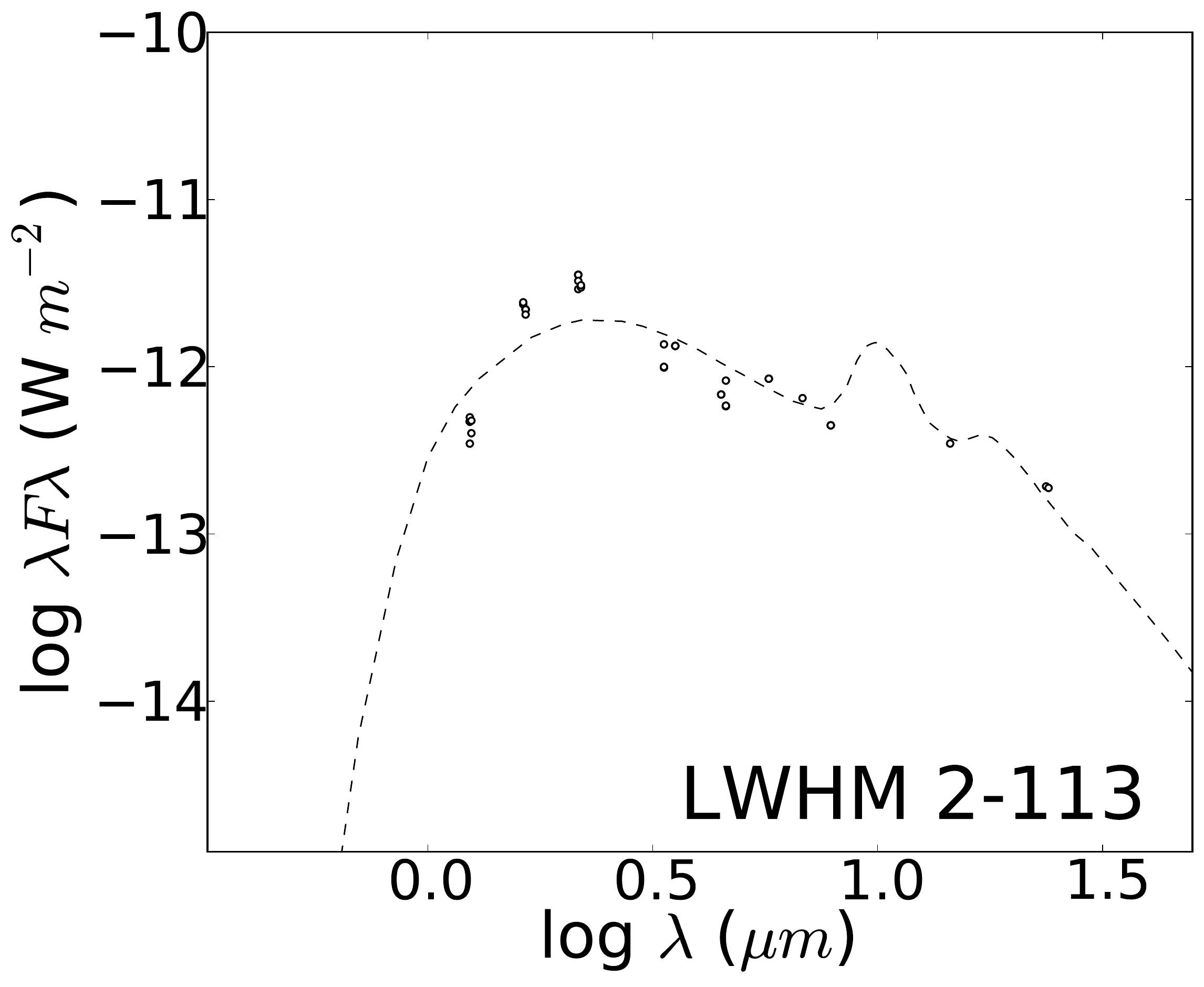}  \\
     \vspace{-0.15cm}
     \includegraphics[width=0.243\textwidth]{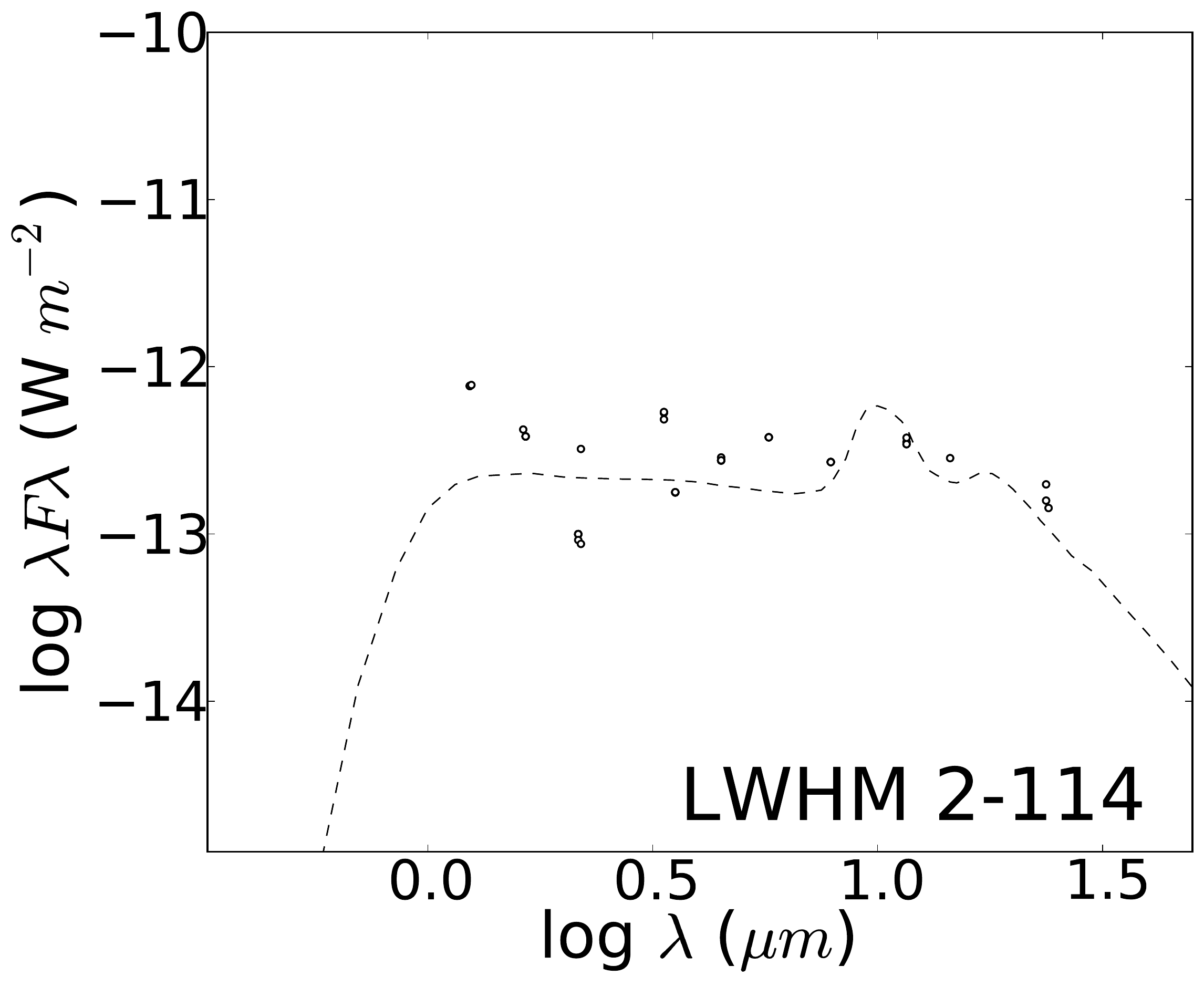}  
     \includegraphics[width=0.243\textwidth]{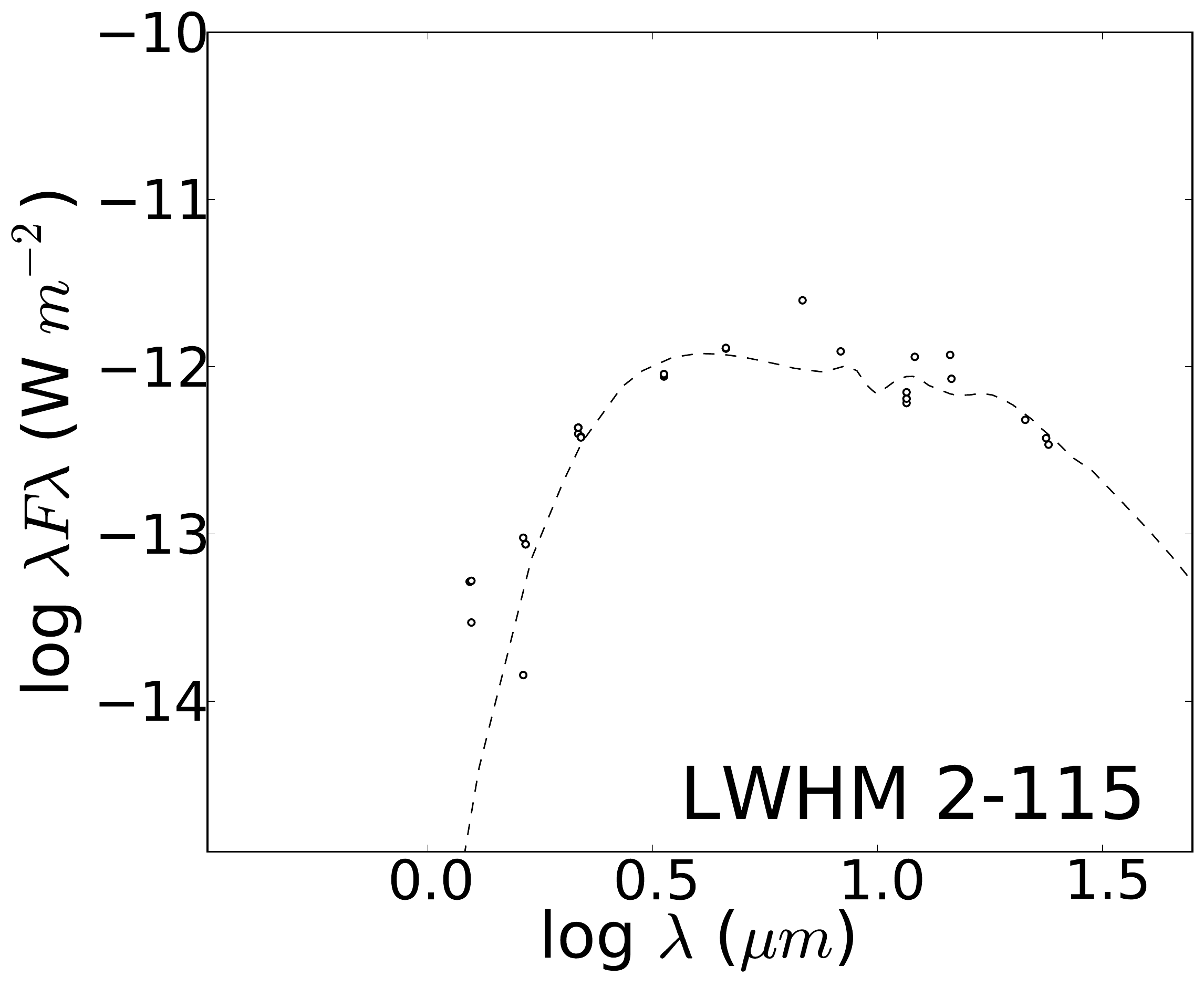}  
     \includegraphics[width=0.243\textwidth]{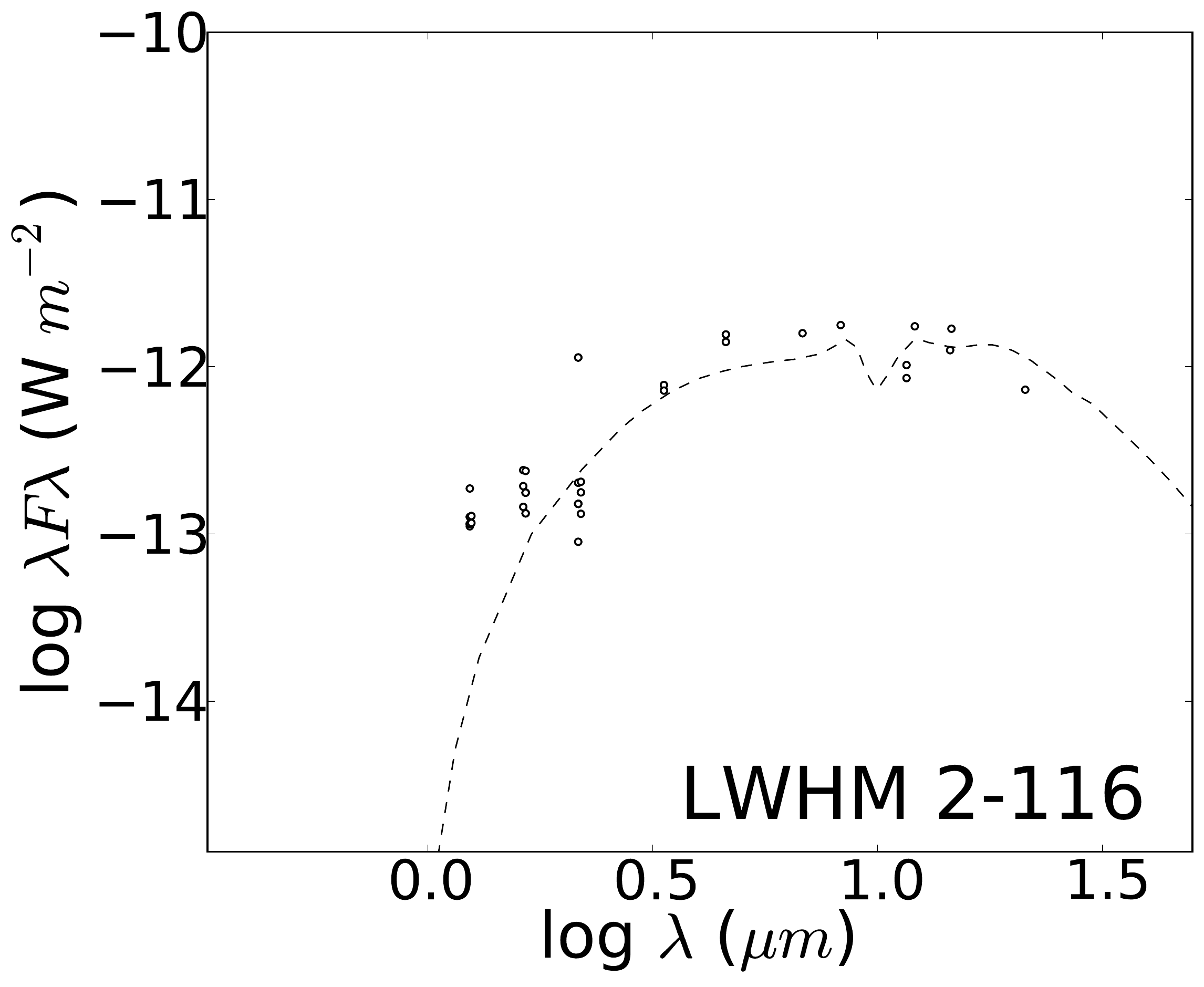}
     \includegraphics[width=0.243\textwidth]{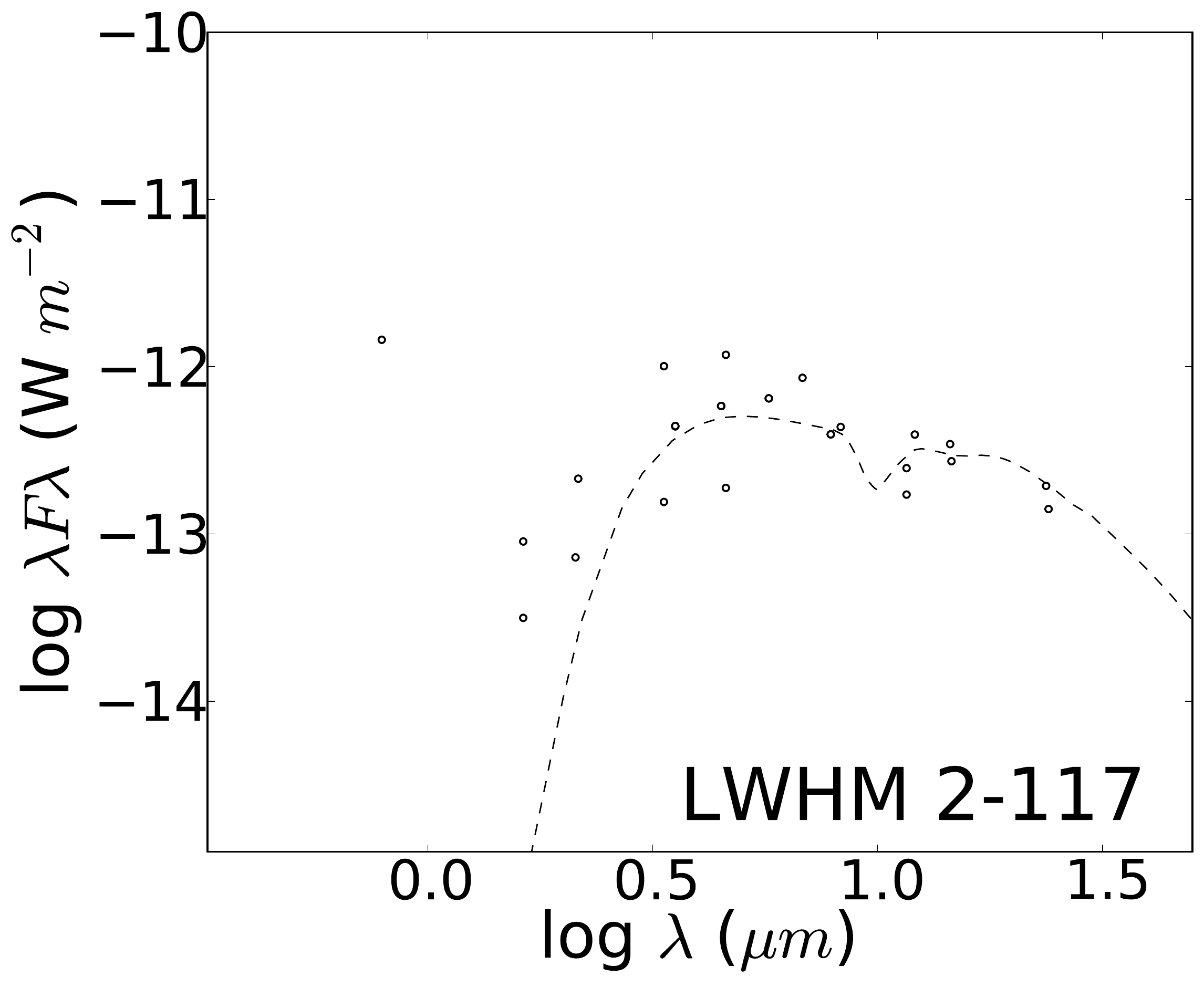}  \\
     \vspace{-0.15cm}
     \includegraphics[width=0.243\textwidth]{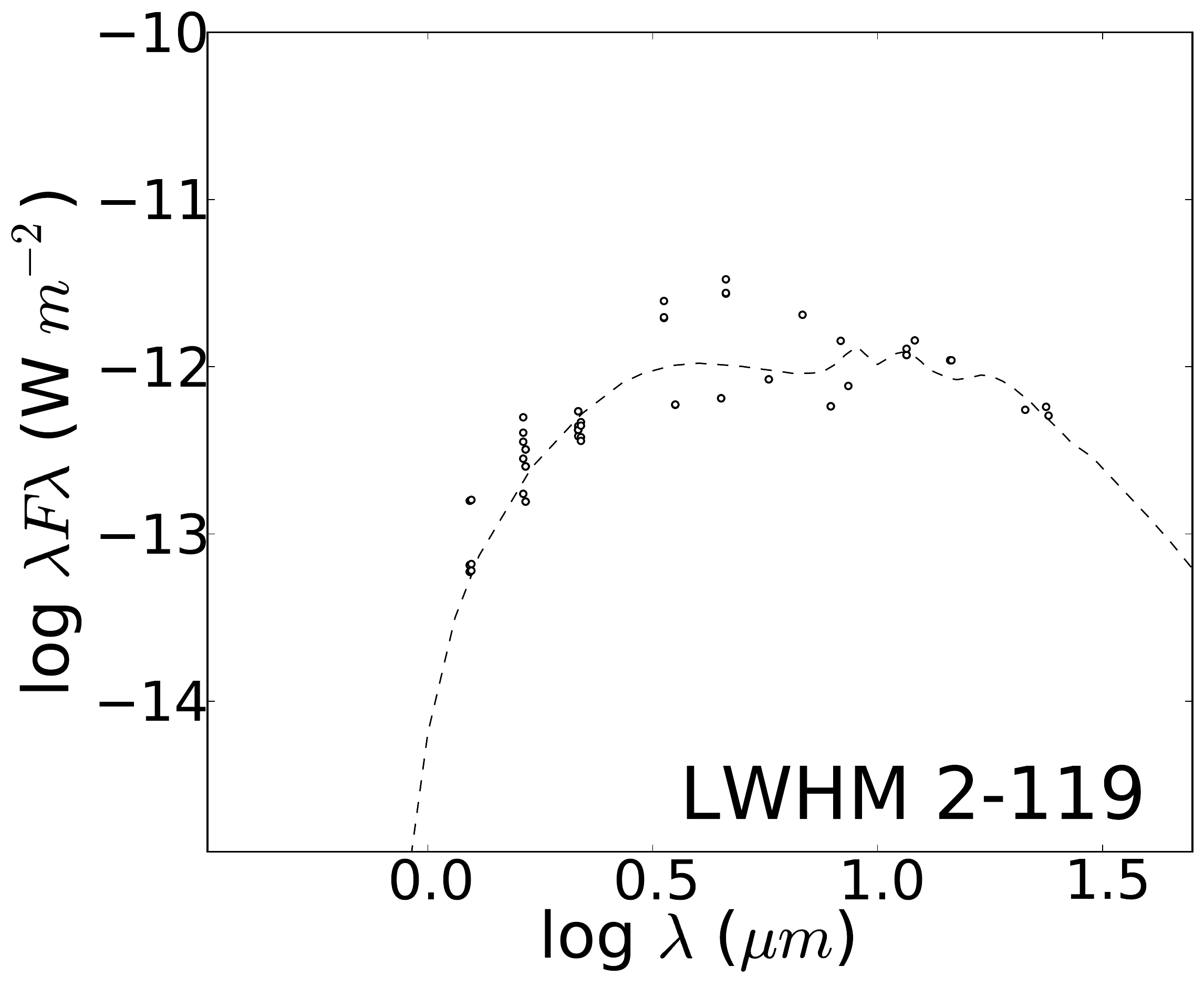}  
     \includegraphics[width=0.243\textwidth]{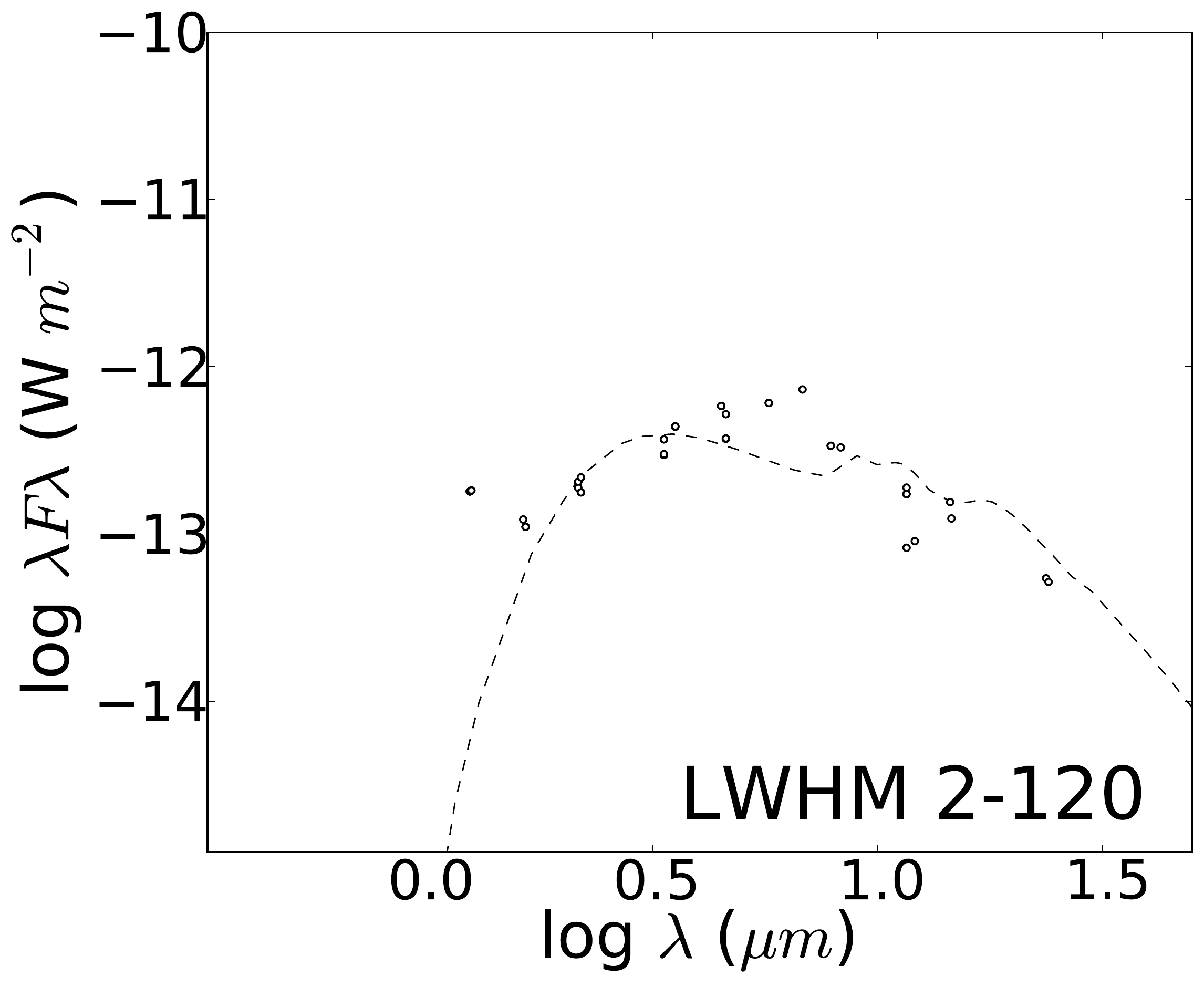}  
     \includegraphics[width=0.243\textwidth]{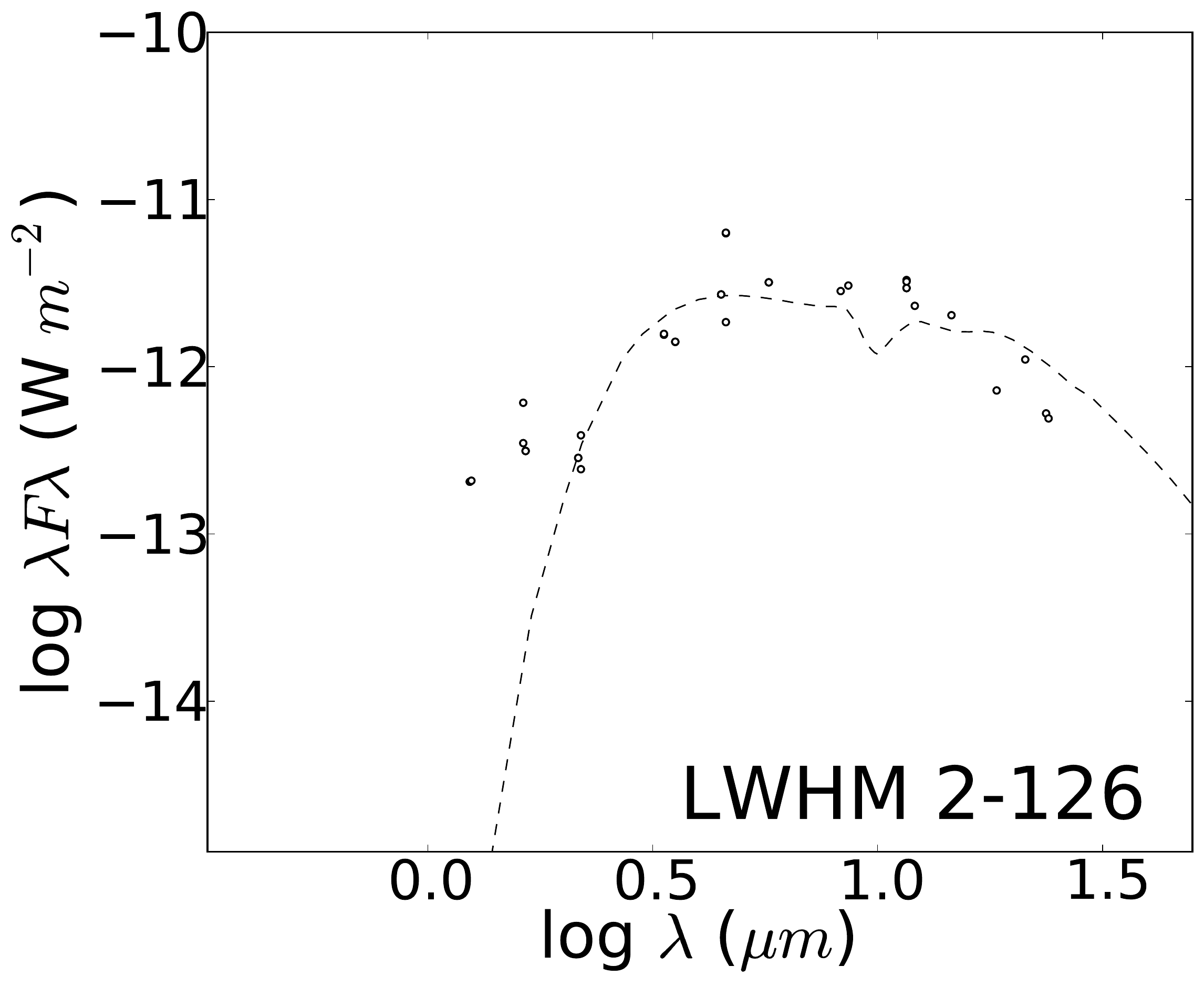}
     \includegraphics[width=0.243\textwidth]{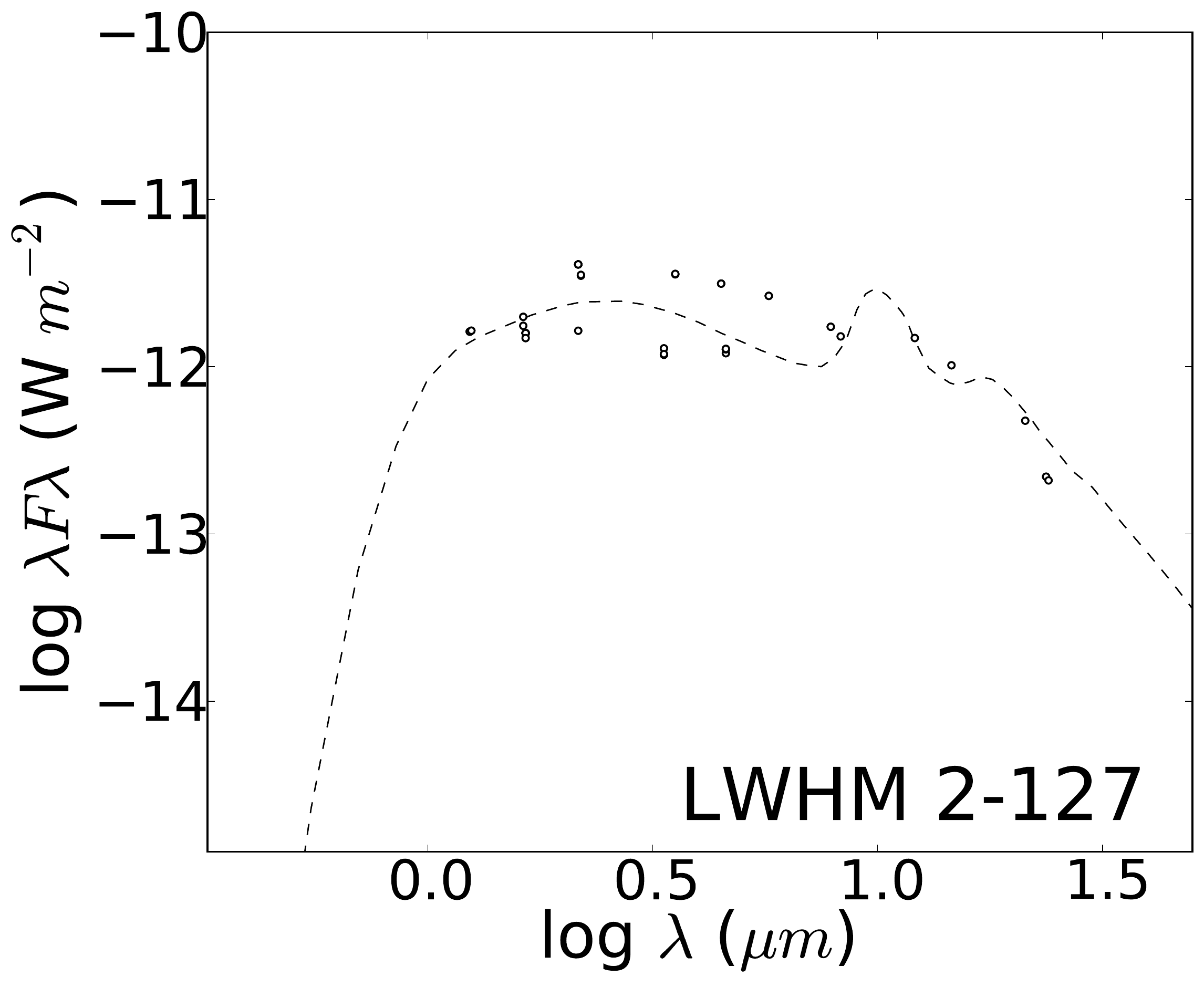}  \\ 

     \end{tabular}
   \end{center}
   \vspace{-0.4cm} 
   \caption{continued}
   \label{dfg} 
   \end{minipage}
   \vspace{1in}
\end{figure*}

\begin{figure*}
  \begin{minipage}[c]{\textwidth}
  \begin{center}
     \begin{tabular}{c}

     \vspace{-0.15cm}
     \includegraphics[width=0.243\textwidth]{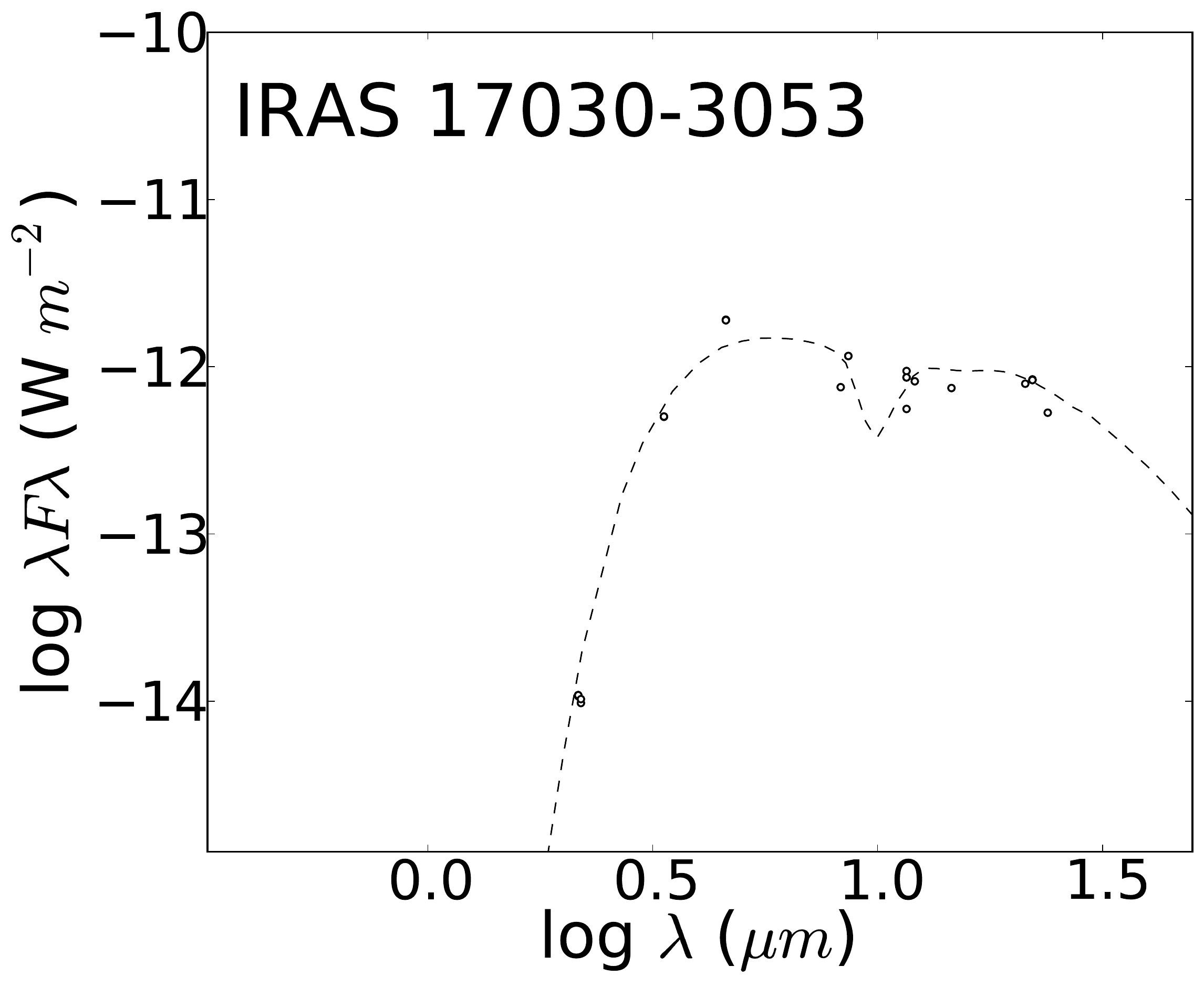}  
     \includegraphics[width=0.243\textwidth]{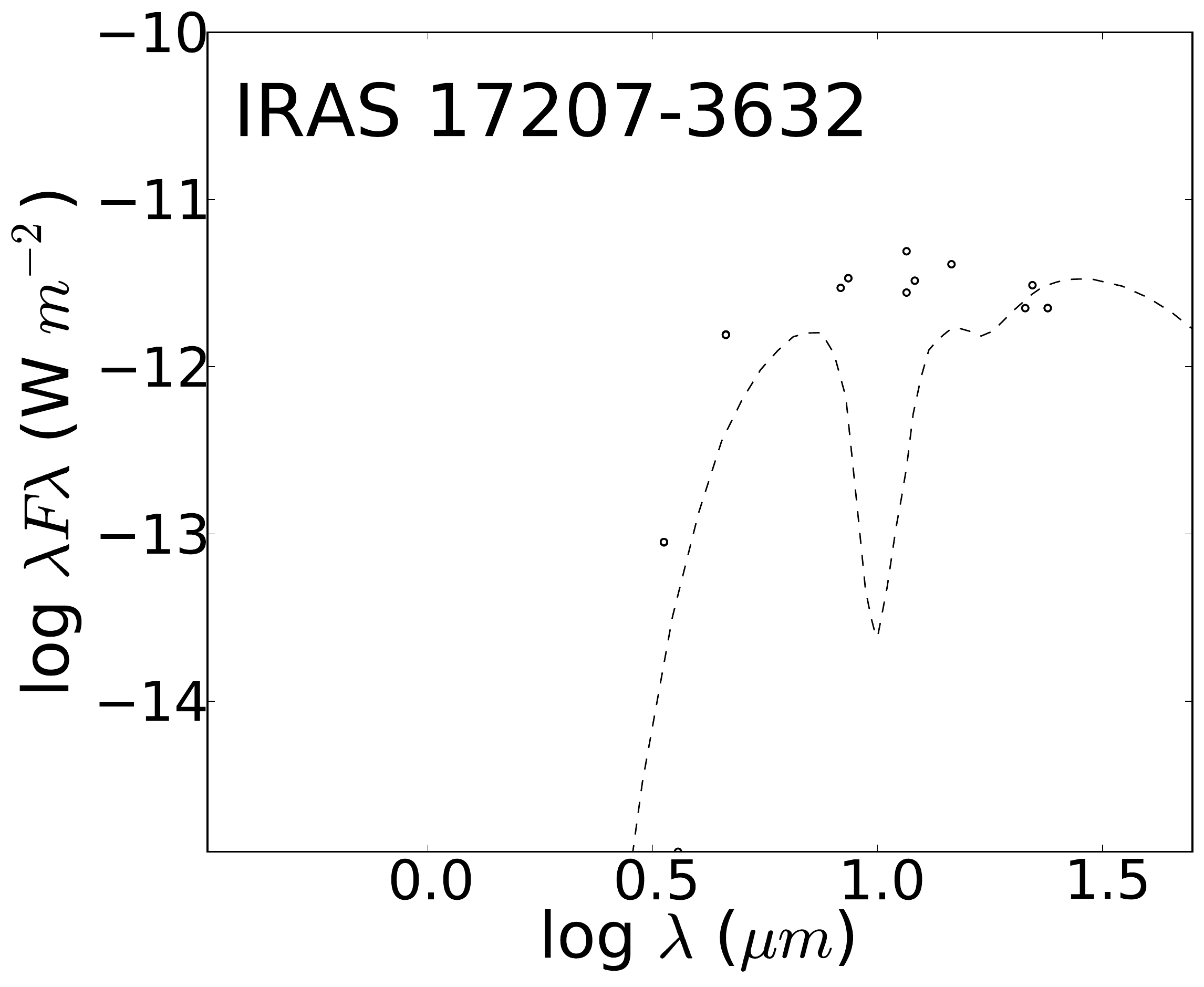}  
     \includegraphics[width=0.243\textwidth]{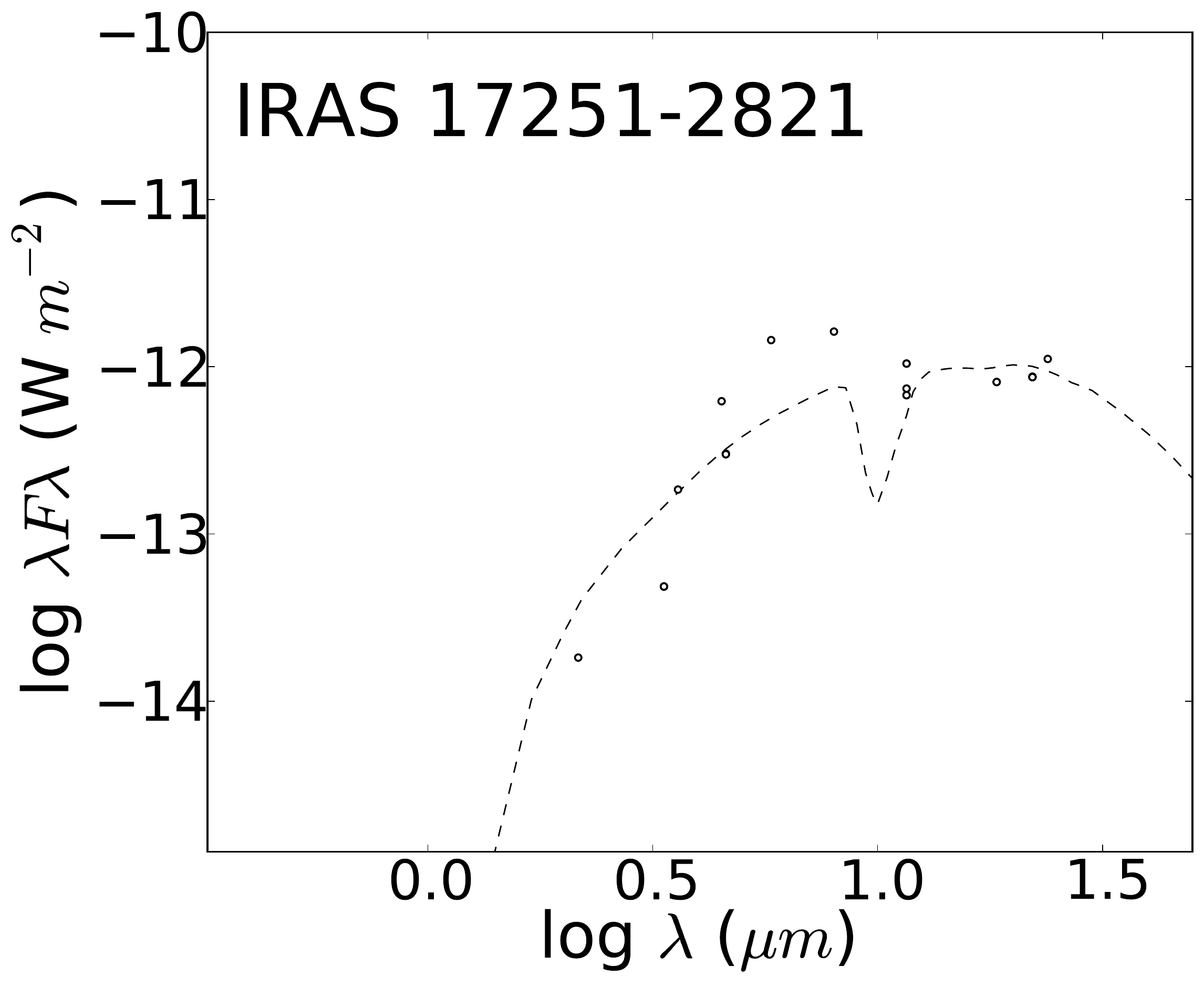}  
     \includegraphics[width=0.243\textwidth]{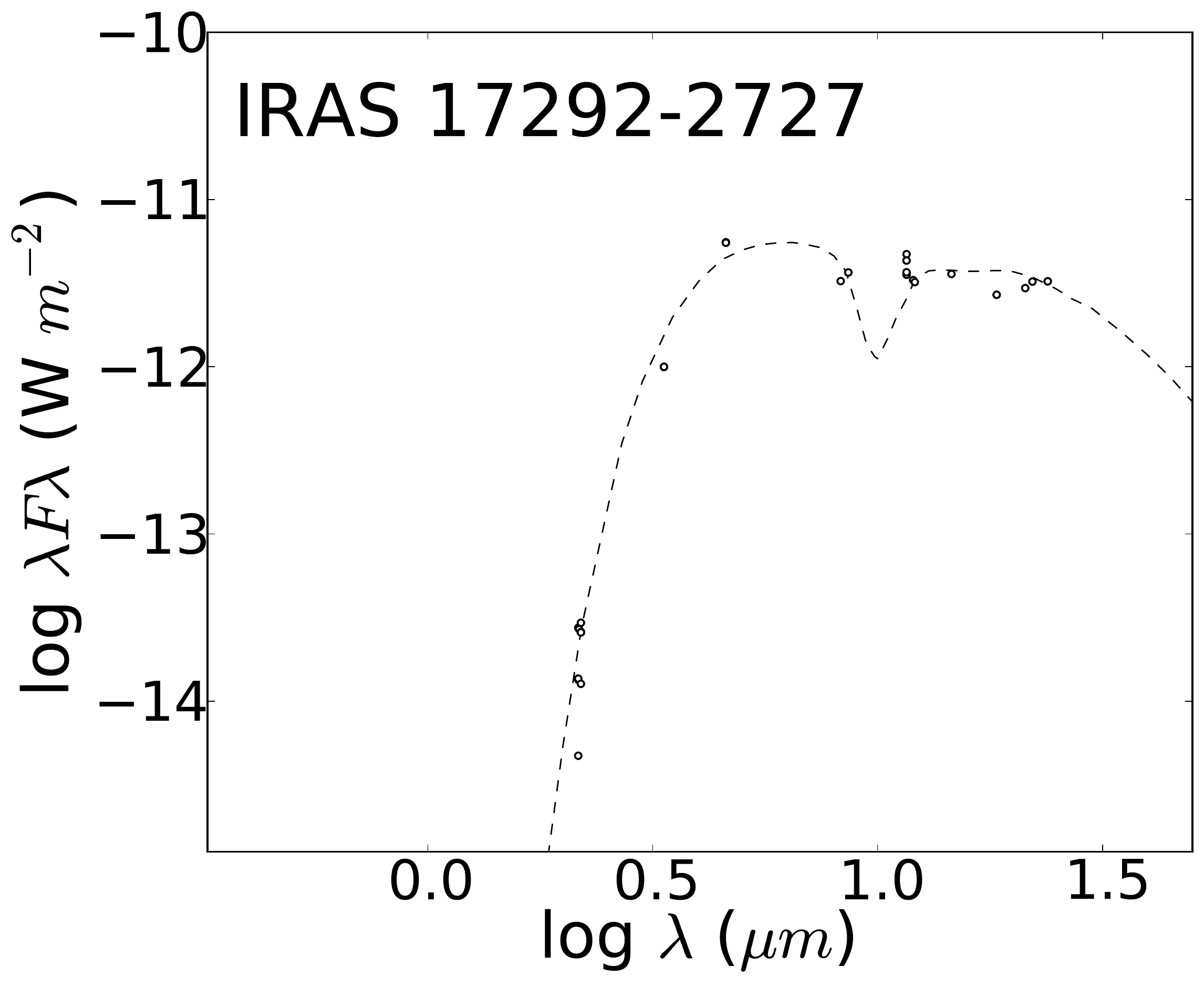} \\
     \vspace{-0.15cm}
     \includegraphics[width=0.243\textwidth]{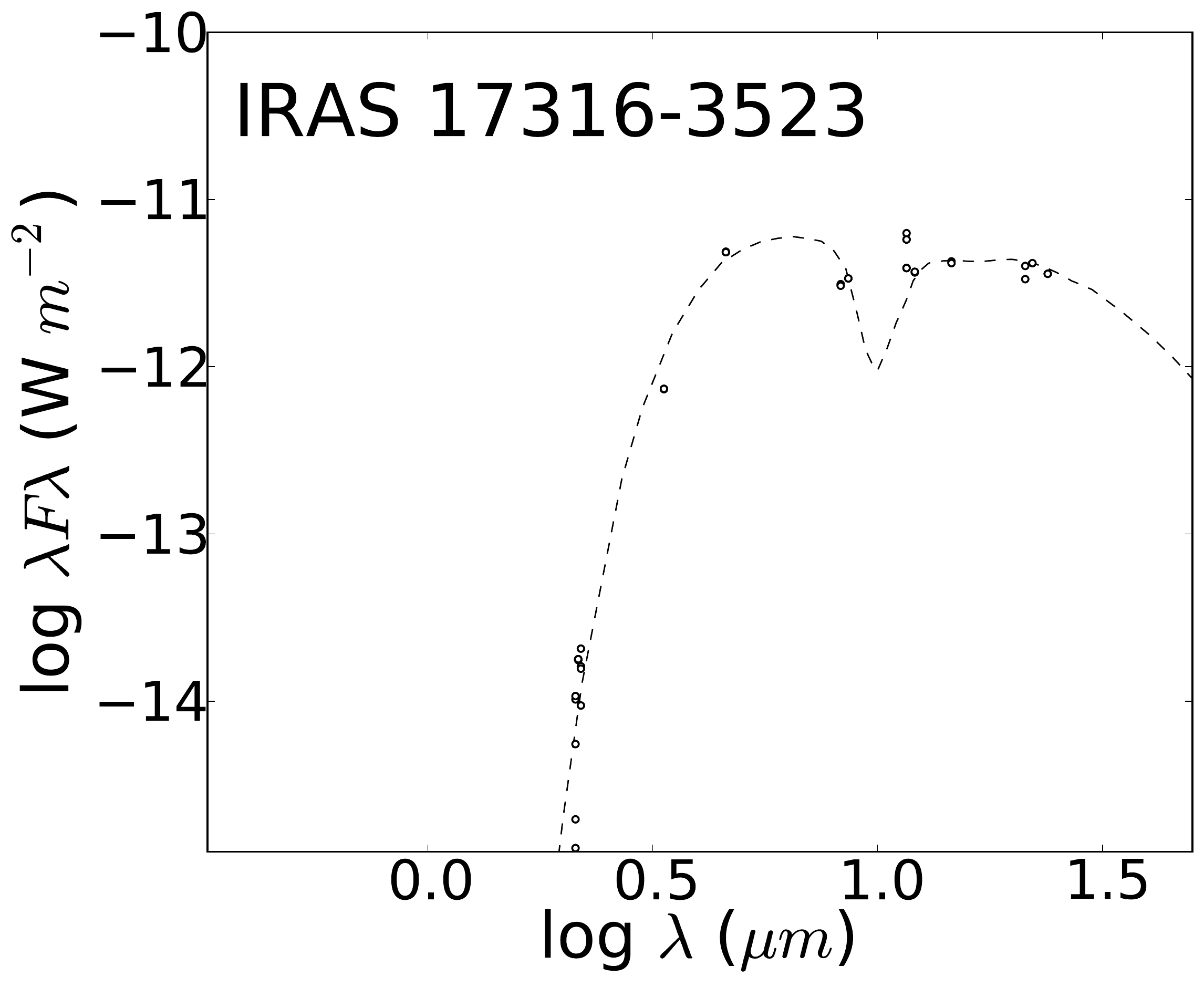}  
     \includegraphics[width=0.243\textwidth]{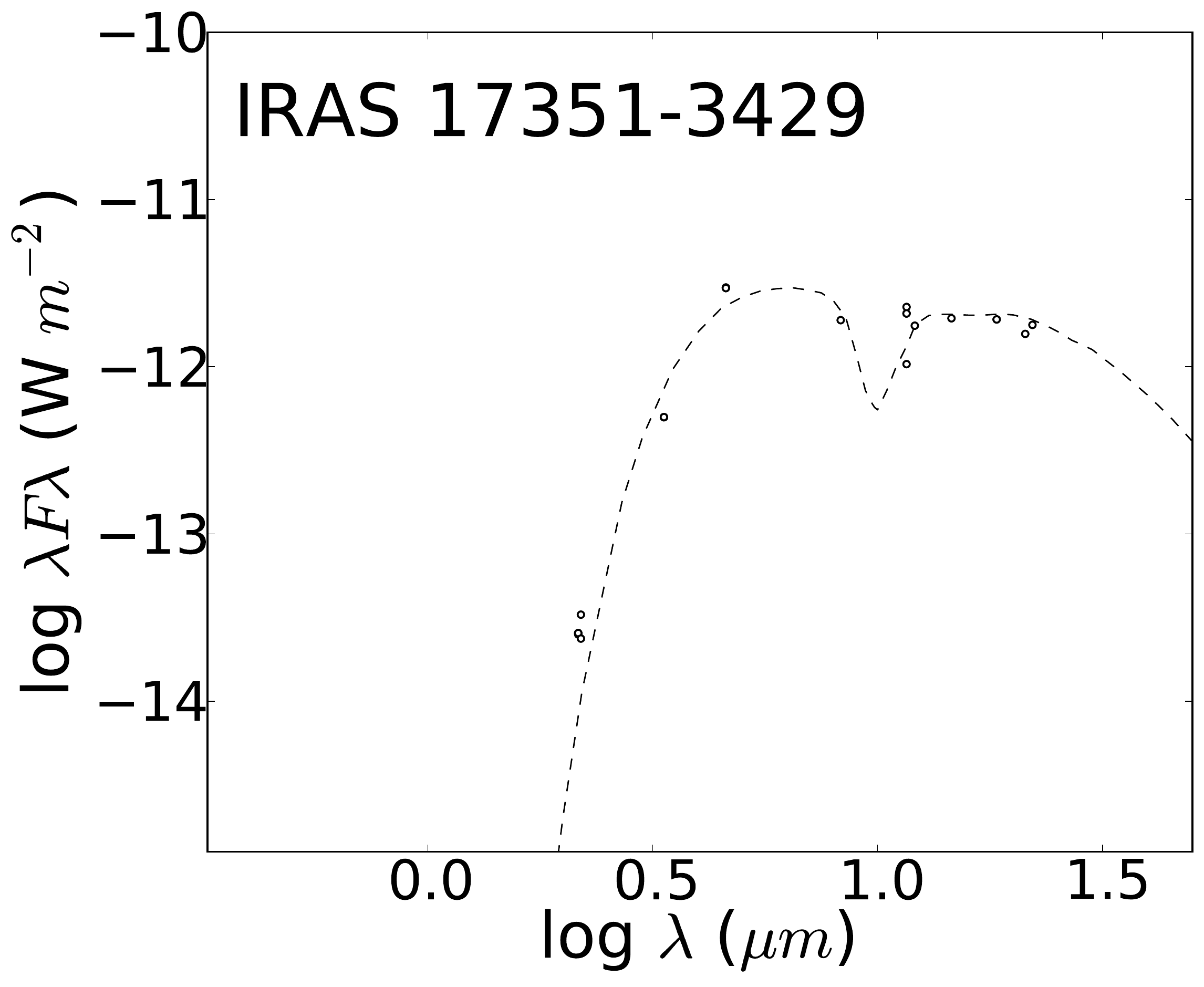}  
     \includegraphics[width=0.243\textwidth]{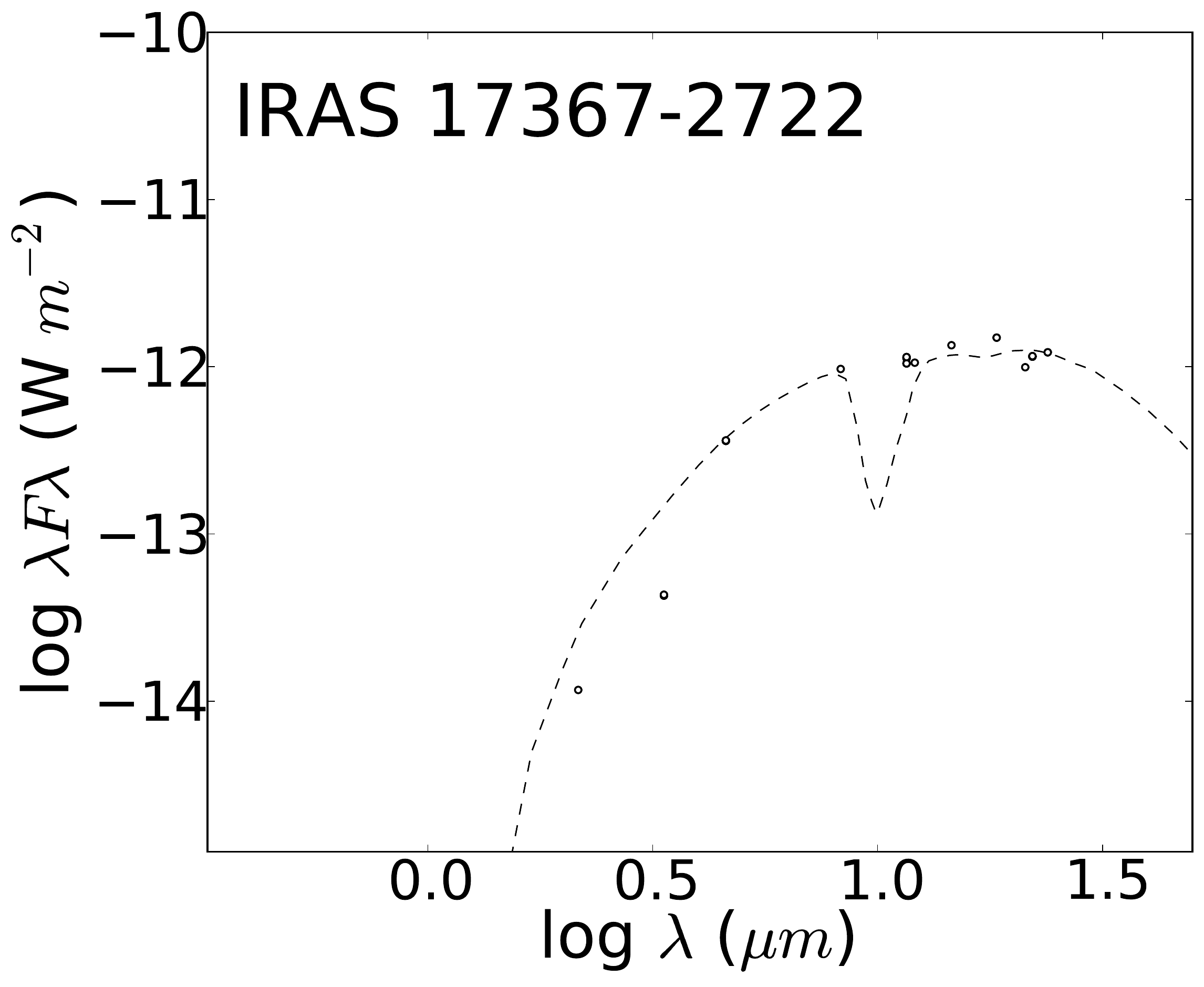} 
     \includegraphics[width=0.243\textwidth]{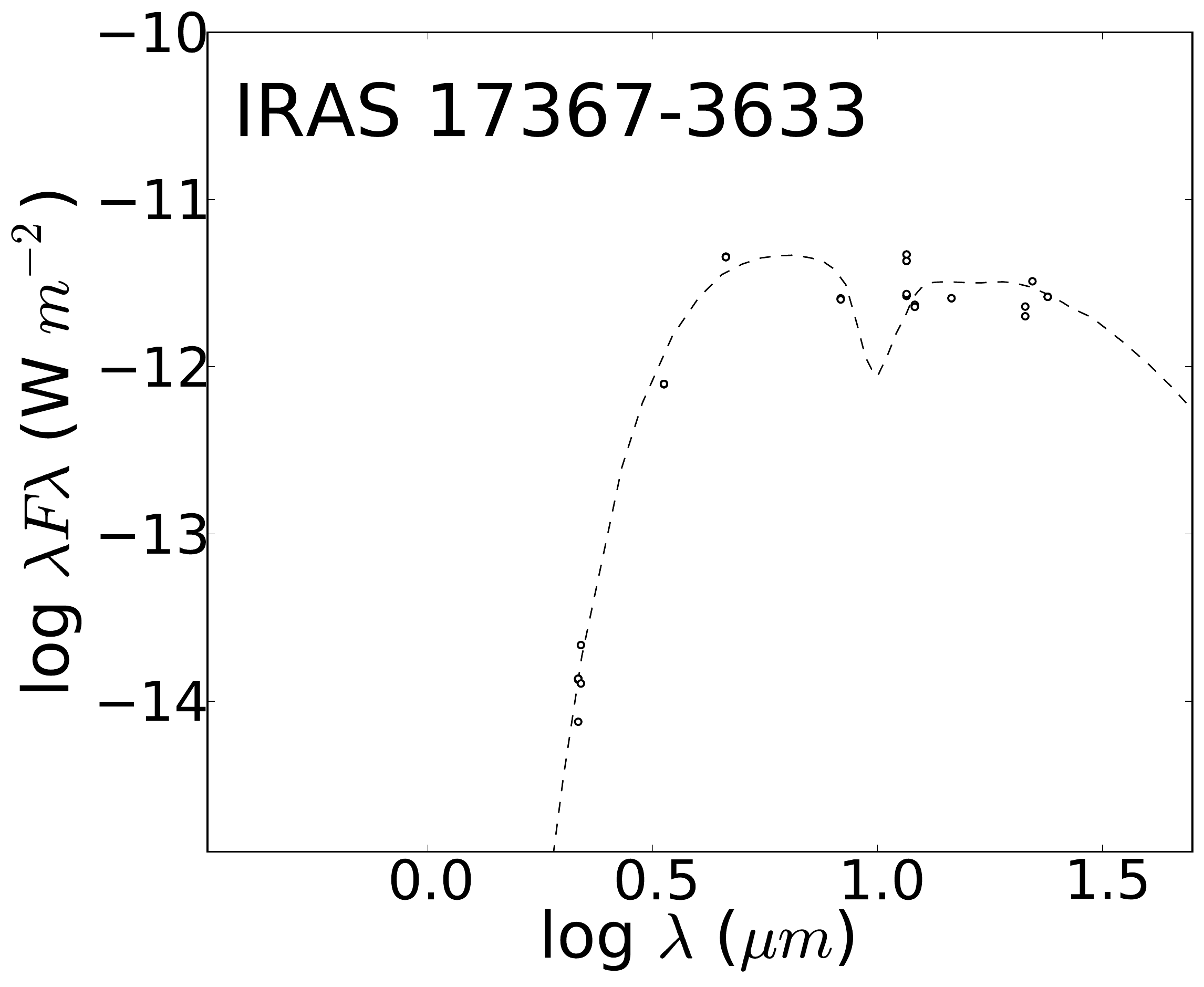} \\
     \vspace{-0.15cm}
     \includegraphics[width=0.243\textwidth]{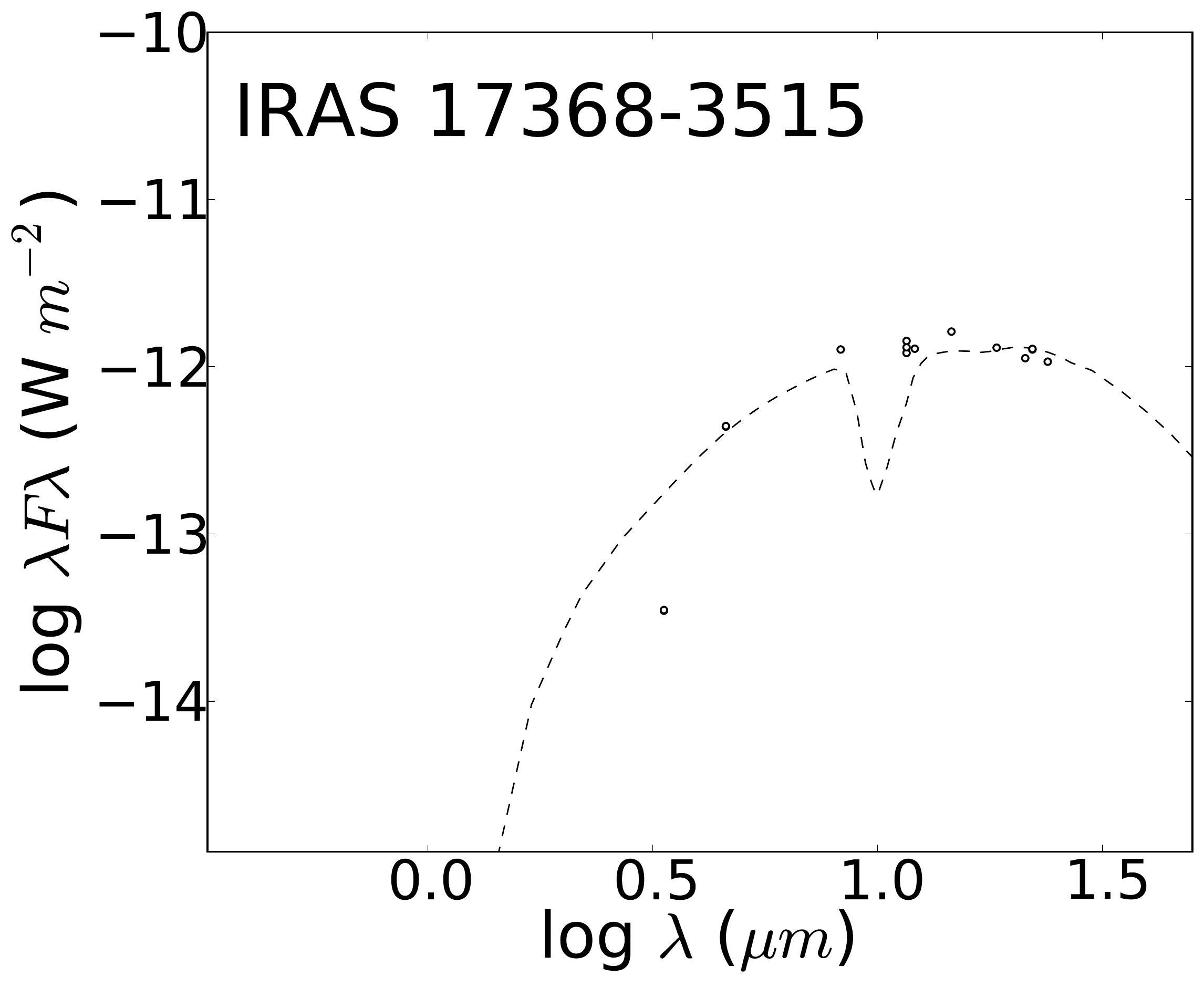}  
     \includegraphics[width=0.243\textwidth]{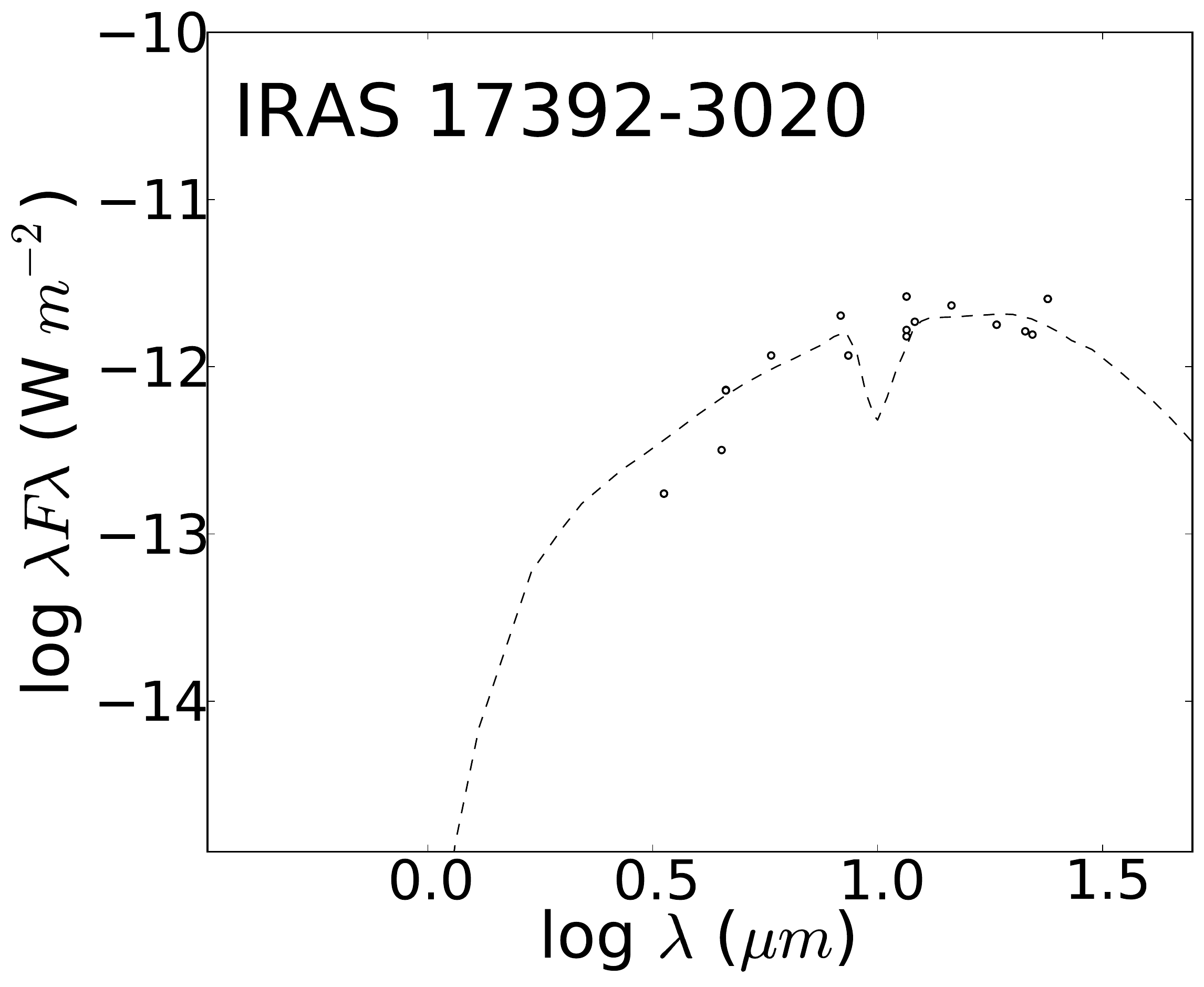}  
     \includegraphics[width=0.243\textwidth]{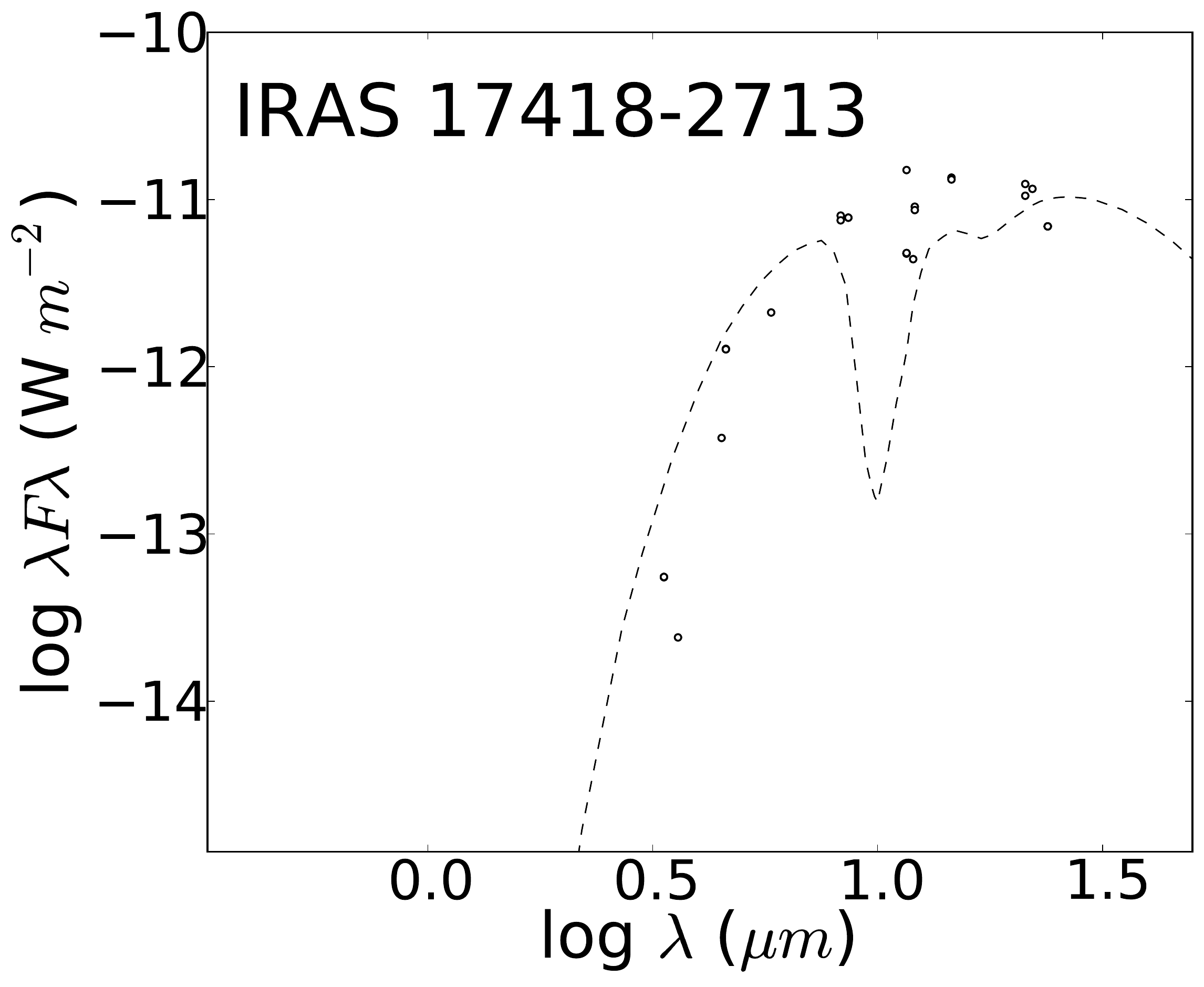} 
     \includegraphics[width=0.243\textwidth]{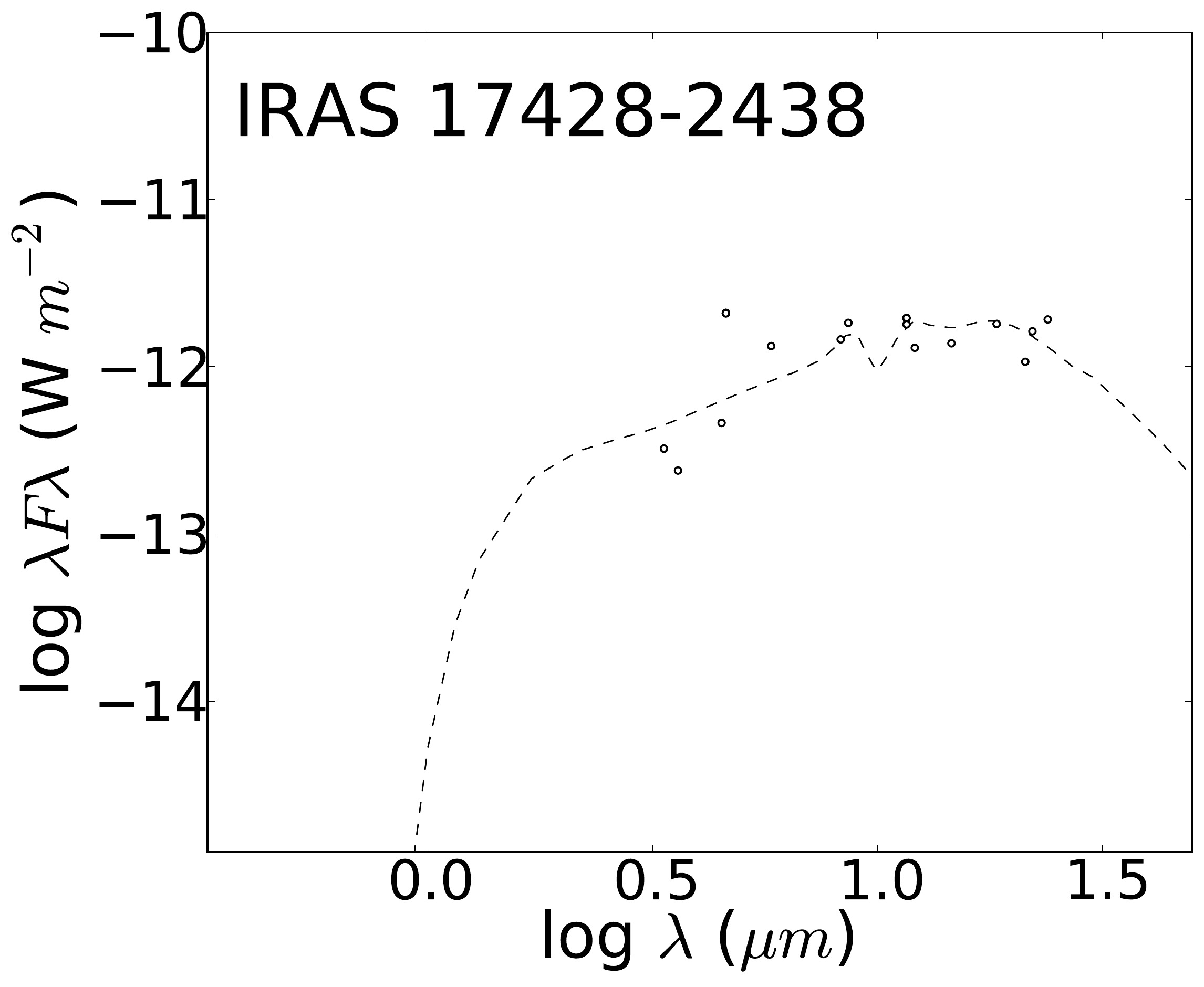} \\
     \vspace{-0.15cm}
     \includegraphics[width=0.243\textwidth]{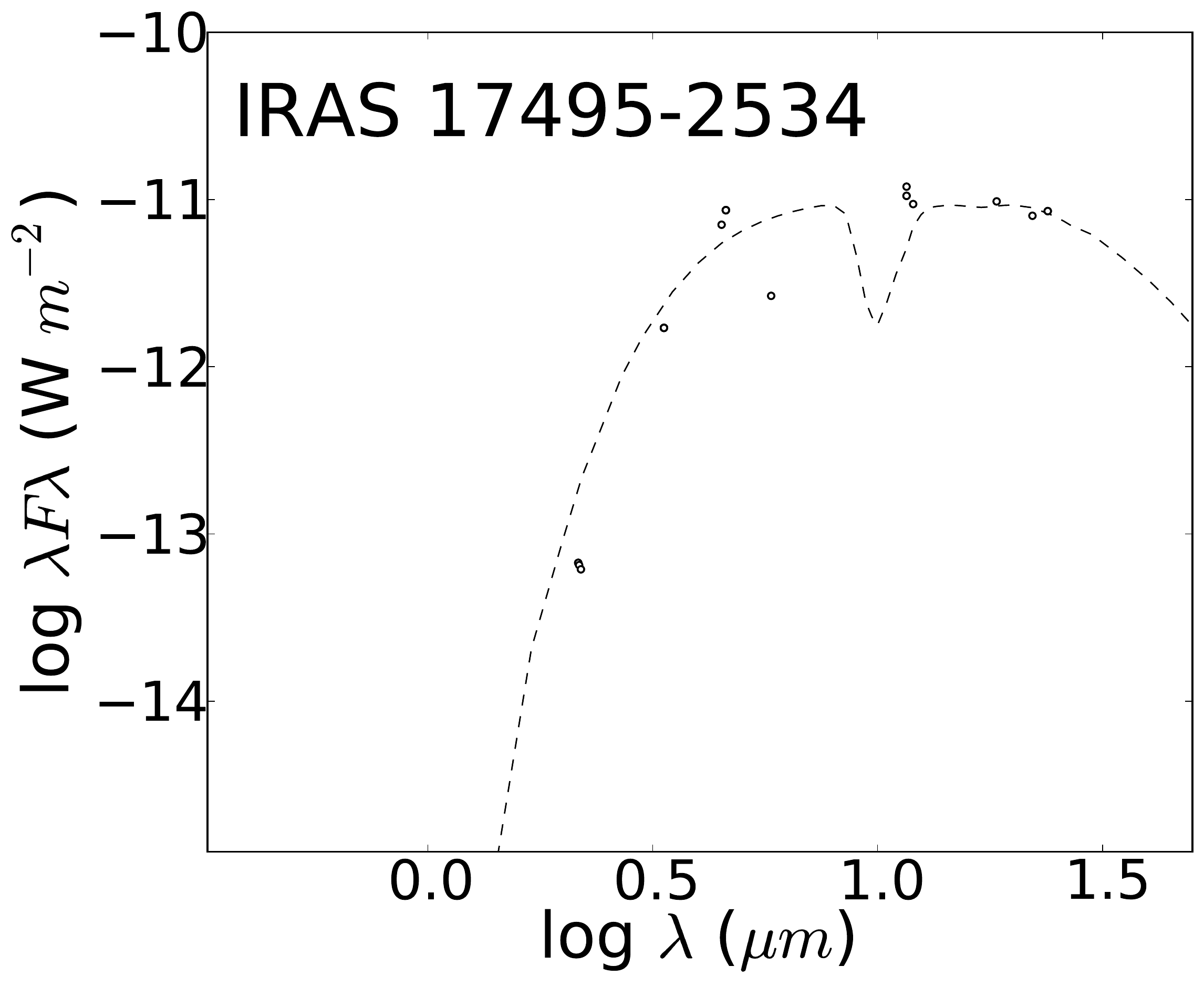}  
     \includegraphics[width=0.243\textwidth]{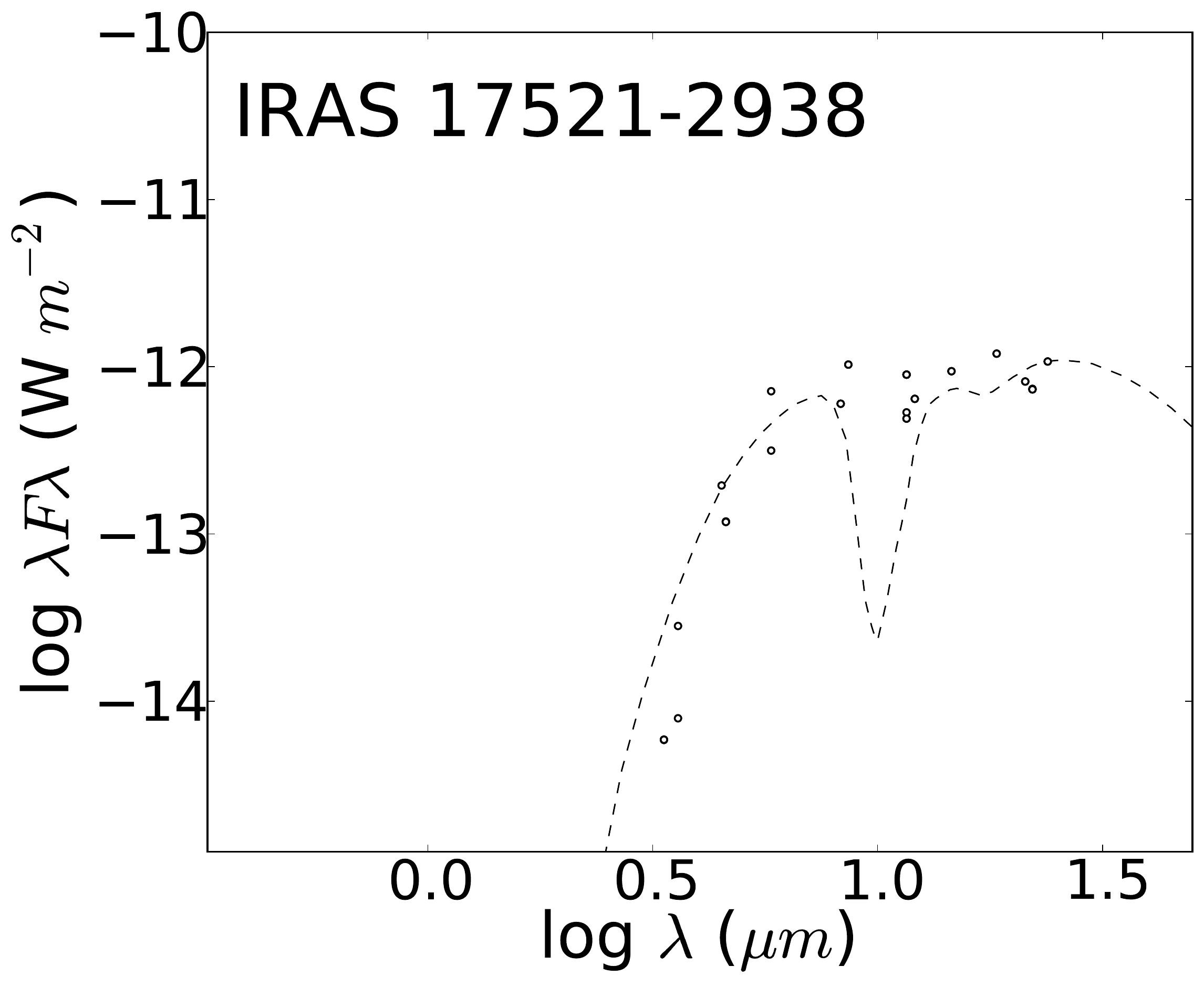}  
     \includegraphics[width=0.243\textwidth]{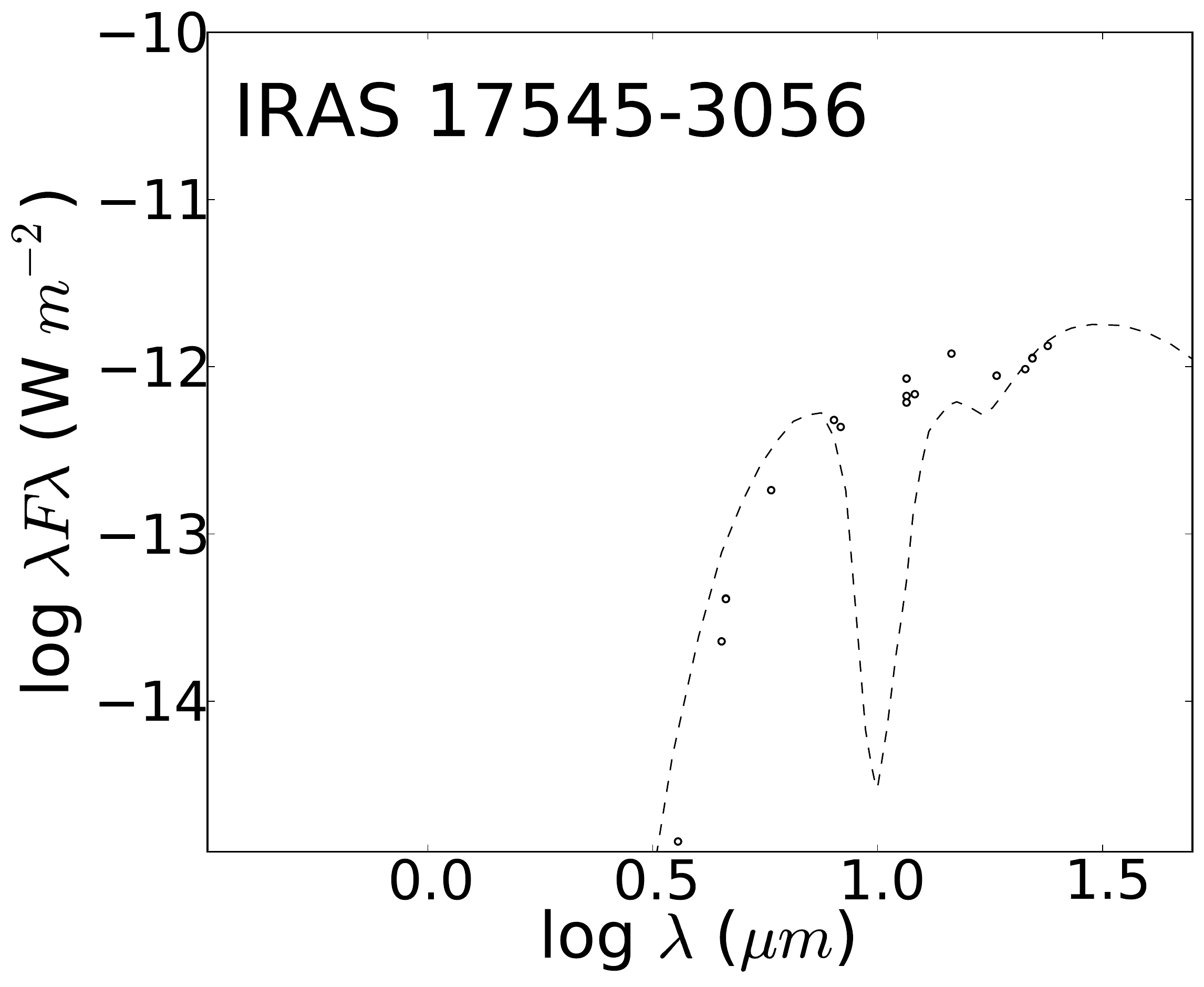} 
     \includegraphics[width=0.243\textwidth]{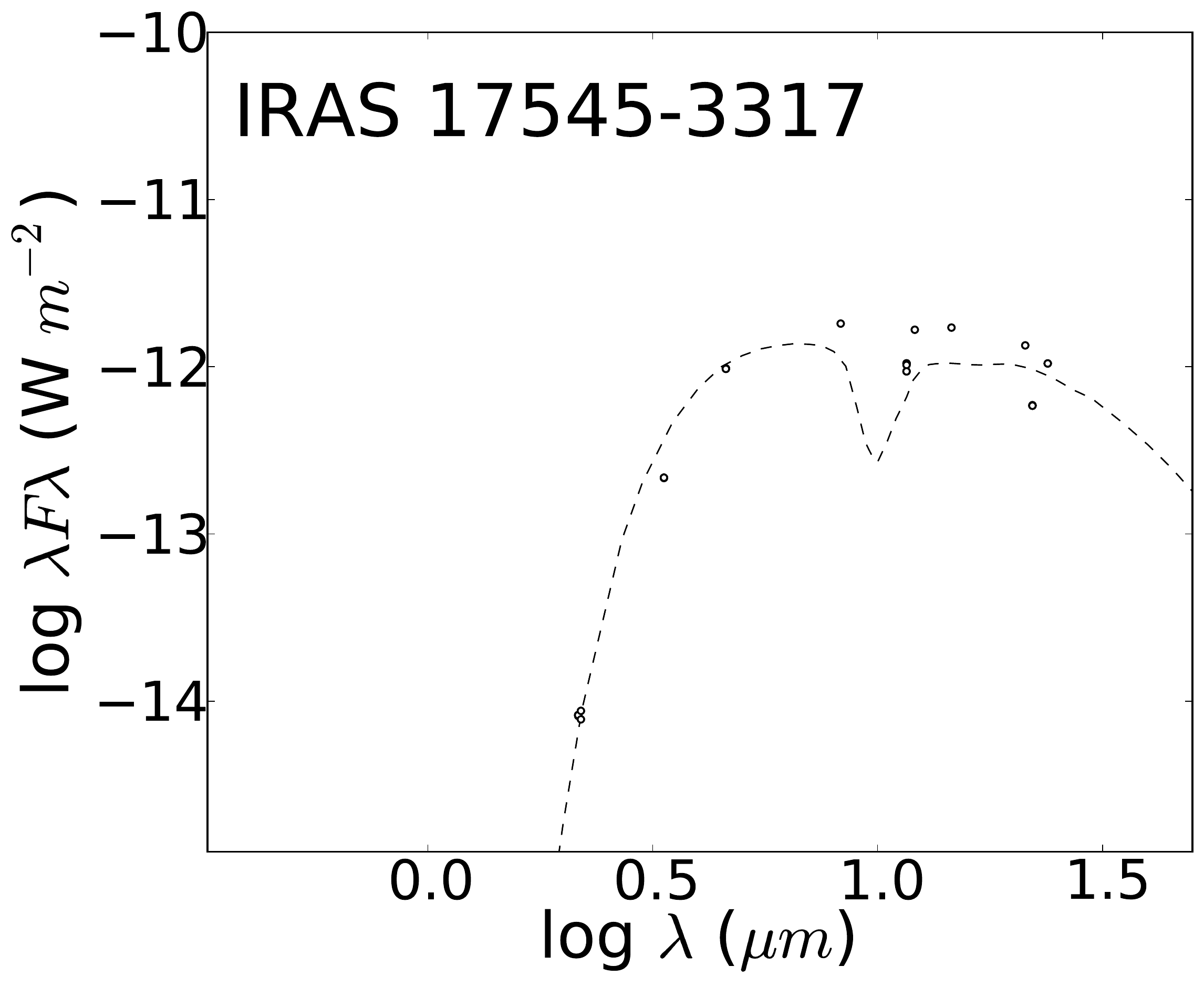} \\

     \includegraphics[width=0.243\textwidth]{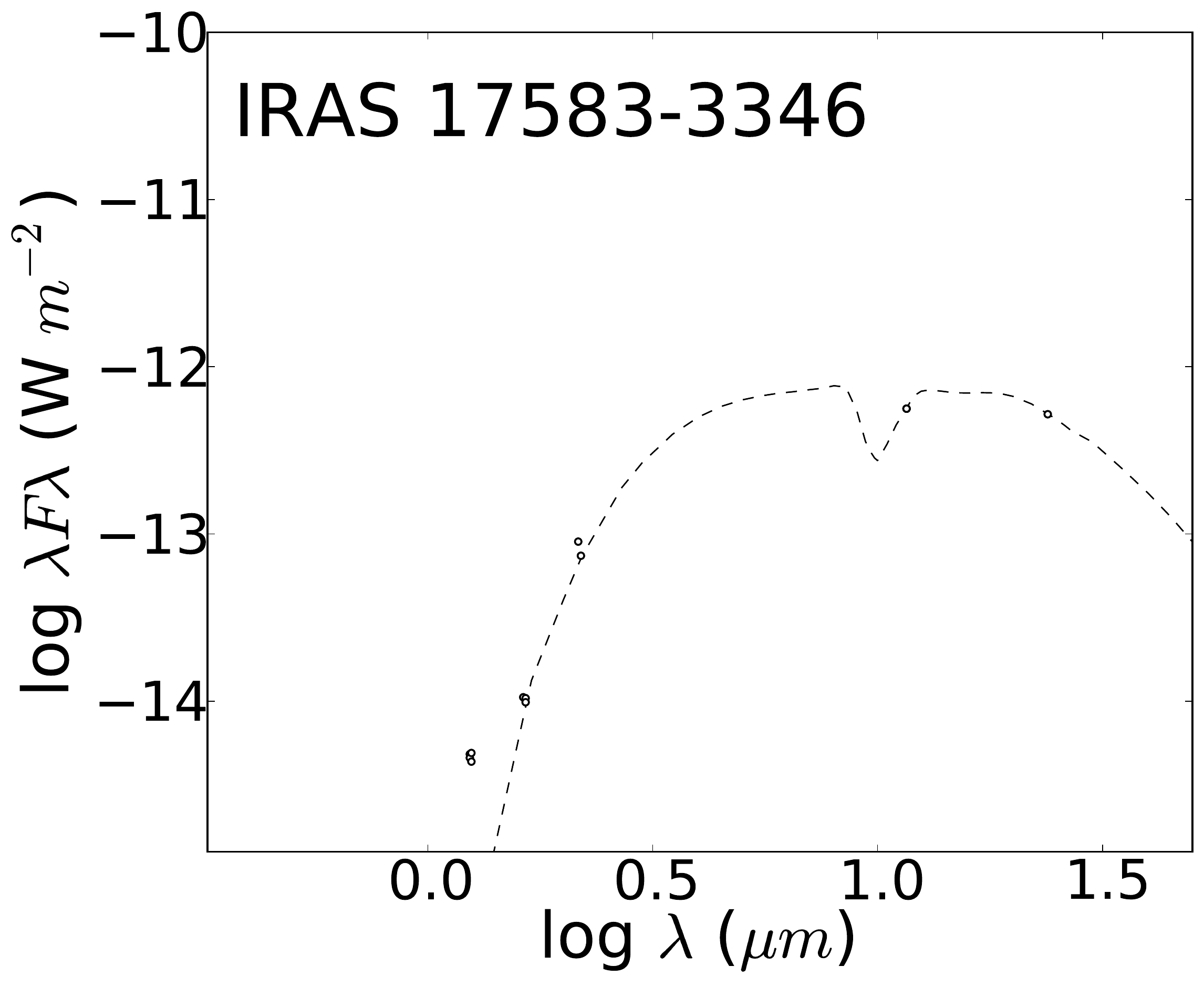}  
     \includegraphics[width=0.243\textwidth]{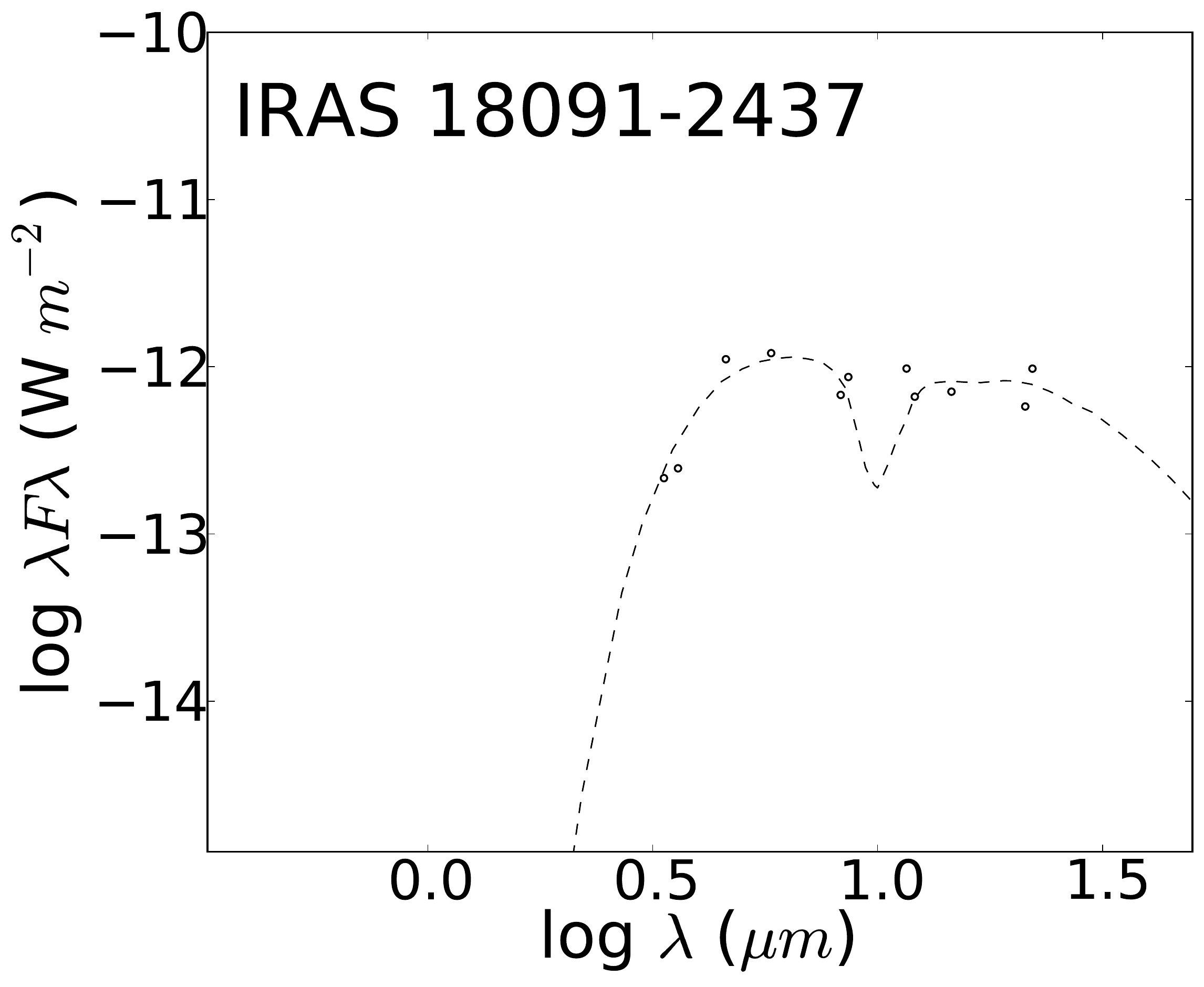}  
     \includegraphics[width=0.243\textwidth]{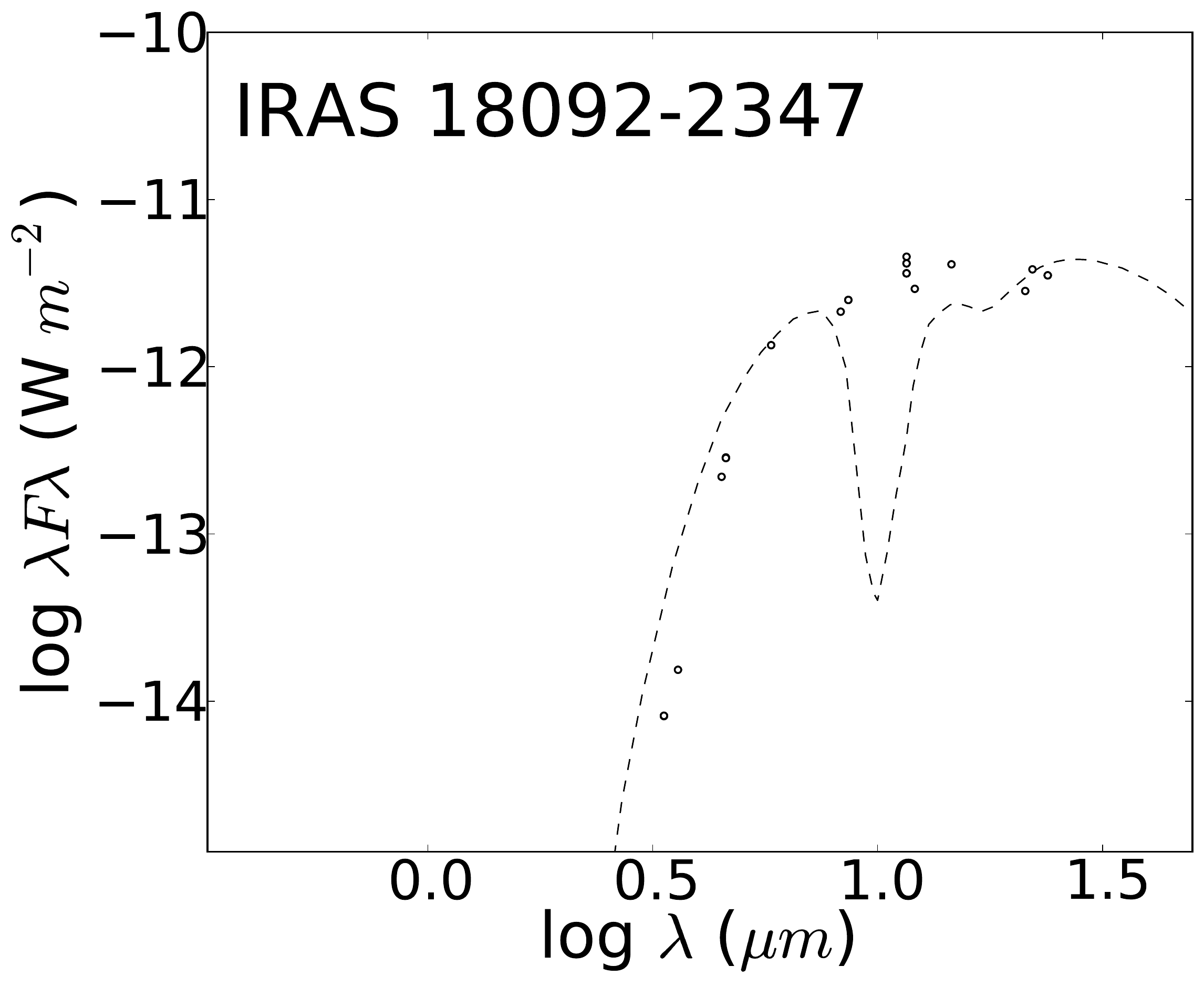} 
     \includegraphics[width=0.243\textwidth]{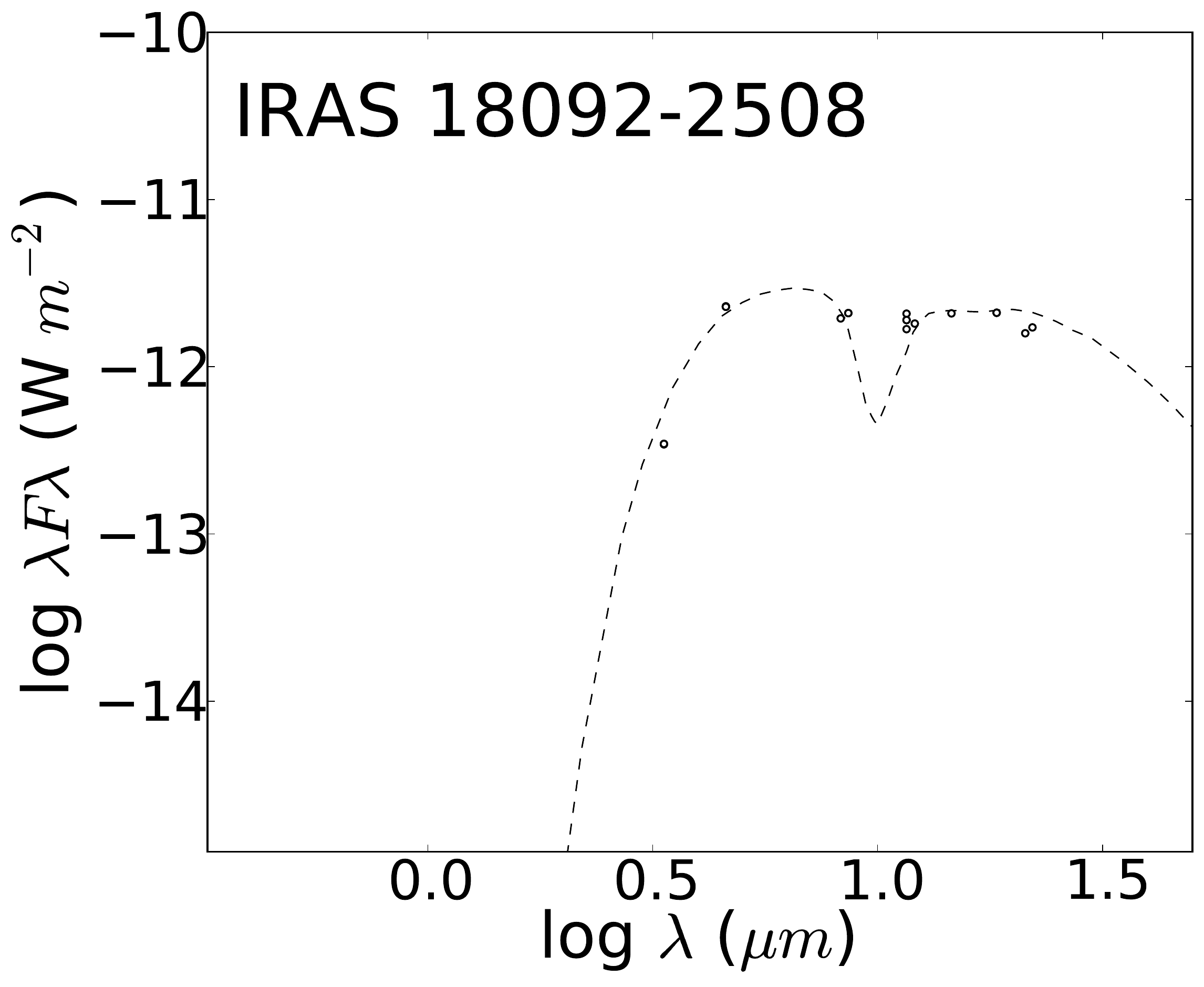} \\ 
     \includegraphics[width=0.243\textwidth]{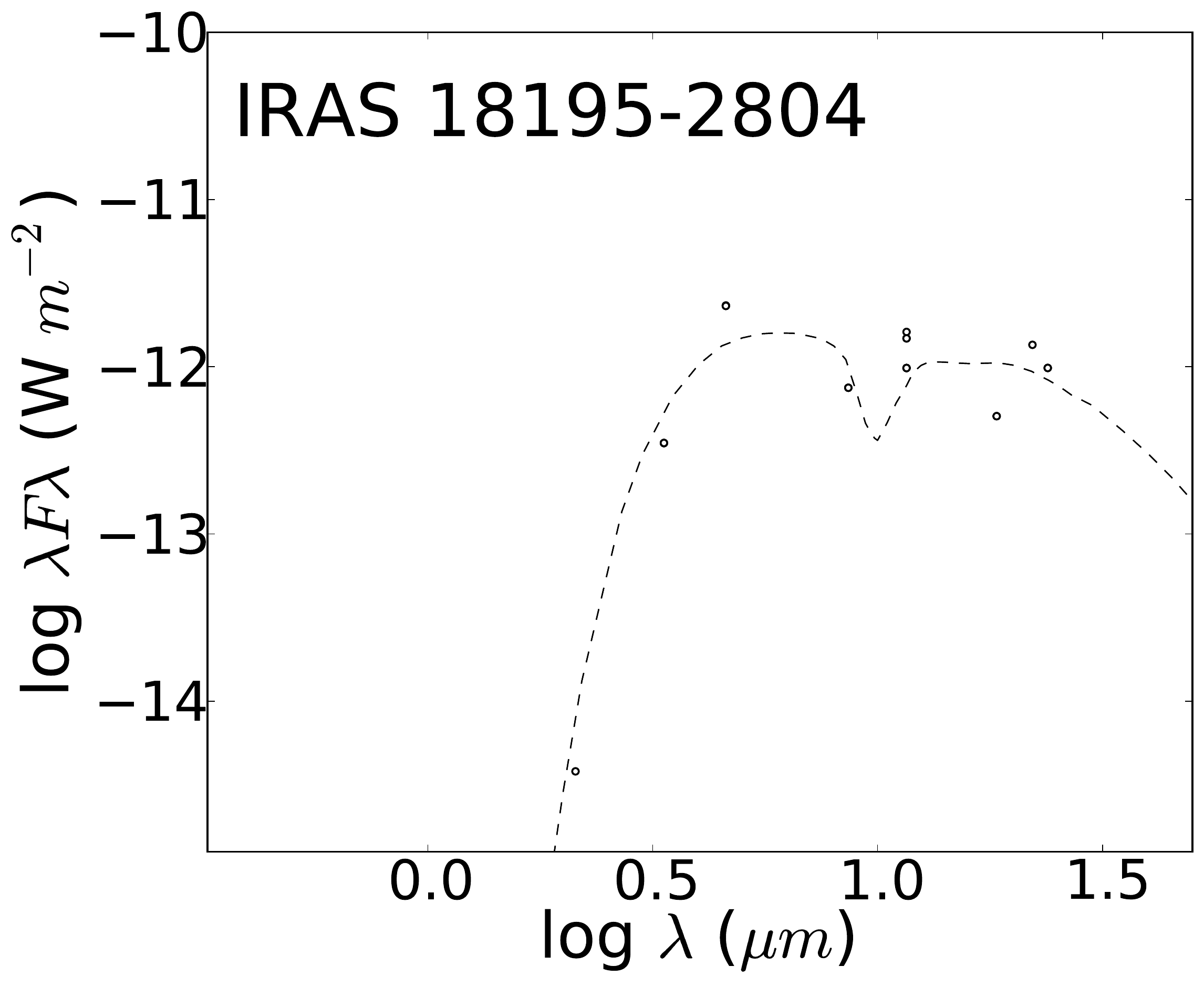} \\
     \end{tabular}
   \end{center}
   \caption{The SED fitting of \textsc{dusty} models to the dereddened photometry from our Galactic Bulge sources.}
   \label{dfj} 
   \end{minipage}
\end{figure*}

\renewcommand{\thefigure}{\arabic{figure}}

\section{Galactic OH maser sources }
\renewcommand{\thefigure}{B\arabic{figure}}
\setcounter{figure}{0}
 \begin{figure*}
 \centering
 \begin{minipage}[c]{\textwidth}
  \centering
  \includegraphics[width=17cm]{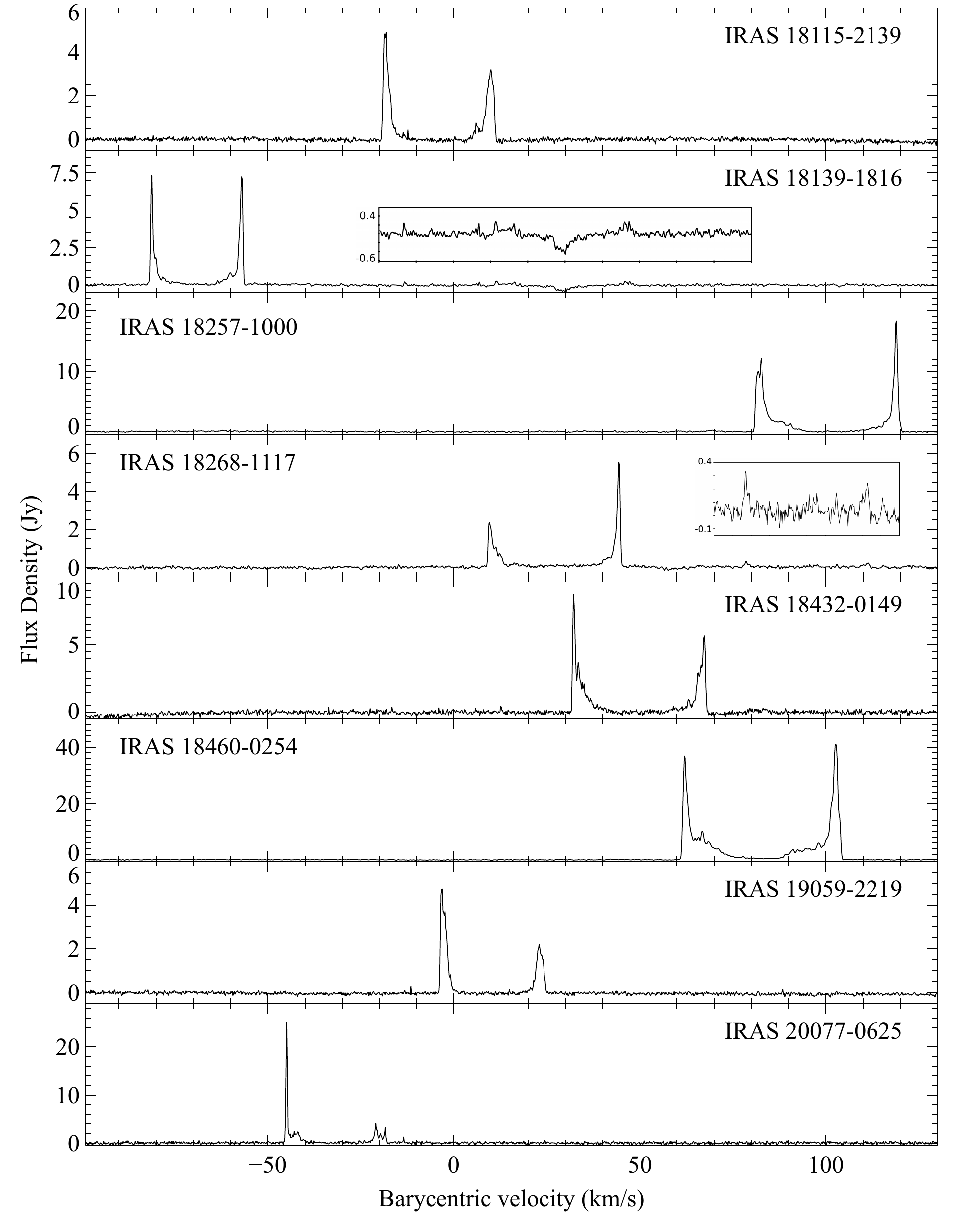}
  \caption{OH 1612-MHz maser emission from our Parkes observations of additional Galactic sources. Sources are from different Galactic samples than those modeled in previous sections.}
 \end{minipage}
 \end{figure*}

We observed eight Galactic maser sources with Parkes in addition to our LMC sources for pointing checks (Table B1). We did not find any noticeable changes in expansion velocity when compared to previous observations, but we did identify several interesting features. From the observation of IRAS 18139$-$1816 we found an absorption feature around 30 km s$^{-1}$. We expect this is a result of absorption within the nearby dust cloud that can be seen in optical images of the region. We also found a secondary maser source during our observation of IRAS 18268$-$1117. This additional maser profile is coming from another source within the $13^\prime$ beam of Parkes. The only potential known sources within around $13^\prime$ are the evolved S-type AGB star, IRAS 18269$-$1111, and the source IRAS 18272$-$1117. The S-type source has a carbon-to-oxygen ratio of 1, making it an unlikely candidate for OH maser emission but still within the realm  of possibility, especially if it has recently undergone an additional thermal pulse and the OH shell corresponds to a time when it was still an M giant .

\section{UVES spectra}
\renewcommand{\thefigure}{C\arabic{figure}}
\setcounter{figure}{0}
We have been able to determine systemic velocities for two of our LMC sources, IRAS 04407$-$7000 and IRAS 04516$-$6902 using UVES reduced spectra from the ESO data archive. Using a cross-spectrum fitting technique we have fit the molecular bandheads of these two sources to that of the UVES spectrum of IRAS 04498$-$6842. This technique has allow us to determine the best fit phase difference and determine the velocity shift. We have used the bright source IRAS 04498$-$6842 as our model, which has a clear maser profile yielding a systemic velocity of 260 km s$^{-1}$. The systemic velocities for IRAS 04407$-$7000 and IRAS 04516$-$6902 seem to fit well between the maser profiles, corroborating the weaker maser peaks and providing more evidence for all three maser profiles. 
\begin{figure*}
 \centering
 \begin{minipage}[c c]{\textwidth}
  \centering
  \includegraphics[width=0.45\textwidth]{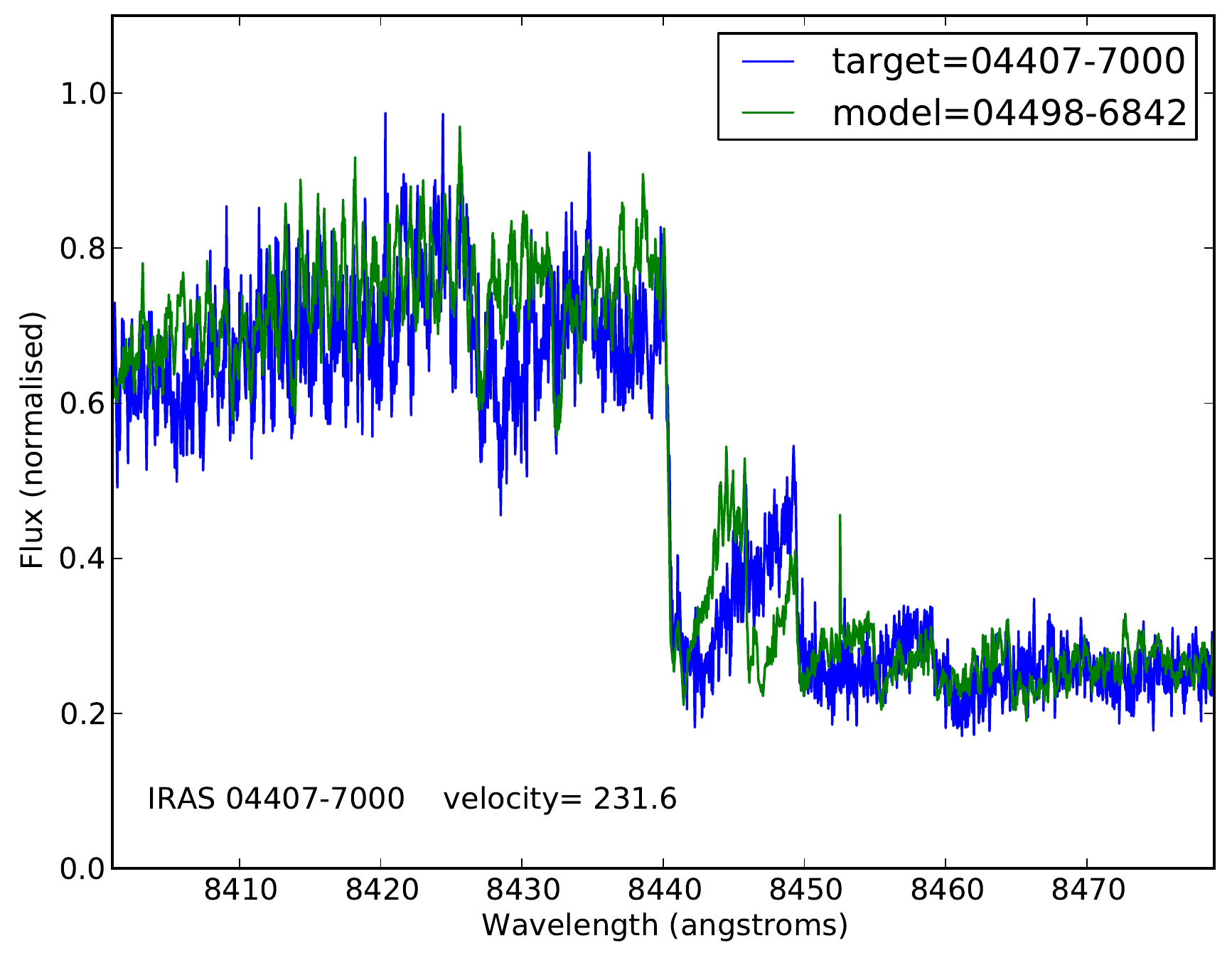} \hspace{1cm}
  \includegraphics[width=0.45\textwidth]{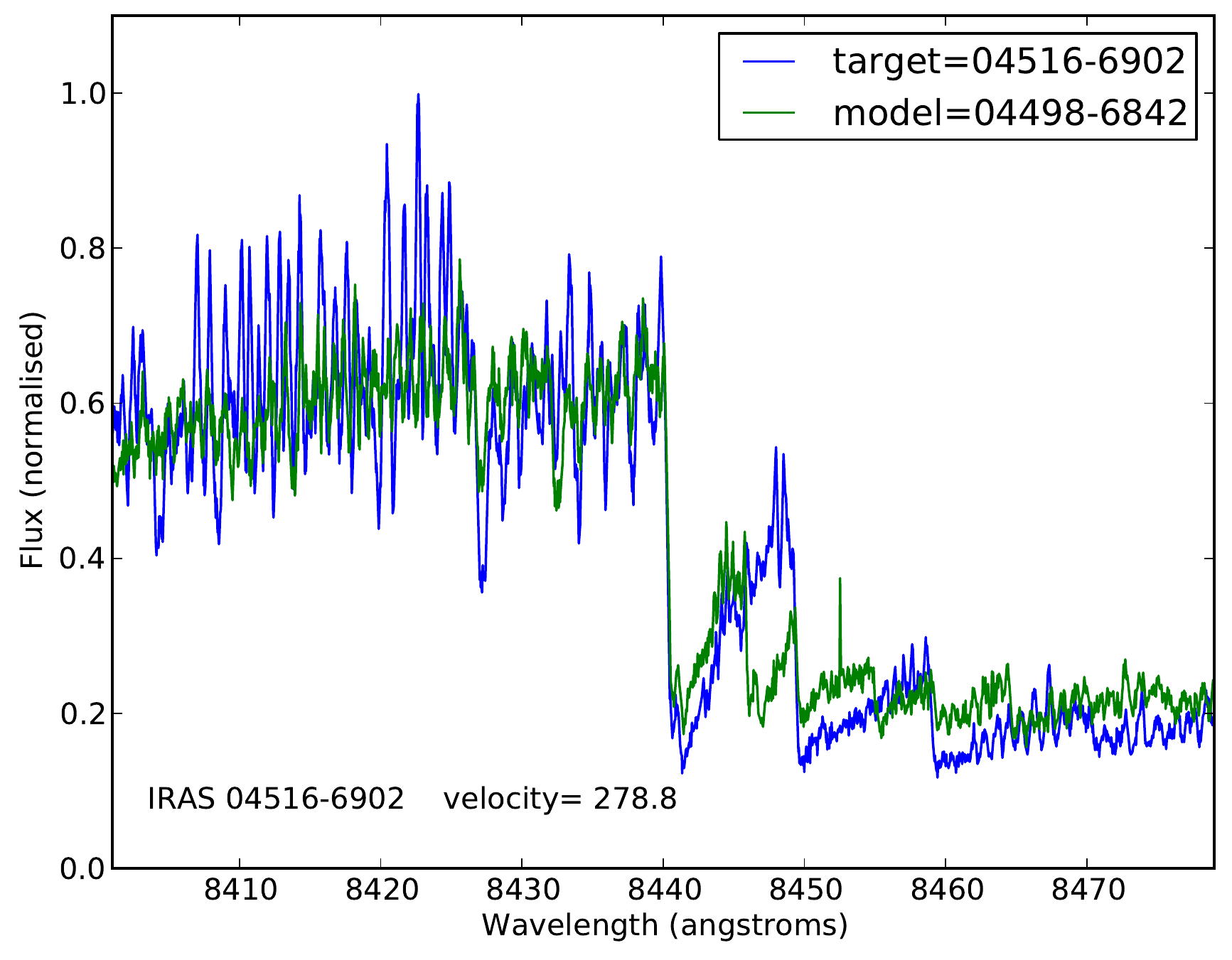} \\
  \caption{The UVES spectra of IRAS 04407$-$7000 and IRAS 04516$-$6902 fit with the UVES spectrum of IRAS 04498$-$6842 using a cross-spectrum fitting technique. The phase shift has been used to calculate systemic velocities for these sources.  }
 \end{minipage}
\end{figure*}

\section{VVV pulsation periods}
We have derived pulsation periods for 7 Galactic Bulge stars from the Jim\'{e}nez-Esteban \& Engels (2015) Galactic Bulge sample of extreme highly reddened OH/IR stars. Aperture photometry was done on K band photometry for all of our Galactic Bulge sources within the tiles of the Vista Variables in the Via Lactea (VVV) survey (Minniti et al. 2010). A calibrator star was used in each field to derive relative flux. Fourier analysis was used to fit the period and a least squares fit was used to calculate the errors. Three of the sources have previously derived pulsation periods from van der Veen \& Habing (1990). Our periods of 690, 781, and 833 days are dramatically different than the values previously derived of 1200, 1500, and 1200 respectively. The inaccuracy of these derived periods has been shown in Whitelock et al. (1991) and now in this work.  

 \renewcommand{\thefigure}{D\arabic{figure}}

 \begin{figure*}
 \vspace{1cm}
 \centering
 \begin{minipage}[c c]{\textwidth}
  \centering
  \includegraphics[width=0.45\textwidth]{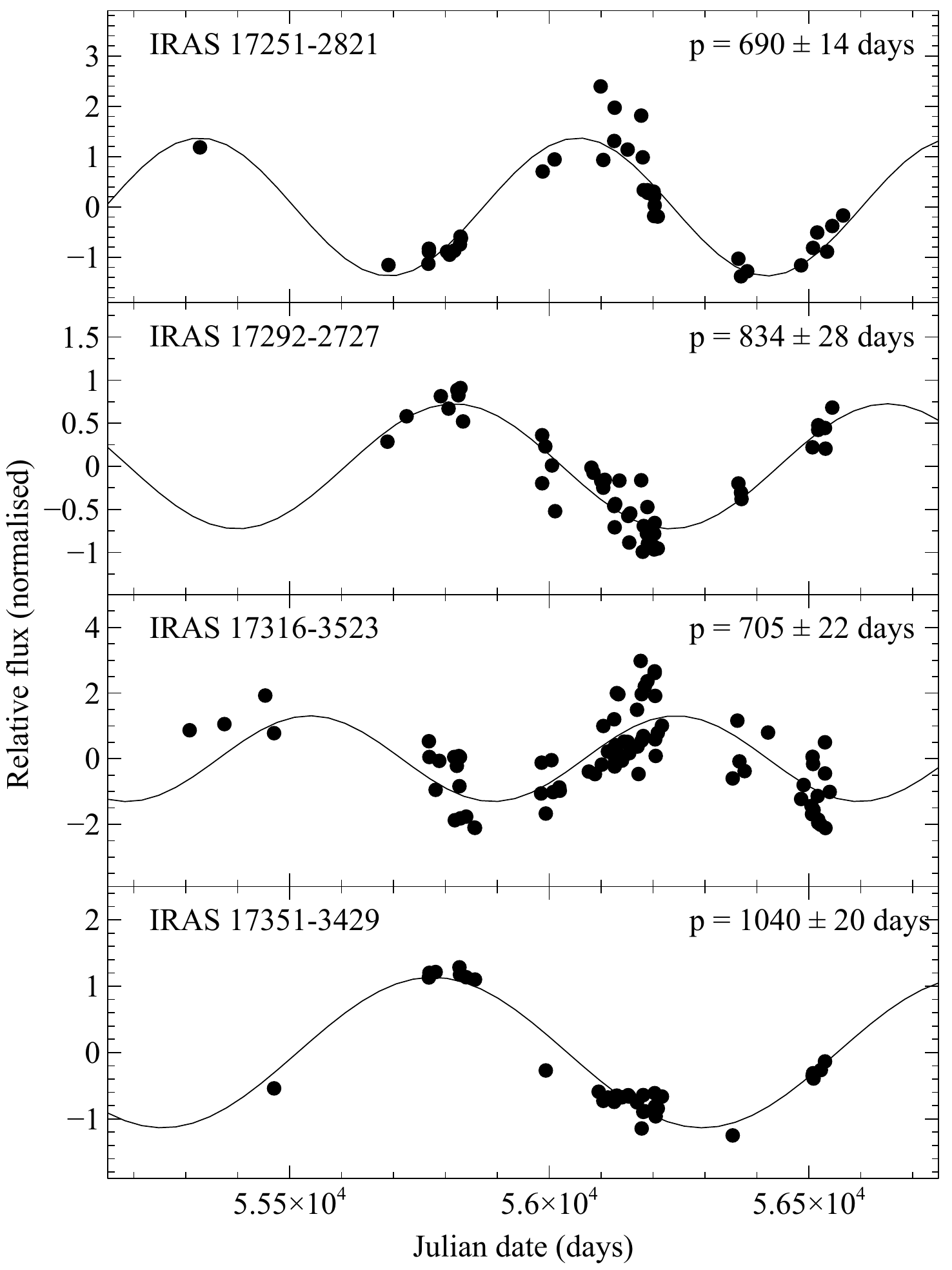} \hspace{1cm}
  \includegraphics[width=0.45\textwidth]{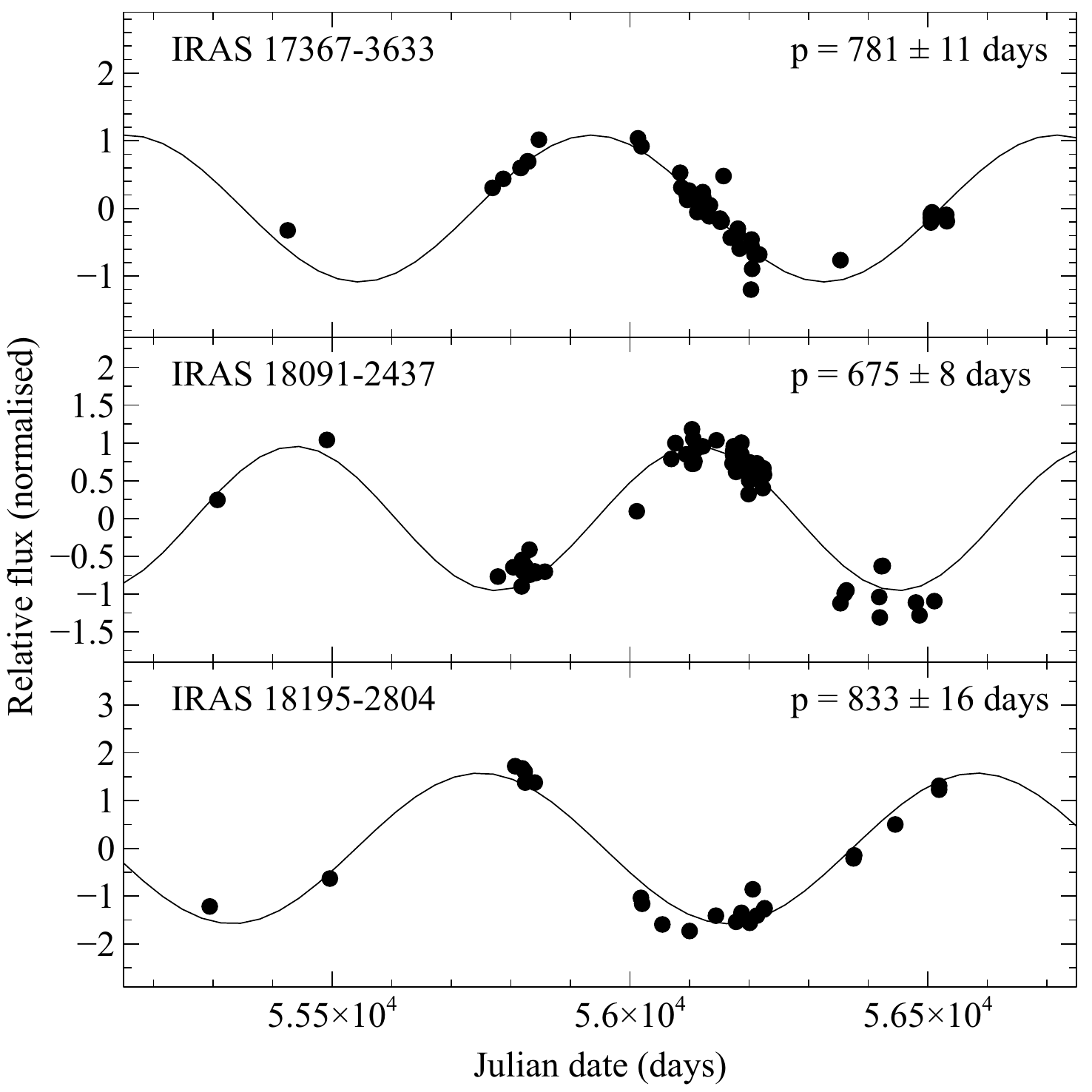}
  \caption{The fit pulsation periods of highly evolved stars from our Galactic Bulge sample. Aperture photometry was done on tiles from the VVV survey and fit using a Fourier analysis fitting technique.}
 \end{minipage}
\end{figure*}  
 
\bsp
\label{lastpage}
\end{document}